\documentclass[11pt,english]{article}
\usepackage{amsmath}
\usepackage[T1]{fontenc}
\usepackage[latin9]{inputenc}
\usepackage{geometry}
\usepackage{setspace}
\usepackage[affil-it]{authblk}
\usepackage[english]{babel}
\usepackage{blindtext}
\usepackage{graphicx,psfrag,epsf}
\usepackage{amssymb}
\usepackage[authoryear]{natbib}
\usepackage{mathtools}
\usepackage[unicode=true,
 bookmarks=false,
 breaklinks=false,pdfborder={0 0 1},colorlinks=true,citecolor = blue]{hyperref}
\usepackage{caption}
\usepackage{subcaption}
\usepackage{xcolor}
\usepackage{bm}
\usepackage{color}
\usepackage{tikz}
\usepackage{stmaryrd}
\usepackage{url}
\usepackage{mathrsfs}  
\usepackage{enumerate}
\usepackage{colortbl}
\usepackage{comment}
\usepackage{gensymb}
\usepackage[tableposition=top]{caption}

\newtheorem{assumption}{Assumption}
\newtheorem{example}{Example}

\newtheorem{proposition}{Proposition}
\newtheorem{theorem}{Theorem}
\newtheorem{remark}{Remark}

\newtheorem{lemma}{Lemma}
\newcommand{\M}{\mathbf}
\newcommand{\lbt}{\llbracket}
\newcommand{\rbt}{\rrbracket}
\newcommand{\B}{\boldsymbol}
\newcommand{\qed}{\hfill \ensuremath{\Box}}
\newcommand{\ind}{\mbox{$\perp\!\!\!\perp$}}

\DeclareMathOperator{\sgn}{sgn}
\DeclareMathOperator{\diag}{diag}
\newcommand{\dd}{{\B\cdot}}
\newcommand{\od}{{\scriptstyle\star}}
 \newcommand\scalemath[2]{\scalebox{#1}{\mbox{\ensuremath{\displaystyle #2}}}}

\newcommand*{\QEDB}{\hfill\ensuremath{\square}}
\newcommand{\bee}{\begin{equation}\begin{aligned}}
	\newcommand{\ee}{\end{aligned}\end{equation}}
\newcommand{\beee}{\begin{eqnarray}}
	\newcommand{\eee}{\end{eqnarray}}
\newcommand{\indep}{\perp\!\!\!\perp}

\newcommand{\logit}{\text{logit}}

\newcommand{\pr}{\mathbb{P}}

\newcommand{\E}{\mathbb{E}}
\newcommand{\LOTE}{\textup{LOTE}}

\newcommand{\nn}{\nonumber}

\newcommand{\T}{\top}

\newcommand{\R}{\mathbb{R}}

\newcommand*\bigcdot{\mathpalette\bigcdot@{.5}}

\usepackage{xr}
\makeatletter
\newcommand*{\addFileDependency}[1]{
  \typeout{(#1)}
  \@addtofilelist{#1}
  \IfFileExists{#1}{}{\typeout{No file #1.}}
}
\makeatother
\newcommand*{\myexternaldocument}[1]{%
    \externaldocument{#1}%
    \addFileDependency{#1.tex}%
    \addFileDependency{#1.aux}%
}
\myexternaldocument{supp}

\makeatletter

\@ifundefined{date}{}{\date{}}
\bibpunct{(}{)}{,}{autheryear}{,}{,}
\allowdisplaybreaks

\usepackage{babel}

\usepackage{babel}

\usepackage{babel}

\makeatother

\usepackage{babel}

\begin{document}
\title{{\huge\textbf{Semiparametric Localized Principal Stratification Analysis with Continuous Strata 
}}}
\author{Yichi Zhang\thanks{Department of Statistics, Indiana University Bloomington, Bloomington, Indiana
47405, U.S.A. Email: yiczhan@iu.edu}$\ \ $and Shu Yang\thanks{Department of Statistics, North Carolina State University, Raleigh, North Carolina
27607, U.S.A. Email: syang24@ncsu.edu} \thanks{Corresponding author.}}
\maketitle
\begin{abstract}
Principal stratification is essential for revealing causal mechanisms involving post-treatment intermediate variables, {\color{black}in real-world applications like surrogate marker evaluation}. Principal stratification analysis with continuous intermediate variables is increasingly common but challenging due to the infinite principal strata and the nonidentifiability and nonregularity of principal causal effects. Inspired by recent research, we resolve these challenges by first using a flexible copula-based principal score model to identify principal causal effect under weak principal ignorability. We then target the local functional substitute of principal causal effect, which  is  statistically regular and can accurately approximate principal causal effect with vanishing bandwidth. We simplify the full efficient influence function of the local functional substitute by considering its oracle-scenario alternative. This leads to a computationally efficient and straightforward estimator for the local functional substitute and principal causal effect with vanishing bandwidth. 
We prove the double robustness of our proposed estimator, and derive its asymptotic normality for inferential purposes. {\color{black}With a vanishing bandwidth, our method attains minimax optimality for the nonparametric estimation of the principal causal effect. With a fixed bandwidth, it achieves semiparametric efficiency in estimating its local functional substitute.} {\color{black} We demonstrate the strong performance of our proposed estimator through simulations and apply it to surrogate analysis of short-term CD4 count in ACTG 175.}

\bigskip{}
\end{abstract}

\section{Introduction}\label{sec:intro}
When an intermediate variable exists between treatment and outcome, identifying and inferring treatment effects becomes more challenging compared to the classical ``covariate-treatment-outcome" mechanism. Principal strata, defined by the joint potential values of the post-treatment intermediate variable under both treatment and control, offer a principled way to understand such underlying causal mechanisms \citep{frangakis2002principal}. Principal stratification involves comparing causal effects between key strata or groupings of the data, often referred to as principal causal effects (PCEs). {\color{black}PCEs are commonly used to evaluate surrogate markers  \citep{frangakis2002principal}. In AIDS studies, for example, post-treatment CD4 count can act as a surrogate for survival time. The PCE surface reflects the average treatment effects on survival across patient strata where CD4 treatment effects are uniform. For CD4 count to be a {\it principal surrogate}, PCEs must align with CD4 effects--non-zero when CD4 effects exist and zero otherwise. Thus, analyzing the PCE surface is crucial for evaluating CD4 count as a surrogate.} PCEs have also been useful in diverse applications, such as   noncompliance and truncation-by-death problems in clinical trials \citep{frangakis2002principal,rubin2006causal} 
and causal mediation analysis \citep{kim2019bayesian}. 

Like potential outcomes, potential intermediate variable values suffer from inherent missing data because at most one potential value is observable. 
Therefore, additional assumptions are necessary to identify PCEs. In this paper, we consider {\em weak principal ignorability} \citep{jo2009use}, an ignorability-type assumption for the intermediate variable analogous to the celebrated treatment ignorability assumption \citep{rosenbaum1983central}. Principal ignorability has been widely adopted in various applications \citep{follmann2000effect, hill2002differential, jo2009use, stuart2015assessing, jiang2020multiply}. For a binary intermediate variable, \citet{jiang2020multiply} establish various identification formulas for PCEs under principal ignorability and an additional monotonicity assumption and derive the semiparametric efficient influence function (EIF). The resulting PCE estimator exhibits desirable  properties, e.g., triple robustness and semiparametric efficiency.
\subsection{From a binary to continuous intermediate variable}
{\color{black}Unlike the noncompliance and truncation-by-death problems, surrogate evaluation often involves a continuous intermediate variable. For example, \citet{follmann2006augmented} and \citet{gilbert2008evaluating} consider immune response measured after randomization as a continuous surrogate outcome for infection in vaccine trials. Similarly, \citet{zigler2012bayesian} evaluate the surrogacy of short-term CD4 count as a continuous predictor for a long-term clinical endpoint in AIDS clinical trials. In causal mediation analysis, the intermediate variable can also be continuous. For instance, \citet{schwartz2011bayesian} investigate the mediating effect of body mass index on cardiovascular disease.} The identification and estimation of PCE with a continuous intermediate variable (Cont.PCE) are  challenging due to  infinitely many principal strata generated by a continuous intermediate variable. {\color{black}Early attempts identify and estimate Cont.PCE through parametric models \citep{gilbert2008evaluating,jin2008principal}. Later works consider more flexible semiparametric models of  continuous principal strata (see, e.g., \citet{bartolucci2011modeling,schwartz2011bayesian,zigler2012bayesian}). However, these methods still depend on correctly specifying the entire underlying distribution (cf. Section~\ref{sec:rw}). 
\par
This paper aims to develop an efficient and flexible semiparametric framework for the identification and estimation of Cont.PCEs under the weak principal ignorability. This work can be seen as an extension of \citet{jiang2020multiply}, which, however, is non-trivial due to
challenges with respect to estimand, identification, and estimation:
\begin{enumerate}
\item[(i).] The Cont.PCE is a non-pathwise differentiable estimand, making classical semiparametric efficiency theory inapplicable for deriving a semiparametric efficient estimator.
\item[(ii).] The monotonicity assumption used to identify PCE in \citet{jiang2021identification} fails to identify PCE when the intermediate variable is continuous and may be implausible in various applications such as surrogate analysis.
\item[(iii).] After addressing (i) and (ii) properly, the resulting full EIF (cf.~Section~\ref{sec:spe}) is still too complex to construct a computationally feasible estimator.
\end{enumerate}}
\subsection{From a perfectly localized estimand to local estimand}\label{intro:pl2l}
The non-pathwise differentiability, as mentioned in (i) above, presents a significant obstacle in developing an efficient estimator for Cont.PCE. To resolve this type of challenge for efficient statistical estimation, \citet{chernozhukov2018biased,chernozhukov2021simple} focus on a specific  non-pathwise differentiable estimand, namely, the \textit{perfectly localized functional} $\vartheta_0$, 
\bee\label{def:perfect}
\vartheta_0 = \lim_{h\rightarrow 0}\vartheta_h,\quad \vartheta_h = \E\big\{ m(W,Y,\gamma^*,k_{h,d_0})\big\},
\ee
where $\vartheta_h$ is the \textit{local functional}    with a fixed and sufficiently small $h > 0$, which is pathwise differentiable.\footnote{Notably, \citet{chernozhukov2021simple}  terms $\vartheta_0$ as the ``local functional'' and $\vartheta_h$ as the ``global functional''. This paper adopts the terminology in \citet[Definition 2.2]{chernozhukov2018biased} to ensure consistence.} Here $Y$ is the  response, $W$ includes other variables, $\gamma^*(\cdot)=\E(Y\mid W = \cdot)$, $m(w,y,\gamma^*,k_{h,d_0})$ is a linear functional of $\gamma^*(\cdot)$, and $k_{h,d_0}(\cdot)$ is a pre-specified smoothing function centered at  a specific value $D = d_0$ in $W$ with bandwidth $h > 0$.  Examples of $\vartheta_0$ include the conditional average treatment effect and  continuous treatment effect. {This paper generalizes the original definition of $\vartheta_0$ by allowing $W$ to contain unobserved variables and proves that Cont.PCE is also a perfectly localized functional under proper identification assumptions (cf. Proposition \ref{po:approximate}).

 Instead of directly estimating the perfectly localized functional, \citet{chernozhukov2018biased,chernozhukov2021simple} alternatively estimate its substitute, the {local functional}  $\vartheta_h$  in \eqref{def:perfect} with small $h > 0$. They further propose a doubly robust estimator using Riesz representation. Switching from $\vartheta_0$ to $\vartheta_h$ as the target estimand has  multiple advantages, as discussed in \citet{chernozhukov2018biased, chernozhukov2021simple}: $\vartheta_h$ is pathwise differentiable and can be  estimated semiparametrically efficiently; $\vartheta_h$ serves as a good approximation of $\vartheta_0$ when $h$ is small, thus efficient estimation of $\vartheta_h$ translates to efficient estimation of $\vartheta_0$ up to a small localization bias; and inference on $\vartheta_h$ is more robust and does not rely on the  localization bias.  Following   \citet{chernozhukov2018biased, chernozhukov2021simple}, this paper aims to construct a computationally and statistically efficient  estimator for the local functional substitute of Cont.PCE (Loc.PCE), {\color{black}and subsequently, estimate Cont.PCE efficiently and accurately as $h\rightarrow 0$ with growing sample size}. However, the paper deals with a more complex setting than \citet{chernozhukov2018biased, chernozhukov2021simple} due to the unobserved variable in $W$ and an additional nuisance function. To address these challenges, a new EIF-invoked estimation strategy is proposed for Loc.PCE. {\color{black} The recent work of \citet{lu2023principal} adopts a different strategy by using a global approximation of the underlying Cont.PCE surface, focusing on a finite-dimensional parameter within a working model. This enables them to derive the corresponding EIF and doubly robust estimators. In contrast, we directly target the localized Cont.PCE and Cont.PCE (c.f.~Section~\ref{sec:rw}). }}

\subsection{Outline and contributions}
The paper is organized as follows. Section \ref{sec:bs} provides the  background for PCE estimation. Section~\ref{sec:example:sa} motivates Cont.PCE estimation using a synthetic example in  surrogate analysis, while Section \ref{sassump:TAignorability} presents the identification assumptions and formula for Cont.PCE. Section \ref{sec:localCont.PCE} introduces Loc.PCE, the local functional substitute of Cont.PCE, as the target estimand {\color{black} for which we develop a semiparametric efficient estimator}. The identification and estimation of Cont.PCE and Loc.PCE involve  three nuisance models: 
\begin{itemize}
\item[] $\mathcal{M}_{{\rm tp}}$:  \textit{treatment probability}, the distribution of treatment given  covariates; 
\item[] $\mathcal{M}_{{\rm ps}}$: \textit{principal score}, the distribution of intermediate variable given covariates \& treatment; 
\item[] $\mathcal{M}_{{\rm om}}$:  \textit{outcome
mean}, the conditional expectation of outcome given all other variables. 
\end{itemize} 
 In Section \ref{sec:spe}, we  derive the EIF of Loc.PCE under nonparametric nuisance functions. However, the resulting EIF is too complex for a computationally viable estimator. Thus, we derive simplified EIFs in two special scenarios: (i) the parametric scenario where $\mathcal{M}_{{\rm ps}}$ is parametrically modeled, and (ii) the oracle scenario where $\mathcal{M}_{{\rm ps}}$ is known. We construct our proposed estimator using the oracle-scenario EIF, which is computationally feasible. 
 In Sections \ref{sec:dbrb} and \ref{sec:lsp}, we show that the proposed estimator is doubly robust and asymptotically normal {\color{black}for Loc.PCE and Cont.PCE.} Interestingly, our estimator attains the semiparametric efficiency bound {\color{black}for Loc.PCE}, under the more general  condition when $\mathcal{M}_{\text{ps}}$ is parametric, despite being derived using the EIF assuming $\mathcal{M}_{\mathrm{ps}}$ is known. We can also construct confidence intervals (CIs) for the inference of {\color{black}Loc.PCE  and Cont.PCE}. In Section~\ref{sec:simu}, simulation results corroborate our theoretical results. {\color{black}In Section~\ref{sec:actg175}, we apply the proposed estimator for the surrogate analysis of short-term CD4 count in ACTG 175, and in Section~\ref{sec:dis} we provide concluding remarks.}
\section{Principal stratification  with continuous strata}\label{sec:pce}

\subsection{Basic setups and Cont.PCE}\label{sec:bs}
Suppose  we have $n$ i.i.d. observations $\{V_i = (X_i,Z_i,M_i,Y_i)\}_{i = 1}^n$. For each sample $V_i$, we denote  the pre-treatment covariates by $X_i\in\mathbb{X}\subseteq\R^{d_X}$, treatment assignment by $Z_i\in\{0,1\}$,  intermediate variable by $M_i\in \mathbb{M}\subseteq \R$, and  outcome by $Y_i\in \mathbb{Y}\subseteq \R$. {\color{black}Our data generation mechanism invokes the  Stable Unit Treatment Value Assumption  \citep{rubin1980randomization}.} In particular, we assume
$$
{(X_i,Z_i,M_{1i},M_{0i},Y_{1i},Y_{0i}) \,\overset{\text{i.i.d.}}{\sim} \,(X,Z,M_1,M_0,Y_1,Y_0),}
$$
where $M_{1i}$ and $M_{0i}$ are potential intermediate variables, and $Y_{1i}$ and $Y_{0i}$ are potential outcomes, such that $M_i = Z_i M_{1i} + (1 - Z_i)M_{0i}$ and $Y_i = Z_i Y_{1i} + (1 - Z_i)Y_{0i}$.  
\par
\cite{frangakis2002principal} use the joint potential values of
 intermediate variables to define the principal stratum
$U=(M_{1},M_{0})$. A 
stratum $U$ captures the information on how the intermediate variable is affected by the
treatment. 
When $M$ is a binary variable, principal stratum $U$ takes  values in
$\{(0,0),(1,0),(0,1),(1,1)\}$, which in noncompliance problems \citep{frumento2012evaluating}
are referred to the never-taker, complier, defier, and always-taker,
respectively.
When $M$ is a continuous variable, stratum $U$  takes  values in
$\R^2$.  
In surrogate analysis, $M_1$ and $M_0$ are the counterfactual surrogate outcomes had treatments been set to $1$ and $0$, respectively, and $M_1 - M_0$ is the surrogate treatment effect.
\par
  For a specific principal stratum $u^* = (m^*_1,m^*_0)\in\R^2$, define PCE as the average causal effect on the primary outcome $Y$ within the principal stratum $U = u^*$,
\begin{equation*} 
{\tau}^*_{u^*} =\E(Y_{1}-Y_{0}\mid U=u^*).
\end{equation*}  
Referring to \citet[$\mathsection$ 2]{jiang2020multiply}, ${\tau}^*_{u^*}$ can have different scientific meanings when $M$ is binary. For continuous $M$,  it is the {Cont.PCE} as we alluded earlier. In the next section, we demonstrate the practical utility of Cont.PCE in evaluating surrogate outcomes \citep{frangakis2002principal}.
 \subsection{Interpretation of Cont.PCE in surrogate analysis}\label{sec:example:sa}
 In clinical trials, obtaining primary clinical outcomes can be time-consuming and expensive. As a result, there is considerable interest in identifying an easy-to-obtain surrogate that can reliably reflect  treatment effects on the primary clinical outcome. Cont.PCE, also known as the causal effect predictiveness (CEP) surface \citep{gilbert2008evaluating}, can provide a comprehensive evaluation of any tested surrogate. In particular, \citet{frangakis2002principal} suggest evaluating the associative and dissociative effects of a surrogate outcome, both of which can be quantified through Cont.PCE.
\begin{itemize}
\item A good surrogate is required to meet {\it causal sufficiency}, indicating that any causal influence of the treatment on the surrogate must be accompanied by an effect on the primary outcome \cite[$\mathsection$ 3.1]{gilbert2008evaluating}. The {associative effect} compares the average difference  between $Y_1$ and $Y_0$ when $M_1 = m_1\neq M_0 = m_0$, which is quantified by $\tau^*_{(m_1,m_0)}$ for any two distinct principal levels $m_1, m_0\in\R$. Therefore, causal sufficiency is satisfied  {\it if} all associative effects ${\tau}^*_{(m_1,m_0)} \neq 0$ when $m_1\neq m_0$.   In some cases, particular interest lies in the {\it one-sided causal sufficiency} of the surrogate \citep{gilbert2008evaluating,vanderweele2013surrogate}, which requires that a positive (resp. negative) treatment effect on the surrogate implies a positive (resp. negative) treatment effect on the primary outcome. For example, our surrogate evaluation for ACTG 175 in Section~\ref{sec:actg175} focuses on this one-sided causal sufficiency. Formally, one-sided causal sufficiency is defined analogously to causal sufficiency above, with ``\( \neq \)'' replaced by ``\( > \)''.
\item {\color{black} Another criterion a good surrogate shall satisfy is {\it causal necessity}, which entails that the absence of a causal impact of the treatment on the surrogate outcome must also imply the absence of any effect on the primary outcome \cite[$\mathsection$ 3.1]{gilbert2008evaluating}. The
{dissociative effect} compares the average difference  between $Y_1$ and $Y_0$  when $M_1= M_0 = m$, which is quantified through $ \tau^*_{(m,m)}$ for any principal level $m\in \R$. Causal necessity can be true {\it only if} dissociative effects ${\tau}^*_{(m,m)} =0$ for all $m\in \R$. In another word, if one observes some ${\tau}^*_{(m,m)} \neq0$, $M$ is evaluated as a surrogate not satisfying causal necessity.}
\end{itemize}{\color{black}
We use  synthetic  examples to demonstrate the usage of Cont.PCE, in which we denote $\Sigma_0 = \begin{psmallmatrix}
1 & 0.25\\
0.25 & 1
\end{psmallmatrix}$ and  generate $X = (X_1,X_0) \sim \mathcal{N}\big\{(0,0)^\T,\Sigma_0\big\}$ and $Z\sim \text{Bern}(1/2)$, independently. Given $X$, we consider two different generating procedures of $(M_1,M_0)$:
$$
\text{P(i). } (M_1,M_0) \sim \mathcal{N}\left\{(X_1,X_0)^\T,\Sigma_0\right\};\quad\text{P(ii). }(M_1,M_0) \sim \mathcal{N}\left\{(0,0)^\T,\Sigma_0\right\}.
$$Finally, we generate the potential outcomes. To facilitate the mediation analysis, we generalize the single-indexed potential outcome notation $Y_z$ to  double-indexed  $Y_{z,m}$. {\color{black} Under the background of surrogate analysis, in principle, the post-treatment surrogate outcome $M$ can be regarded as an additional continuous treatment that has causal effect on the primary outcome $Y$.  Conceptually, $Y_{z,m}$ is the potential clinical outcome if the treatment $Z$ and surrogate outcome $M$ were simultaneously manipulated to some levels $z$ and $m$, respectively \citep{robins1992identifiability,mealli2012refreshing,baccini2017bayesian}. The single-indexed and double-indexed potential outcome notations can be linked through $
Y_z = Y_{z,M_z}$ for $z \in \{0,1\}$ \cite[$\mathsection$ 2.1]{forastiere2018principal}.} For each $z\in\{0,1\}$ and $m\in\R$, with given $X$, we generate the potential outcome $Y_{z,m}$ independently according to the following distributions: 
$$
\mathrm{P(i).}\, Y_{z,m} \sim m/2 + \mathcal{N}(X_{z} + 0.5X_{1 - z},1);\quad\mathrm{P(ii).}\, Y_{z,m} \sim \mathcal{N}(X_{z} + 0.5X_{1 - z},1).
$$ Figure~\ref{fig:s12} visualizes the CEP surfaces in P(i) where $\color{black}\tau_u^*=0.75(m_1-m_0)$ and in P(ii) where $\tau_u^*=0$. Therefore, P(i) and P(ii) generate two types of surrogates: in P(i), $M_1 - M_0$ has a positive correlation with $Y_1 - Y_0$, and $M$ is both causally necessary and  sufficient to be a good surrogate for $Y$, while in P(ii) $M$ has no association with $Y$ and thus is uninformative. 
\begin{figure}[!t]\centering
\begin{subfigure}{0.3\linewidth} 
\includegraphics[width=\linewidth]{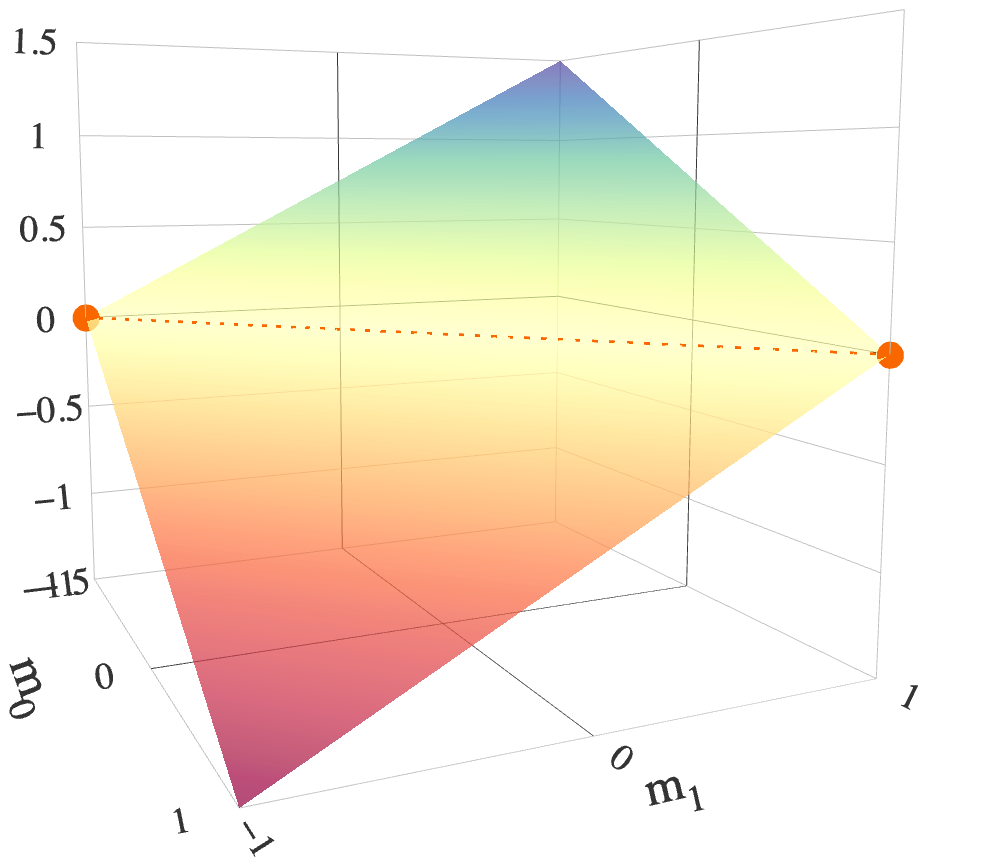}
\caption{$\tau^*_u$ under Procedure (i)}
\end{subfigure}  
 \quad \quad\quad\quad\quad\quad\quad\quad
\begin{subfigure} {0.3\linewidth} 
\includegraphics[width=\linewidth]{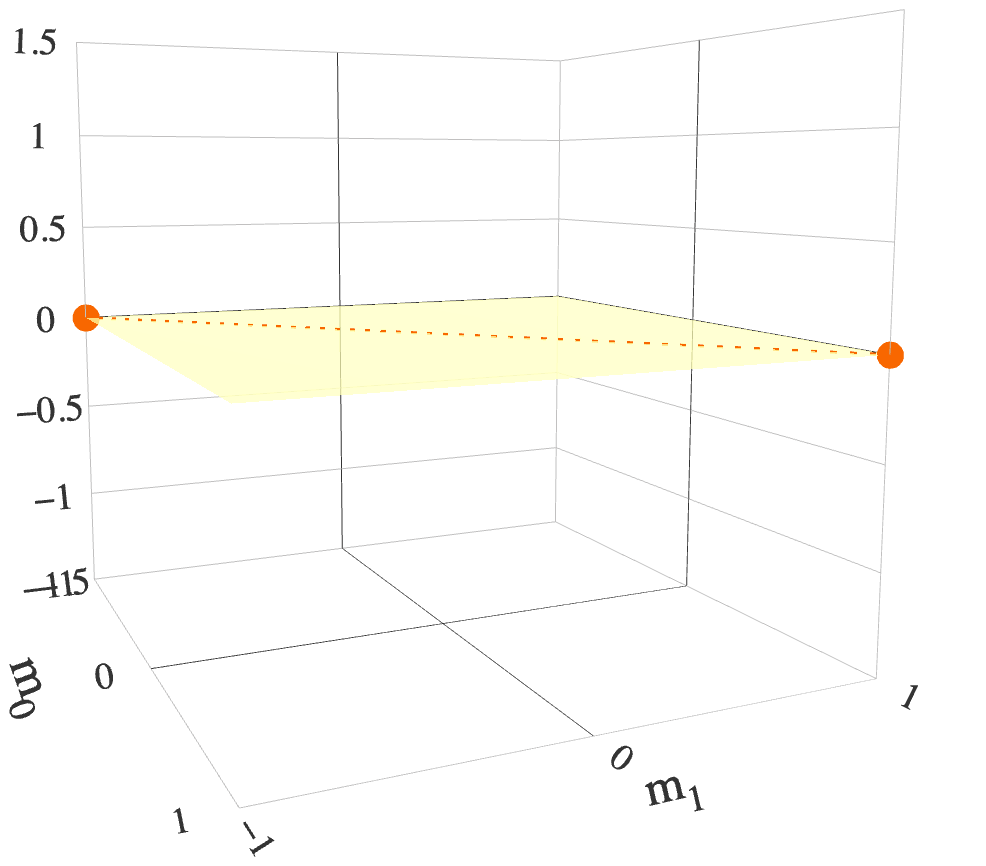}
\caption{$\tau^*_u$ under Procedure (ii)}
\end{subfigure}
\caption{Cont.PCEs under Procedure (i) and (ii). The X- and Y- axes represent $m_1$ and $m_0$, respectively, and Z-axis represents the corresponding $\tau^*_u$. The orange dashed line represents $\tau^*_u = 0$ when $m_1 = m_0$.}
\label{fig:s12}
\end{figure}

Mediation analysis is another widely recognized approach for handling a continuous intermediate variable. Through our synthesized examples, we demonstrate that the conventional mediation analysis may not be suitable for evaluating the surrogate and, thus, highlight the need for Cont.PCE estimation. Mediation analysis focuses on how $M$  plays a role in the causal pathway $Z\rightarrow M\rightarrow Y$. The natural indirect effect (NIE) and natural direct effect (NDE) are $\text{NIE} = \E(Y_{1,M_1} - Y_{1,M_0})$ and $\text{NDE} = \E(Y_{1,M_0} - Y_{0,M_0})
$, respectively. By fixing $Z$ at the treatment level ($Z = 1$), NIE measures the average causal effect of  counterfactual mediators. By fixing the mediator at the control level ($M = M_0$), NDE measures the average causal effect of  treatments  \citep{imai2010identification}. In the synthetic examples, we can deduce that $Y_{1,M_1} - Y_{1,M_0}\sim  \mathcal{N}\{(M_1 - M_0)/2,2\}$ in P(i) and $Y_{1,M_1} - Y_{1,M_0}\sim  \mathcal{N}(0,2)$ in P(ii), while $Y_{1,M_0} - Y_{0,M_0}\sim \mathcal{N}\{(X_1 - X_0)/2,2\}$ in both P(i) and P(ii). Then by the zero means of $(X_1,X_0)$ and $(M_1,M_0)$, we have $\text{NIE} = \text{NDE}  = 0$ in both P(i) and P(ii), which make NIE and NDE uninformative for the surrogate evaluation of $M$. However, the strong and positive correlation between $M_1 - M_0$ and $Y_1 - Y_0$  actually makes $M$ a good surrogate in P(i), which can be captured by the Cont.PCE surface.}
\begin{remark}\label{dcr}
The mediation analysis and principal stratification are generally not comparable due to their conceptual difference \citep{rubin2004direct,vanderweele2011principal,lipkovich2022using}.  When the intermediate variable cannot be intervened, or causal pathway $Z\rightarrow M\rightarrow Y$ does not exist, the principal stratification is a more reasonable framework to deal with intermediate variables \citep{rubin2004direct,vanderweele2011principal}. In mediation analysis, Cont.PCE can also be used to evaluate the direct effect; Non-zero $\tau^*_{(m,m)}$ for some $m\in\R$, imply that $Z$ affects $Y$ through some causal pathways other than $Z\rightarrow M\rightarrow Y$ \citep{mattei2011augmented}. 
\end{remark}
\begin{remark}\label{rk:other}
{\color{black}Another branch of research analyzes the surrogate variable through the estimation of proportion of the treatment effect (PTE) \citep{wang2002measure,parast2016robust,agniel2021evaluation,wang2020model,han2022identifying}. PTE summarizes how much of the treatment effect on $Y$ is explained by its effect on $M$, offering a different perspective on surrogacy. We compare the identification assumptions of these works with ours in Remark~\ref{rk:compare}.}
\end{remark}
\vspace{-.5cm}
\paragraph{\bf Additional notations}  Before we present our identification results, we introduce some new notations for ease of exposition. Denote $f(V = v)$ the density function of $V = (X,Z,M,Y)$. 
Define $\pi_z(x)=\pr(Z=z\mid X = x)$ as
the treatment probability, also known as the propensity score, $e_{u}(x)=f\big(U= u\mid X = x\big)$
the generalized principal score, which is the density function
of principal stratum $u$ given the covariates, and $\mu_{z}(x,m)=\E(Y\mid X = x,Z=z,M=m)$ the outcome mean. When there is no conflict of notations, we  abbreviate  $\pi_z(x)$ and $\mu_{z}(x,m)$ as $\pi$ and $\mu$. Denote $\|\cdot\|_{\infty}$ as the functional sup-norm. Let  $e_{u}=\E\{e_{u}(X)\} = f(U = u)$
denote the marginalized principal score. 
 We use $f(\cdot)$ and $F(\cdot)$ to represent the probability density function (PDF) and the conditional cumulative distribution function (CDF), respectively. We  also denote
$
f_{zm}(x) = f(M = m\mid X = x,Z = z )$ and $F_{zm}(x) = \pr(M \leq m\mid X = x,Z = z ).
$
\subsection{Nonparametric identification}\label{sassump:TAignorability}\label{sec:Nonparametric-identification}
We first consider the treatment ignorability for both  the intermediate variable and
outcome, which is an extension of the classic treatment ignorability \citep{rubin1974estimating}. 
\begin{assumption}[Treatment ignorability]\label{assump:TAignorability}
$Z\ind(M_{0},M_{1},Y_{0},Y_{1})\mid X$. \end{assumption}
Assumption~\ref{assump:TAignorability} precludes unobservable confounders that affect both the treatment and intermediate variable and those that impact both the treatment and outcome. In observational studies, its plausibility
relies on whether or not  {\color{black}observed covariates include all confounders that affect the treatment as well as the outcome and intermediate variables.} Next, we extend the weak principal ignorability assumption \citep{jiang2020multiply} to the continuous-$M$ scenario. Weak principal ignorability is
an ignorability-type assumption for the principal
strata similar to the treatment ignorability. 
\begin{assumption}[Weak principal ignorability]\label{assump:weak-pi} 
$\E(Y_z\mid X, U = u) = \E(Y_z\mid X, M_z = m_z)$ for any $z \in\{0,1\}$ and $ u =(m_1,m_0)\in\R^2$. 
\end{assumption}
{\color{black}Assumption~\ref{assump:weak-pi} is a relaxed, expectation-based version  of the general weak principal ignorability (see, e.g., \citet[Assumption 12]{mattei2023assessing}) that  assumes $Y_z\ind M_{1 - z}\mid X, M_z$, $\forall z\in\{0,1\}$. Briefly, weak principal ignorability implies that given $X$ and $M_{1}$ (resp., $M_0$), the distribution of $Y_1$ (resp., $Y_0$) does not depend on the intermediate variable under the opposite treatment $M_0$ (resp., $M_1$) \citep{mattei2023assessing}. Assumption~\ref{assump:weak-pi} is generally weaker than such implication{\color{black}; see Remark~\ref{rk:compare}.} In particular, it  requires that, for both $z=0$ and $z=1$, the expected value of the potential outcome $Y_z$ should be the same across principal strata when the value of $M_z$ is fixed at a specific level $m_z\in \R$, and $M_{1 - z}$ can vary over any level $m_{1 - z}\in \R$, given observed covariates. Under Assumption \ref{assump:TAignorability}, Assumption \ref{assump:weak-pi} is equivalent to
\bee\label{asm1:further}
\E(Y_z\mid X, Z = z, U =  u) = \E(Y_z\mid X, Z = z, M = m_z), \text{ for any } z\in\{0,1\}\text{ and }(m_1,m_0)\in\R^2,
\ee 
which further implies that $\E(Y_z\mid X, U = u)$ can be  simplified to an observable conditional expectation $\E(Y\mid X, Z = z, M = m_z)$. {\color{black} If $X$ captures all $Z$-$Y$ and $M$-$Y$ confounders,
it suffices for the two ignorability assumptions to hold. This unconfoundedness
is implied by sequential ignorability \citep{imai2010identification},
while our assumptions are generally weaker; see Remark~\ref{rk:compare}.} 
For the surrogate analysis, the surrogate outcome under the opposite treatment does not have a direct causal effect on the primary outcome \citep{parast2016robust}, then weak principal ignorability is plausible if $X$ collects all the $Z$-$Y$ and $M$-$Y$ confounders.} 

\begin{remark}\label{rk:compare}
{In mediation analysis, the strong principal ignorability \citep{forastiere2018principal} and sequential ignorability \citep{imai2010identification} are two standard assumptions for the identification of causal mediation effects. {\color{black}Some other surrogate identifying conditions have been proposed  in  \citet{parast2016robust,wang2020model} for PTE estimation, under the randomized clinical trail (RCT).  All assumptions above, however, directly imply and thus are stronger than Assumption~\ref{assump:weak-pi}; see Section \ref{compare:sec} in the Supplementary File for   detailed discussions. }}
\end{remark}

By Assumption~\ref{assump:TAignorability}, the conditional density of $M_z$ given $X$ can be identified as
$$
f(M_{z} = m_z\mid X = x)=f(M = m_z\mid X = x, Z=z),
$$ for any $x\in\R^{d_X},z\in\{0,1\}, m_z\in \R$. However, the joint density $e_u(x) = f(M_{1} = m_1,M_{0} = m_0\mid X = x)$ is still not identifiable due to the fundamental
problem that $M_{1}$ and $M_{0}$ cannot be jointly observed. Moreover, the monotonicity assumption, as used in the binary-$M$ scenario \citep{jiang2020multiply}, is inadequate  to identify $e_u(x)$ in the continuous-$M$ scenario. To overcome this issue, we propose a copula-based approach \citep{bartolucci2011modeling,kim2020health} to model the association between $M_1$ and $M_0$ conditional on $X$. Our approach encompasses several popular association models, including the monotonicity and equipercentile equating model \citep{efron1991compliance,jin2008principal} as special cases; see, e.g., \citet[$\mathsection$ 4.1]{jiang2021identification} and \citet{nelsen2007introduction}.  
\begin{assumption}[Copula models of $e_u(x)$]\label{am:cp}Let $F(M_z= \cdot\mid X = \star)$ be the CDF of $M_z$ given $X$, 
$c_\rho(\cdot,\star)$ be the specified copula function, and $\rho$ be the {\color{black}selected} correlation between $M_1$ and $M_0$ given $X$. \label{fm1m0x} Assume
$
e_u(x) = f(M_{1} = m_1,M_{0} = m_0\mid X = x) =c_\rho\big\{F(M_1 = m_1\mid X = x), F(M_0 = m_0\mid X = x)\big\}\prod_{z\in\{0,1\}}f(M_z =m_z\mid X = x),
$
 for any $x\in\R^{d_X}, z\in\{0,1\},u=(m_1,m_0)\in\R^2$. 
\end{assumption}
Sklar's theorem \citep{sklar1959fonctions} guarantees any regular $e_u(x)$ can be represented via Assumption \ref{fm1m0x} with correctly specified $c_{\rho}(\cdot,\star)$. Some popular  models of $c_{\rho}(\cdot,\star)$ include  Gaussian copula, Clayton copula, 
and Farli-Gumbel-Morgenstern copula \citep{jaworski2010copula}. We assume  the copula function $c_\rho(\cdot,\star)$ and correlation parameter $\rho$ are correctly specified {\color{black}for our theoretical analysis}. 
\begin{remark}[Choices of sensitivity  parameters]\label{rk:sen}{\color{black} As sensitivity parameters, the true $c_{\rho}(\cdot,\star)$
and $\rho$ cannot be inferred from the data, and the robustness of
the proposed method to varying choices depends on the dataset. Practitioners
should perform sensitivity analysis by varying $c_{\rho}(\cdot,\star)$
and $\rho$ to assess whether the estimation results and conclusions
remain robust.

Nonetheless, when $\mathcal{M}_{\mathrm{om}}$
follows a linear model, our proposed method demonstrates enhanced
robustness to different sensitivity parameter choices. Theorem~\ref{prop:linear:triple}
in the Supplementary File shows that, under mild regularity conditions,
as long as $\mathcal{M}_{\mathrm{om}}$ is linear and correctly specified,
the proposed estimator can correctly identify $\tau_{u^{*}}^{*}$
for any choice of $c_{\rho}(\cdot,\star)$ and $\rho$; see Section~\ref{sec:extrarobust}
for further discussion. This result implies that when the true outcome
model is linear or approximately linear and correctly specified, the
estimator remains consistent and insensitive to diverse choices of
$c_{\rho}(\cdot,\star)$ and $\rho$.
} 
\end{remark}
\par
The following theorem provides the identification result for Cont.PCE.
\begin{theorem}[Identification of Cont.PCE]\label{thm:iddd} {\color{black}Under Assumptions \ref{assump:TAignorability} and \ref{assump:weak-pi}, Cont.PCE can be expressed as
\bee\label{contpce:express}
\tau_{u^*}^* = \E\big\{\mu_{1}(X,m_1^*) - \mu_{0}(X,m_0^*)\mid U = u^*\big\}.
\ee
}Suppose Assumptions~\ref{assump:TAignorability}--\ref{am:cp} hold. Given any principal level $u^*=(m_1^*,m_0^*)\in\R^{2}$ with $e_{u^*} > 0$,  $\tau^*_{u^*}$ can be identified as
\bee\label{tau:idd}
\tau^*_{u^*}  = \frac{\E\big[\big\{\mu_{1}(X,m^*_1) - \mu_{0}(X,m^*_0)\big\}e_{u^*}(X)\big]}{e_{u^*}},
\ee 
where $e_{u^*}(X) = c_\rho\big\{F_{1m^*_1}(X), F_{0m^*_0}(X)\big\}\prod_{z\in\{0,1\}}f_{zm^*_z}(X)$.
\end{theorem}
{\color{black}In Theorem \ref{thm:iddd}, we emphasize that $\tau_{u^*}^*$ in \eqref{contpce:express} is defined by conditioning on the unobserved $U$. This distinguishes our problem from other localized functional estimation problems discussed by  \citet{chernozhukov2018biased, chernozhukov2021simple}. For example, while CATE can be expressed through conditional outcome means similar to \eqref{contpce:express}, it conditions on an observable variable $X$. This highlights the need to model the joint distribution of $U$ in Assumption \ref{am:cp} for   $\tau_{u^*}^*$ identification.} In the identification formula \eqref{tau:idd}, under the treatment and weak principal ignorability, we have
$\mu_{1}(X,m^*_1) - \mu_{0}(X,m^*_0) = \E(Y_1 - Y_0\mid X,U = u^*)$, which is the conditional Cont.PCE given  $X$. The weight $e_{u^*}(X)/e_{u^*}$, by Bayes' Theorem, is the density ratio of $X$ conditional and unconditional on $U = u^*$; Here  $e_{u^*}$ is the normalizing constant of $e_{u^*}(X)$, and they can be identified by Assumptions \ref{assump:TAignorability} and \ref{am:cp}. Therefore, our identification formula marginalizes the conditional Cont.PCE over the distribution of $X$ given $U = u^*$, and thereby giving Cont.PCE within stratum $U= u^*$.

\subsection{Target estimand: from Cont.PCE to Loc.PCE}\label{sec:localCont.PCE}
We  show that Cont.PCE is a perfectly localized estimand \citep{chernozhukov2018biased}. Specifically, there exists a sequence  of local estimands indexed by $h>0$, i.e., Loc.PCE (cf.~\eqref{def:tau}), that converges to Cont.PCE as $h\rightarrow 0$.  Define the normalized kernel function centered at $u^* \in \R^2$ as
$k_{u^*}(u)=k_{h,\mathcal{K}(\cdot,\cdot),u^*}\{u = (m_1,m_0)\} =h^{-2}\mathcal{K}\{h^{-1}(m_1 - m_1^*),h^{-1}(m_0 - m_0^*)\},$ where $\mathcal{K}(\cdot,\cdot)$ is any pre-specified  two-dimensional kernel function and $h > 0$ is a fixed bandwidth. Define Loc.PCE  as
\begin{equation}\label{def:tau}
\tau_{u^*} = \tau_{u^*}\{\mathcal{K}(\cdot,\cdot),h\}
 =  \frac{\E\big[\big\{\mu_{1}(X,M_1) - \mu_{0}(X,M_0)\big\}k_{u^*}(U)\big]}{\E\{k_{u^*}(U)\}},
\end{equation}
where we drop the dependence of $\tau_{u^*}$ on  $\mathcal{K}(\cdot,\cdot)$ and $h$ for brevity. 
\par
Proposition \ref{po:approximate} verifies that Cont.PCE is a perfectly localized functional and Loc.PCE is its local functional substitute under the identification assumptions (c.f.~Section~\ref{sassump:TAignorability}). To enhance the understanding, we demonstrate  how to verify the conditional average treatment effect as a perfectly localized functional in Section \ref{sec:cate:exp} in  Supplementary File; see also \citet[Examples 3.1--3.3]{chernozhukov2021simple} for more  examples.
\begin{proposition}\label{po:approximate}Suppose  Assumptions~\ref{assump:TAignorability}--\ref{am:cp} and the regularity conditions in Section \ref{sec:po1:con} in the  Supplementary File hold. For  any $u^*\in\R^2$ such that $e_{u^*} > 0$, we have,
\bee\label{bias:res}|\tau_{u^*} - \tau^*_{u^*}| = \mathcal{O}(h^2).
\ee 
In \eqref{def:perfect}, let $W = (X,Z,U)$, $\gamma^*(x,z,u)    = \mu_{z}(x,m_z)$, $D = U$, $d_0 = u^*$,   $m(w,y,\gamma^*,k_{h}) = \{\gamma^*(x,1,u) - \gamma^*(x,0, u)\}\big[\E\{k_{h}(U)\}\big]^{-1}{k_{h}(u)}$, $\vartheta_0 = \tau^*_{u^*}$, and $\vartheta_h =\tau_{u^*}$. Then \eqref{bias:res} implies that $\tau^*_{u^*}$ is a perfectly localized functional, and $\tau_{u^*}$ is the associated local functional substitute.
\end{proposition}
In light of Proposition~\ref{po:approximate} and following \citet{chernozhukov2018biased}, we will focus on Loc.PCE as our target estimand to develop an efficient estimator. We begin by deriving the identification result for Loc.PCE, similar to Theorem~\ref{thm:iddd}. For ease of exposition, given   any function $h_{u}(\mathcal{T})$ with  $\mathcal{T}$ being any variables other than $u$, we define  $\llbracket\cdot\rrbracket_{u^*}$ as a kernel smoother of $h_u(\mathcal{T})$ around $u = u^*$, 
\bee\label{def:lu}
\big\llbracket h_{(\dd,\od)}(\mathcal{T})\big\rrbracket_{u^*} = \int_{\R^2} h_{u}(\mathcal{T})k_{u^*}(u)du.
\ee
Here $(\dd,\od)$ represents the two-dimensional vector $u$ in $h_u(\mathcal{T})$ are integrated out.
\begin{theorem}[Identification of Loc.PCE]\label{thm:idd}
Suppose Assumptions~\ref{assump:TAignorability}--\ref{am:cp} hold. Given $u^*\in\R^{2}$, kernel function $k_{u^*}(\cdot)$ and bandwidth $h > 0$, then $\tau_{u^*}$ can be identified as
\bee\label{tau:id}
\tau_{u^*} = \frac{\E\big\llbracket\big\{\mu_{1}( X,\dd) - \mu_{0}(X,\od)\big\}e_{\dd\od}(X)\big\rrbracket_{u^*}}{\E\big\llbracket e_{\dd\od}(X)\big\rrbracket_{u^*}},
\ee where $e_{u^*}(X) = c_\rho\big\{F_{1m^*_1}(X), F_{0m^*_0}(X)\big\}\prod_{z\in\{0,1\}}f_{zm^*_z}(X)$.
\end{theorem}
Comparing \eqref{tau:idd} and \eqref{tau:id}, it becomes clear   that Loc.PCE acts as a locally kernel-smoothed approximation of Cont.PCE. 

\section{EIFs and an oracle EIF-invoked  estimator}\label{sec:spe}

We adopt a semiparametric approach similar to the binary-$M$ case in \citet{jiang2020multiply} to propose a principled estimator for Loc.PCE, {\color{black}and subsequently, for Cont.PCE when $h\rightarrow 0$ as $n\rightarrow \infty$}. We begin by deriving EIFs under two scenarios, where $\mathcal{M}_{{\rm tp}}$ and $\mathcal{M}_{{\rm om}}$ are both nonparametric, and $\mathcal{M}_{{\rm ps}}$ is either nonparametric or follows a parametric model,
\bee\label{paraps}
f_{zm}(x\mid \beta) = f(M = m\mid X = x,Z = z,\beta).
\ee Here $\beta\in \R^{d_\beta}$ is the parameter and \ $f_{zm}(x) = f_{zm}(x\mid \beta^*)$. The resulting EIFs are complex, posing challenges in constructing a computationally viable estimator. We will elaborate on this issue later. To simplify the EIF and facilitate computation, we further derive the EIF under an oracle scenario, where the true $\mathcal{M}_{\mathrm{ps}}$ is known. 
\par
\begin{theorem}\label{eif:tau}
Suppose  Assumptions~\ref{assump:TAignorability}--\ref{fm1m0x}
hold, and $\mathcal{M}_{{\rm tp}}$ and $\mathcal{M}_{{\rm om}}$ are nonparametric.
\begin{itemize}
\item[(i)] Suppose $\mathcal{M}_{{\rm ps}}$ is  known. The  EIF of $\tau_{u^*}$ is
\bee\label{eifeuknown}
\phi_{{\rm kn}}(x,z,m,y)  \ =  \   
\underbrace{\frac{ {\xi}_{1,u^*}(x,z,m)\{y - \mu_{z}(x,m)\}}{\E\big\lbt e_{\dd\od}(X)\big\rbt_{u^*}}}_{\mathrm{(I)}} + \underbrace{\frac{ \big\lbt\big\{\mu_{1}(x,\dd) - \mu_{0}(x,\od)\big\}e_{\dd\od}(x)\big\rbt_{u^*}
 - \tau_{u^*}\big\llbracket e_{\cdot \star}(x)\big\rrbracket_{u^*}}{\E\big\lbt e_{\dd\od}(X)\big\rbt_{u^*}}}_{\mathrm{(II)}},
\ee
 where $\xi_{1,u^*}(x,z,m) = \{\pi_z(x) f_{zm}(x)\}^{-1} {(-1)^{z+1}{\gamma^{(z)}_{1,u^*}}(x,m)}$, $\gamma_{1,u^*}^{(1)}(x,m)= \int_{\R} k_{u^*}(m,m_0)$ $e_{(m
,m_0)}(x)dm_{0}$, and $\gamma_{1,u^*}^{(0)}(x,m)= \int_{\R} k_{u^*}(m_1,m)e_{(m_1
,m)}(x)dm_{1}$.
\item[(ii)] Suppose $\mathcal{M}_{{\rm ps}}$ is unknown but follows a  parametric model \eqref{paraps}.  Let $s_{zm}(x\mid \beta^*) = \partial \log f_{zm}(x\mid \beta)/\partial \beta \mid_{\beta = \beta^*}$ be the score function of \eqref{paraps}, and suppose information matrix $I_{\beta^*} = \E \{s_{ZM}(X\mid \beta^*)s^\T_{ZM}(X\mid \beta^*)\}$ is positive-definite.  The EIF of $\tau_{u^*}$ is
$
\phi_{{\rm p}}(x,z,m,y)
 =   \phi_{{\rm kn}}(x,z,m,y) + \tilde{\phi}_{\mathrm{p}}(x,z,m)
$, where $\tilde{\phi}_{\mathrm{p}}(x,z,m)$ is defined in \eqref{tphip}.

\item[(iii)]Suppose $\mathcal{M}_{{\rm ps}}$ is unknown and nonparametric. The EIF of $\tau_{u^*}$ is 
$
\phi_{{\rm np}}(x,z,m,y)
 = \phi_{{\rm kn}}(x,z,m,y) + \tilde{\phi}_{\mathrm{np}}(x,z,m)
$, where $\tilde{\phi}_{\mathrm{np}}(x,z,m)$ is defined in \eqref{tphinp}.
\end{itemize}
\end{theorem}
The expressions for $\tilde{\phi}_{\mathrm{p}}(x,z,m)$ 
and $\tilde{\phi}_{\mathrm{np}}(x,z,m)$ are complex and do not provide additional insights and thus are deferred to \eqref{tphip} and  \eqref{tphinp} in the Supplement File.
We discuss the connections and comparisons among the EIFs. In $\phi_{{\rm np}}(\cdot)$, function $\phi_{{\rm kn}}(\cdot)$ comprises two orthogonal functionals, (I) and (II) (cf. \eqref{eifeuknown}), which belong to the orthogonal tangent spaces spanned by the scores of $f(y\mid x,z,m)$ and $f(x)$, respectively, and function $\tilde{\phi}_{\mathrm{np}}(\cdot)$  belongs to the tangent space spanned by the score of $f(m\mid x,z)$. To derive EIFs $\phi_{\textrm{p}}(\cdot)$ and $\phi_{\textrm{kn}}(\cdot)$, we project $\phi_{{\rm np}}(\cdot) = \phi_{\mathrm{kn}}(\cdot) + \tilde{\phi}_{\mathrm{np}}(\cdot)$ onto the corresponding tangent spaces under Scenarios (i) and (ii), respectively. Compared with Scenario (iii), under Scenarios (i) and (ii), only the assumption on $f(m\mid x,z)$  changes, and $\phi_{\mathrm{kn}}(\cdot)$ is   orthogonal to the corresponding tangent spaces spanned by the scores of $f(m\mid x,z)$. Therefore,  $\phi_{\mathrm{kn}}(\cdot)$ remains the same after projection. The projection of $\tilde{\phi}_{\mathrm{np}}(\cdot)$ is zero  under Scenario (i), because the tangent space spanned by a known $f(m\mid x,z)$ is null, and the projection of $\tilde{\phi}_{\mathrm{np}}(\cdot)$ is $\tilde{\phi}_{\mathrm{p}}(\cdot)$  under Scenario (ii). Under Scenarios (i)--(iii), the corresponding semiparametric efficiency bounds (SEBs) are $\mathrm{SEB}_{\star} = \E\{\phi^2_{\star}$ $(X,Z,M,Y)\}$, $\star\in\{{\rm kn},{\rm p},{\rm np}\}$. The additional terms in $\phi_{{\rm p}}(\cdot)$  and $\phi_{{\rm np}}(\cdot)$ amplify the variances in the SEBs for Scenarios  (ii) and (iii). This is because,  under Scenario (i), we have complete information about $\mathcal{M}_{{\rm ps}}$, which results in the smallest efficiency bound. In particular, as the knowledge about $\mathcal{M}_{\mathrm{ps}}$ reduces from Scenario (i) to Scenario (iii), we have $\mathrm{SEB}_{\mathrm{kn}}\leq \mathrm{SEB}_{\mathrm{p}} \leq \mathrm{SEB}_{\mathrm{np}}$. 
\par
The EIFs in Theorem~\ref{eif:tau} can be used to construct semiparametrically efficient estimators via classical semiparametric theory \citep{bickel1993efficient}, but the complexity of $\phi_{\mathrm{p}}(\cdot)$ and $\phi_{\mathrm{np}}(\cdot)$ makes it impractical to generate computationally feasible estimators. For instance, 
accurately computing  $\tilde{\phi}_{\mathrm{np}}(\cdot)$ and $\tilde{\phi}_{\mathrm{p}}(\cdot)$ can be time-consuming due to the need for   numerical integraions with intricate integrands, and   
this can be further exacerbated when using Bootstrap to construct confidence intervals. Additionally, obtaining derivatives such as $c^{(1)}_{\rho}(\cdot,\star)$ and $s_{zm}(x\mid \beta^*)$ in $\tilde{\phi}_{\mathrm{p}}(\cdot)$ and $\tilde{\phi}_{\mathrm{np}}(\cdot)$ requires additional effort from the researchers. 

\par As a result, our proposed estimator $\hat\tau_{u^*}$ is based on the oracle EIF, which is the simplest among the three. Importantly, the whole $\hat\tau_{u^*}$ only requires computing  numerical integrations   {\color{black}with simple integrands like $\lbt\hat{e}_{\dd\od}(X_i)\rbt_{u^*}$,} and does not require analytically calculating any derivatives. Consider the empirical version of ${\phi}_{{\rm kn}}(x,z,m,y)$,
\bee\nonumber
\hat{\phi}_{{\rm kn}}(x,z,m,y)  
 = \frac{\hat{{\xi}}_{1,u^*}(x,z,m)\{y - \hat{\mu}_{z}(x,m)\} + \big\lbt\big\{\hat{\mu}_{1}(x,\dd) - \hat{\mu}_{0}(x,\od)\big\}\hat{e}_{\dd\od}(x)\big\rbt_{u^*}
 - \hat{\tau}_{u^*}\big\lbt\hat{e}_{\dd\od}(x)\big\rbt_{u^*}}{\mathbb{P}_n\big\lbt\hat{e}_{\dd\od}(X)\big\rbt_{u^*}} , 
\ee
where the nuisance functions  $\{\hat{\pi}_z(x),\hat{f}_{zm}(x), \hat{\mu}_{z}(x,m)\}$ are trained by any off-the-shelf estimation algorithms, and other quantities are approximated by the plug-in estimators: 
\begin{itemize}
\item[(i)] $\hat{F}_{zm_z}(x)=\int_{-\infty}^{m_z}\hat{f}_{zm}(x)dm$, $\hat{e}_u(x) = c_\rho\{\hat{F}_{1m_1}(x), \hat{F}_{0m_0}(x)\}\prod_{z\in\{0,1\}}\hat{f}_{zm_z}(x)$; 
\item[(ii)] $\hat{\gamma}_{1,u^*}^{(1)}(m,x)=\int_{\R} k_{u^*}(m,m_0)$ $\hat{e}_{(m
,m_0)}(x)dm_{0}$, $\hat{\gamma}_{1,u^*}^{(0)}(m,x)=\int_{\R} k_{u^*}(m_1,m)\hat{e}_{(m_1,m
)}(x)dm_{1}$; 
\item[(iii)] $\hat{{\xi}}_{1,u^*}(x,z,$ $m)  = (-1)^{z+1}\{\hat{\pi}_z(x) \hat{f}_{zm}(x)\}^{-1}$ ${{\hat{\gamma}_{1,u^*}}^{(z)}(m,x)}$. 
\end{itemize}
Solving  $\hat{\tau}_{u^*}$ from the estimating equation $\mathbb{P}_n\big\{\hat{\phi}_{{\rm kn}}$ $(X,Z,M,Y)\big\} = 0$, we obtain the proposed estimator,
\bee\label{form:hattau}
\hat{\tau}_{u^*}
  =  \frac{\mathbb{P}_n\Big[\hat{{\xi}}_{1,u^*}(X,Z,M)\{Y - \hat{\mu}_{Z}(X,M)\}
+\big\lbt\big\{\hat\mu_{1}( X,\dd) - \hat\mu_{0}(X,\od)\big\}\hat{e}_{\dd\od}(X)\big\rbt_{u^*}\Big]}{\mathbb{P}_n\big\lbt\hat{e}_{\dd\od}(X)\big\rbt_{u^*}},
\ee
for Loc.PCE, and subsequently, for Cont.PCE when $h\rightarrow 0$ as $n\rightarrow \infty$.
\begin{remark}[Computational complexity]\color{black}
The computational
complexity of our estimator involves two components: (i) estimating
nuisance functions and (ii) calculating $\hat{\tau}_{u^{*}}$. The
complexity of estimating nuisance functions depends on the methods
used. Once estimated, the primary computational cost of $\hat{\tau}_{u^{*}}$
arises from numerically evaluating the integrals $\lbt\hat{e}_{\dd\od}(X_{i})\big\rbt_{u^{*}}$,
$\lbt\{\hat{\mu}_{1}(X_{i},\dd)-\hat{\mu}_{0}(X_{i},\od)\}\hat{e}_{\dd\od}(X_{i})\rbt_{u^{*}}$
and $\hat{\gamma}_{1,u^{*}}^{(Z_{i})}(M_{i},X_{i})$ for all $i\in[n]$.
For $\hat{\tau}_{u^{*}}$ to satisfy its asymptotic properties, these
integrals must achieve approximation errors of $o(n^{-1/2})$. For
example, with a Gaussian kernel and under mild regularity conditions,
our theoretical results show that computing these integrals with $o(n^{-1/2})$
errors requires $\mathcal{O}(n^{2})$ time (ignoring log factors)
when nuisance functions are estimated using simpler methods, such
as the parametric strategy introduced in Section~{\color{black}6}. Using more
complex estimators (e.g., kernel regression or neural networks) may
increase this complexity. For detailed discussions and theoretical
results, see Section~{\color{black}S4} of the Supplementary File.
\end{remark}

\section{Theoretical properties}\label{sec:sp}
\subsection{Limit double robustness}\label{sec:dbrb}
Denote the functional limits $\bar{\pi}_z(x)$ and $\bar{\mu}_{z}(x,m)$, such that $\hat{\pi}_{z}(x) \rightarrow \bar{\pi}_{z}(x)$, $\hat{\mu}_{z}(x,m)\rightarrow \bar{\mu}_{z}(x,m)$ as $n\rightarrow \infty$. Recalling \eqref{form:hattau}, {\color{black}when $\mathcal{M}_{{\rm ps}}$ is correctly specified,} the probabilistic limit of $\hat{\tau}_{u^*}$ is
\bee\nonumber
\bar\tau_{u^*} = \frac{\E\Big[\bar\xi_{1,u^*}(X,Z,M)\{Y - \bar{\mu}_{Z}(X,M)\} 
+\big\lbt\big\{\bar\mu_{1}(X,\dd) - \bar\mu_{0}(X,\od)\big\}{e}_{\dd\od}(X)\big\rbt_{u^*}\Big]}{\E\big\lbt{e}_{\dd\od}(X)\big\rbt_{u^*}},
\ee
 where $\bar\xi_{1,u^*}(x,z,m) =  \{\bar\pi_z(x) f_{zm}(x)\}^{-1}{(-1)^{z+1}\gamma_{1,u^*}^{(z)}(x,m)} $ and {\color{black} $\gamma_{1,u^*}^{(z)}(x,m)$ is defined similar with $\hat\gamma_{1,u^*}^{(z)}(x,m)$ by replacing $\hat{e}_u(x)$ with ${e}_u(x)$}. Theorem \ref{thm:dbrb} below justifies the limit double robustness of  $\hat{\tau}_{u^*}$ when $\mathcal{M}_{{\rm ps}}$ is correctly specified.
\begin{theorem}\label{thm:dbrb}
Suppose  Assumptions~\ref{assump:TAignorability}--\ref{am:cp}
hold, and $\mathcal{M}_{{\rm ps}}$ is correctly specified.  If either $\bar{\pi} = \pi$ or $\bar{\mu} = \mu$,  we have $\bar\tau_{u^*} = \tau_{u^*}${\color{black},  and moreover,  $\lim_{h\rightarrow 0}\bar{\tau}_{u^*} = \tau_{u^*}^*$ under  the conditions of Proposition~\ref{po:approximate}}.
\end{theorem}
Intuitively, with nonparametric $\mathcal{M}_{\text{tp}}$ and $\mathcal{M}_{\text{om}}$, our EIF-invoked estimator enjoys the Neyman orthogonality \citep{chernozhukov2017double} with respect to these two nuisance functions, which leads to the double robustness property as shown in Theorem \ref{thm:dbrb}.
 \begin{remark}\label{rk:triplerobust}
 When $M$ is binary, \citet{jiang2020multiply}  illustrate the triple robustness of their EIF-based PCE estimator; i.e., their estimator is consistent as long as any pair of the three models in $(\mathcal{M}_{{\rm tp}},\mathcal{M}_{{\rm ps}},\mathcal{M}_{{\rm om}})$ is correctly specified. The crucial reason why their estimator is triply robust while ours is doubly robust is that the underpinning identification assumptions are different. Their monotonicity assumption provides a neat  identification PCE formula, enabling triple robustness. However, when $M$ is continuous, the monotonicity fails to identify Cont.PCE and can be implausible in real-world problems such as surrogate evaluation. Due to the nonlinearity  of copula  $c_{\rho}(\cdot,\star)$, copula-based identification requires correct specification of $\mathcal{M}_{\text{ps}}$, losing robustness against its misspecification. If $\mathcal{M}_{\text{ps}}$  is incorrectly specified in $e_u(x)$,  the bias for $\bar{\tau}_{u^*}$ {\color{black}generally }cannot be eliminated through correct specifications of ${\mu}_{z}(x,m)$ and ${\pi}_z(x)$.

{\color{black}Nevertheless, in the special case where \(\mathcal{M}_{\mathrm{om}}\) follows a linear model, Theorem~\ref{prop:linear:triple} in Section~\ref{sec:extrarobust} of the Supplementary File demonstrates that \(\hat{\tau}_{u^{*}}\) exhibits robustness surpassing triple robustness for Cont.PCE estimation. Specifically, under the condition that \(\mathcal{M}_{\mathrm{om}}\) is linear and all identification assumptions (Assumptions~\ref{assump:TAignorability}-\ref{am:cp}) along with mild regularity conditions hold, our proposed estimator consistently identifies and estimates the true Cont.PCE  with vanishing $h$, in the following cases:  
(i) \(\mathcal{M}_{\mathrm{om}}\) is correctly specified, or  
(ii) \(\mathcal{M}_{\mathrm{tp}}\) and \(\mathcal{M}_{\mathrm{ps}}\) are both correctly specified.}
\end{remark} 

\subsection{Rate-double robustness and optimality}\label{sec:lsp}
Desirable asymptotic properties of $\hat{\tau}_{u^*}$ are shown under the  semiparametric setting that $\mathcal{M}_{{\rm ps}}$ is parametrically modeled through \eqref{paraps}, while $\mathcal{M}_{{\rm tp}}$ and $\mathcal{M}_{{\rm om}}$ are nonparametrically modeled. Examples of \eqref{paraps} include the linear model, additive errors nonlinear model and generalized linear model  \citep{boos2013essential}. With fixed $h > 0$, we first derive the asymptotic results of $\hat{\tau}_{u^*}$ towards the Loc.PCE $\tau_{u^*}$. 
Assumption \ref{am:emp} below imposes some typical regularity conditions for the nuisance functions and their estimators; see, e.g.~\citet{kennedy2016semiparametric}.
\begin{assumption}\label{am:emp}
(i) Either $\bar{\mu} = \mu$ or $\bar{\pi} = \pi$; (ii) $\|\hat{\mu} - \bar{\mu}\|_{\infty}, \|\hat{\pi} - \bar{\pi}\|_{\infty} = o_{\mathbb{P}}(1)$; (iii)  $\mathbb{P}(|\hat{\pi} - {\pi}|^2) = \mathcal{O}(r^2_\pi)$ and $\mathbb{P}(|\hat{\mu} - {\mu}|^2)= \mathcal{O}(r^2_\mu)$ with $r_{\pi}, r_{\mu} \precsim 1$ as $n\rightarrow \infty$; (iv) Functions like $\hat{\pi}(x),\bar{\pi}(x),\hat{\mu}_{z}(x,m),\bar{\mu}_{z}(x,m)$ and $\hat{\mu}_{a}(x,m_a),\bar{\mu}_{a}(x,m_a)$ for $a = 0,1$, are  contained in uniformly bounded function classes with finite uniform entropy integrals (see Section \ref{pf:lm:emp} in the Supplementary File for details), and $\hat{\pi}^{-1}$, $\bar{\pi}^{-1}$ are  uniformly bounded.
\par { If one exploits the cross-fitting strategy for $\hat{\tau}_{u^*}$, (iv) can be further dropped, and our theoretical results remain valid; see, e.g., \citet{robins2008higher} and \citet{chernozhukov2018double} for details.}
\end{assumption} 
\begin{theorem}[Loc.PCE inference]\label{thm:consist} Suppose Assumptions \ref{assump:TAignorability}--\ref{am:emp} and other  regularity conditions in Section \ref{sec:otherreg} hold. In addition, assume the parametric model of $\mathcal{M}_{{\rm ps}}$  in \eqref{paraps} is correctly specified and $\hat{\beta}$ is a regular $\sqrt{n}$-consistent estimator of $\beta^*$.
\begin{enumerate}
\item[I.](Rate-double robustness) As $n\rightarrow \infty$, we have
\bee\label{main:rate}
\hat{\tau}_{u^*} - \tau_{u^*} = \mathcal{O}_{\mathbb{P}}\big( n^{-1/2} + r_\pi r_\mu\big).
\ee
\item[II.](Asymptotic normality) Furthermore, suppose $r_{\pi}r_{\mu} = o(n^{-1/2})$, and $\hat{\beta}$ is asymptotically linear, i.e., there exists some $\psi(x,z,m)$ such that $\E\{\psi(X,Z,M)\} = 0$, $\E\|\psi(X,Z,M)\|^2 < +\infty$, and
$
\sqrt{n}(\hat{\beta} - \beta^*) = n^{-1/2}\sum_{i = 1}^n\psi(X_i,Z_i,M_i) + o_{\mathbb{P}}(1).
$
Then, with $\bar{\sigma}_{u^*}^2$  defined in \eqref{sigmaform}, we have
\bee\label{lim:tau}
\sqrt{n}  (\hat{\tau}_{u^*} - \tau_{u^*})\rightsquigarrow \mathcal{N}(0,\bar\sigma^2_{u^*}).
\ee
\item[III.] (Semiparametric efficiency) Since   $h$ is fixed,  $\tau_{u^*}$ is a regular estimand. Then we can use the classic semiparametric statistics theory to justify the optimality of $\hat{\tau}_{u^*}$ \citep{bickel1993efficient}. Suppose all conditions above hold. Further assume \ $\bar\pi = \pi$, $\bar\mu = \mu$, and $\hat{\beta}$ is a semiparametrically efficient estimator for $\beta^*$ such that $\psi(x,z,m) = I_{\beta^*}^{-1}s_{zm}(x\mid \beta^*)$, where $s_{zm}(x\mid \beta^*)$ and $I_{\beta^*}$ are defined in Theorem \ref{eif:tau}. Then we have $\sigma_{u^*}^2 = \mathrm{SEB}_{\mathrm{p}}=\E\{\phi^2_{{\rm p}}(X,Z,M,Y)\}$.  Thus, by  Theorem~\ref{eif:tau}(ii),  
 $\hat{\tau}_{u^*}$  is {semiparametrically efficient}. 
\end{enumerate}
{}
\end{theorem}
The error rate   \eqref{main:rate}  in Theorem~\ref{thm:consist}(I) illustrates the  rate-double robustness of our proposed estimator. The asymptotic linearity condition for $\hat{\beta}$  can be attained by commonly-used  methods, e.g., the likelihood-based or quasi-likelihood-based estimators, and generalized method of moments  \citep{boos2013essential,mccullagh2019generalized}. To attain the optimal root-$n$ convergence of $\hat{\tau}_{u^*}$, we only require $\mu_{z}(x,m)$ and $\pi_z(x)$ to be well estimated, e.g., $r_{\pi},r_{\mu}\precsim n^{-1/4}$. When one of $\mu_{z}(x,m)$ and $\pi_z(x)$ can be estimated parametrically (i.e., $r_{\pi} \text{ or }r_{\pi}\precsim n^{-1/2}$), $\hat{\tau}_{u^*}$ is root-$n$ consistent even the other one  is not correctly specified.  For asymptotic normality   \eqref{lim:tau} in  Theorem~\ref{thm:consist}(II), the condition   $r_{\pi}r_{\mu} = o_{\mathbb{P}}(n^{-1/2})$ is standard in the double machine learning  literature \citep{chernozhukov2018double}. Under the condition of Theorem~\ref{thm:consist}(II), we can   construct the CI for Loc.PCE.   A closed-form  estimator for the asymptotic variance $\sigma_{u^*}^2$ is feasible, which however is complicated due to the complex form of $\sigma_{u^*}^2$ (c.f.~\eqref{sigmaform}). Thus a standard nonparametric bootstrap procedure \citep{efron1981nonparametric} is recommended to estimate ${\sigma}_{u^*}^2$; see, e.g.,  \citet[$\mathsection$ IV]{imbens2004nonparametric} and \citet{jiang2020multiply} for the use of bootstrap  asymptotic variance estimation in different problems.
\par
Theorem \ref{thm:consist}(III) further justifies the optimality of $\hat{\tau}_{u^*}$. It  requires correct specification of both $\mathcal{M}_{{\rm tp}}$ and $\mathcal{M}_{{\rm om}}$, and that $\hat{\beta}$ is semiparametrically efficient, which holds when $\hat{\beta}$ is a regular MLE. These conditions are necessary for achieving semiparametric efficiency in doubly robust estimators, such as the augmented inverse propensity weighted estimator \citep{hahn1998role}.

{\color{black}
Finally, we  derive nonparametric convergence results  of $\hat{\tau}_{u^*}$ for Cont.PCE estimation. In particular, the  asymptotic rate for $\hat{\tau}_{u^*}-{\tau}_{u^*}$ in \eqref{main:rate} can be generalized to include $h$, when we let $h\rightarrow 0$ as $n\rightarrow \infty$. Combing such  results with the asymptotic bias of ${\tau}_{u^*} - {\tau}^*_{u^*} $ in Proposition \ref{po:approximate}, we show the  convergence rate and rate-double robustness of $\hat{\tau}_{u^*} - {\tau}^*_{u^*}$. Similarly, we generalize the result in Theorem~\ref{thm:consist}(II)  and show the asymptotic normality of $\hat{\tau}_{u^*}-{\tau}^*_{u^*}$.
\begin{theorem}[Cont.PCE inference]\label{hto0}
Suppose Assumptions \ref{assump:TAignorability}--\ref{am:emp} and   regularity conditions in Section \ref{sec:po1:con}  and Section \ref{sec:otherreg} hold.  We have  
\bee\label{rate:cont.pce}
\hat{\tau}_{u^*} - \tau^*_{u^*} = \mathcal{O}_{\mathbb{P}}(h^{-1}n^{-1/2}  + h^2 + h^{-1}r_\pi r_\mu).
\ee
Moreover, suppose the additional conditions in Theorem \ref{thm:consist}{\rm(II)} hold and we choose any $h = o(n^{-1/6})$. Then with $\tilde\sigma^2_{u^*}$  defined in \eqref{tildesigma} in the Supplementary File, we have
\bee\label{normality:contpce}
\sqrt{n}h(\hat{\tau}_{u^*} - \tau^*_{u^*})\rightsquigarrow \mathcal{N}(0,\tilde\sigma^2_{u^*}).
\ee
\end{theorem}
In \eqref{rate:cont.pce}, we derive the nonparametric convergence rate of $\hat{\tau}_{u^*}$ towards $ \tau^*_{u^*}$. The three terms on the right-hand side of \eqref{rate:cont.pce} represent the nonparametric variance, nonparametric bias, and the effect of nuisance estimation errors, respectively. Specifically, under the rate-doubly robust condition $r_{\pi}r_{\mu} = \mathcal{O}(n^{-1/2})$, we have  $\hat{\tau}_{u^*} - \tau^*_{u^*} = \mathcal{O}_{\mathbb{P}}(n^{-1/3}) $ by choosing $h\asymp n^{-1/6}$, which is  minimax optimal for the pointwise estimation of a two-dimensional and twice-differentiable nonparametric regression function $\tau_{u}^*$ \citep{tsybakov2008introduction}.  See the regularity conditions for $\tau_{u}^*$ in Section~\ref{sec:po1:con}. With an undersmoothing selection of $h > 0$ and well-estimated nuisance functions such that $r_\pi r_\mu = \mathcal{O}(n^{-1/2})$, we further achieve the asymptotic normality in \eqref{normality:contpce}. Similar to the Loc.PCE, we can use nonparametric bootstrap to estimate $\tilde{\sigma}_{u^*}^2$ and thereby construct CI for  $\tau^*_{u^*}$.
}
{\color{black}\begin{remark}[Nonparametric estimation of  $\mathcal{M}_{\mathrm{ps}}$]\label{rk:np}
In general, suppose $\hat{f}_{zm}(x)$ is  nonparametrically estimated with
$\E\{\hat{f}_{zM}(X) - {f}_{zM}(X)\}^2 = \mathcal{O}_{\mathbb{P}}(r^2_{f}) = o_{\mathbb{P}}(1)$ for $z = 0,1$, and  $r_\pi r_{\mu} \rightarrow 0$. Following the proofs of Theorem \ref{thm:consist} and Theorem~\ref{hto0}, we can show  the rates in \eqref{main:rate} and \eqref{rate:cont.pce} become $\mathcal{O}_{\mathbb{P}}\big(r_f +r_\pi r_\mu\big)$ and $\mathcal{O}_{\mathbb{P}}\big(h^{-1}r_f + h^{-1}r_\pi r_\mu + h^2 \big)$, respectively, and thus $\hat{\tau}_{u^*}$ remains consistent for $\tau_{u^*}$, and furthermore, for $\tau^*_{u^*}$ when we select $r_f\prec h\prec 1$ as $n\rightarrow \infty$.
 {\color{black}}
\end{remark}}
\section{Related works}\label{sec:rw}

\paragraph{Joint modeling of continuous principal strata.}  Pioneering works have proposed flexible methods to model the joint distribution of continuous $M_1$ and $M_0$. \citet{bartolucci2011modeling}  use copula to model the joint distribution of $(M_1,M_0)$  with parametric outcome model for randomized experiment. They  propose a likelihood approach to estimate and perform sensitivity analysis with respect to $\rho$. \citet{schwartz2011bayesian} further adopt the Dirichlet process mixture model to model $(M_1,M_0)\mid X$. \citet{zigler2012bayesian} develop a fully Bayesian approach, and assume $(M_1,M_0)\mid X$ follows a joint conditional Gaussian model with a known $\rho$, which corresponds to a Gaussian linear regression model for  \eqref{paraps} with a Gaussian copula  $c_{\rho}(\cdot,\star)$. 
Unlike previous works that rely on concrete modeling assumptions without assuming principal ignorability, our proposed estimator under weak principal ignorability is more robust and flexible. It does not necessarily require parametric assumptions for all nuisance functions and remains consistent even if $\mathcal{M}_{\rm{tp}}$ or $\mathcal{M}_{\rm{om}}$ is misspecified.
 \paragraph{Semiparametric estimation of Cont.PCE under principal ignorability.}   {\color{black}During the revision of this paper, another paper uploaded to arXiv \citep{lu2023principal} also investigated the semiparametric estimation of Cont.PCE under principal ignorability. To address the non-regularity of Cont.PCE, the authors proposed a global projection-based approach, projecting the Cont.PCE surface onto a general parametric working model and subsequently studying its semiparametric efficient estimation. In contrast, we estimate a local smoothing-based substitute of Cont.PCE, referred to as Loc.PCE. The fundamental distinction between these two methods arises from the inherent differences between local smoothing-based and global projection-based estimators in approximating and estimating non-regular estimands. In particular, there exists a bias-variance trade-off between the two approaches in estimation. Assuming all regularity conditions hold and all nuisance functions are well estimated, when the parametric working model in \citet{lu2023principal} correctly specifies Cont.PCE, both methods consistently estimate Cont.PCE, with the projection-based approach achieving a sharper root-\(n\) convergence rate. However, if the working model in \citet{lu2023principal} misspecifies Cont.PCE, their method introduces a non-vanishing misspecification bias but remains consistent with its global projection. In contrast, our method does not require the correct specification of a parametric working model and remains consistent with the exact Cont.PCE. An overview of \citet{lu2023principal}, along with a comprehensive comparison of the two papers from multiple perspectives--including approaches for non-regular estimand approximation, bias-variance trade-offs in estimation, computational complexity, and application foci--is provided in Section~\ref{sec:compare} of the Supplementary~File.}
\paragraph{\color{black}Local and localized functional estimation.}
\citet{chernozhukov2018biased,chernozhukov2021simple} propose a general Riesz-representation-invoked method which estimates  local functionals semiparametrically efficiently \citep[Theorem 4.2]{chernozhukov2018biased}, and thereby approximates the corresponding perfectly localized functionals accurately. 
They consider the minimal Riesz representer and outcome mean model as  nuisance functions, and 
 demonstrate the double robustness and  semiparametric efficiency of their estimator. However, their framework cannot be applied to estimate Loc.PCE and Cont.PCE, because they assume  all variables in $(W,Y)$ (cf.~\eqref{def:perfect}) are observable, while under this paper's setting, principal strata $U$ in $W$  contain one unobserved variable. Moreover, for identification, we need to  deal with three nuisance functions instead of two, which introduces extra difficulties in estimation.  To resolve  these challenges while achieving the same  desirable properties as \citet{chernozhukov2018biased,chernozhukov2021simple}, we propose our estimator through a new oracle EIF-invoked strategy. Although $\hat{\tau}_{u^*}$ is derived through the EIF under the oracle $\mathcal{M}_{\mathrm{ps}}$-known scenario, it can still attain the  semiparametric efficient bound $\rm{SEB}_{{\rm p}}$ when $\mathcal{M}_{\mathrm{ps}}$ follows a parametric model \eqref{paraps} (c.f.~Theorem \ref{thm:consist}(III)). 

\paragraph{Leveraging surrogates for causal effect identification and estimation.}In this paper, we focus on evaluating the effectiveness of certain surrogates in assessing the treatment effects on the primary outcome. Beyond identifying informative surrogates from data, another line of research investigates how to leverage these surrogates to identify the long-term treatment effect or enhance its estimation efficiency. 
For instance, \citet{athey2019surrogate} study the estimation of the long-term average treatment effect (ATE) using multiple surrogates and two separate datasets, where treatments and long-term outcomes are observed in different datasets. By exploiting surrogates observed in both datasets, they develop a semiparametric efficient estimator of the ATE under the statistical surrogacy condition. More recently, \citet{kallus2024role} propose a semiparametric efficient estimator of the ATE without assuming the statistical surrogacy condition when treatments are observed in both datasets. Given the extensive literature on causal effect identification and estimation using surrogates, we do not attempt to provide a comprehensive review here. Interested readers are referred to \citet[Section~1.2]{kallus2024role} for a comprehensive review.

\section{Simulation study}\label{sec:simu}
\begin{figure}[!t]
\centering
\includegraphics[height=5.9cm]{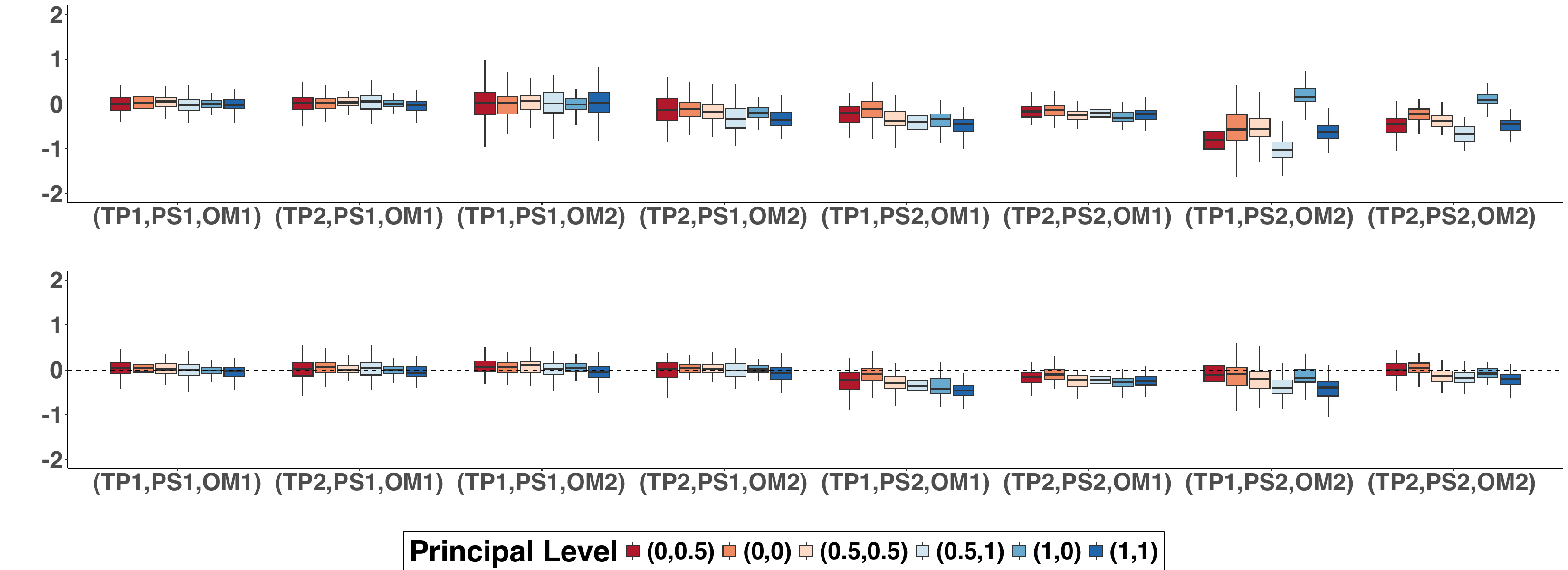}
\caption{\color{black}The box-plots of $\hat{\tau}_{u}-\tau^*_{u}$ over $100$ MC rounds when $n = 2000$. X-axis represents different data generation mechanisms. For example, (TP2,PS1,OM1) means  data are generated from $(\mathcal{F}_{{\rm tp}}^{(2)},\mathcal{F}_{{\rm ps}}^{(1)},\mathcal{F}_{{\rm ps}}^{(1)})$. Different colors represent different principal levels $u$.  The top and bottom panels present the results when nuisance functions are estimated through  parametric and semiparametric strategies, respectively.}\label{fig:para}
\end{figure}
{We perform numerical experiments to assess the large-sample properties of $\hat{\tau}_{u}$ on estimating Cont.PCE. The data generation process involves the following steps: first, we generate $X = (X_1,X_2,X_3)^\T$ from the standard normal distribution $\mathcal{N}(0,I_3)$. Second, we generate $Z$ from $\mathcal{F}_{{\rm tp}}$ given $X$. Third, we generate $M$ from $\mathcal{F}_{{\rm ps}}$ given $X$ and $Z$. Finally, we generate $Y$ from $\mathcal{F}_{{\rm om}}$ given all the other variables. We use the standard Gaussian copula $c_{\rho}(\cdot,\star)$ with $\rho = 0.5$. We consider two comparative distributions for each of the three distributions $\mathcal{F}_{{\rm tp}},\mathcal{F}_{{\rm ps}},\mathcal{F}_{{\rm om}}$ as follows:
\begin{itemize}{\color{black}
\item $\mathcal{F}^{(1)}_{{\rm tp}} = \text{Bern}\{\logit(  X_2 + X_3)\}$ and $\mathcal{F}^{(2)}_{{\rm tp}} = \text{Bern}\{\logit(X_1^2/2 + X^3_2/2)\}$;
\item $\mathcal{F}^{(1)}_{{\rm ps}} = \mathcal{N}(X_1/2 + X_2/2 + X_3/2+Z,0.5^2)$ and $\mathcal{F}^{(2)}_{{\rm ps}} = \mathcal{N}\{X_1 + Z(X^2_1 +X^3_1/2) , 0.5^2\}$;
\item  $\mathcal{F}^{(1)}_{{\rm om}} = \mathcal{N}(X_1 +X_3 +   Z X_1+Z + M/2,{\color{black}0.5^2})$ and $\mathcal{F}^{(2)}_{{\rm om}} = \mathcal{N}\{X_2 +Z(X_1 + {\color{black}X^2_1} +{\color{black}X^3_1/5})+ M/2,{\color{black}0.5^2}\}$.}
\end{itemize}}
{\color{black}Based on the identification formula \eqref{tau:idd}, we generate $10^6$ samples, $\{(X_i,Z_i,M_i,Y_i)\}_{i = 1}^{10^6}$, under the corresponding data generation setting, and approximate $\tau_u^*$ as
$
\tau^*_{u} \approx\big[\sum_{i = 1}^{10^6}\big\{\mu_{1}(X_i,m_1) - \mu_{0}(X_i,m_0)\big\}e_u(X_i)\big]/\big[\sum_{i = 1}^{10^6}e_u(X_i)\big],
$
where $e_u(X)$ can be specified by Assumption \ref{am:cp}.}
\subsection{Simulation: the consistency and double robustness}\label{sec:cd}
{We  simulate data through eight different data generation mechanisms, with $(\mathcal{F}_{{\rm tp}},\mathcal{F}_{{\rm ps}},\mathcal{F}_{{\rm om}})$ $= (\mathcal{F}^{(a)}_{{\rm tp}},\mathcal{F}^{(b)}_{{\rm ps}},\mathcal{F}^{(c)}_{{\rm om}})$, $a,b,c\in \{1,2\}$. In addition, we choose  $n \in \{500,1000,2000\}$ and $\color{black}h = 0.15 n^{-1/6}$ following the optimal rate in Theorem \ref{hto0}. We consider both a parametric and a semiparametric strategy for estimating the nuisance functions.
\begin{table}[t]
 
\caption{Empirical coverage rates (rounded to three decimal places) of $\hat{\tau}_{u}$'s $95\%$  CIs over 300 Monte Carlo rounds, when all nuisance functions are trained via linear models. The triplet $abc$ in the first row represents scenario $(\mathcal F_{\rm tp}^{(a)} , \mathcal F_{\rm ps}^{(b)} , \mathcal F_{\rm om}^{(c)} )$ for $a,b,c\in\{1,2\}$.}
\label{tb:realdata1}\label{tb:1}\centering\scalebox{0.9}{
\setlength{\tabcolsep}{14pt}
 {\begin{tabular}{c|c c c c c c c c}
\hline
$u$& { 111 } & { 211 }  & { 112 } & { 212 } & { 121 } & { 221 } & { 122 }& { 222 }
\\
\hline
${\bf(0,0)}$ & $0.960$ & $0.953$& $ 0.957$& $ 0.907$& $ 0.850$& $ 0.920 $ & $0.917$& $ 0.940$
\\
${\bf(0,0.5)}$ & $ 0.947$ & $ 0.970$ & $ 0.947$ & $ 0.943$ & $ 0.703$ & $ 0.917 $ & $0.727 $ & $0.743$
\\
${\color{black}\bf(1,0)}$& $ 0.970 $&$0.937$&$ 0.947$&$ 0.887$&$ 0.510$&$ 0.777$&$ 0.863$&$ 0.903$
\\
${\bf(0.5,0.5)}$ & $0.953$& $ 0.953$& $ 0.947$& $ 0.913 $& $0.653$& $ 0.783$& $ 0.887 $& $0.777$
\\
${\bf(0.5,1)}$ &  $0.963 $ &  $0.970$ &  $ 0.970$ &  $ 0.890$ &  $ 0.700$ &  $ 0.820 $ &  $0.450 $ &  $0.437 $
\\
${\bf(1,1)}$ & $ 0.940$ &  $ 0.960 $ &  $0.963 $ &  $0.773 $ &  $0.637$ &  $ 0.793$ &  $ 0.580$ &  $ 0.553$
\\
\hline

\end{tabular}}}
\end{table}

\paragraph{\bf The parametric strategy}{\color{black} We estimate \(\pi_{z}(x)\) using logistic regression, treating \(Z\) as the outcome and \(X\) as the covariates. We estimate \(\mu_{z}(x,m)\) using linear regression, with \(Y\) as the outcome and \(X, Z, M\), along with their interaction effects \(XZ = (X_{1}Z, X_{2}Z, X_{3}Z)\), as the covariates.} To estimate $f_{zm}(x)$, we specify the parametric model in \eqref{paraps} as
$
f_{zm}\big\{x\mid \beta =(\ell,\sigma^2)\big\} = \varphi\{m\mid    \ell^\T\begin{psmallmatrix}
x
\\
z
\end{psmallmatrix},\sigma^2\},
$
where $\ell$ is a vector in $\R^4$ and $\varphi(\cdot\mid \iota, \omega)$ denotes the Gaussian density function with mean $\iota$ and variance $\omega$. We estimate $\ell$ and $\sigma^2$ through the linear regression $M\sim(X_1,X_2,X_3,Z)$, where $\hat{\ell}$ denotes the vector of regression coefficients and $\hat{\sigma}^2$ represents the empirical residual variance. Hence, $\mathcal{M}_{\mathrm{ps}}$ is correctly specified if and only if $\mathcal{F}_{{\rm ps}} = \mathcal{F}_{{\rm ps}}^{(1)}$.

\paragraph{The semiparametric strategy} We estimate ${\pi}_z(x)$ and ${\mu}_{z}(x,m)$ through an ensemble learning algorithm, namely \texttt{SuperLearner} \citep{van2007super}, which combines various machine learning methods including linear regression, multivariate adaptive regression splines, single-layer neural networks, and recursive regression trees. The estimation of $\mathcal{M}_{{\rm ps}}$ is the same as the parametric strategy, where we estimate $\ell$ and $\sigma^2$ through the linear regression $M\sim(X_1,X_2,X_3,Z)$, and $\hat{\ell}$ is the vector of regression coefficients while $\hat{\sigma}^2$ is the empirical residual variance. 

\par
\
\par
{We evaluate the performance of $\hat{\tau}_{u}$ on Cont.PCE estimation at six different levels of $u$, namely, $u = (0,0)$, $(0,0.5)$, $\color{black}(1,0)$, $(0.5,0.5)$, $(0.5,1)$, $(1,1)$, and demonstrate the estimation errors of $\hat{\tau}_{u}$ across $100$ Monte Carlo (MC) iterations in Figure \ref{fig:para} when $n = 1000$. The complete simulation results with different sample sizes $n \in \{500,1000,2000\}$ are contained in Figure \ref{fig:completesimu} in the Supplementary File. Our numerical results indicate: (1) When employing a parametric nuisance strategy, the estimation biases of  $\hat{\tau}_{u} - \tau^*_u$ are close to zero for all tested $u$, provided that $\mathcal{F}_{{\rm ps}} = \mathcal{F}^{(1)}_{{\rm ps}}$ and either $\mathcal{F}_{{\rm tp}} = \mathcal{F}^{(1)}_{{\rm tp}}$ or $\mathcal{F}_{{\rm om}} = \mathcal{F}^{(1)}_{{\rm om}}$, and the estimation variances are particularly small when all nuisance functions are correctly specified, i.e., $(\mathcal{F}_{{\rm tp}},\mathcal{F}_{{\rm ps}},\mathcal{F}_{{\rm om}}) = (\mathcal{F}^{(1)}_{{\rm tp}},\mathcal{F}^{(1)}_{{\rm ps}},\mathcal{F}^{(1)}_{{\rm om}})$; (2) When using a semiparametric nuisance strategy, the estimation biases are close to zero and the estimation variances are particularly small whenever $\mathcal{F}_{{\rm ps}} = \mathcal{F}^{(1)}_{{\rm ps}}$, as $\mathcal{M}_{{\rm tp}}$ and $\mathcal{M}_{{\rm om}}$ are always correctly specified by $\mathtt{SuperLearner}$; (3) Under aforementioned settings, the estimation errors  become smaller as $n$ increases. These findings confirm the consistency and rate-double robustness of $\hat{\tau}_{u}$ as  shown in Theorem~\ref{hto0}.  
\subsection{Simulation: the limiting distribution}
We verify the asymptotic normality of $\hat{\tau}_{u}$ by  $n = 700$ samples and calculating the bootstrap standard deviation of $\hat{\tau}_{u}$ using 100 bootstrap samples, which serves as the nonparametric Bootstrap estimator of ${\sigma}_u$. {\color{black}To validate the asymptotic normality in Theorem \ref{hto0}, we choose undersmoothing $\color{black}h = 0.1n^{-1/5}$.} We obtain the 95$\%$ bootstrap CIs  and calculate the empirical coverage rate (ECR)  over 300 MC rounds, to check if the CIs capture the true $\tau^*_{u}$.   All nuisance functions are estimated via the parametric strategy (c.f.~Section~\ref{sec:cd}) for tested $u$. As shown in Table \ref{tb:1}, when $(\mathcal{F}_{{\rm tp}},\mathcal{F}_{{\rm ps}},\mathcal{F}_{{\rm om}}) = (\mathcal{F}^{(1)}_{{\rm tp}},\mathcal{F}^{(1)}_{{\rm ps}},\mathcal{F}^{(1)}_{{\rm om}})$ and thus all nuisances can be estimated under the root-$n$ rates,  ECRs are close to $0.95$ for all tested $u$ because of the asymptotic normality of $\hat{\tau}_{u}$ (cf.~\eqref{lim:tau}). When $(\mathcal{F}_{{\rm tp}},\mathcal{F}_{{\rm ps}},\mathcal{F}_{{\rm om}}) = (\mathcal{F}^{(2)}_{{\rm tp}},\mathcal{F}^{(1)}_{{\rm ps}},\mathcal{F}^{(1)}_{{\rm om}}) \text{ or }(\mathcal{F}^{(1)}_{{\rm tp}},\mathcal{F}^{(1)}_{{\rm ps}},\mathcal{F}^{(2)}_{{\rm om}})$, we have $r_{\pi}r_\mu \precsim n^{-1/2}$, as one of $\mu_{z}(x,m)$ and $\pi_z(x)$ is misspecified while the other can be estimated parametrically, which is of the  boundary case of condition  $r_{\pi}r_\mu =o( n^{-1/2})$ for \eqref{lim:tau}.  Nevertheless, 
in these cases, the nuisance function estimation biases are still negligible enough, and  ECRs are close to 0.95 for all tested $u$.

\section{Surrogate analysis of short-term CD4 count in ACTG 175}\label{sec:actg175}
\begin{figure}[t]
\centering
\begin{subfigure}[b]{0.3\textwidth}
\centering
\includegraphics[width=1\textwidth]{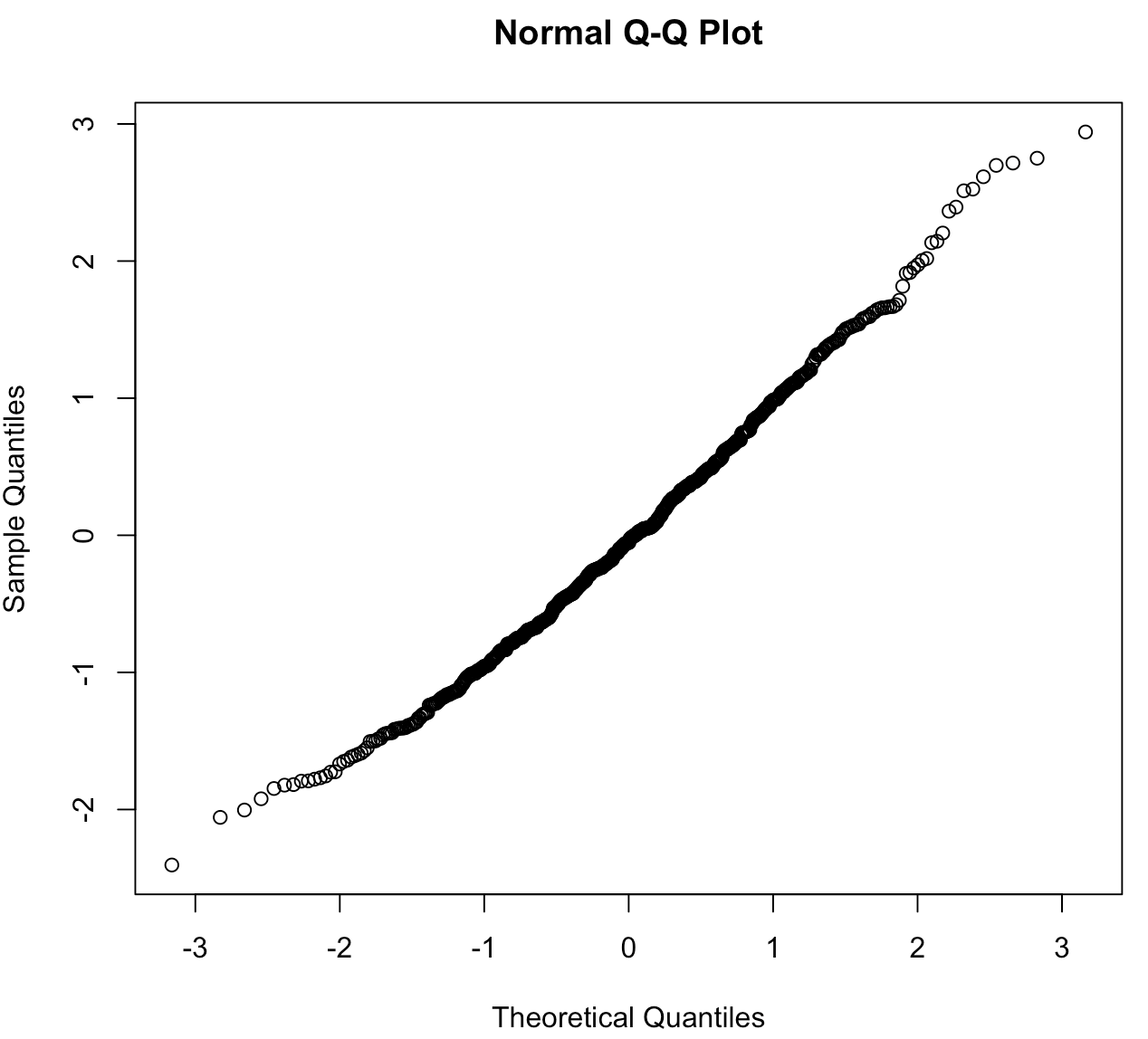}
\caption{The Q-Q plot of the residuals for  linear regression  $M\sim X,Z$.}
\end{subfigure}
\quad\quad\quad\quad\quad\quad\quad\quad
\begin{subfigure}[b]{0.3\textwidth}
\centering
\includegraphics[width=1\textwidth]{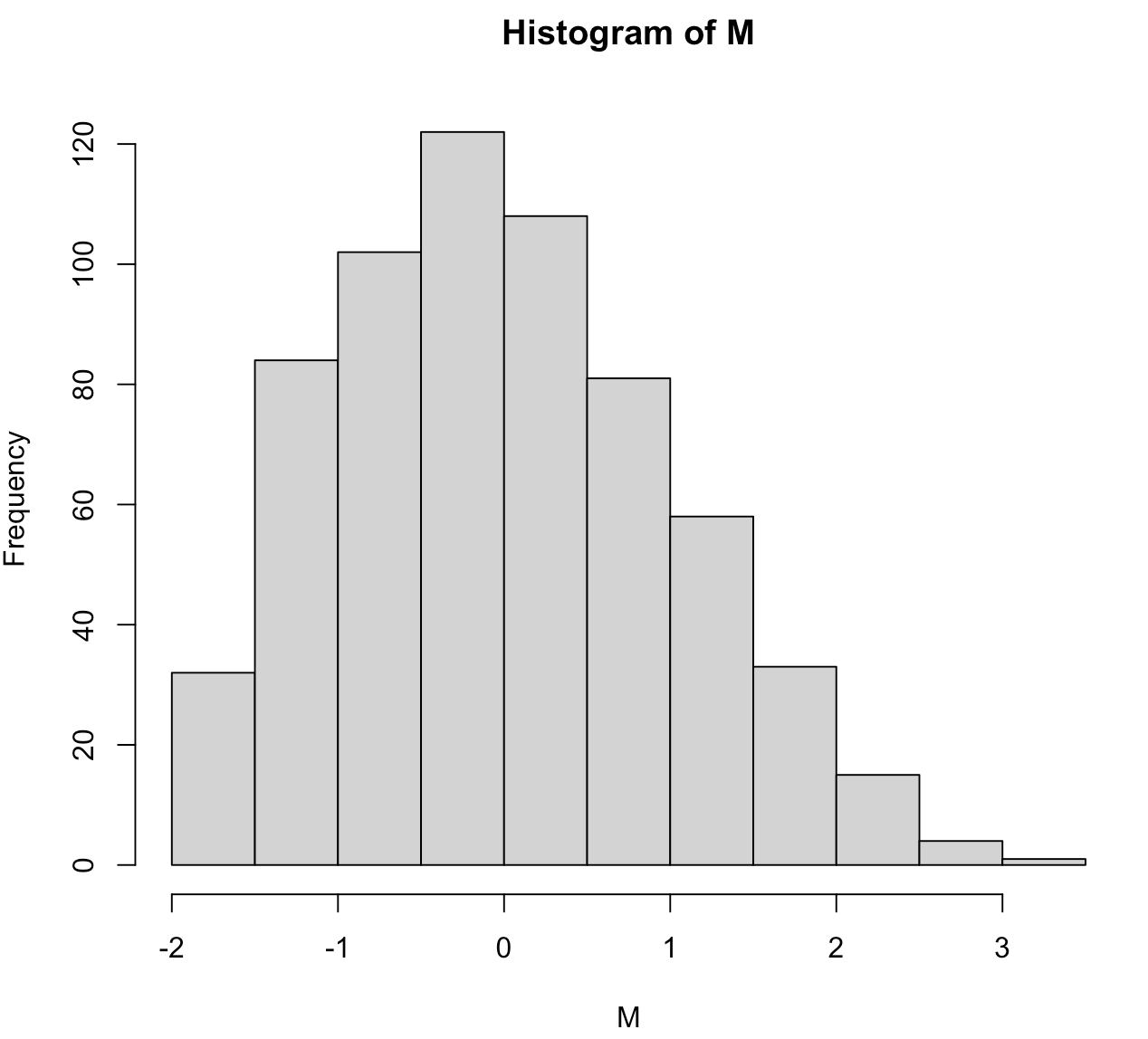}
\caption{The histogram of standardized $M_i$ among all samples.}
\end{subfigure}
\caption{ \color{black} Model diagnosis for surrogate analysis in ACTG 175. Panel(a) shows the Q-Q plot of linear regression residuals when training $\mathcal{M}_{\mathrm{ps}}$ in  the ACTG 175 data analysis. The approximate linear pattern implies that our regular linear regression modeling of $M\mid X,Z$ is reasonable for nuisance training. Panel(b) is the histogram of all standardized $M_i$, roughly ranging from $-2$ to $3$.}
\label{fig:actg175}
\end{figure}
ACTG 175 \citep{hammer1996trial} is one of the first randomized clinical trials to demonstrate the superiority of combination therapy with zidovudine and didanosine over monotherapy with zidovudine or didanosine for patients infected with human immunodeficiency virus type I (HIV-I). Each patient was randomly assigned a specific therapy. It is of essential interest to investigate whether a short-term endpoint measured several weeks after randomization can be a valid surrogate for a long-term clinical endpoint \citep{burzykowski2005evaluation}. In this study, we explore this question using data from ACTG175 contained in the R package \texttt{speff2trial}, focusing on patients treated with zidovudine only ($Z = 0$) or zidovudine + didanosine ($Z= 1$). To ensure weak principal ignorability, we collect various covariates in $X$, including age, weight, hemophilia, homosexual activity, history of intravenous drug use, Karnofsky score, non-zidovudine antiretroviral therapy prior to initiation, zidovudine use in the 30 days prior to treatment initiation, race (white/non-white), gender, antiretroviral history, and symptomatic indicator. We select the short-term CD4 count at $20\pm 5$ weeks as $M$ and the long-term CD4 count at $96\pm 5$ weeks as $Y$, both of which reflect patients' immune functions in different periods. To handle censoring and outliers, we exclude patients censored two years after randomization and patients with $M_i$ larger than the $99$th percentile or smaller than the $1$st percentile among all $M_i$. Ultimately, our study focuses on 640 patients.
\begin{figure}[!t]
\centering
\begin{subfigure}[b]{0.32\textwidth}
\centering
\includegraphics[width=1\textwidth]{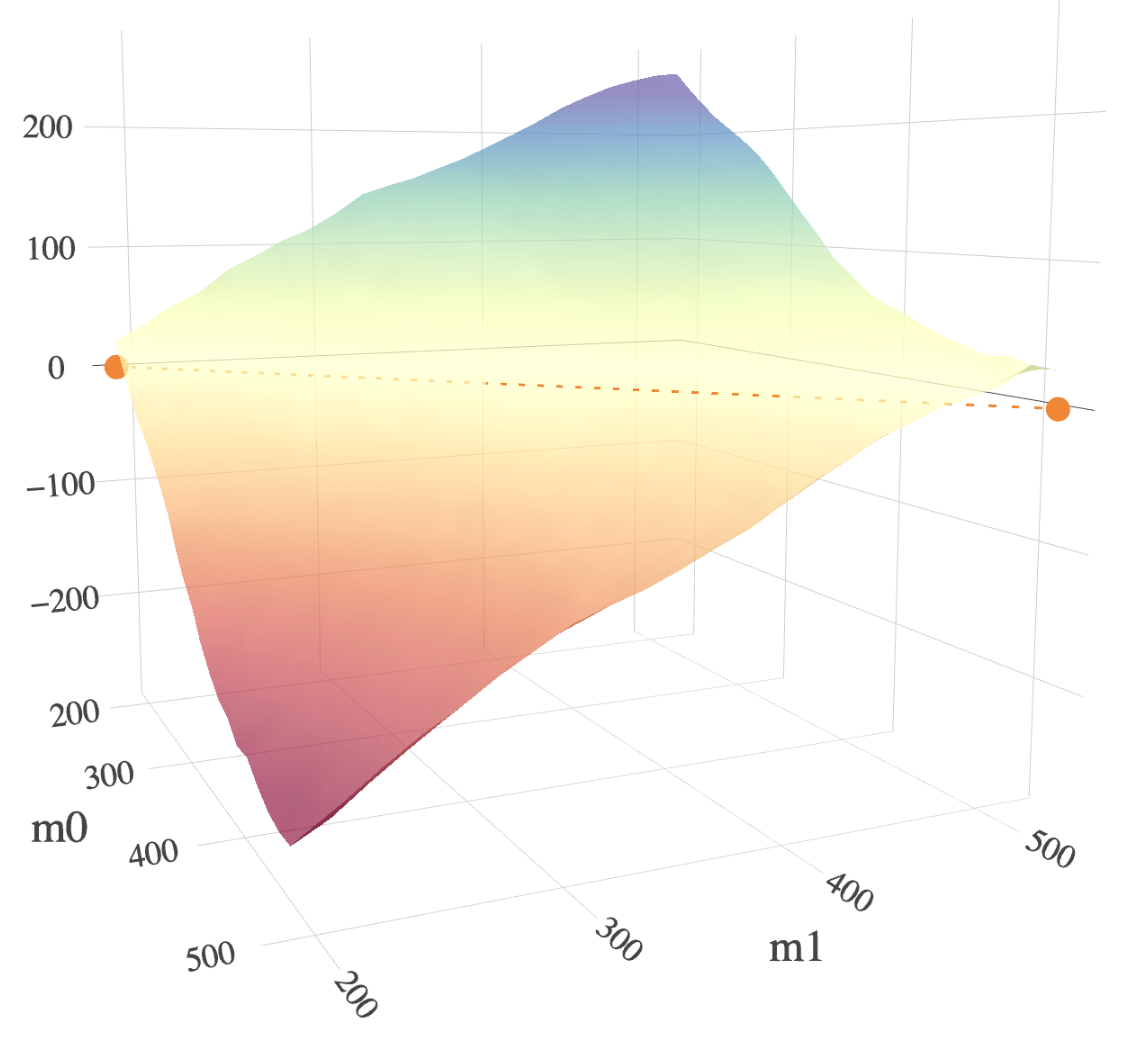}
\includegraphics[width=1\textwidth]{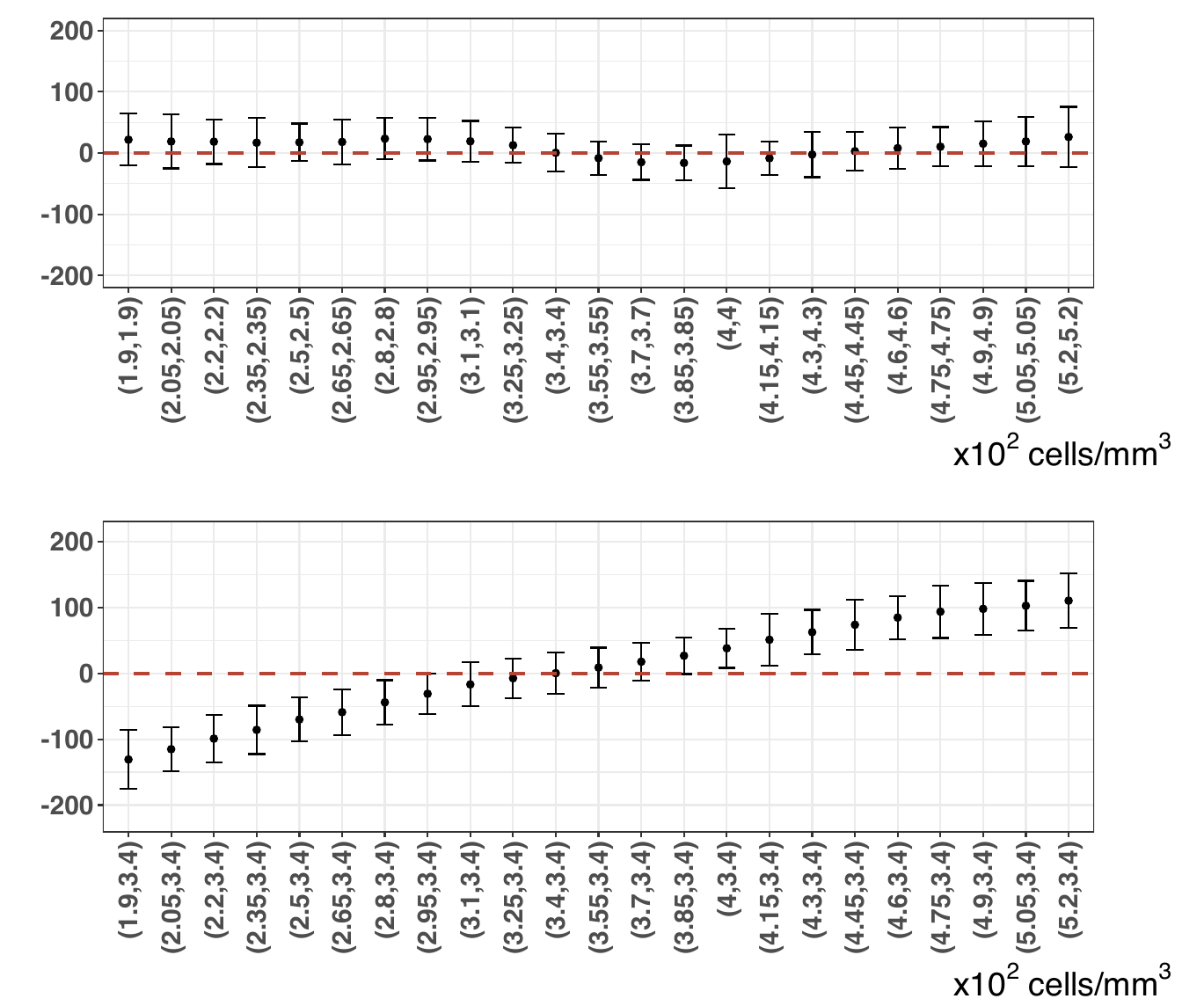}
\caption{$\rho=0.5,h = 0.25$}
\end{subfigure}
\begin{subfigure}[b]{0.32\textwidth}
\centering
\includegraphics[width=1\textwidth]{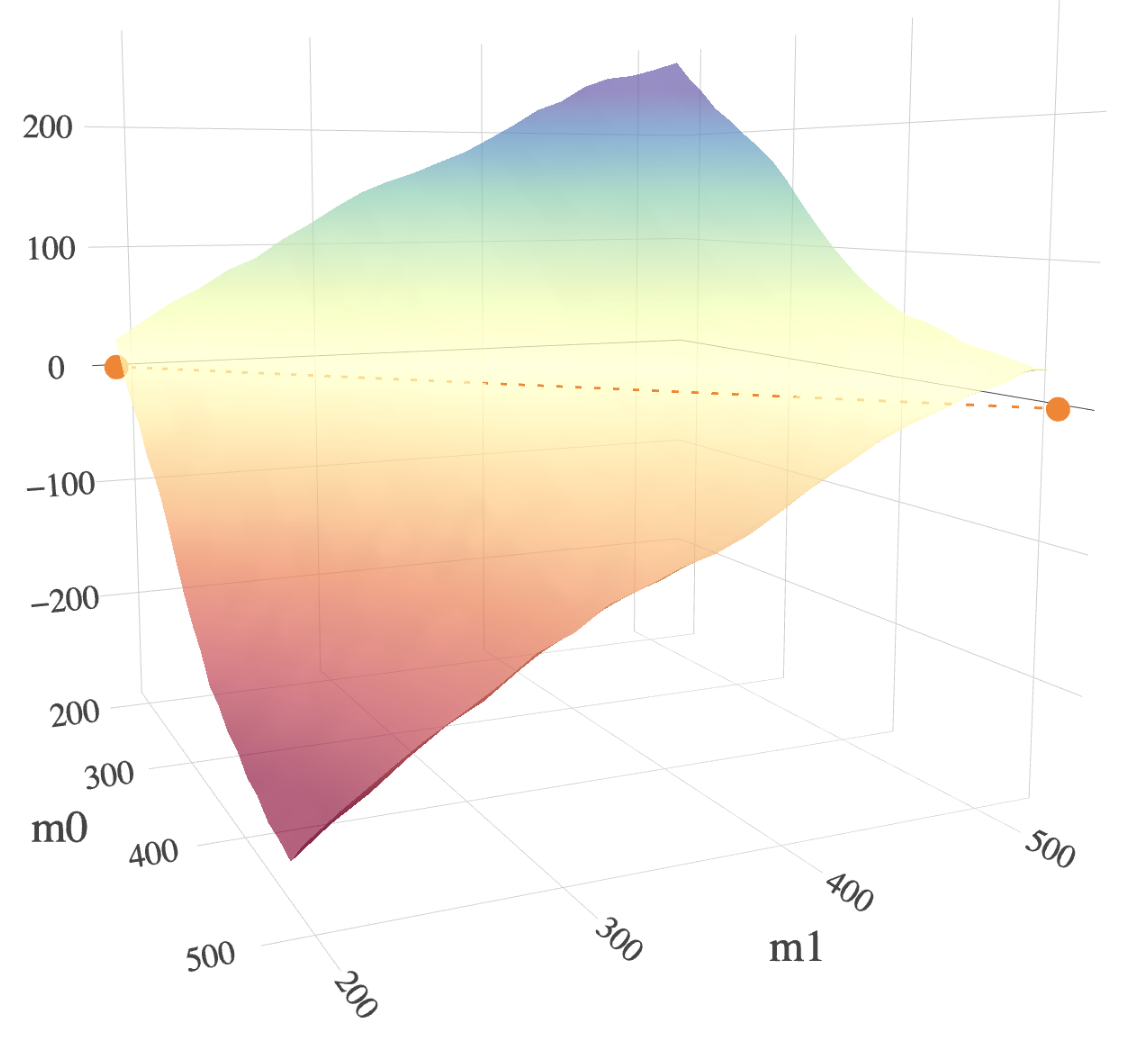}
\includegraphics[width=1\textwidth]{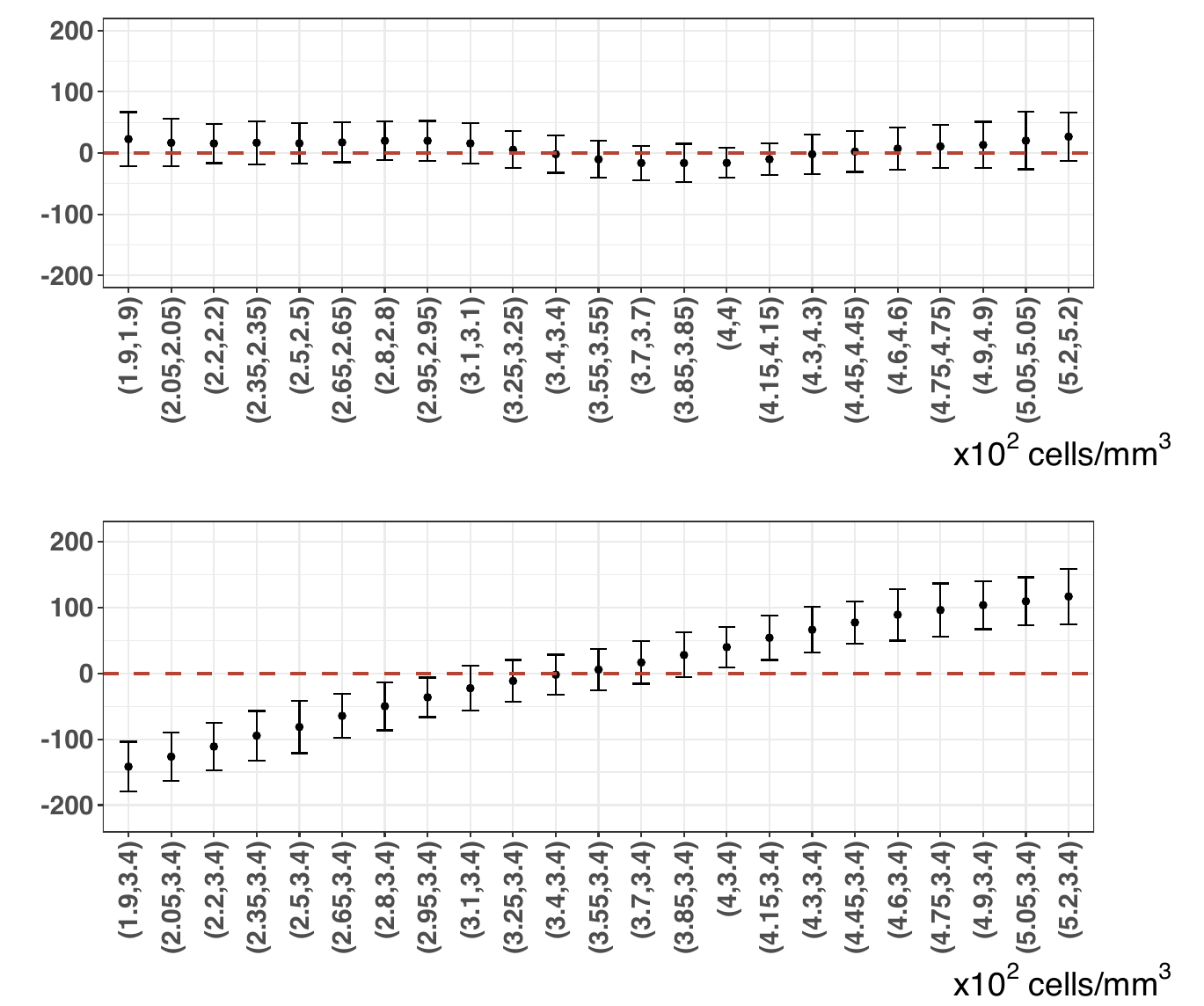}
\caption{$\rho=0,h = 0.25$}
\end{subfigure}
\begin{subfigure}[b]{0.32\textwidth}
\centering
\includegraphics[width=1\textwidth]{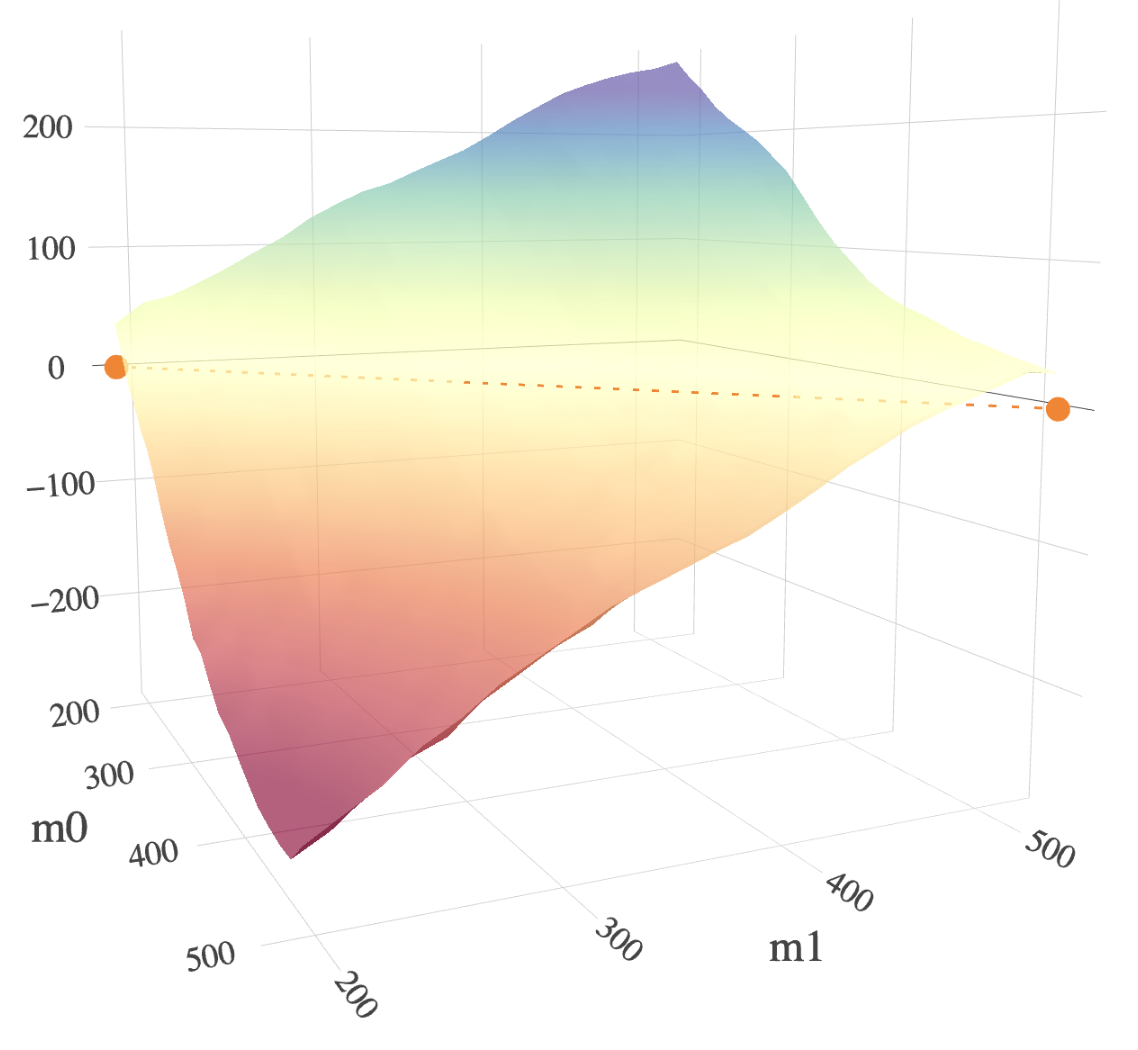}
\includegraphics[width=1\textwidth]{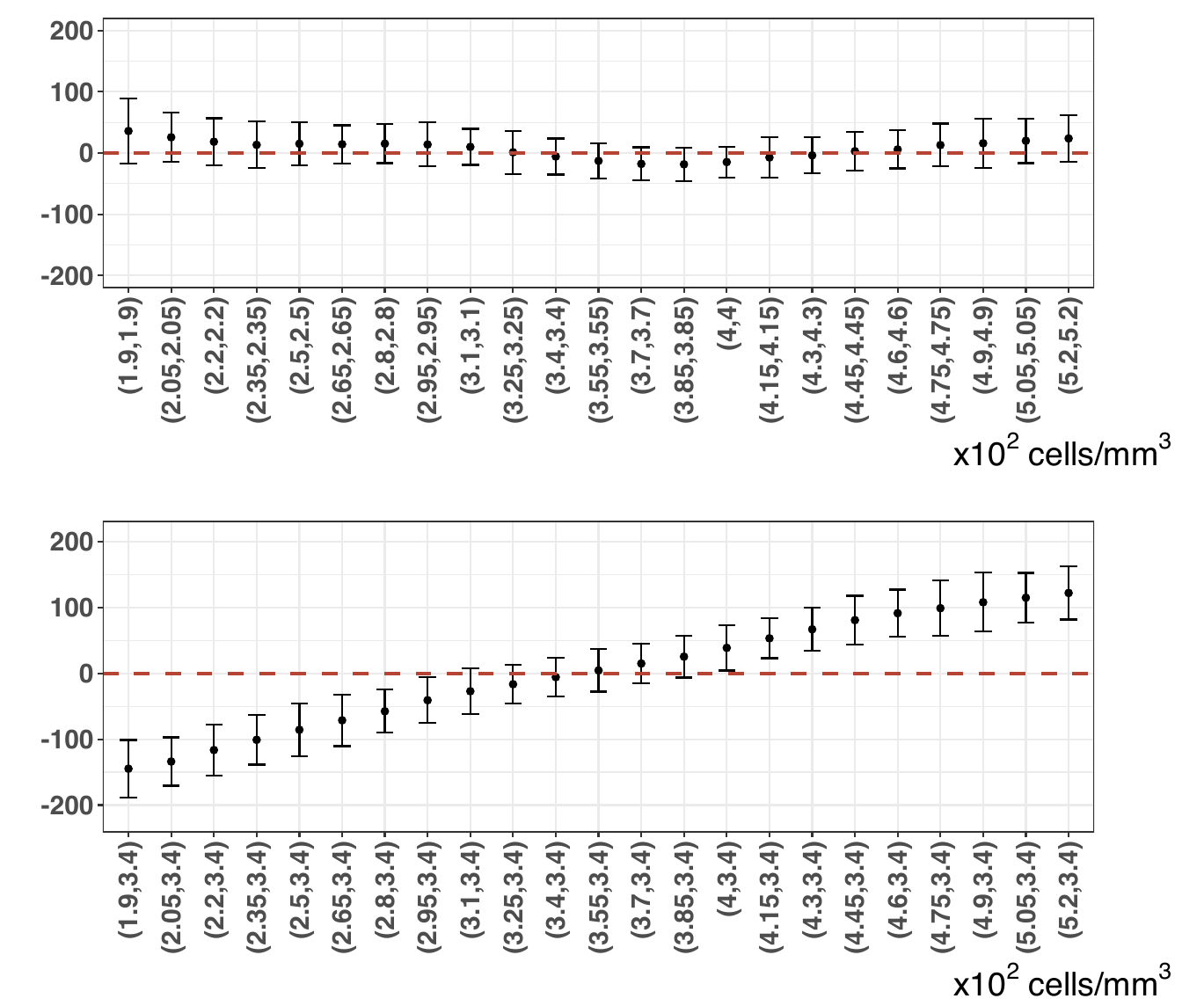}
\caption{$\rho=-0.5,h = 0.25$}
\end{subfigure}
\caption{\color{black}Each subfigure in the first row shows that the  surface of $\hat{\tau}_{u}$, and each   subfigure in the second row shows the $95\%$ bootstrap CIs of $\hat{\tau}_{u}$,  with  $h = 0.25$ (for the standardized $M$),  $\rho \in \{0.5,0,-0.5\}$, and $u\in\big[190\text{ cells/mm}^3,520\text{ cells/mm}^3\big]^2$. In each second-row subfigure, the X-axis represents different levels of $u$. Specifically, the top panel of each second-row subfigure, shows the $95\%$ bootstrap CIs of $\hat{\tau}_{u}$ with varying $u$ when $m_1 = m_0$, while the bottom panel  shows the $95\%$ bootstrap CIs of $\hat{\tau}_u$ with varying $m_1$, and $m_0$ is fixed at the level of $340\text{ cells/mm}^3$.}
\label{fig:cd4main}
\end{figure}
\par
The semiparametric parametric strategy for nuisance training, as outlined in Section \ref{sec:cd}, is employed in this study. Both $\mathcal{M}_{\mathrm{tp}}$ and $\mathcal{M}_{\mathrm{om}}$ are trained via $\mathtt{SuperLearner}$ combining various machine learning algorithms.  The Q-Q plot of $\mathcal{M}_{\mathrm{ps}}$ estimation suggests that a linear regression model for $\mathcal{M}_{\mathrm{ps}}$  is appropriate, as demonstrated in Figure \ref{fig:actg175}(a). Our approach utilizes a Gaussian kernel  $\mathcal{K}(\cdot,\cdot)$, and a Gaussian copula $c_\rho(\cdot,\star)$. To evaluate the estimator's sensitivity under different scenarios of $(M_1, M_0)$'s correlation, we investigate three values of $\rho = -0.5,0,0.5$. All continuous variables are standardized during the implementation of the proposed method, and are rescaled back when presenting the results. The standardized $M_i$ values fall approximately within the range of $-2$ and $3$, as depicted in Figure \ref{fig:actg175}(b). Thus, we have chosen two relatively small bandwidths $h = 0.25,0.5$ to ensure that $\tau_{u}$ can approximate $\tau^*_{u}$ effectively. In practice, we recommend specifying $h$ to be a small fraction of the standard deviation of the observed $M_i$ values. 
\par
For $h \in\{ 0.25,0.5\}$ and $\rho \in\{ 0.5,0,-0.5\}$, the estimated Cont.PCE surface and $95\%$ confidence intervals are reported in Figures~\ref{fig:cd4main} and \ref{fig:cd4main:3}, with both $m_1$ and $m_0$ ranging from $190 \text{ cells/mm}^3$ to $520\text{ cells/mm}^3$, which roughly correspond to the $10$th and $85$th percentiles among all $M_i$.  The estimated Cont.PCE surface exhibits a pattern similar to the synthetic example in Figure~\ref{fig:s12}(a) of an ideal surrogate, where the estimated dissociative effects are close to zero, and the estimated associative effects are positive (negative) when $m_1 > m_0$ ($m_1 < m_0$). {\color{black}This   suggests the causal sufficiency and necessity of short-term CD4 count.} In particular, we show the $95\%$ Bootstrap CIs with $100$ resamples of several different $\hat{\tau}_u$ with $m_1 = m_0$ {\color{black} in top panels of the second-row subfigures in Figures \ref{fig:cd4main} and \ref{fig:cd4main:3}}. {\color{black}Most of them are not} significantly different from zero. On the other hand, the $95\%$ CIs of $\hat{\tau}_u$ with different $m_1$ while fixing $m_0$ at $340\text{ cells/mm}^3$ {\color{black} in bottom panels of the second-row subfigures in Figures \ref{fig:cd4main} and \ref{fig:cd4main:3}}, demonstrate that ${\tau}^*_u$ becomes significantly larger as $m_1 - m_0$ increases. 

\begin{figure}[t]
\centering
\begin{subfigure}[b]{0.32\textwidth}
\centering
\includegraphics[width=1\textwidth]{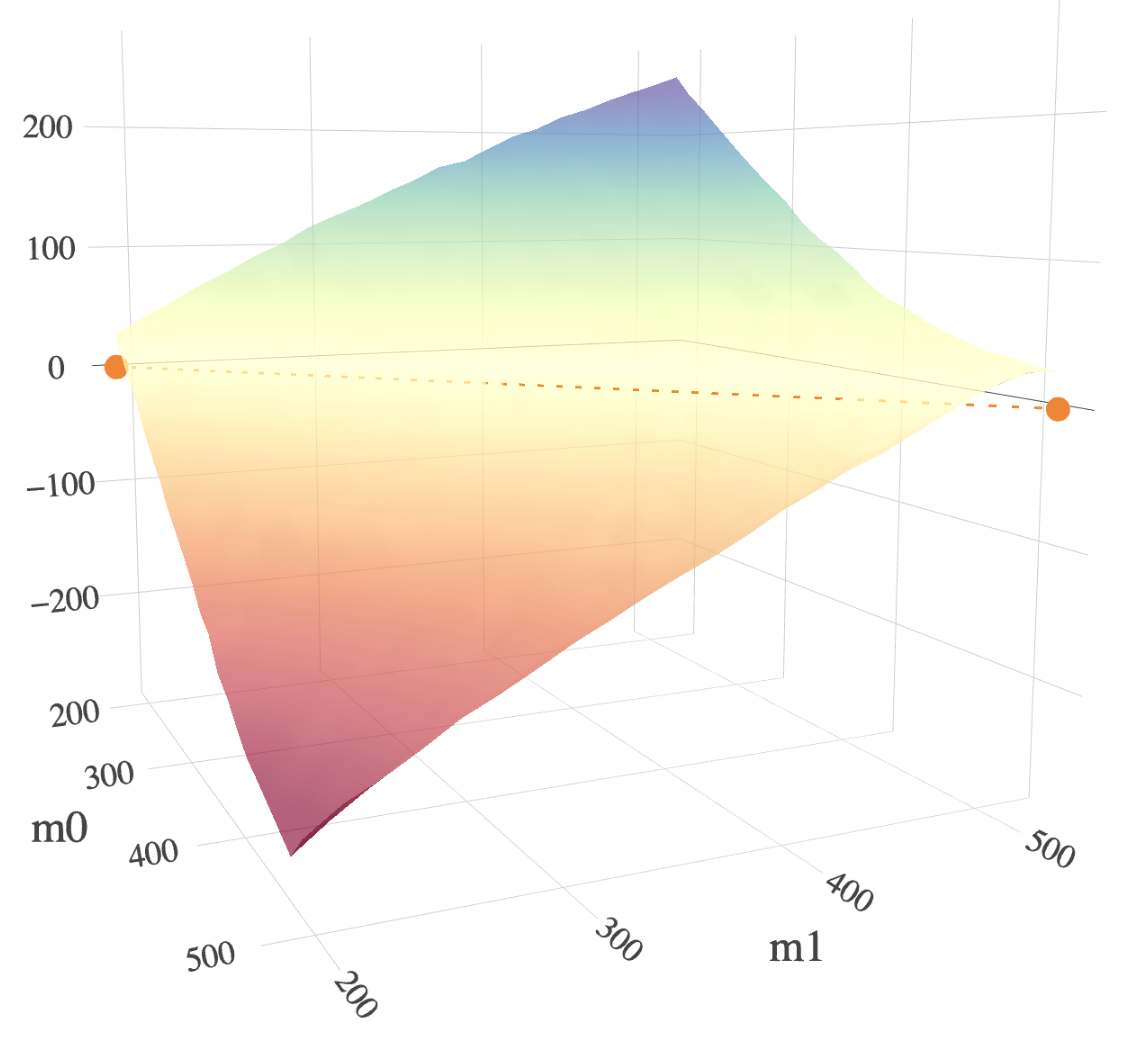}
\includegraphics[width=1\textwidth]{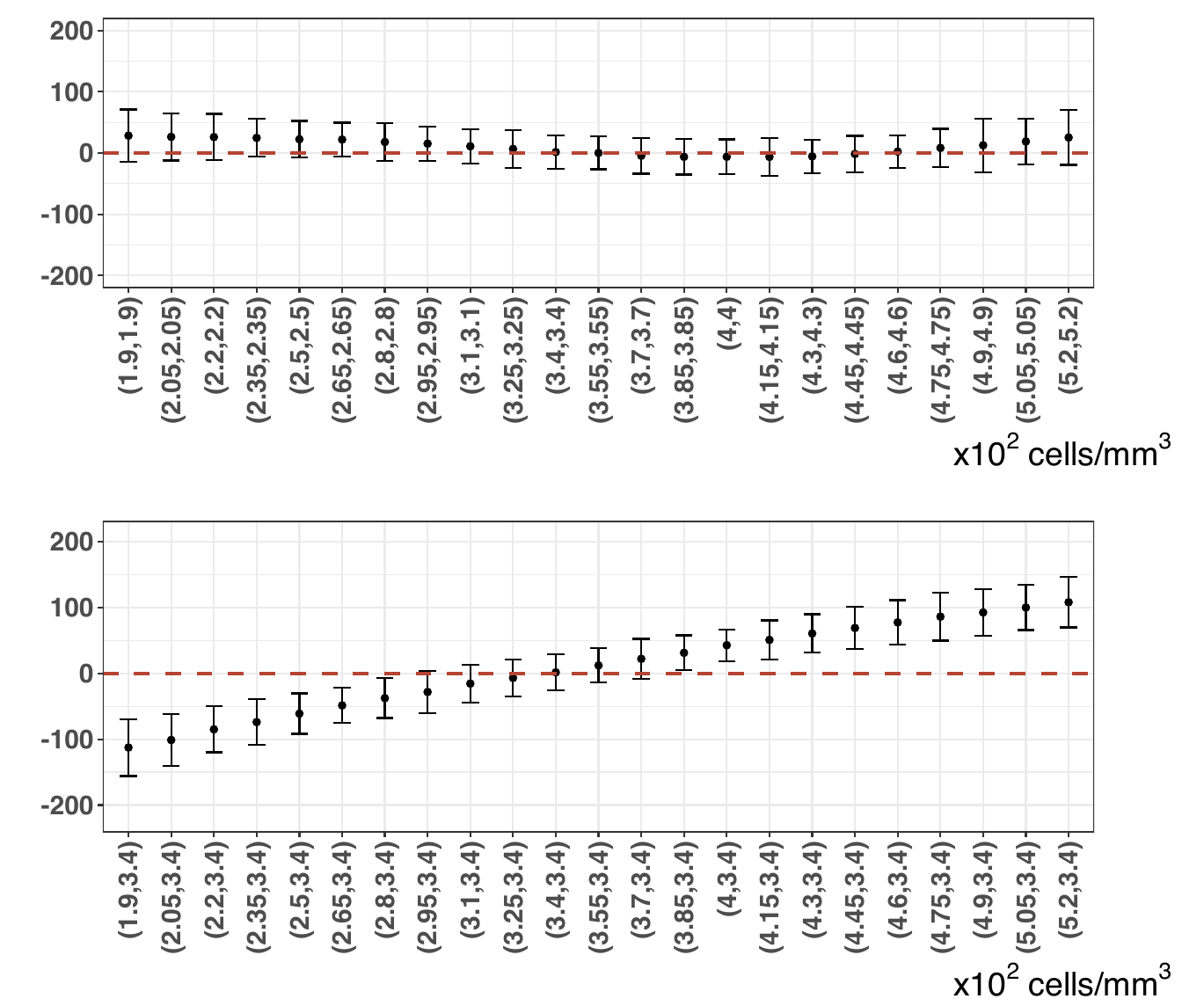}
\caption{$\rho=0.5,h = 0.5$}
\end{subfigure}
\begin{subfigure}[b]{0.32\textwidth}
\centering
\includegraphics[width=1\textwidth]{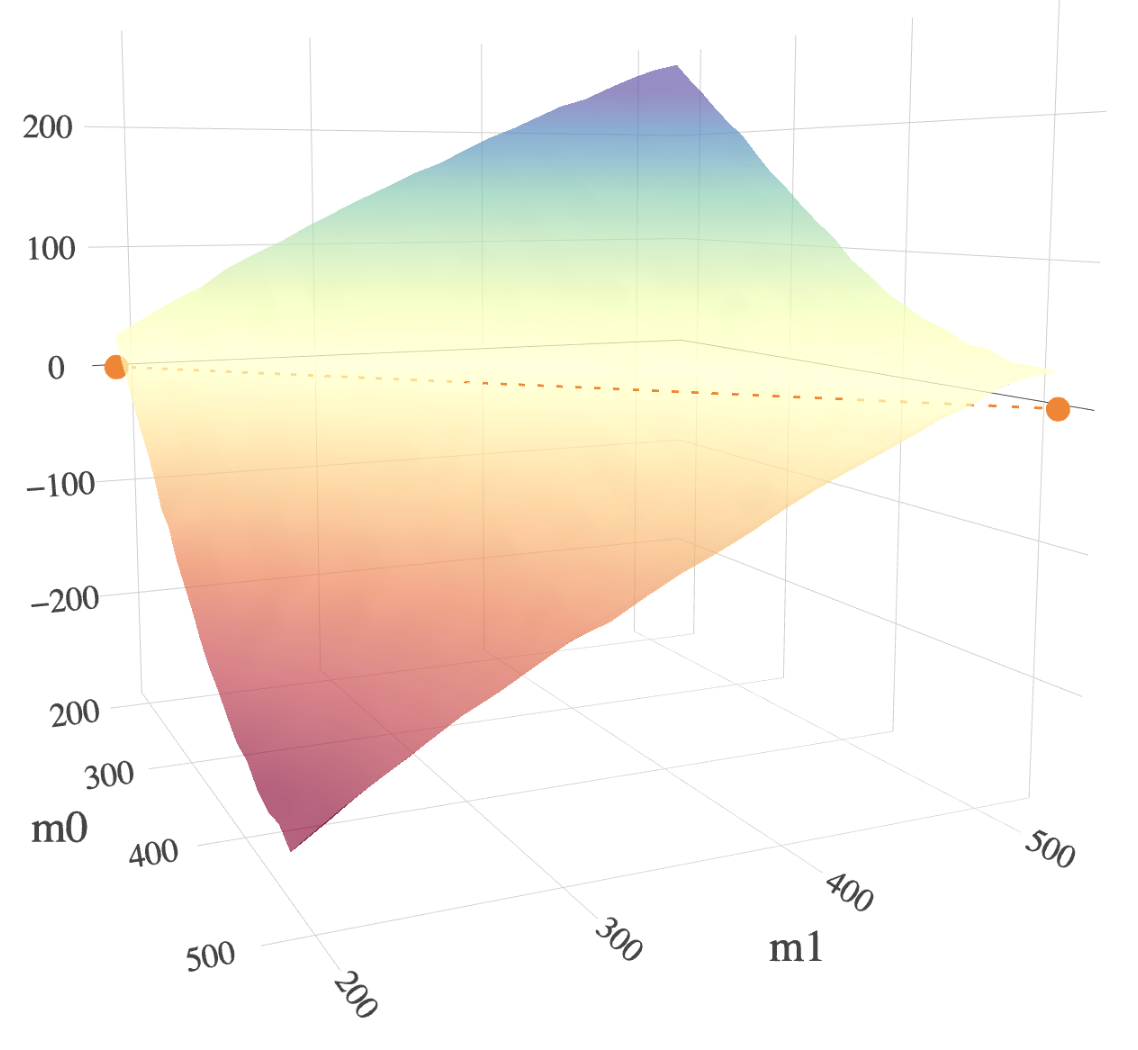}
\includegraphics[width=1\textwidth]{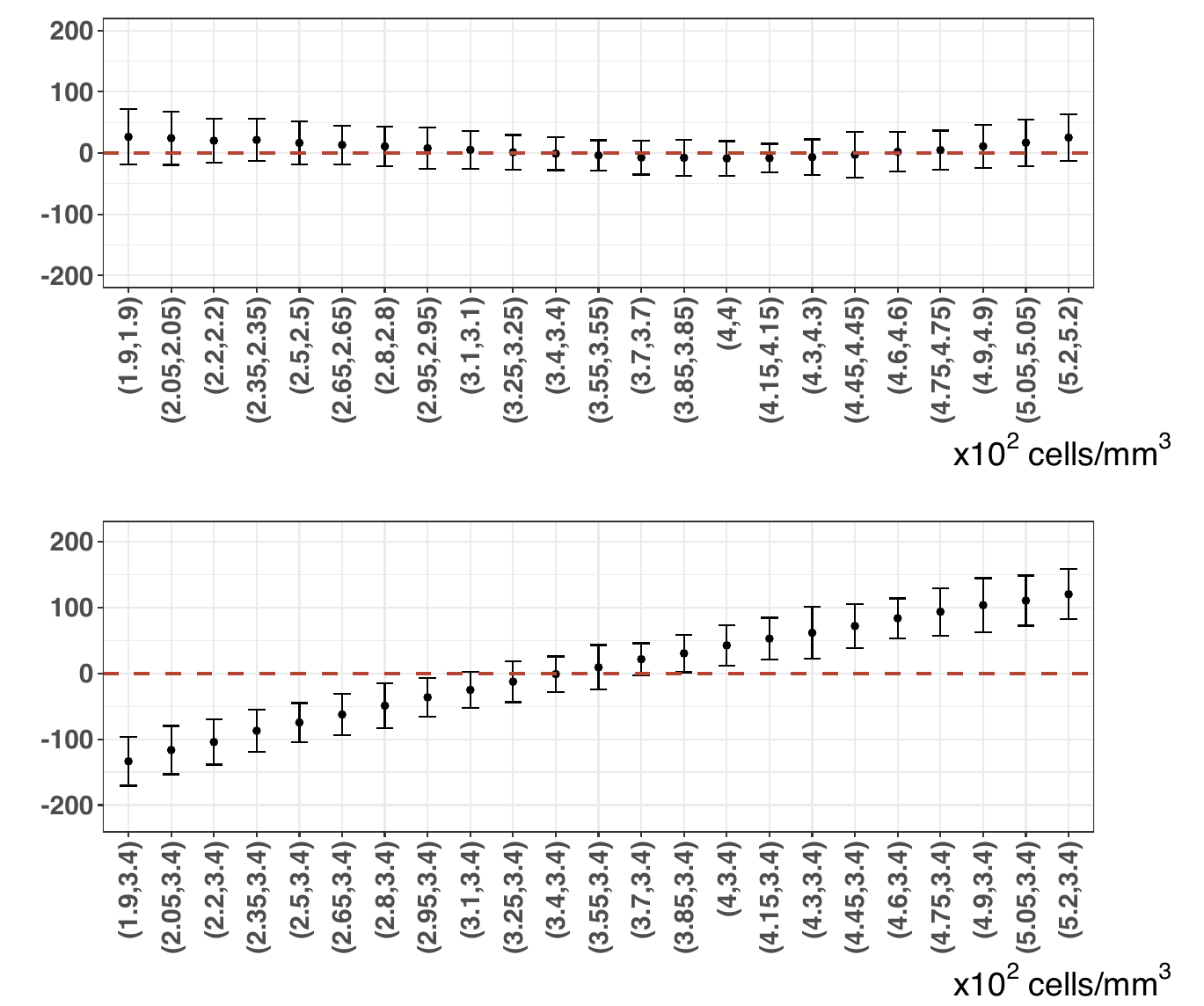}
\caption{$\rho=0,h = 0.5$}
\end{subfigure}
\begin{subfigure}[b]{0.32\textwidth}
\centering
\includegraphics[width=1\textwidth]{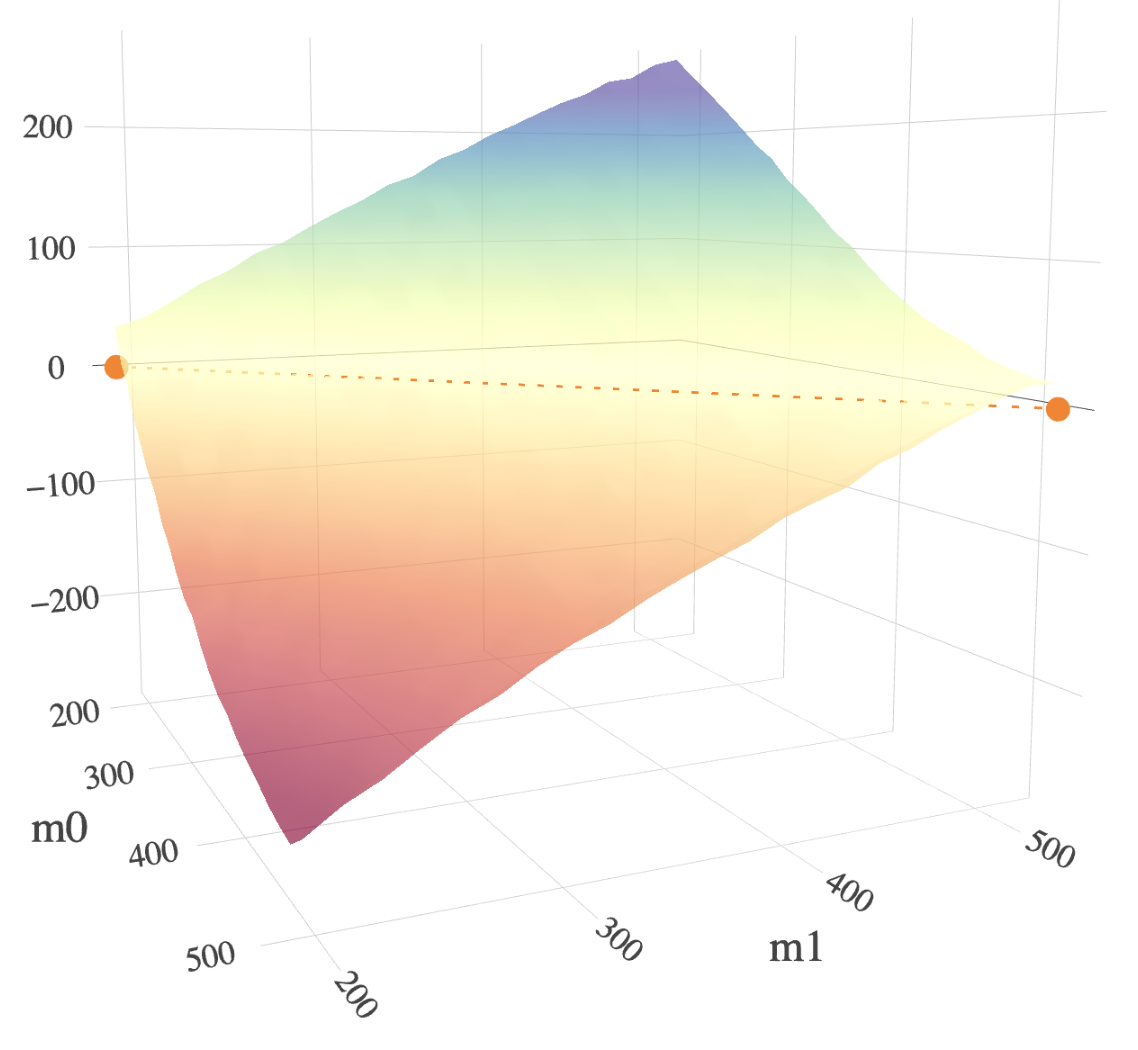}
\includegraphics[width=1\textwidth]{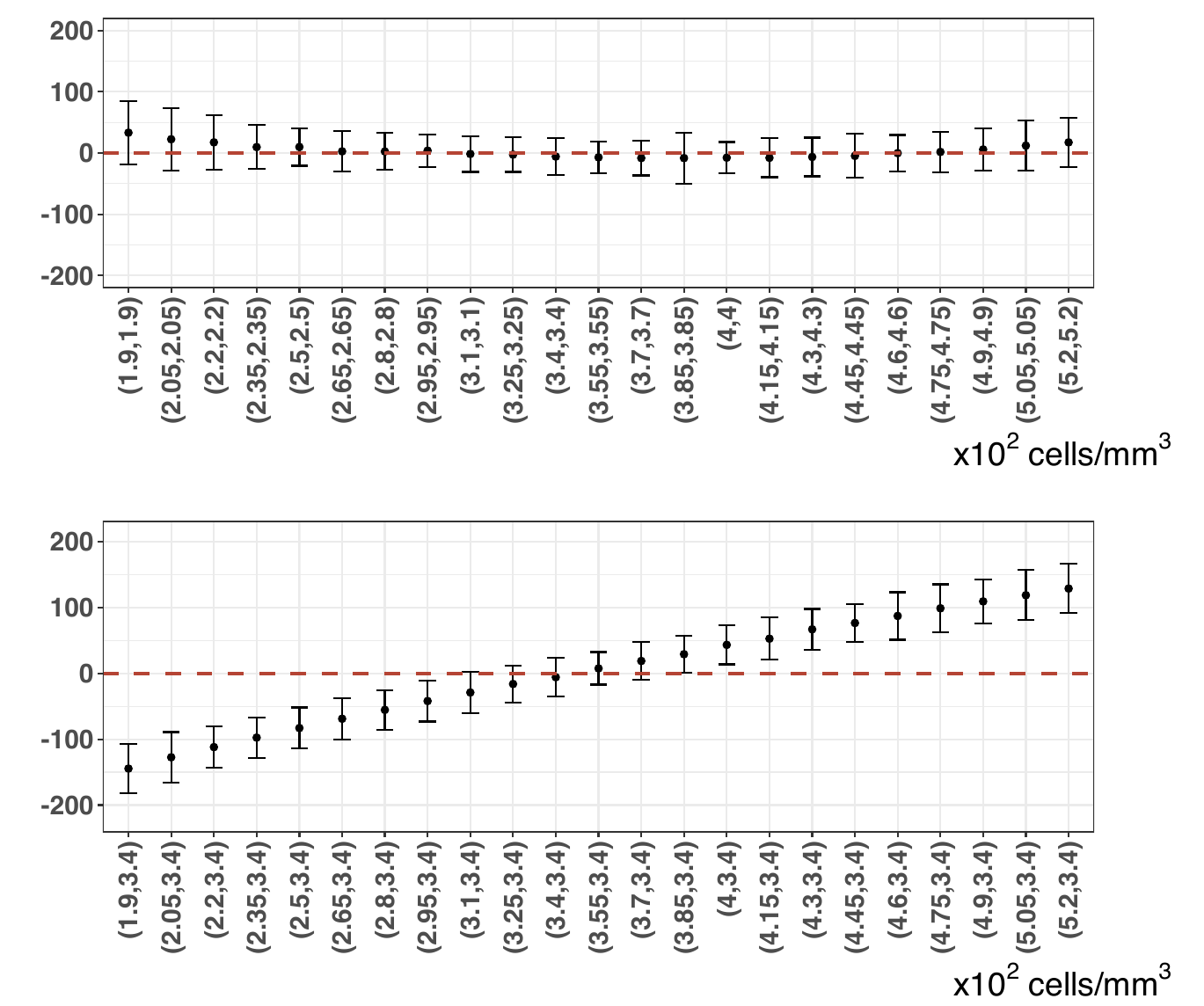}
\caption{$\rho=-0.5,h = 0.5$}
\end{subfigure}
\caption{\color{black}The  surface of $\hat{\tau}_{u}$ and the  bootstrap CIs of $\hat{\tau}_{u}$  with  $h = 0.5$,  $\rho \in \{0.5,0,-0.5\}$, and $u\in\big[190\text{ cells/mm}^3,520\text{ cells/mm}^3\big]^2$. The   explanations of figures are similar to Figure~\ref{fig:cd4main}.}
\label{fig:cd4main:3}
\end{figure}
 
\begin{figure}[t]
\centering
\includegraphics[height=7cm]{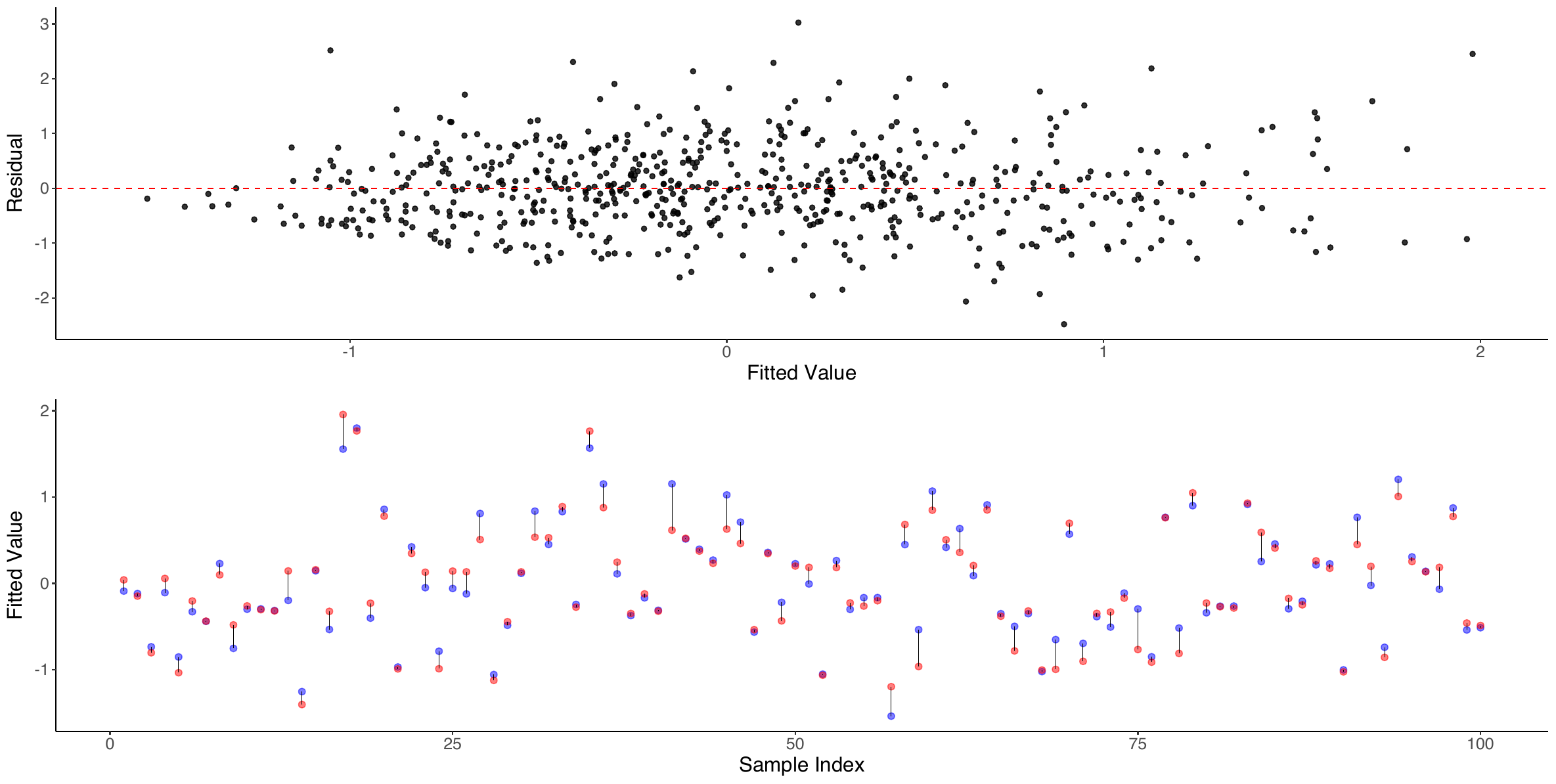}
\caption{\color{black} The top panel is the residuals-vs.-fitted-values plot for the outcome model trained via linear regression. Black dots represent individual samples. X-axis represents the linear fitted value and the Y-axis represents the residual for each sample.
The bottom panel compares the fitted values of the first 100 samples ($i = 1, \dots, 100$) obtained from outcome models trained using linear regression or $\mathtt{SuperLearner}$. The X-axis represents the sample index $i$,  blue dots indicate the fitted values of $\hat{\mu}^{\mathrm{(l)}}_{Z_i}(X_i, M_i)$, and red dots indicate the fitted values of  $\hat{\mu}^{\mathrm{(s)}}_{Z_i}(X_i, M_i)$. For both panels, all continuous variables are standardized.}\label{fig:regression:cehck}
\end{figure} 
\begin{figure}
\centering
\begin{subfigure}[b]{0.24\textwidth}
\centering
\includegraphics[width=1\textwidth]{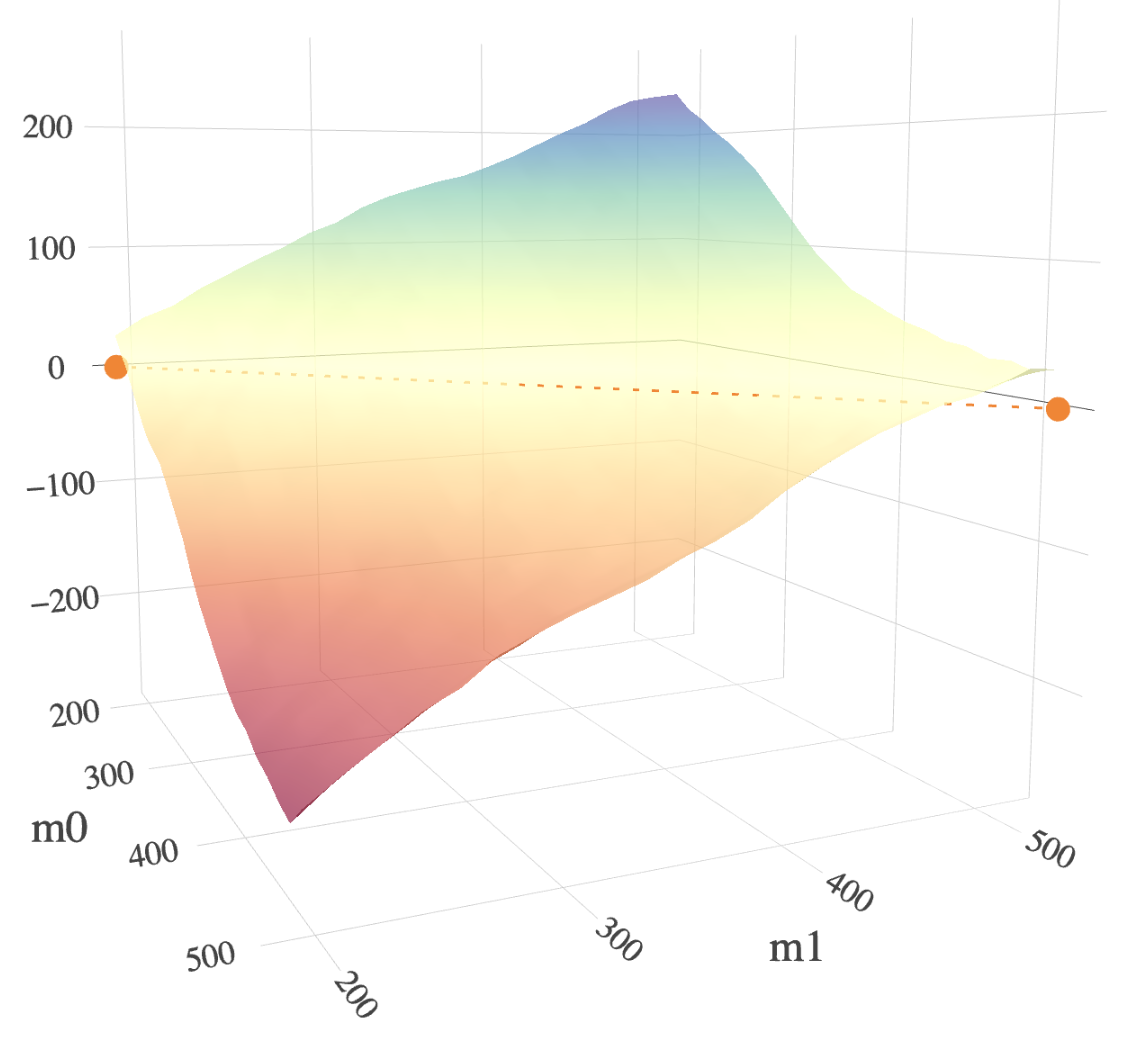}
\includegraphics[width=1\textwidth]{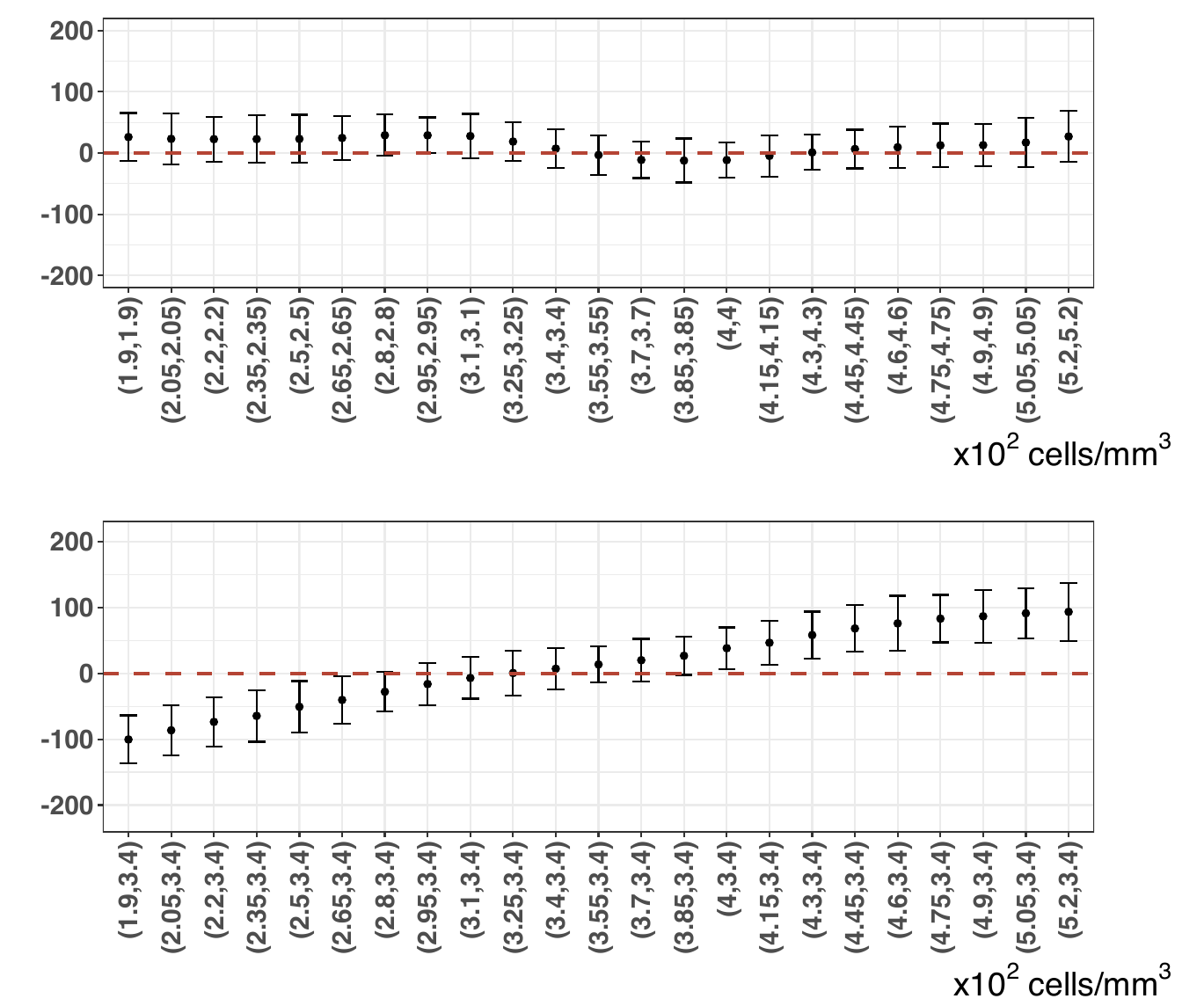}
\caption{$\rho=0.8,h = 0.25$}
\end{subfigure}
\begin{subfigure}[b]{0.24\textwidth}
\centering
\includegraphics[width=1\textwidth]{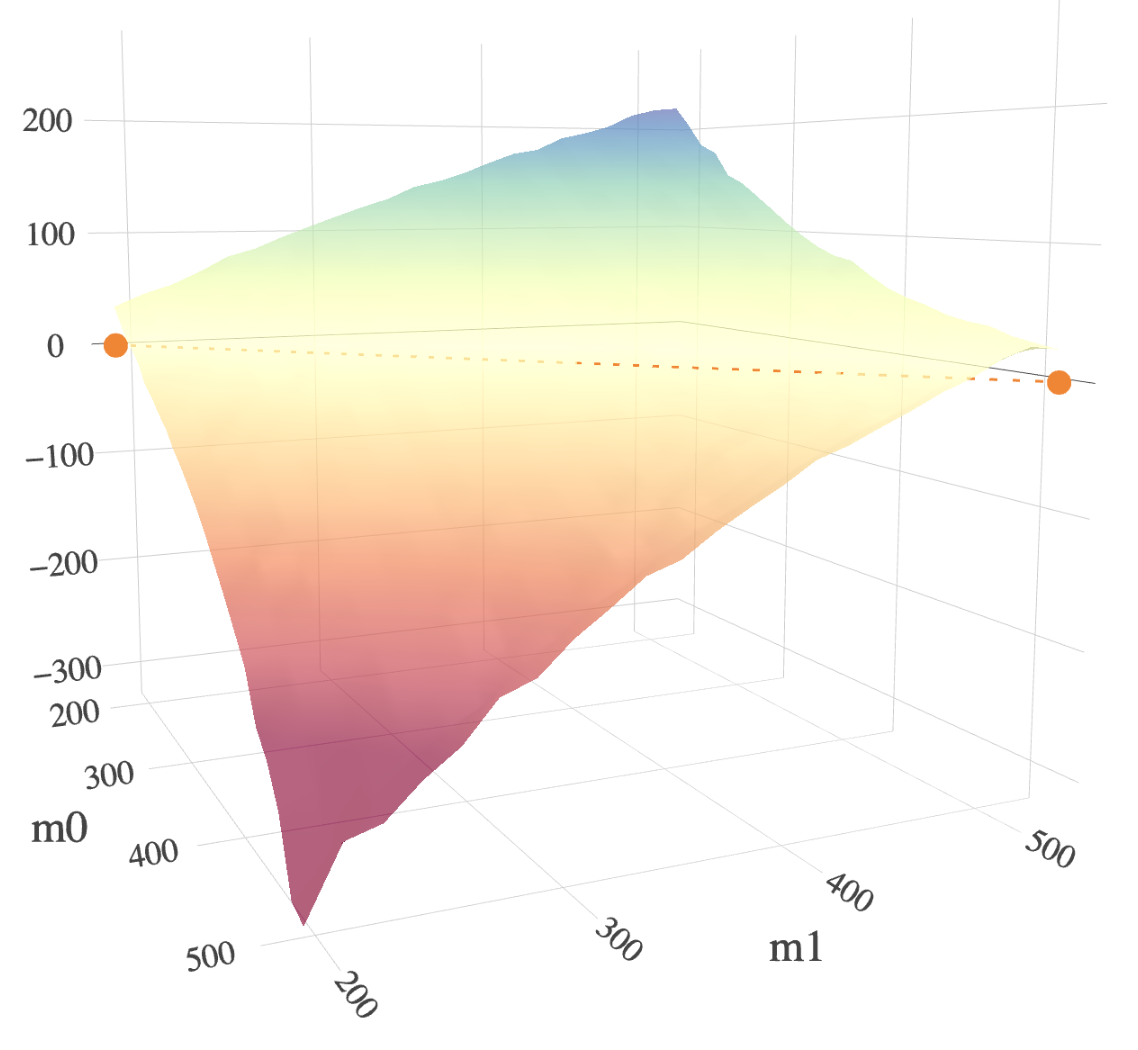}
\includegraphics[width=1\textwidth]{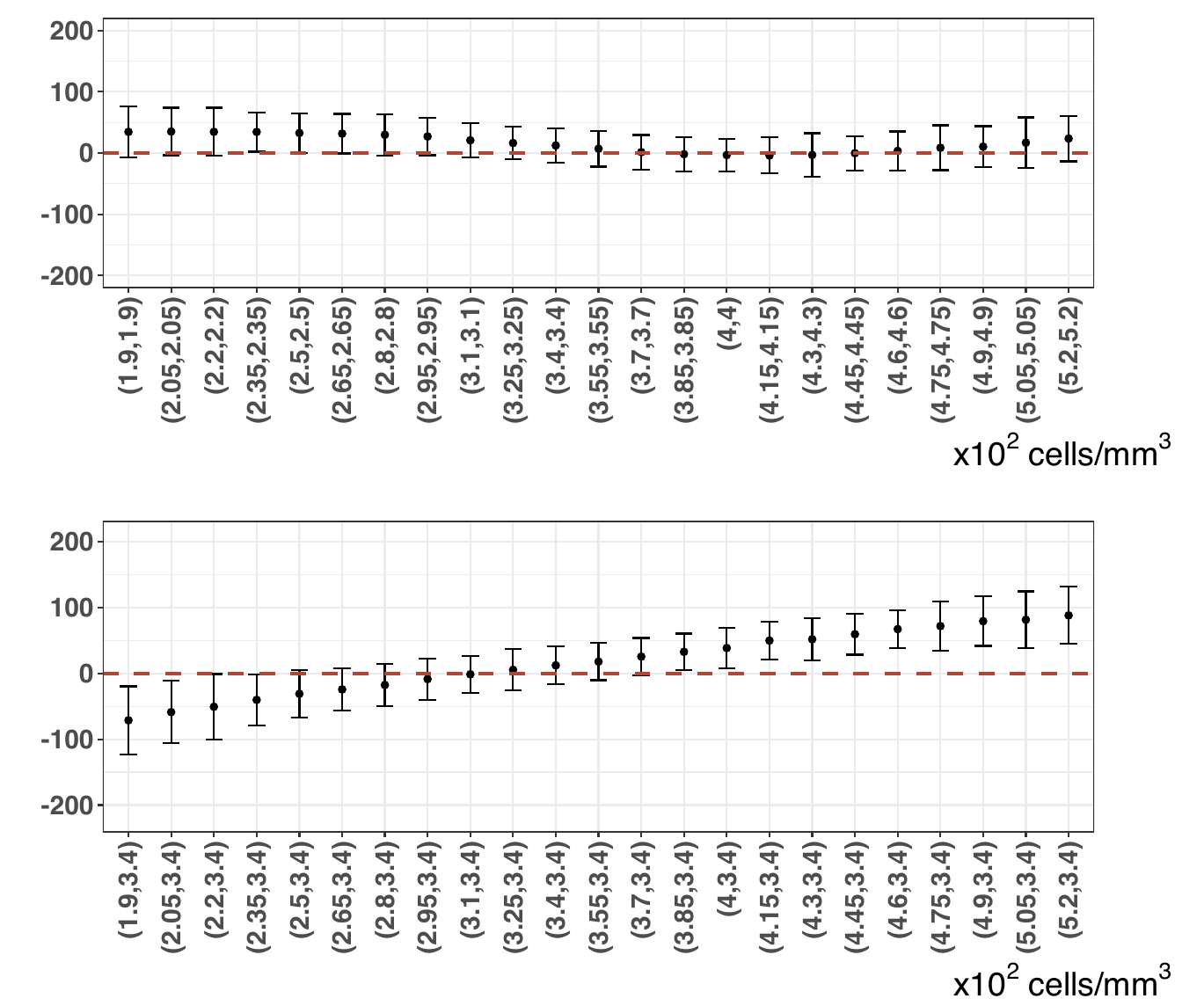}
\caption{$\rho=0.8,h = 0.5$}
\end{subfigure}
\begin{subfigure}[b]{0.24\textwidth}
\centering
\includegraphics[width=1\textwidth]{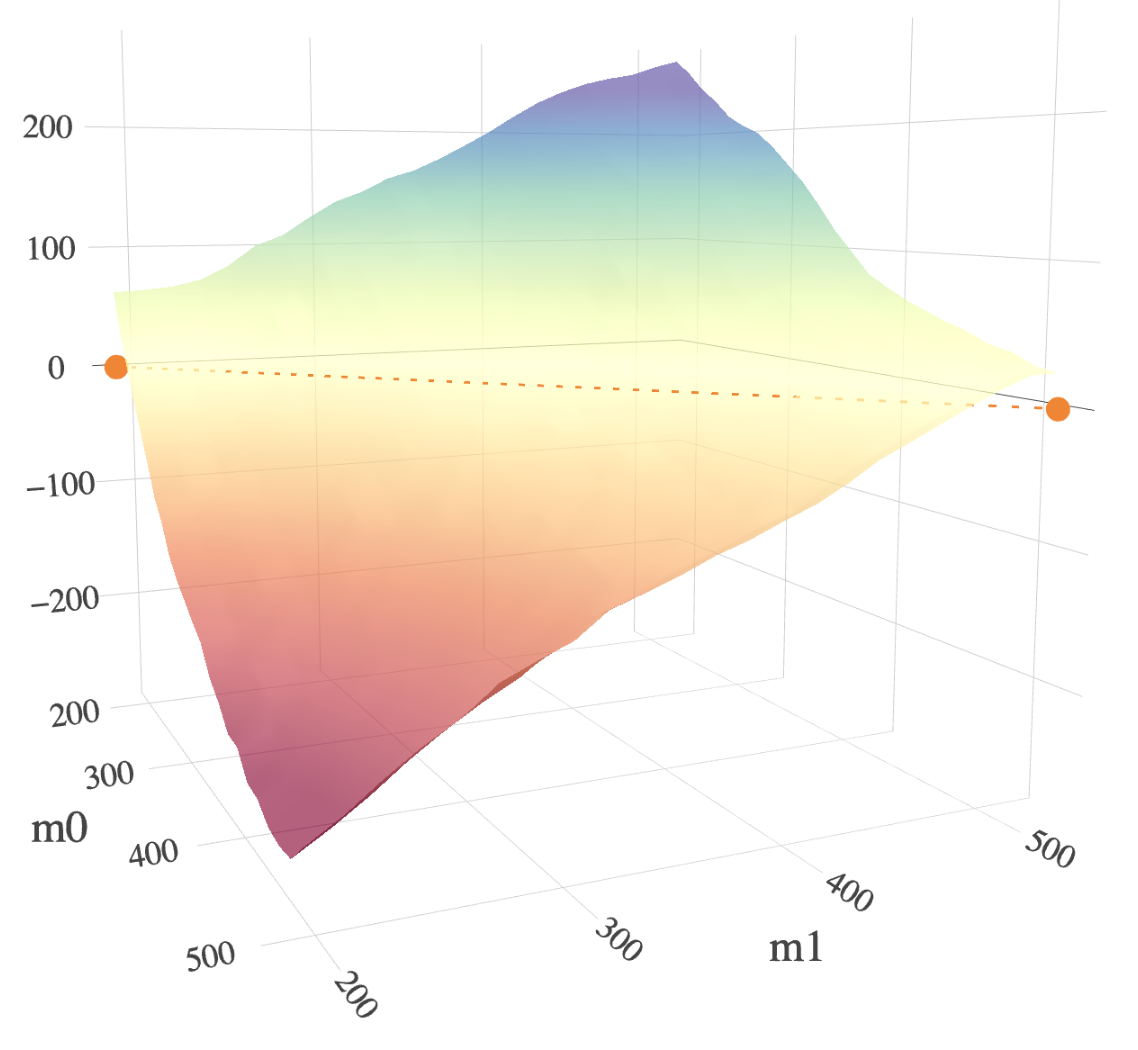}
\includegraphics[width=1\textwidth]{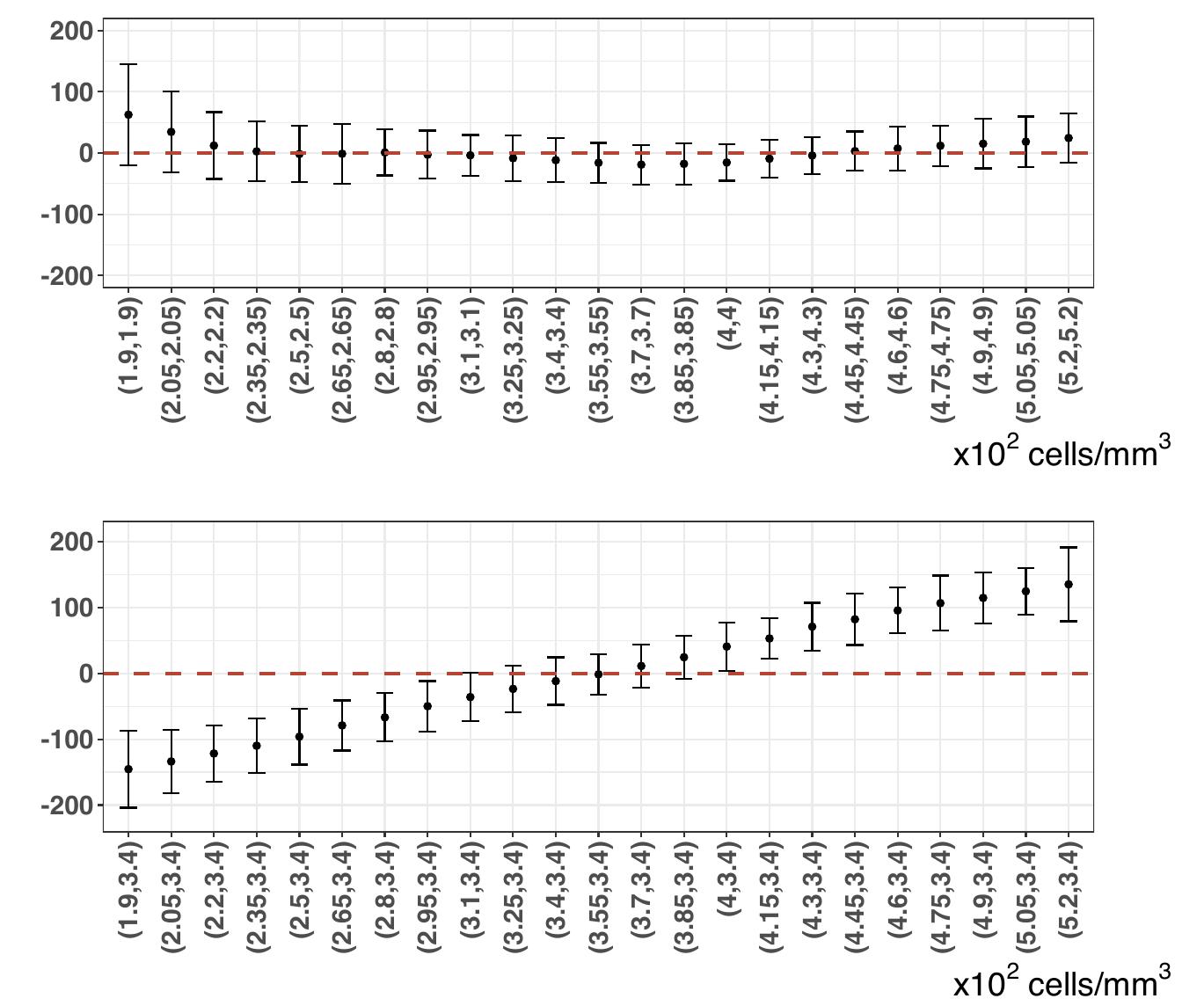}
\caption{$\rho=-0.8,h = 0.25$}
\end{subfigure}
\begin{subfigure}[b]{0.24\textwidth}
\centering
\includegraphics[width=1\textwidth]{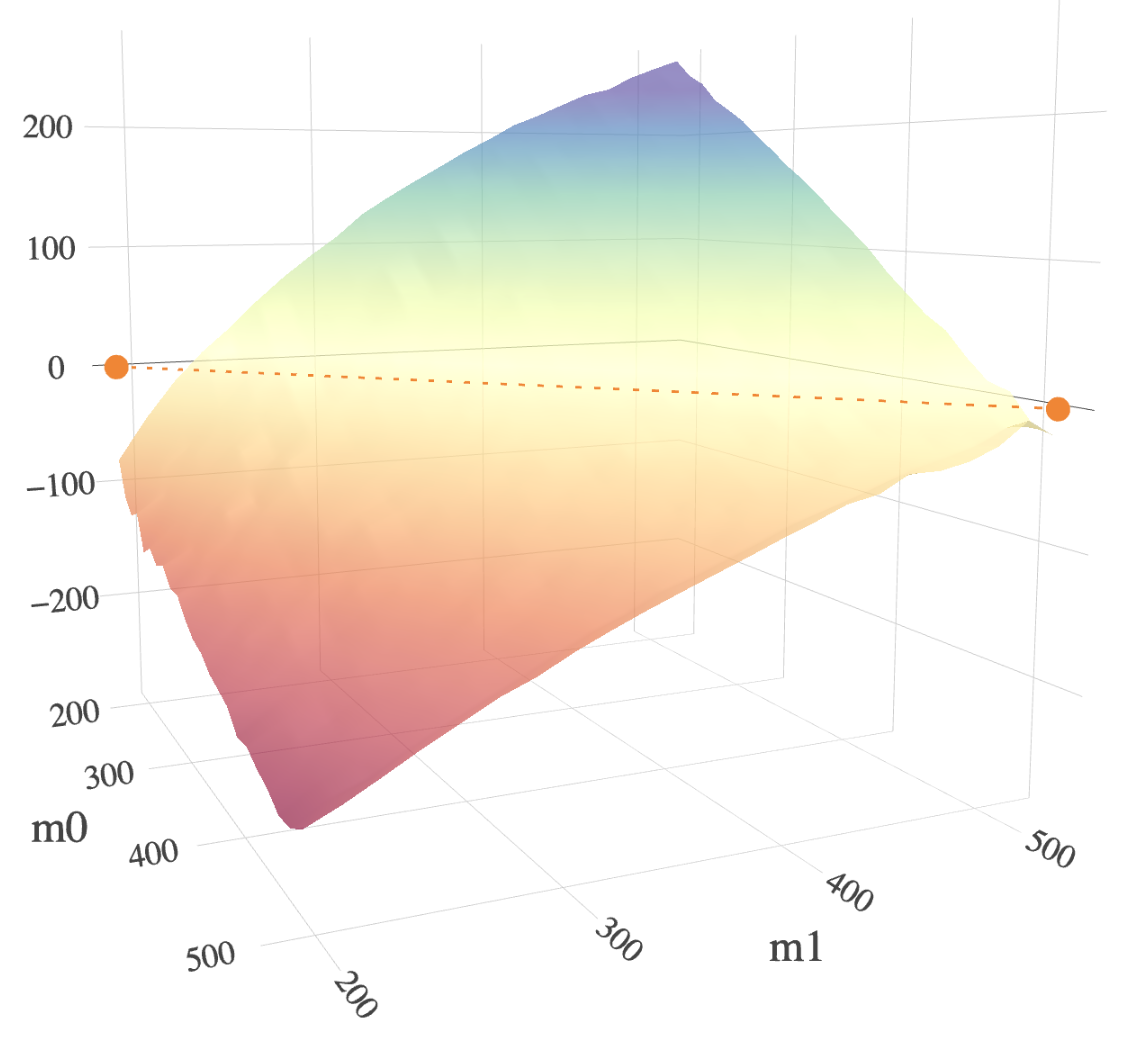}
\includegraphics[width=1\textwidth]{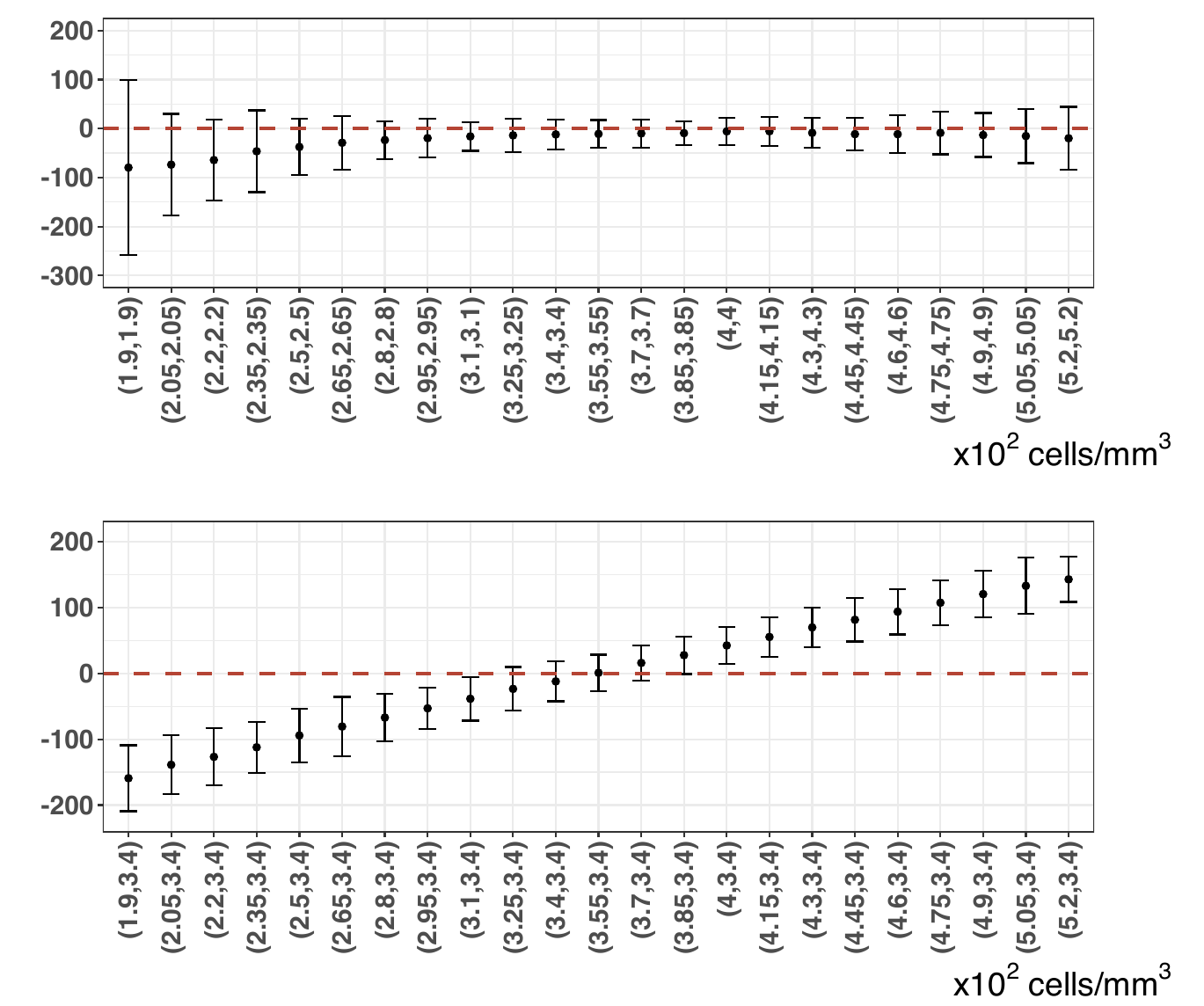}
\caption{$\rho=-0.8,h = 0.5$}
\end{subfigure}
\caption{\color{black}The  surface of $\hat{\tau}_{u}$ and the  bootstrap CIs of $\hat{\tau}_{u}$  with  $h \in \{0.25,0.5\}$,  $\rho \in \{-0.8,0.8\}$, and $u\in\big[190\text{ cells/mm}^3,520\text{ cells/mm}^3\big]^2$. The   explanations of figures are similar to Figure~\ref{fig:cd4main}.}
\label{fig:cd4main:4}
\end{figure}

{\color{black}Figures~{\color{black}4} and {\color{black}5}
demonstrate that our estimation results are robust to varying $\rho$.
As discussed in Remark~{\color{black}4} and Section~{\color{black}S5}
of the Supplementary File, if the true outcome model is linear in
$(X,Z,M)$ and correctly specified in estimation, our estimator achieves
consistency regardless of $\rho$, making the results insensitive
to its choice.

To assess the linearity of the outcome model, we fitted it using linear
regression to obtain $\hat{\mu}_{z}^{(\mathrm{l})}(x,m)$. The top
panel of Figure~\ref{fig:regression:cehck} shows a residuals-vs.-fitted-values
plot, where residuals are evenly distributed around the $y=0$ line,
suggesting near linearity. We also fitted the outcome model nonparametrically
using $\mathtt{SuperLearner}$ (excluding linear regression) to obtain
$\hat{\mu}_{z}^{(\mathrm{s})}(x,m)$. The bottom panel compares the
fitted values for $100$ samples under $\hat{\mu}_{z}^{(\mathrm{l})}(x,m)$
and $\hat{\mu}_{z}^{(\mathrm{s})}(x,m)$, showing close alignment,
further supporting near linearity.

This analysis explains the robustness of our results to $\rho$. However,
as shown in Figure~\ref{fig:cd4main:4}, exploring extreme $\rho\in\{-0.8,0.8\}$
reveals some sensitivity, particularly at $\rho=0.8$, indicating
the true model may not be perfectly linear. Despite this, the   confidence intervals 
still highlight the causal sufficiency and necessity of short-term
CD4 counts, confirming that our conclusions remain robust even under
sensitivity analysis.}

 To the best of authors' knowledge, this is the first time that the short-term CD4 count is evaluated as a surrogate under the principal stratification framework. Previous works assess PTE (cf. Remark~\ref{rk:other}) and suggest not a large proportion of the treatment effect on the  clinical endpoint can be explained by  the short-term CD4 count \citep{hughes1998cd4,agniel2021evaluation}. Nevertheless, our observations imply that, from the perspective of association strength and prediction power, the short-term CD4 count can  still be a desirable principal surrogate of long-term CD4 for the considered population \citep{gilbert2008evaluating}. {\color{black}Another reason why the short-term CD4 count serves as a reasonably good surrogate in our analysis is that, by trimming the dataset to exclude all patients censored within two years post-randomization due to reasons such as death or loss to follow-up, our surrogate analysis focuses on a healthier and more adherent patient population. This reduces instances that could weaken the association between short-term and long-term CD4 counts, such as non-compliance, viral resistance, and therapy adjustments in response to disease progression \citep{hughes1998cd4}. Consequently, within the considered population, the short-term CD4 count serves as a strong surrogate for the long-term CD4 count.}

\section{Discussion}\label{sec:dis}
{To efficiently estimate the principal causal effect  with continuous strata, we extend the  local functional estimation framework in   \citet{chernozhukov2018biased,chernozhukov2021simple}. Our new oracle EIF-invoked estimator serves as a valuable complement to the existing literature for resolving computation and estimation difficulties in this scenario.
%


For future work, we suggest three directions. 
Firstly, it is interesting to examine the relationship between the EIF and the Riesz representation \citep{chernozhukov2018biased,chernozhukov2021simple} of a target estimand for constructing estimators with desirable statistical properties. 
While the Riesz representation is identical to the EIF for some regular estimands such as the average treatment effect, it can be challenging to determine for more complex problems. We hypothesize that the EIF can offer insights into the Riesz representation in general, and plan to establish a formal connection in our future research.
Secondly, some key identification assumptions, such as weak principal ignorability, are not verifiable based on the observed data, necessitating sensitivity analysis to assess the robustness of the proposed estimator against their violations. Similar to \citet{jiang2020multiply},  sensitivity models can be introduced to handle possible breaches of these critical assumptions.   {\color{black}Finally, when $h$ is vanishing as $n$ grows, non-asymptotic analysis of Cont.PCE estimation has been further shown in Theorem \ref{hto0}. Thus, a data-driven algorithm for optimal and adaptive $h$ selection, is also of interest for future research.}
}

\bibliographystyle{abbrvnat}
\bibliography{refs}
\newpage
\def\spacingset#1{\renewcommand{\baselinestretch}%
{#1}\small\normalsize} \spacingset{1}
\spacingset{1.5} 
\begin{center}
\textbf{\Large{}{}Supplementary Material for ``Semiparametric Localized Principal Stratification Analysis with Continuous Strata''}
{\Large{} }{\Large\par}
\par\end{center}
\renewcommand\thesection{S\arabic{section}}
\renewcommand{\thefigure}{S\arabic{figure}}
\renewcommand{\thetable}{S\arabic{table}}
\renewcommand{\theequation}{S\arabic{equation}}
\renewcommand{\thelemma}{S\arabic{lemma}}
\renewcommand{\thetheorem}{S\arabic{theorem}}
\renewcommand{\thecondition}{S\arabic{condition}}
\renewcommand{\theremark}{S\arabic{remark}}
\renewcommand{\theproposition}{S\arabic{proposition}}
\renewcommand{\theassumption}{S\arabic{assumption}}
\pagenumbering{arabic} 
\renewcommand*{\thepage}{S\arabic{page}}

\setcounter{lemma}{0} 
\global\long\def\thelemma{\textup{S}\arabic{lemma}}%
\setcounter{equation}{0} 
\global\long\def\theequation{S\arabic{equation}}%
\setcounter{section}{0} 
\global\long\def\thesection{S\arabic{section}}%
\global\long\def\thesubsection{S\arabic{section}.\arabic{subsection}}%
The Supplementary File is organized as follows. Section \ref{sec:fulleif} presents short-hand notations used throughout the Supplementary File. Section \ref{compare:sec} contains technical details of Remark \ref{rk:compare}. An illustrative example demonstrating the verification of CATE as a perfectly localized estimand is presented in Section \ref{sec:cate:exp}. {\color{black} Section~\ref{sec:compute} presents the computational complexity of computing all involved   numerical integrations for our proposed estimator. 
 Section~\ref{sec:extrarobust} showcases     the enhanced robustness of our proposed estimator regarding nuisance function specification and sensitivity parameter selection when $\mathcal{M}_{\mathrm{om}}$ follows a linear model. Section~\ref{sec:compare} contains a detailed comparison between this paper and \citet{lu2023principal}.} Sections \ref{pf:po:approximate}--\ref{pf:lm:emp} include all lemmas and proofs. Section \ref{sec:add:simu} reports additional results and figures {\color{black}for our simulations}. 
\section{Summary of notations}\label{sec:fulleif} 
For simplicity of presentation, we summarize the short-hand notations used in the proofs. 
\begin{itemize}
\item $\mathbb{I}(\mathscr{E})$ is the indicator function such that $\mathbb{I}(\mathscr{E}) = 1$ when $\mathscr{E}$ happens and $0$ otherwise.
\item $\xi_{1,u^*}(x,z,m) =  \{\pi_z(x) f_{zm}(x)\}^{-1}{(-1)^{z+1}{\gamma_1}^{(z)}(x,m)}$;
\item $\gamma_{1,u^*}^{(1)}(x,m)= \int_{\R} k_{u^*}(m,m_0)e_{(m
,m_0)}(x)dm_{0}$;
\item $\gamma_{1,u^*}^{(0)}(x,m)= \int_{\R} k_{u^*}(m_1,m)e_{(m_1,m)}(x)dm_{1 }$;
\item $\tilde\xi_{a,u^*} (x,z,m) = \xi_{a,u^*}(x,z,m) - \E\big\{\xi_{a,u^*}(X,Z,M) \mid X = x,Z = z\big\}$ for $a = 2,3$;
\item $\xi_{2,u^*}(x,z,m) = \{\pi_{z}(x)\}^{-1}\int_{\R^2}1(m \leq m_{z})\{\mu_{1}(x,m_1) - \mu_{0}(x,m_0)\} k_{u^*}(u)c_{\rho}^{(z)}\big\{F_{1m_1}(x), F_{0m_0}(x)\big\}$ $\{\prod_{z\in\{0,1\}}f_{zm_z}(x)\}du$; 
\item $\xi_{3,u^*}(x,z,m) = \{\pi_{z}(x)\}^{-1}{\gamma}_{2,u^*}^{(z)}(x,m)$; 
\item $\gamma_{2,u^*}^{(1)}(x,m) = \int_{\R}\{\mu_{1}(x,m) - \mu_{0}(x,m_0)\} k_{u^*}(m,m_0)c_{\rho}\{F_{1m}(x), F_{0m_0}( x)\}f_{0m_{0}}(x)dm_{0}$;
\item $\gamma_{2,u^*}^{(0)}(x,m) = \int_{\R}\{\mu_{1}(x,m_1) - \mu_{0}(x,m)\} k_{u^*}(m_1,m)c_{\rho}\{F_{1m_1}(x), F_{0m}(x)\}f_{1m_{1}}(x)dm_{1}$;

\item $\tilde\xi'_{a,u^*} (x,z,m) = \xi'_{a,u^*}(x,z,m) - \E\big\{\xi'_{a,u^*}(X,Z,M) \mid X = x,Z = z\big\}$ for $a = 2,3$;
\item $\xi'_{2,u^*}(x,z,m) = \{\pi_{z}(x)\}^{-1}\int_{\R^2}1(m \leq m_{z})k_{u^*}(u)c_{\rho}^{(z)}\big\{F_{1m_1}(x), F_{0m_0}(x)\big\}\{\prod_{z\in\{0,1\}}f_z(m_z\mid x)\} du$; 
\item $\xi'_{3,u^*}(x,z,m) = \{\pi_{z}(x)\}^{-1}{\gamma'^{(z)}_{2,u^*}}(x,m)$; 
\item ${\gamma'^{(1)}_{2,u^*}}(x,m) = \int_{\R}k_{u^*}(m,m_0)c_{\rho}\{F_{1m}(x), F_{0m_0}(x)\}f_{0m_{0}}(x)dm_{0}$;
\item $\gamma'^{(0)}_{2,u^*}(x,m) = \int_{\R}k_{u^*}(m_1,m)c_{\rho}\{F_{1m_1}(x), F_{0m}(x)\}f_{1m_{1}}(x)dm_{1}$;
\item $\mathcal{A}^*_{1,u^*}$ and  $\mathcal{A}^*_{2,u^*}$ are defined as follows:
\beee\nn
{\mathcal{A}^*_{1,u^*}} & \scalemath{0.8}{=} & \scalemath{0.8}{ \sum_{z = 0,1}\mathbb{E}\Big[\int_{\R^2}k_{u^*}(u)c_{\rho}\big\{{F}_{1m_1}( X),{F}_{0m_0}( X)\big\} f_{(1 - z)m_{1- z}}( X)\{f'_{zm_z}( X)\}^\T du\Big]}
\\\nonumber
&&\scalemath{0.8}{+\sum_{z = 0,1}\mathbb{E}\Big[\int_{\R^2}k_{u^*}(u)\Big[c_{\rho}^{(z)}\big\{{F}_{1m_1}( X),{F}_{0m_0}(X)\big\}{f}_{1m_1}( X){f}_{0m_0}( X)\big\{F'_{zm_z}(X)\big\}^\T \Big]du\Big],}
\\\nonumber
{\mathcal{A}^*_{2,u^*}} & \scalemath{0.8}{=} &  \scalemath{0.8}{\sum_{z = 0,1}\mathbb{E}\Big[\int_{\R^2}k_{u^*}(u)\big\{\mu_{1}(X,m_1) - \mu_{0}(X,m_0)\big\}
c_{\rho}\big\{{F}_{1m_1}( X),{F}_{0m_0}(X)\big\} f_{(1 - z)m_{1- z}}( X)\{f'_{zm_z}( X)\}^\T du\Big]}
\\\nonumber
&&\scalemath{0.8}{+\sum_{z = 0,1}\mathbb{E}\Big[\int_{\R^2}k_{u^*}(u)\big\{\mu_{1}(X,m_1) - \mu_{0}(X,m_0)\big\} c_{\rho}^{(z)}\big\{{F}_{1m_1}( X),{F}_{0m_0}(X)\big\}{f}_{1m_1}( X){f}_{0m_0}( X)\big\{F'_{zm_z}(X)\big\}^\T du\Big];}
\eee
\item $\mathcal{A}_{1,u^*}$ and $\mathcal{A}_{2,u^*}$ are defined as follows:
\beee\nonumber
{\mathcal{A}_{1,u^*}} & \scalemath{0.8}{=} & \scalemath{0.8}{ \sum_{z = 0,1}\mathbb{E}\Big[\int_{\R^2}k_{u^*}(u)c_{\rho}\big\{{F}_{1m_1}( X),{F}_{0m_0}( X)\big\} f_{(1 - z)m_{1- z}}( X)\{f'_{zm_z}( X)\}^\T du\Big]}
\\\label{def:A1}
&&\scalemath{0.8}{+\sum_{z = 0,1}\mathbb{E}\Big[\int_{\R^2}k_{u^*}(u)\Big[c_{\rho}^{(z)}\big\{{F}_{1m_1}( X),{F}_{0m_0}(X)\big\}{f}_{1m_1}( X){f}_{0m_0}( X)\big\{F'_{zm_z}(X)\big\}^\T \Big]du\Big],}
\\\nonumber
{\mathcal{A}_{2,u^*}} & \scalemath{0.8}{=} &  \scalemath{0.8}{\sum_{z = 0,1}\mathbb{E}\Big[\int_{\R^2}k_{u^*}(u)\big\{\bar\mu_{1}(X,m_1) - \bar\mu_{0}(X,m_0)\big\}
c_{\rho}\big\{{F}_{1m_1}( X),{F}_{0m_0}(X)\big\} f_{(1 - z)m_{1- z}}( X)\{f'_{zm_z}( X)\}^\T du\Big]}
\\\nonumber
&&\scalemath{0.8}{+\sum_{z = 0,1}\mathbb{E}\Big[\int_{\R^2}k_{u^*}(u)\big\{\bar\mu_{1}( X,m_1) - \bar\mu_{0}(X,m_0)\big\} c_{\rho}^{(z)}\big\{{F}_{1m_1}( X),{F}_{0m_0}(X)\big\}{f}_{1m_1}( X){f}_{0m_0}( X)\big\{F'_{zm_z}(X)\big\}^\T du\Big] }
\\
\label{def:A2}
&&\scalemath{0.8}{ + \sum_{j = 1,2,3}\mathbb{E}\Bigg[\Big[\frac{(-1)^{Z+1}}{\bar{\pi}_Z(X)}\big\{Y - \bar{\mu}_{Z}(X,M)\big\}\Big]\delta_{X,Z,M}^{(j)}\Bigg];}
\eee
\item $\tilde{\phi}_{\mathrm{p}}(x,z,m)$ and $\tilde{\phi}_{\mathrm{np}}(x,z,m)$ are defined as follows:
\beee\label{tphip}
\tilde{\phi}_{\mathrm{p}}(x,z,m)&=&\scalemath{0.8}{\frac{\mathcal{A}^*_{2,u^*} - \tau_{u^*}\mathcal{A}^*_{1,u^*}}{\E\big\lbt e_{\dd\od}(X)\big\rbt_{u^*}} I_{\beta^*}^{-1}s_{zm}(x\mid \beta^*),}
\\\label{tphinp}
\tilde{\phi}_{\mathrm{np}}(x,z,m) & =&  \scalemath{0.8}{\frac{\tilde{\xi}_{2,u^*}(x,z,m) + \tilde{\xi}_{3,u^*}(x,z,m) - \tau_{u^*}\{\tilde\xi'_{2,u^*}(x,z,m) + \tilde\xi'_{3,u^*}(x,z,m)\} 
 }{\E\big\lbt e_{\dd\od}(X)\big\rbt_{u^*}};}
\eee
\item $\bar\sigma_{u^*}^2$ is defined as follows:
\beee\nn
\bar\sigma_{u^*}^2 
&=& \E\Bigg[\frac{{\bar{\xi}}_{1,u^*}(X,Z,M)\{Y - \bar{\mu}_{Z}(X,M)\}
+\big\lbt\big\{\bar\mu_{1}( X,\dd) - \bar\mu_{0}(X,\od)\big\}{e}_{\dd\od}(X)\big\rbt_{u^*} - \tau_{u^*}\big\lbt e_{\dd\od}(X)\big\rbt_{u^*}}{\big\lbt{e}_{\dd\od}(X)\big\rbt_{u^*}}
\\\label{sigmaform}
&&+ \frac{\mathcal{A}_{2,u^*} - \tau_{u^*}\mathcal{A}_{1,u^*}}{\big\lbt{e}_{\dd\od}(X)\big\rbt_{u^*}} \psi(X,Z,M)\Bigg]^2;
\eee
\item $\delta^{(1)}_{X,1,M,u^*}$--$\delta^{(3)}_{X,1,M,u^*}$ are defined as follows:
\beee\nn
\delta^{(1)}_{X,1,M,u^*}&=&\{F'_{1M}( X\mid\beta^*)\}^\T \int_{\R}k_{u^*}(M,m_0)c_{\rho}^{(1)}\big\{F_{1M}(X),F_{0m_0}(X)\big\}f_{0m_0}( X)dm_0,
\\\nonumber
\delta^{(2)}_{X,1,M,u^*}&=&\int_{\R}k_{u^*}(M,m_0)c_{\rho}^{(0)}\big\{F_{1M}( X),F_{0m_0}( X)\big\}\{F'_{0m_0}( X\mid\beta^*)\}^\T f_{0m_0}( X)dm_0,
\\\label{haha}
\delta^{(3)}_{X,1,M,u^*} &=& \int_{\R}k_{u^*}(M,m_0)c_{\rho}\big\{F_{1M}(X),F_{0m_0}( X)\big\}\{f'_{0m_0}( X\mid \beta^*)\}^\T dm_0;
\eee
\item $\delta^{(1)}_{X,0,M,u^*}$--$\delta^{(3)}_{X,0,M,u^*}$ are defined as follows:
\beee\nn
\delta^{(1)}_{X,0,M,u^*}&=&\{F'_{0M}( X\mid\beta^*)\}^\T \int_{\R}k_{u^*}(m_1,M)c_{\rho}^{(0)}\big\{F_{1m_1}(X),F_{0M}(X)\big\}f_{1m_1}( X)dm_1,
\\\nonumber
\delta^{(2)}_{X,0,M,u^*}&=&\int_{\R}k_{u^*}(m_1,M)c_{\rho}^{(1)}\big\{F_{1m_1}( X),F_{0M}( X)\big\}\{F'_{1m_1}( X\mid\beta^*)\}^\T f_{1m_1}(X)dm_1,
\\\label{haha2}
\delta^{(3)}_{X,0,M,u^*} &=& \int_{\R}k_{u^*}(m_1,M)c_{\rho}\big\{F_{1m_1}(X),F_{0M}( X)\big\}\{f'_{1m_1}( X\mid \beta^*)\}^\T dm_1.
\eee
\end{itemize}

\section{Comparison with identification assumptions in mediation analysis }\label{compare:sec}
 Recall the setting of mediation analysis in Section \ref{sec:example:sa}. In particular, the  potential outcome notation $Y_z$ and potential outcome notation $Y_{z,m}$ in mediation analysis are linked by  
\beee\label{link}
Y_z &=& Y_{z,M_z}, \quad z= 0,1,
\eee with no conflict of  settings; See \citet[ \textsection2.1]{forastiere2018principal}. We show that Assumption~\ref{assump:weak-pi} is weaker than the following two common identification assumptions in mediation analysis.
\begin{itemize} 
\item The \textit{strong principal ignorability} assumes 
\beee\label{spia}
Y_{z,m}\indep U\mid X&\text{for any}& z = 0,1
\eee
and $m\in\R$. When $M$ is binary, \citet{forastiere2018principal} show that this assumption is equivalent to the sequential ignorability under monotonicity. When $M$ is continuous,  it implies Assumption \ref{assump:weak-pi} as 
\beee\nonumber
\E\{Y_z\mid X, U = (m_1,m_0)\} &=& \E\{Y_{z,m_z}\mid X, U = (m_1,m_0)\}\quad\text{(by \eqref{link} and $M_z = m_z$)} 
\\\nonumber
&=& \E(Y_{z,m_z}\mid X)\quad\text{(by \eqref{spia})}
\\\nonumber
&=& \E(Y_{z,m_z}\mid X, M_z = m_z) \quad\text{(by \eqref{spia})}
\\\nonumber
&=& \E(Y_{z,M_z}\mid X, M_z = m_z) = \E(Y_{z}\mid X, M_z = m_z)\quad\text{(by \eqref{link})}.
\eee
 \item  The \textit{sequential ignorability} \citep{imai2010identification,tchetgen2012semiparametric} assumes  for any $z,z' \in \{0,1\}$,  $m\in\mathbb{R}$:  \begin{eqnarray}\label{newlabel:aa}
 &&\text{(i). }\big(Y_{z',m}, M_z\big)\indep Z \mid X; 
 \\\label{newlabel:aa2}
 &&\text{(ii). } Y_{z',m} \indep M\mid Z = z ,X,
 \end{eqnarray} which imply Assumption \ref{assump:weak-pi}. This is because for any $z\in\{0,1\}$ and $(m_1,m_0)\in\R^2$, 
\begin{eqnarray*}
\E(Y_{z}\mid  X, M_z = m_z) & = & \E(Y_{z,m_z}\mid  X, Z = 1-z, M_z = m_z)\quad\text{(by \eqref{link}\text{ and }\eqref{newlabel:aa})}
\\ 
& = &\E(Y_{z,m_z}\mid  X, Z = 1-z, M_z = m_z,M = m_{1-z})\quad\text{(by \eqref{newlabel:aa2})}
\\
& = & \E(Y_{z,m_z}\mid  X, Z = 1-z, M_z = m_z,M_{1-z} = m_{1-z})
\\
& = &  \E\{Y_{z,m_z}\mid  X, U = (m_1,m_0)\} \quad\text{(by \eqref{newlabel:aa})}
\\
&=&\E\{Y_{z,M_z}\mid  X, U = (m_1,m_0)\} = \E\{Y_{z}\mid  X, U = (m_1,m_0)\}.
\end{eqnarray*} 
\item \citet{parast2016robust} and \citet{wang2020model} consider the RCT setting where $Z$ is randomly assigned and no covariates $X$ are collected. The identification assumption of \citet{parast2016robust} assumes 
\bee\label{surrogate1}
Y_1 \ind M_0 \mid M_1 \text{ and }Y_0 \ind M_1 \mid M_0.
\ee
The identification assumption of \citet{wang2020model} assumes
\bee\label{surrogate2}
(Y_1,M_1)\ind (Y_0,M_0).
\ee
Under either \eqref{surrogate1} or \eqref{surrogate2}, one can see that $Y_z\mid M_z = m_z$ and $Y_z\mid U = u$ share the same distribution and thus $\E(Y_{z}\mid  M_z = m_z) = \E\{Y_{z}\mid  U = (m_1,m_0)\}$, which verifies Assumption \ref{assump:weak-pi}.
\end{itemize} 
\section{Verification of CATE as a perfectly localized functional}\label{sec:cate:exp}
{\color{black}In \eqref{def:perfect}, let $W = (X,Z)$, $D = X$,  and $x_0$ be a specific individual level of $X$. Let $\gamma^*(x,z) = \E(Y\mid X = x,Z = z)$ and $k_h(x\mid x_0)$ be some regular kernel centered at $x_0$. 
Denote $$m(w,y,\gamma^*,k_h) = k_h(x\mid x_0)\{\gamma^*(x,1) - \gamma^*(x,0)\}.$$ We show that CATE with $X = x_0$, namely $\vartheta_0 = \E(Y_1 - Y_0\mid X = x_0)$, can be represented through  \eqref{def:perfect} and thus is a perfectly localized function. In particular, under the assumption of selection on observables, we have
$
\vartheta_0 = \E\{\gamma^*(X,1) - \gamma^*(X,0)\mid X = x_0\} = \lim_{h\rightarrow 0}\E[k_h(X\mid x_0)\{\gamma^*(X,1) - \gamma^*(X,0)\}] = \lim_{h\rightarrow 0}\vartheta_h
$, which complies with \eqref{def:perfect}.}
{\color{black}\section{Computational complexity of numerical integrations}\label{sec:compute}
In this section, using the standard Gaussian kernel as an example, we analyze the computational complexity of the numerical integrations required for our proposed estimator in \eqref{form:hattau}. All integrations can be computed with a computational complexity bounded by \(\mathcal{O}(n^2r_n)\) (ignoring logarithmic factors), ensuring the numerical approximation errors are controlled at \(o(n^{-1/2})\). 

Here, \(r_n\) represents the maximum computational cost for evaluating a single \(\hat{e}_{u}(x)\) or \(\hat{\mu}_{z}(x,m)\) given specific inputs, after all nuisance function estimators have been trained. Specifically, we define  
\bee\label{def:rn}
r_n = \max\left\{\sup_{(x,u)\in\mathbb{X}\times \mathbb{M}^2}\text{CT}\{\hat{e}_u(x)\},\sup_{(x,z,m)\in\mathbb{X}\times\{0,1\}\times \mathbb{M}}\text{CT}\{\hat{\mu}_z(x,m)\}\right\},
\ee
where \(\text{CT}\{g(v)\}\) denotes the computational time required to calculate the function output \(g(v)\) with a given input \(v\).  We assume that the number of parameters in the estimators \(\hat{e}_u(x)\) and \(\hat{\mu}_{z}(x,m)\) grows with the sample size \(n\), implying that the computational complexity of calculating a single \(\hat{e}_u(x)\) or \(\hat{\mu}_{z}(x,m)\) can be viewed as a quantity of \(n\).
\par
In the following, we present three examples to illustrate the different orders of \(r_n\) when \(\hat{e}_u(x)\) and \(\hat{\mu}_z(x,m)\) are estimated using different statistical methods. For simplicity, we assume that all involved random variables are uniformly bounded by a constant.
\begin{example}\label{exp:1}
Suppose we estimate $f_{zm}(x)$ through the Gaussian linear model such that $\hat{f}_{zm}(x)
  = \varphi\{m\mid \hat{\ell}^\T \begin{psmallmatrix}x\\z\end{psmallmatrix} ,\hat{\sigma}^2\},
$
where $\hat\ell$ and $\hat{\sigma}^2$ are estimated coefficients in $\R^{d_X + 1}$ and $\R$, respectively, and $\varphi(\cdot\mid \iota, \omega)$ denotes the Gaussian density function with mean $\iota$ and variance $\omega$. Then for any given $x\in\mathbb{X}$ and $u\in\mathbb{M}^2$, all $\hat{f}_{1m_1}(x),\hat{f}_{0m_0}(x),\hat{F}_{1m_1}(x),\hat{F}_{0m_0}(x)$ can be calculated with   $\mathcal{O}(1) $ complexities following the Gaussian linear model, and thus we can calculate the corresponding $\hat{e}_u(x) = c_\rho\{\hat{F}_{1m_1}(x), \hat{F}_{0m_0}(x)\}\prod_{z\in\{0,1\}}\hat{f}_{zm_z}(x)$ with   $\mathcal{O}(1)$ complexity, and thus  $$\sup_{(x,u)\in\mathbb{X}\times \mathbb{M}^2}\mathrm{CT}\{\hat{e}_u(x)\}  = \mathcal{O}(1).$$
\end{example}
\begin{example}\label{exp:2}
Suppose we estimate $\mu_z(x,m)$ through the linear regression such that $  \hat{\mu}_z(x,m) = \hat{\ell}'^\T\begin{psmallmatrix}x\\z\\m
\end{psmallmatrix}$, where $\hat{\ell}'$ is a vector of coefficients in $\R^{d_X + 2}$, and assume $d_X$ is fixed with $n$. Then we can calculate   $\hat{\mu}_z(x,m)$, for any given $x\in\mathbb{X}, z\in\{0,1\},m\in\mathbb{M}$, with $\mathcal{O}(1)$ complexity, and thus $$\sup_{(x,u)\in\mathbb{X}\times \mathbb{M}^2}\mathrm{CT}\{\hat{e}_u(x)\}  = \mathcal{O}(1).$$
\end{example}
\begin{example}\label{exp:3}
Suppose we estimate $\mu_z(x,m)$ through the multivariate Nadaraya--Watson estimator \citep{nadaraya1964estimating,watson1964smooth}:  
$$
 \hat{\mu}_z(x,m) = \frac{\sum_{i = 1}^n\mathcal{K}_{d_X + 2}\{(x_i - x,z_i-z,m_i - m)/h\}y_i}{\sum_{i = 1}^n\mathcal{K}_{d_X + 2}\{(x_i - x,z_i-z,m_i - m)/h\}},
$$
where $h > 0$ is a pre-specified bandwidth, $\mathcal{K}_{d_X + 2}(\cdot)$ is any kernel function mapping from $\R^{d_X + 2}$ to $\R$, and for concreteness, here we consider $\mathcal{K}_{d_X + 2}(\cdot)$ to be the  $(d_X + 2)$-dimensional standard Gaussian kernel. Then for any given  $x\in\mathbb{X}, z\in\{0,1\},m\in\mathbb{M}$, one can see that calculating $ \hat{\mu}_z(x,m)$   needs $\mathcal{O}(n)$ complexity, and thus 
$$
\sup_{(x,z,m)\in\mathbb{X}\times\{0,1\}\times \mathbb{M}}\mathrm{CT}\{\hat{\mu}_z(x,m)\} = \mathcal{O}(n).
$$
\end{example}
 When we estimate $\hat{e}_u(x)$  and   $\hat{\mu}_{z}(x,m)$ following Example~\ref{exp:1} and Example~\ref{exp:2}, respectively, we have $r_n = \max\{\mathcal{O}(1),\mathcal{O}(1)\} = \mathcal{O}(1)$. When we estimate $\hat{e}_u(x)$  and   $\hat{\mu}_{z}(x,m)$ following Example~\ref{exp:1} and Example~\ref{exp:3}, respectively, we have $r_n = \max\{\mathcal{O}(1),\mathcal{O}(n)\} = \mathcal{O}(n)$. 

 Theorem~\ref{thm:complex} demonstrates the concrete computational complexity bound of calculating all numerical integrations for our proposed estimator, with    approximation errors of all numerical integrations controlled by $o(n^{-1/2})$.
  
\begin{theorem}\label{thm:complex}
Consider the following regularity conditions.   For some universal constant $C > 0$, assume for any $i\in[n]$,  all $\hat{e}_u(X_i)$, $\partial\hat{e}_u(X_i)/\partial m_1$, $\partial\hat{e}_u(X_i)\partial m_0$, and $\{\hat\mu_{1}(X_i,m_1) - \hat\mu_{0}(X_i,m_0)\}\hat{e}_u(X_i)$ are uniformly bounded by $C$ over $u\in\R^2$, and assume
\bee\label{regular:int:condition}
\int_{\R^2}\left|\frac{\partial^2g(u)}{\partial m_1\partial m_0}\right|du,\,\int_{\R^2}\left|\frac{\partial g(u)}{\partial m_1 }\right|du,\,\int_{\R^2}\left|\frac{\partial g(u)}{ \partial m_0}\right|du,\,\sup_{m_1\in\R}\int_{\R}\left|\frac{\partial g(u)}{ \partial m_0}\right|dm_0,\,\sup_{m_0\in\R}\int_{\R}\left|\frac{\partial g(u)}{ \partial m_1}\right|dm_1 \leq C,
\ee
for   $g(u) \in\{\hat{e}_u(X_i),\{\hat\mu_{1}(X_i,m_1) - \hat\mu_{0}(X_i,m_0)\}\hat{e}_u(X_i)\}$. Let $h \succsim n^{-1/2}$,      $\mathcal{K}$ be the  standard Gaussian kernel, and $r_n$  is defined in \eqref{def:rn}.
\par
The event of calculating all the numerical integrations in  $\{\lbt\hat{e}_{\dd\od}(X_i)\big\rbt_{u^*}\}_{i = 1}^n$,  $\{\lbt\{\hat\mu_{1}( X_i,\dd) - \hat\mu_{0}(X_i,\od) \}$ $\hat{e}_{\dd\od}(X_i)\rbt_{u^*}\}_{i = 1}^n$ and $\{\hat{\gamma}_{1,u^*}^{(Z_i)}(M_i,X_i)\}_{i = 1}^n$ such that all approximation errors of the  numerical integrations are  bounded by $o(n^{-1/2})$, can be addressed by a procedure with computational complexity bounded by 
$
\mathcal{O}\{(\log n)^{2 + \epsilon}n^2r_n\}
$, for any fixed $\epsilon > 0$.
\end{theorem}
The proof of Theorem~\ref{thm:complex} is deferred to Section~\ref{sec:pf:ni}. Below, we briefly discuss the conditions in Theorem~\ref{thm:complex}. In general, the regularity conditions on \(\hat{e}_u(X_i)\) and \(\{\hat{\mu}_1(X_i,m_1) - \hat{\mu}_0(X_i,m_0)\} \hat{e}_u(X_i)\) in Theorem~\ref{thm:complex} can be satisfied if \(\hat{e}_u(x)\) is derived from a smooth two-dimensional density family and \(\hat{\mu}_z(x,m)\) is uniformly bounded with a bounded first-order derivative with respect to \(m\) for all \((x,z,m) \in \mathbb{X} \times \{0,1\} \times \mathbb{M}\).  
For example, suppose \(c_\rho(\cdot)\) is a Gaussian copula with \(\rho\) strictly bounded away from \(-1\) and \(1\), and all nuisance functions are estimated using the same parametric strategy described in Section~\ref{sec:cd} for our numerical experiments. Then, the regularity conditions are satisfied. Additionally, Theorem~\ref{hto0} establishes that \(h \succsim n^{-1/2}\) is necessary to ensure that \(\hat{\tau}_{u^*}\) remains a consistent nonparametric estimator of \(\tau_{u^*}^*\). Thus, the requirement \(h \succsim n^{-1/2}\) in Theorem~\ref{thm:complex} is also mild.

Theorem~\ref{thm:complex}   suggests that if \(\hat{e}_u(x)\) and \(\hat{\mu}_z(x,m)\) are estimated using relatively simple machine learning algorithms with \(r_n = \mathcal{O}(1)\), the numerical integrations can be computed in approximately \(\mathcal{O}(n^2)\) time (ignoring logarithmic factors), making the computation quite efficient. However, employing more complex estimation methods for \(\hat{e}_u(x)\) and \(\hat{\mu}_z(x,m)\) can increase computation time.   For instance, in Examples~\ref{exp:1} and \ref{exp:3}, we have \(r_n = \mathcal{O}(1)\), resulting in a nearly \(\mathcal{O}(n^2)\) complexity. However, if \(\hat{e}_u(x)\) follows Example~\ref{exp:1} while \(\hat{\mu}_z(x,m)\) follows Example~\ref{exp:3}, then \(r_n = \mathcal{O}(n)\), increasing the total complexity to approximately \(\mathcal{O}(n^3)\).  From a practical perspective, selecting computationally efficient machine learning algorithms for nuisance function estimation, leading to a smaller \(r_n\) can enhance the overall computational efficiency of our proposed estimator.

\section{Enhanced robustness under linear $\mathcal{M}_{\mathrm{om}}$}\label{sec:extrarobust} 
In this section, we present Theorem~\ref{prop:linear:triple}, which shows the enhanced robustness of our proposed estimator with respect to   nuisance function specification and sensitivity parameter selection when $\mathcal{M}_{\mathrm{om}}$ follows a linear model.  The implications of Theorem~\ref{prop:linear:triple} can be summarized by the following two key takeaway messages.
\begin{enumerate}
\item[(i)] Suppose $\mathcal{M}_{\mathrm{om}}$ is linear, and all  identification assumptions (Assumptions~\ref{assump:TAignorability}-\ref{am:cp}) and mild regularity conditions hold. Our proposed estimator has the robustness surpassing the triple robustness. That is, whenever $\mathcal{M}_{\mathrm{om}}$ is correctly specified, or both $\mathcal{M}_{\mathrm{tp}}$  and $\mathcal{M}_{\mathrm{ps}}$ are correctly specified, our proposed estimator identifies the true Cont.PCE correctly with vanishing $h$.
\item[(ii)] Suppose  $\mathcal{M}_{\mathrm{om}}$ is linear, and two causal identification assumptions (Assumptions~\ref{assump:TAignorability}-\ref{assump:weak-pi})     and mild regularity conditions hold. When $\mathcal{M}_{\mathrm{om}}$ is correctly specified in estimation, even if one does not choose the true sensitivity parameters $c_\rho(\cdot,\star)$ and $\rho$, our proposed estimator can still identify  the true Cont.PCE correctly  with vanishing $h$.  This    indicates that when the outcome model is linear or nearly linear, the choice of sensitivity parameters can have a non-sensitive effect on the estimation bias of our proposed estimator towards the true Cont.PCE.
\end{enumerate}
\begin{theorem}\label{prop:linear:triple} 
Suppose  Assumptions~\ref{assump:TAignorability}--\ref{assump:weak-pi}
hold, and $\mathcal{M}_{\mathrm{om}}$ follows a linear model: $$\mu_z(x,m) = \lambda^\T(1,x^\T,z,m)\text{ for some }\lambda = (\lambda_1,\cdots,\lambda_{d_X + 3})^\T. $$ Denote the functional limit $\bar{f}_{zm}(x)$ such that $\hat{f}_{zm}(x)\rightarrow \bar{f}_{zm}(x)$ as $n\rightarrow \infty$, and $\bar{F}_{zm}(x)= \int_{(-\infty,m]}\bar{f}_{zm'}(x)dm'$. The selected copula function and correlation parameter for our proposed estimator, are   $\bar{c}_{\bar\rho}(\cdot,\star)$ and $\bar{\rho}$, respectively. Thus the specified principal model is
\bee\nonumber
\bar{e}_u(x) = \bar{c}_{\bar \rho}\big\{\bar F(M_1 = m_1\mid X = x),\bar  F(M_0 = m_0\mid X = x)\big\}\prod_{z\in\{0,1\}}\bar f(M_z =m_z\mid X = x).
\ee 
The true principal score model is $
e_u(x) = f(M_{1} = m_1,M_{0} = m_0\mid X = x) =c_\rho\big\{F(M_1 = m_1\mid X = x), F(M_0 = m_0\mid X = x)\big\}\prod_{z\in\{0,1\}}f(M_z =m_z\mid X = x)
$ following Assumption~\ref{am:cp}. Also suppose the regularity conditions in Proposition~\ref{po:approximate} hold, and $\bar{e}_u(x)$ satisfies  Assumption~\ref{ppa:ii} similarly  with $e_u(x)$.
\par
When the true copula function and correlation parameter are replaced by $\bar{c}_{\bar\rho}(\cdot,\star)$ and $\bar{\rho}$  for our proposed estimator $\hat{\tau}_{u^*}$ in  \eqref{form:hattau},    the probabilistic limit of $\hat{\tau}_{u^*}$ becomes 
\bee\nonumber
\check\tau_{u^*} = \frac{\E\Big[\check\xi_{1,u^*}(X,Z,M)\{Y - \bar{\mu}_{Z}(X,M)\} 
+\big\lbt\big\{\bar\mu_{1}(X,\dd) - \bar\mu_{0}(X,\od)\big\}\bar{e}_{\dd\od}(X)\big\rbt_{u^*}\Big]}{\E\big\lbt{\bar e}_{\dd\od}(X)\big\rbt_{u^*}},
\ee
where 
$\check\xi_{1,u^*}(x,z,m) =  \{\bar\pi_z(x) \bar f_{zm}(x)\}^{-1}{(-1)^{z+1}\bar\gamma_{1,u^*}^{(z)}(x,m)} $, and   $\bar{\gamma}_{1,u^*}^{(1)}(m,x)=\int_{\R} k_{u^*}(m,m_0)$ $\bar{e}_{(m
,m_0)}(x)dm_{0}$, $\bar{\gamma}_{1,u^*}^{(0)}(m,x)=\int_{\R} k_{u^*}(m_1,m)\bar{e}_{(m_1,m
)}(x)dm_{1}$. Then when either $\bar{\mu} = \mu$ or $(\bar{\pi},\bar{e}) = (\pi,e)$, we have $$\lim_{h\rightarrow 0}\check\tau_{u^*} = {\tau}^*_{u^*}.$$
\end{theorem}
The proof of Theorem~\ref{prop:linear:triple} is deferred to Section~\ref{sec:pf:linear}.}
\section{Comparison with \citet{lu2023principal}}\label{sec:compare}
 In this section, we provide a comprehensive comparison of our work with \citet{lu2023principal}. To facilitate comparison, we first provide a brief overview of \citet{lu2023principal} using our paper's notation. We then detail the major differences of two papers, emphasizing the fundamentally distinct approaches in approximating and estimating Cont.PCE: the local   smoothing-based approach versus the global projection-based approach.  

\paragraph{Overview of the method in \citet{lu2023principal}}
To ensure self-containment, we first provide a brief overview of \citet{lu2023principal}. Their main paper focuses on the identification and estimation of the average potential outcomes:  
\[
m_{z}(u) = \E(Y_{z} \mid U = u),
\]
for \( z = 0,1 \) and \( u \in \mathbb{R}^{2} \). The Cont.PCE is connected to the average potential outcomes via  
\[
{\tau}_{u}^{*} = \E(Y_{1} - Y_{0} \mid U = u) = m_{1}(u) - m_{0}(u).
\]  
Similar to Cont.PCE, the average potential outcomes are also non-pathwise differentiable. \citet{lu2023principal} propose projecting \( m_{z}(u) \) onto parametric working models \( f_{z}(u; \eta) \) for \( z \in \{0,1\} \), where \( \eta \in \mathbb{R}^{q} \) represents the working model parameter.

Specifically, they approximate \( m_{1}(u) \) and \( m_{0}(u) \) using functional forms \( f_{1}(u;\eta_{1}) \) and \( f_{0}(u;\eta_{0}) \), respectively, where
\bee\label{response:eta10}
\eta_1 = \arg\min_{\eta\in\R^q} \E\Big[w_1(U)\{m_1(U) - f_1(U;\eta)\}^2\Big], \quad
\eta_0 = \arg\min_{\eta\in\R^q} \E\Big[w_0(U)\{m_0(U) - f_0(U;\eta)\}^2\Big],
\ee
with user-specified weight functions \( w_1(u) \) and \( w_0(u) \), e.g., setting \( w_1(u) = w_0(u) = 1 \) for all \( u \in \mathbb{M}^{2} \). This reduces the problem of approximating \( m_{z}(u) \) to estimating \( \eta_{z} \).

\citet[$\mathsection$3, $\mathsection$4]{lu2023principal} develop various identification and estimation methods for \( \eta_{z} \), including the EIF-based estimator \( \hat{\eta}_{z,\mathrm{eif}} \), obtained as the solution to:
\bee\label{Lu:EQ}
\sum_{i = 1}^n{D}_{z,\mathrm{eif}}(Y_i,M_i,Z_i,X_i;\hat{\eta}_{z,\mathrm{eif} },\hat{\pi},\hat{e},\hat{\mu}_1) = 0,
\ee
where \( {D}_{z,\mathrm{eif}}(\cdot) \) is detailed in \citet[Lemma~3]{lu2023principal}. The semiparametric efficiency of \( \hat{\eta}_{z,\mathrm{eif}} \) is established in \citet[$\mathsection$5]{lu2023principal} under the doubly robust condition, ensuring that nuisance functions  can be well approximated.

For a general linear working model:
\bee\label{response:linear}
f_z(u;\eta) = \eta^\T g(u),\quad z\in\{0,1\}, 
\ee
where \( g(u) \) is a pre-specified vector-valued function and \( \eta \) is the unknown parameter. \citet[$\mathsection$4.2]{lu2023principal} provide closed-form solutions for \( \hat{\eta}_{1,\mathrm{eif}} \) and \( \hat{\eta}_{0,\mathrm{eif}} \) from \eqref{Lu:EQ}.

As noted in Remark~1 of \citet{lu2023principal}, one can also project Cont.PCE directly onto a parametric working model \( f_{\tau}(u;\eta) \):
\[
\eta_{\tau}=\arg\min_{\eta\in\R^{q}}\E\left[w_{\tau}(U)\{\tau_{U}^{*}-f_{\tau}(U;\eta)\}^{2}\right],
\]
where \( w_{\tau}(u) \) is a user-specified weight function. \citet{lu2023principal} derive the EIF of \( \eta_{\tau} \) in the general case. For linear working models with identical weight functions, \citet[Corollary S1]{lu2023principal} show that the EIF of \( \eta_{1}-\eta_{0} \) coincides with that of \( \eta_{\tau} \). Thus, the closed-form EIF-based estimators \( \hat{\eta}_{1,\mathrm{eif}} \) and \( \hat{\eta}_{0,\mathrm{eif}} \) yield a closed-form EIF-based estimator for \( \eta_{\tau} \) as \( \hat{\eta}_{1,\mathrm{eif}}-\hat{\eta}_{0,\mathrm{eif}} \).

{\paragraph{Considered scenario for comparisons}  
To simplify and clarify our comparisons, we focus on the scenario where all working models are identical and adopt the general linear model in \eqref{response:linear}, with all weight functions also being identical. Specifically, we define the considered working model as  
\bee\nonumber
f(u;\eta) = \eta^\T g(u),
\ee  
for a fixed vector-valued function \( g(u):\mathbb{R}^2\to\mathbb{R}^{q} \), where \( q \) is a fixed positive integer, and we denote the corresponding weight function by \( w(u) \). Accordingly, we assume  
\[
f(u;\eta) = f_1(u;\eta) = f_0(u;\eta) = f_\tau(u;\eta), \quad w(u) = w_1(u) = w_0(u) = w_\tau(u).
\]  
Under this setting, the efficient estimator for the Cont.PCE projection parameters proposed by \citet{lu2023principal} is given by  
\[
\hat{\eta}_{\tau,\mathrm{eif}} = \hat{\eta}_{1,\mathrm{eif}} - \hat{\eta}_{0,\mathrm{eif}},
\]  
which we compare against our proposed Cont.PCE estimator.}

{\color{black}\paragraph{I. Difference in  non-regular estimand approximation and estimation}  
To address the non-pathwise differentiability of Cont.PCE, the two papers adopt different approaches to approximate the target causal estimands using regular substitutive estimands. Both approaches develop semiparametric efficient estimators for these substitutive estimands, thereby approximating causal quantities such as Cont.PCE and average potential outcomes. We summarize the details of these two approaches below.

\begin{itemize}
\item Our approach considers Loc.PCE \( \tau_{u} \) as the regular substitutive estimand, which provides a local functional approximation of Cont.PCE \( \tau^*_{u} \) via a kernel-based method. By focusing on any specific principal strata \( u \), we develop a semiparametric efficient estimator for Loc.PCE. If  
\[
\tau_u\rightarrow \tau_u^* \quad \text{as} \quad h \to 0,  
\]
then our estimator serves as a nonparametric estimator for Cont.PCE.  

\item \citet{lu2023principal} instead consider the working model projection \( f (u;\eta_\tau) \), which provides a global approximation of Cont.PCE. They develop the EIF-based estimator \( \hat{\eta}_{\tau,\mathrm{eif}} \) for the substitutive estimand \( \eta_{\tau} \), and \( f (u;\hat{\eta}_{\tau,\mathrm{eif}}) \) serves as an efficient estimator for the global projection \( f (u;\eta_\tau) \). If the working model \( f \) correctly specifies Cont.PCE, then \( f (u;\hat{\eta}_{\tau,\mathrm{eif}}) \) also serves as an efficient estimator for Cont.PCE.  
\end{itemize} 
	The key distinction between our method and \citet{lu2023principal} lies in the fundamental difference between local smoothing-based and global projection-based estimation for non-regular estimands. Both approaches, along with fully model-based estimation when the non-regular estimand is specified by a parametric model, involve bias-variance trade-offs. We first compare the estimation properties of our method and \citet{lu2023principal}, then introduce the fully model-based approach and discuss the bias-variance trade-offs among the three methods in the Cont.PCE estimation context.
\begin{itemize}
\item As a local smoothing-based estimator, our proposed estimator \( \hat{\tau}_u \) achieves semiparametric efficiency for Loc.PCE with an \( \mathcal{O}(n^{-1/2}) \) convergence rate and further estimates Cont.PCE with an \( \mathcal{O}(n^{-1/3}) \) rate, under the regularity conditions of Theorem~\ref{hto0} and the doubly robust condition \( r_{\pi}r_{\mu} = \mathcal{O}(n^{-1/2}) \).  

\item As a global projection-based estimator, \( \hat{\eta}_{\tau,\mathrm{eif}} \) achieves \( \mathcal{O}(n^{-1/2}) \) convergence for \( \eta_\tau \) under the regularity conditions in \citet[Theorem~5]{lu2023principal} and the doubly robust assumption. Consequently, \( f(u;\hat{\eta}_{\tau,\mathrm{eif}}) \) efficiently estimates the global projection of Cont.PCE with an \( \mathcal{O}(n^{-1/2}) \) rate under \eqref{response:linear}. However, if the working model does not correctly specify Cont.PCE, a non-vanishing bias will persist between \( f(u;\hat{\eta}_{\tau,\mathrm{eif}}) \) and \( \tau_u^* \) as \( n \to \infty \), leading to inconsistency in estimating Cont.PCE. The usefulness of \( f(u;\hat{\eta}_{\tau,\mathrm{eif}}) \) then depends on the alignment between the working model and the true Cont.PCE.  

\item If the working model in \eqref{response:linear} correctly specifies Cont.PCE, then \( f(u;{\eta}_{\tau}) = \tau^*_u \), and \( f(u;\hat{\eta}_{\tau,\mathrm{eif}}) \) achieves an \( \mathcal{O}(n^{-1/2}) \) convergence rate for Cont.PCE. However, in this case, the projection-based estimator does not generally attain the semiparametric efficiency bound. Specifically, the asymptotic variance bound of \( \hat{\eta}_\tau \) in \citet[Theorem~5]{lu2023principal} depends on the weight function \( w(u) \), whereas the semiparametric efficiency bound under the parametric model assumption of Cont.PCE does not. To achieve the semiparametric efficiency bound under the parametric assumption that $\tau_u^*$ can be correctly specified by the parametric working model $f(u;\eta)$, one can further develop a fully model-based estimator for \( \tau_u^* \) through the semiparametric statistics framework.  
\end{itemize}
The three estimation approaches: local smoothing-based, global projection-based and fully model-based, exhibit different bias-variance tradeoffs in estimation:
\begin{itemize}
\item The local smoothing-based estimator does not rely on a parametric model for Cont.PCE, avoiding model misspecification bias. However, it may be less efficient with a slower convergence rate  than the other two methods, when the Cont.PCE is indeed correctly specified by the parametric working model.  
\item The global projection-based estimator balances efficiency and robustness. It can be more efficient than the local smoothing-based estimator when the parametric model assumption holds, while still providing an interpretable approximation of the Cont.PCE projection when the assumption does not hold.  
\item The fully model-based estimator is the most efficient when the parametric model assumption holds but loses interpretability if the model is misspecified, as it may not converge to some meaningful projection of Cont.PCE; see also \citet[$\mathsection$3.1]{kennedy2019robust}.  
\end{itemize}

\begin{remark}  
If we let \( g(u) \) in \eqref{response:linear} be a set of basis functions (e.g., power series, B-spline basis, or wavelet basis; see \citet{eubank1999nonparametric}) and allow its dimension \( q \) to grow as \( n \to \infty \), then \( f(u;\hat{\eta}_{\tau,\mathrm{eif}}) \) may serve as a nonparametric spline estimator for Cont.PCE. This enables nonparametric estimation of the Cont.PCE surface without requiring that \eqref{response:linear} correctly specifies \( \tau_u^* \) for any finite \( q \). However, the convergence rate of \( f(u;\hat{\eta}_{\tau,\mathrm{eif}}) \) will generally be slower than \( \mathcal{O}(n^{-1/2}) \) due to the increasing dimension \( q \).   The difference between our proposed estimator and such a spline estimator parallels the classic difference between kernel and spline estimators in nonparametric regression. Specifically, our proposed estimator is point-wise, and its accuracy relies on the point-wise smoothness of the Cont.PCE surface. In contrast, the accuracy of the spline estimator relies on the global approximation error of the spline basis towards the  Cont.PCE surface.
\end{remark}} 
\paragraph{II. Difference in computations} Recall that $\hat{\eta}_{\tau,\mathrm{eif}} = \hat{\eta}_{1,\mathrm{eif}} - \hat{\eta}_{0,\mathrm{eif}}$. As discussed in the overview, the EIF-based estimator $\hat{\eta}_{z,\mathrm{eif}}$ for $z \in \{1,0\}$ is obtained from \eqref{Lu:EQ}, with its computational complexity depending on the working model specification. When $f_z$ is a constant or follows the general linear model \eqref{response:linear}, \citet{lu2023principal} derive   closed-form solutions for $\hat{\eta}_{z,\mathrm{eif}}$. For instance, under the general linear model \eqref{response:linear} for $f_1$, they show that
\bee\label{def:1:eif}
\hat{\eta}_{1,\text{eif}} = \frac{\sum_{i=1}^n \hat{A}_{i2}}{\sum_{i=1}^n \hat{A}_{i1}},
\ee
 with
\bee\nonumber
\hat{A}_{i1} = &\frac{Z_i}{\hat{\pi}_1(X_i)} \left[ 
\frac{\int \hat{G}_1(M_i, m_0, X_i; \hat{\mu}_1) \, \mathrm{d}m_0}{\hat{f}_{1M_i}(X_i)} 
- \iint \hat{G}_1(m_1, m_0, X_i; \hat{\mu}_1) \hat{r}_u(m_1,m_0,M_i, X_i) \, \mathrm{d}m_1 \mathrm{d}m_0
\right] 
\\
&+ \frac{1 - Z_i}{ \hat{\pi}_0(X_i)} \left[ 
\frac{\int \hat{G}_1(m_1, M_i, X_i; \hat{\mu}_1) \, \mathrm{d}m_1}{\hat{f}_{0M_i}(X_i)} 
- \iint \hat{G}_1(m_1, m_0, X_i; \hat{\mu}_1) \hat{r}_v(m_1, m_0, M_i, X_i) \, \mathrm{d}m_1 \mathrm{d}m_0
\right]
\\
&+ \iint \hat{G}_1(m_1, m_0, X_i; \hat{\mu}_1) \, \mathrm{d}m_1\mathrm{d}m_0 
+ \frac{Z_i}{\hat{\pi}_1(X_i)} 
 \frac{\int\hat{G}_1(M_i, m_0, X_i; 1) \, \mathrm{d}m_0}{\hat{f}_{1M_i}( X_i)} \{Y_i - \hat{\mu}_1(M_i, X_i)\},
\ee 
\bee\nonumber
\hat{A}_{i2} = &\frac{Z_i}{\hat{\pi}_1(X_i)} \left[
 \frac{\int\hat{G}_1(M_i, m_0, X_i; g_1) \, \mathrm{d}m_0}{\hat{f}_{1M_i}( X_i)}
- \iint \hat{G}_1(m_1, m_0, X_i; g_1) \hat{r}_u(m_1, m_0, M_i,X_i) \, \mathrm{d}m_1 \mathrm{d}m_0
\right]
\\
&+ \frac{1 - Z_i}{ \hat{\pi}_0(X_i)} \left[
\int \frac{\hat{G}_1(m_1, M_i, X_i; g_1) \, \mathrm{d}m_1}{\hat{f}_{0M_i}( X_i)}
- \iint \hat{G}_1(m_1, m_0, X_i; g_1) \hat{r}_v(m_1, m_0, M_i, X_i) \, \mathrm{d}m_1 \mathrm{d}m_0
\right]\\
&+ \iint \hat{G}_1(m_1, m_0, X_i; g_1) \, \mathrm{d}m_1 \mathrm{d}m_0;
\ee
the definitions of some additional functions like $\hat{G}_1(\cdot)$ and $\hat{r}_u(\cdot)$ can be found in \citet[$\mathsection$4.2]{lu2023principal}, and similar results also hold for $\hat{\eta}_{0,\mathrm{eif}}$. Comparing \eqref{form:hattau} and \eqref{def:1:eif}, we observe that both approaches require computing \(\mathcal{O}(n)\) two-dimensional numerical integrations. Thus, the computational complexity analysis of numerical integrations, similar to that presented in Section~\ref{sec:compute} in the revision, applies to \(\hat{\eta}_{z,\mathrm{eif}}\) as well. Overall, the computational complexities of obtaining \(\hat{\eta}_{z,\mathrm{eif}}\) and \(\hat{\tau}_{u}\) are comparable. 

When evaluating Cont.PCE \(\tau^*_{u}\) across a large number of different principal strata, the method proposed by \citet{lu2023principal} can be computationally more efficient. This is because \(\hat{\eta}_{1,\mathrm{eif}}\) and \(\hat{\eta}_{0,\mathrm{eif}}\) need to be computed only once, and the linear model \(f(u;\hat{\eta}_{1,\mathrm{eif}} - \hat{\eta}_{0,\mathrm{eif}})\) is generally straightforward to evaluate for each \(u\). In contrast, our proposed estimator \(\hat{\tau}_u\) must be computed separately for each \(u\), increasing the computational cost.

For non-linear working models for \(f_1\) and \(f_0\), \eqref{Lu:EQ} may not have a closed-form solution, requiring numerical methods such as Newton's method to approximate \(\hat{\eta}_{1,\mathrm{eif}}\) and \(\hat{\eta}_{0,\mathrm{eif}}\). In contrast, our proposed estimator always has a closed-form solution, as shown in \eqref{form:hattau}, and does not require pre-specification of working models, making it very easy to implement in practical applications. 
\paragraph{III. Difference in the application foci}Our paper has a distinct application motivation and focus--surrogate evaluation--differentiating it from \citet{lu2023principal}, whose real data analysis primarily focuses on causal median analysis. For instance, Section~\ref{sec:example:sa} of our paper provides a detailed discussion on the connection between Cont.PCE and key concepts in surrogate evaluation, including associative and dissociative effects, causal sufficiency, and causal necessity. We further illustrate how Cont.PCE surfaces can be leveraged for surrogate evaluation through a simple and intuitive numerical example in Section~\ref{sec:example:sa}.  
Beyond the conceptual framework, our real data analysis also centers on the surrogate evaluation of short-term CD4 count in ACTG 175, providing a comprehensive empirical investigation.

\section{Proof of Proposition \ref{po:approximate}}\label{pf:po:approximate}
\allowdisplaybreaks
\subsection{Regularity conditions}
\label{sec:po1:con}
We  list the three regularity conditions for Proposition \ref{po:approximate} as follows.
\begin{assumption}\label{ppa:i}
The second-order partial derivative of $\mu_{z}(x,m)$ with respect to $m$ is uniformly bounded for all $x\in\mathbb{X}$, $m_z\in\R^2$ and $z = 0,1$.
\end{assumption}
\begin{assumption}\label{ppa:ii}
All second-order partial derivatives of $f\{(M_1,M_0) = u\mid X = x\} = e_u(x)$  with respect to $u$, are uniformly bounded for all $x \in\mathbb{X}$, $u\in\R^2$ and $z = 0,1$.
\end{assumption}
\begin{assumption}\label{ppa:iii}
$\int_{\R^2}\mathcal{K}(\delta_1,\delta_0)d\delta_1d\delta_0 = 1$, $\int_{\R^2}\delta_z\mathcal{K}(\delta_1,\delta_0)d\delta_1d\delta_0 = 0$, and $\int_{\R^2}|\delta_1|^{c_1}|\delta_0|^{c_0}\mathcal{K}(\delta_1,\delta_0)$ $d\delta_1d\delta_0 < +\infty$ for all integers $c_1,c_0$ such that $c_1 + c_0\leq 4$ and $z = 0,1$.
\end{assumption}
The bounded second-order partial derivatives of $\mu_{z}(x,m)$ and $e_u(x)$ in Assumptions \ref{ppa:i} and \ref{ppa:ii} are standard in the nonparametric kernel regression literature. Assumption \ref{ppa:iii} for $\mathcal{K}(\cdot,\cdot)$  is also mild and can be typically satisfied by the two-dimensional Gaussian Kernel, box kernel,  Epanechnikov kernel, among others. See e.g. \citet{fan1993local} and \citet{wasserman2006all}. 
\subsection{Main proof of Proposition \ref{po:approximate}}
Some technical arguments are standard in the nonparametric kernel regression literature for bias control, and thus we will omit the corresponding details; See, e.g. \citet{fan1993local} for more discussions. For the ease of exposition, denote $\Delta_{u}(x) = \mu_{1}(x,m_1) - \mu_{0}(x,m_0)$ for any $u = (m_1,m_0)\in\R^2$ and $x\in\R^{d_X}$. By definition, one has
\beee\nn
\tau_{u^*}  &= &\int_{\R^{d_X + 2}}\Delta_u(x)\frac{k_{u^*}(u)}{\E\big\{k_{u^*}(U)\big\}}e_u(x)f(x)dm_1dm_0dx
\\\nn
&= &\int_{\R^{d_X + 2}}\Delta_{u^* + (\delta_1,\delta_0)}(x)\frac{k_{u^*}\{u^* + (\delta_1,\delta_0)\}}{\E\big\{k_{u^*}(U)\big\}}e_{u^* + (\delta_1,\delta_0)}(x)f(x)d\delta_1d\delta_0dx
\\\nn
&= &\int_{\R^{d_X + 2}}\Delta_{u^* + h\delta'}(x)\frac{k_{u^*}\{u^* + h\delta'\}}{\E\big\{k_{u^*}(U)\big\}}e_{u^* + h\delta'}(x)f(x)h^2d\delta'_1d\delta'_0dx
\\\nn
&= &\int_{\R^{d_X + 2}}\Delta_{u^* + h\delta'}(x)\frac{\mathcal{K}(\delta_1',\delta_0')}{\E\big\{k_{u^*}(U)\big\}}e_{u^* + h\delta'}(x)f(x)d\delta'_1d\delta'_0dx
\\\label{po:limit:1}
&= &\int_{\R^{d_X}}\frac{f(x)}{\E\big\{k_{u^*}(U)\big\}}\int_{\R^2}\Delta_{u^* + h\delta'}(x)\mathcal{K}(\delta_1',\delta_0')e_{u^* + h\delta'}(x)d\delta'_1d\delta'_0dx,
\eee
where we denote $\delta' = (\delta_1',\delta_0')$. By Taylor expansion, one has for any given $x\in\mathbb{X}$,
\beee\nn
\Delta_{u^* + h\delta'}(x)& = &\mu_{1}(x,m_1^*) + h\delta_1'\frac{\partial\mu_{1}(x,m_1)}{\partial m_1}\Big|_{m_1 =m_1^*} + \frac{(h\delta_1')^2}{2}\frac{\partial^2\mu_{1}(x,m_1)}{\partial^2 m_1}\Big|_{m_1 =m_1^* + h_1\delta_1'}
\\\label{po:limit:2}
&&-\mu_{0}(x,m_0^*) - h\delta_0'\frac{\partial\mu_{0}(x,m_0)}{\partial m_0}\Big|_{m_0 =m_0^*} - \frac{(h\delta_0')^2}{2}\frac{\partial^2\mu_{0}(x,m_0)}{\partial^2 m_0}\Big|_{m_0 =m_0^* + h_0\delta_0'}
\eee
where $|h_z|\leq h$ and $\sgn(h_z) = \sgn(h)$ for $z = 0,1$; $\text{sgn}(\cdot)$ is the sign function. Similarly, we have,
\beee\label{po:limit:3}
e_{u^* + h\delta'}(x) &=& e_{u^*}(x) + h\delta_1'\frac{\partial e_u(x)}{\partial m_1}\Big|_{u = u^*} + h\delta_0'\frac{\partial e_u(x)}{\partial m_0}\Big|_{u = u^*}
\\\nn
&&+\frac{(h\delta_1')^2}{2}\frac{\partial^2 e_u(x)}{\partial^2 m_1}\Big|_{u = u^* + h_{10}\delta'}+\frac{(h\delta_0')^2}{2}\frac{\partial^2 e_u(x)}{\partial^2 m_0}\Big|_{u = u^* + h_{10}\delta'}+\frac{h^2\delta_1'\delta_0'}{2}\frac{\partial^2 e_u(x)}{\partial m_1\partial m_0}\Big|_{u = u^* + h_{10}\delta'},
\eee
for some $|h_{10}|\leq h$ and $\sgn(h_{10}) = \sgn(h)$. Substituting the right-hand side of \eqref{po:limit:1} with \eqref{po:limit:2} and \eqref{po:limit:3}, we have
\beee\nn
\tau_{u^*} &=& \int_{\R^{d_X}}\frac{f(x)}{\E\big\{k_{u^*}(U)\big\}}\Delta_{u^*}(x)e_{u^* }(x)\int_{\R^2}\mathcal{K}(\delta_1',\delta_0')d\delta'_1d\delta'_0dx
\\\nn
&&+h\cdot\Bigg\{\int_{\R^{d_X}}\frac{f(x)}{\E\big\{k_{u^*}(U)\big\}}\Delta_{u^*}(x)\frac{\partial e_u(x)}{\partial m_1}\Big|_{u = u^*}\underbrace{\int_{\R^2}\delta_1'\mathcal{K}(\delta_1',\delta_0')d\delta'_1d\delta'_0}_{ = 0}dx +\cdots
\\\nn
&&-\int_{\R^{d_X}}\frac{f(x)}{\E\big\{k_{u^*}(U)\big\}}e_{u^*}(x)\delta_0'\frac{\partial\mu_{0}(x,m_0)}{\partial m_0}\Big|_{m_0 = m_0^*}\underbrace{\int_{\R^2}\delta_0'\mathcal{K}(\delta_1',\delta_0')d\delta'_1d\delta'_0}_{ = 0}dx\Bigg\}
\\\nn
&&+\Bigg\{h^2\int_{\R^{d_X}}\frac{f(x)}{\E\big\{k_{u^*}(U)\big\}}\Big\{\frac{\partial e_u(x)}{\partial m_1}\frac{\partial\mu_{1}(x,m_1)}{\partial m_1}\Big\}\Big|_{u =u^*}\int_{\R^2}(\delta_1')^2\mathcal{K}(\delta_1',\delta_0')d\delta'_1d\delta'_0dx + \cdots
\\\nn
&& - \frac{h^4}{4}\int_{\R^{d_X}}\frac{f(x)}{\E\big\{k_{u^*}(U)\big\}}\int_{\R^2}\frac{\partial^2\mu_{0}(x,m_0)}{\partial^2 m_0}\Big|_{m_0 =m_0^* + h_0\delta_0'}\frac{\partial^2 e_u(x)}{\partial m_1\partial m_0}\Big|_{u = u^* + h_{10}\delta'}\delta_1'(\delta_0')^3\mathcal{K}(\delta_1',\delta_0')d\delta'_1d\delta'_0dx\Bigg\}
\\\nn
&=&\int_{\R^{d_X}}\frac{f(x)}{\E\big\{k_{u^*}(U)\big\}}\Delta_{u^*}(x)e_{u^* }(x)dx + h R_1 + h^2R_2
\\\label{po:go:6}
& =& \int_{\R^{d_X}}\frac{f(x)}{\E\big\{k_{u^*}(U)\big\}}\Delta_{u^*}(x)e_{u^*}(x)dx + \mathcal{O}(h^2),
\eee
where the last equality is because that, by the regularity conditions and the fact that $[\E\{k_{u^*}(U)\}]^{-1} = \mathcal{O}(1)$ (see \eqref{xikubounds} in the following), we can derive $R_1 = 0$ and $R_2 = \mathcal{O}(1)$ when $h\rightarrow 0$ with standard arguments for controlling the bias of the kernel nonparametric regression \citep{fan1993local}. 
\par
Next, we show $|1/\E\{k_{u^*}(U)\} - 1/e_{u^*}| = \mathcal{O}(h^2)$. By the Taylor expansion, one has
\beee\label{po:limit:5}
\Big|\frac{1}{\E\{k_{u^*}(U)\}} - \frac{1}{e_{u^*}}\Big| &= &\frac{1}{\xi^2}\big|\E\{k_{u^*}(U)\} - e_{u^*}\big|
\eee
for some $\xi$ between $\E\{k_{u^*}(U)\}$ and $e_{u^*}$. Similar to \eqref{po:limit:1}--\eqref{po:limit:3}, one has
\beee\nn
\E\{k_{u^*}(U)\} & = & \E\big[\E\{k_{u^*}(U)\mid X\}\big]
\\\nn
&=& \int_{\R^{d_X}}f(x)\int_{\R^2}\frac{1}{h^2}\mathcal{K}\Big(\frac{m_1 - m_1^*}{h},\frac{m_0 - m_0^*}{h}\Big)e_u(x)dm_1dm_0dx
\\\nn
&=&\int_{\R^{d_X}}f(x)\int_{\R^2}\mathcal{K}\big(\delta_1',\delta_0'\big)e_{u^* + h\delta'}(x)d\delta_1'd\delta_0'dx
\\\nn
&=&\int_{\R^{d_X}}f(x)e_{u^*}(x)\int_{\R^2}\mathcal{K}\big(\delta_1',\delta_0'\big)d\delta_1'd\delta_0'dx
\\\label{s10000}
&&+h\cdot\Bigg\{\int_{\R^{d_X}}f(x)\frac{\partial e_u(x)}{\partial m_1}\Big|_{u = u^*}\underbrace{\int_{\R^2}\delta_1'\mathcal{K}\big(\delta_1',\delta_0'\big)d\delta_1'd\delta_0'dx}_{=0} 
\\\nn
&&+ \int_{\R^{d_X}}f(x)\frac{\partial e_u(x)}{\partial m_0}\Big|_{u = u^*}\underbrace{\int_{\R^2}\delta_0'\mathcal{K}\big(\delta_1',\delta_0'\big)d\delta_1'd\delta_0'dx}_{=0}\Bigg\}
\\\nn
&& + h^2\cdot\Bigg\{\int_{\R^{d_X}}\frac{f(x)}{2}\int_{\R^2}\frac{\partial^2 e_u(x)}{\partial^2 m_1}\Big|_{u = u^* + h_{10}\delta'}(\delta_1')^2\mathcal{K}\big(\delta_1',\delta_0'\big)d\delta_1'd\delta_0'dx + \cdots
\\\nn
&&+\int_{\R^{d_X}}\frac{f(x)}{2}\int_{\R^2}\frac{\partial^2 e_u(x)}{\partial m_1\partial m_0}\Big|_{u = u^* + h_{10}\delta'}\delta_1'\delta_0'\mathcal{K}\big(\delta_1',\delta_0'\big)d\delta_1'd\delta_0'dx\Bigg\}
\\\nn
&=&\int_{\R^{d_X}}f(x)e_{u^*}(x)dx + \mathcal{O}(h^2)
\\\label{euh}
&=&e_{u^*} + \mathcal{O}(h^2).
\eee
Thus $|\E\{k_{u^*}(U)\} - e_{u^*}| = \mathcal{O}(h^2) = o(1)$ when $h\rightarrow 0$, which implies $\xi \rightarrow e_{u^*}$. Thus when $h\rightarrow 0$, we have,
\beee\label{xikubounds}
\xi^{-1}\precsim 1 &\text{ and  } &[\E\{k_{u^*}(U)\}]^{-1}\precsim 1,
\eee
as $e_{u^*} > 0$ is fixed. Then \eqref{po:limit:5}--\eqref{xikubounds}  imply,
\beee\label{po:go:7}
\Big|\frac{1}{\E\{k_{u^*}(U)\}} - \frac{1}{e_{u^*}}\Big| &=& \mathcal{O}(h^2).
\eee
\par
Recalling \eqref{po:go:6}, one has
\beee\nn
\tau_{u^*} 
&=& \int_{\R^{d_X}}\Delta_{u^*}(x)\{e_{u^* }(x)/e_{u^*}\}f(x)dx
  + \Big[\frac{1}{\E\big\{k_{u^*}(U)\big\}} - \frac{1}{e_{u^*}}\Big]\int_{\R^{d_X}}\Delta_{u^*}(x)e_{u^* }(x)f(x)dx  + \mathcal{O}(h^2)
\\\nn
&=&  \int_{\R^{d_X}}\Delta_{u^*}(x)f(X = x\mid U  = u^*)dx  + \underbrace{\Big[\frac{1}{\E\big\{k_{u^*}(U)\big\}} - \frac{1}{e_{u^*}}\Big]e_{u^*}\int_{\R^{d_X}}\Delta_{u^*}(x)f(X = x\mid U  = u^*)dx}_{= R_3}
\\\label{po:go:10}
&& + \mathcal{O}(h^2),
\eee
where we note $\{e_{u^*}(x)/e_{u^*}\}f(x) = f\{(M_1,M_0) = u^*\mid X = x\}f(X = x)/f\{(M_1,M_0) = u^*\} = f\{X = x\mid U  = u^*\}$. Furthermore, under Assumptions \ref{assump:TAignorability}--\ref{assump:weak-pi}, one has
\beee\nn
\int_{\R^{d_X}}\Delta_{u^*}(x)f\{X = x\mid (M_1,M_0)  = u^*\}dx  & = & \E\big\{\mu_{1}(X,m^*_1) - \mu_{0}(X,m^*_0)\mid U = u^*\big\}
\\\label{heheh}
& = &\tau_{u^*}^*,
\eee
by an argument same as \eqref{id1112} in the Proof of Theorem \ref{thm:iddd}. Combining  \eqref{po:go:7}--\eqref{heheh}, one has
\beee\nn
|R_3| &\leq& \Big|\frac{1}{\E\big\{k_{u^*}(U)\big\}} - \frac{1}{e_{u^*}}\Big|\cdot e_{u^*}\int_{\R^{d_X}}\Delta_{u^*}(x)f\{X = x\mid U  = u^*\}dx
\\\nn
& = &\Big|\frac{1}{\E\big\{k_{u^*}(U)\big\}} - \frac{1}{e_{u^*}}\Big|\cdot e_{u^*}\tau_{u^*}^*
\\\label{po:go:11}
& = &\mathcal{O}(h^2).
\eee
Combining \eqref{po:go:10}--\eqref{po:go:11}, we  have $\tau_{u^*} = \tau_{u^*}^* + \mathcal{O}(h^2)$ and thus
$
\big|\tau_{u^*} - \tau_{u^*}^*\big| = \mathcal{O}(h^2)
$ as desired.

\qed
\section{Proofs of main theorems}
Throughout the proofs, we will use $f(\cdot)$ to denote the probability
density function for continuous random variables.  We will use the law of
total expectation repeatedly and will mark the steps as $\LOTE$. We first recall some notations in the main text, 
\[
\begin{array}{rclrcl}
\pi_z(X) & = & \pr(Z=z\mid X), & e_{u}(X) & = & f(U=u\mid X),
\\
e_{u} & = & \E\{e_{u}(X)\}= f(U=u),  & \mu_{z}(X,m) & = & \E(Y\mid X,Z=z,M=m),\\
f_{zm}(X) & = & f(M=m\mid X,Z=z), & F_{zm}( X) & = & \pr(M\leq m\mid X, Z = z),
\end{array}
\]
for $z=0,1$, $m\in \R$ and $u=(m_{1},m_{0})$. Under Assumptions~\ref{assump:TAignorability}~and \ref{fm1m0x}, we have 
\begin{eqnarray*}
e_{u}(X) & = & c_\rho\big\{F(M_1 = m_1\mid X), F(M_0 = m_0\mid X)\big\}\prod_{z\in\{0,1\}}f(M_z = m_z\mid X)
\\
& = & c_\rho\big\{F_{1m_1}(X), F_{0m_0}(X)\big\}\prod_{z\in\{0,1\}}f_{zm_z}( X).
\end{eqnarray*}
%

\label{app:balancing} \QEDB
\allowdisplaybreaks
\subsection{Proof of Theorem \ref{thm:iddd}}
We first introduce the following lemma from \citet[Lemma S3]{jiang2020multiply}.
\begin{lemma}\label{ab}
Let $A$ and $B$ be two random variables with densities $f_A(\cdot)$ and $f_B(\cdot)$. For any function $h(\cdot)$ with $\E\{h(A)\} < \infty$, we have
\beee\nn
\E\{h(A)\} &=& \E\Bigg\{\frac{f_A(B)}{f_B(B)}h(B)\Bigg\}.
\eee
\end{lemma}
 By the law of the expectation (LOTE), we have
\begin{eqnarray}\nonumber
\tau^*_{u^*} & = & \E\Big(Y_1 - Y_0\mid U = u^*\Big)
\\\nonumber
& = & \E\Big\{\E\big(Y_1 - Y_0\mid X, U = u^*\big)\mid U = u^*\Big\}
\\\nonumber
& = & \E\Big[\E\{Y_1\mid X, U = u^*\} - \E\{Y_0\mid X, U = u^*\}\mid U  = u^*\Big] 
\\\nonumber
& = & \E\Big[\E\{Y_1\mid X, M_1 = m_1^*\} - \E\{Y_0\mid X, M_0 = m_0^*\}\mid U  = u^*\Big] \quad\text{(Assumption \ref{assump:weak-pi})}
\\\nonumber
& = & \E\Big[\E\{Y_1\mid X,Z = 1, M_1 = m_1^*\} - \E\{Y_0\mid X,Z = 0, M_0 = m_0^*\}\mid U  = u^*\Big] \quad\text{(Assumption \ref{assump:TAignorability})}
\\\nonumber
& = & \E\Big\{\E(Y\mid X,Z = 1, M = m_1^*) - \E(Y\mid X,Z = 0, M = m_0^*)\mid U  = u^*\Big\}
\\\label{id1112}
& = &\E\big\{\mu_{1}(X,m^*_1) - \mu_{0}(X,m^*_0)\mid U = u^*\big\}
\\\label{id111}
& = & \E\Big[\frac{f(X\mid U = u^*)}{f(X)}\{\mu_{1}(X,m^*_1) - \mu_{0}(X,m^*_0)\}\Big]
\\\nn
& = & \E\Big[\frac{f( U = u^*\mid X)f(X)}{f(U = u^*)f(X)}\{\mu_{1}(X,m^*_1) - \mu_{0}(X,m^*_0)\}\Big]
\\\nn
& = & \E\Big[\frac{e_{u^*}(X)}{e_{u^*}}\{\mu_{1}(X,m^*_1) - \mu_{0}(X,m^*_0)\}\Big],
\end{eqnarray}
where in \eqref{id111} we use Lemma \ref{ab} with $A = X\mid U = u^*$ and $B = X$. By Assumptions \ref{assump:TAignorability} and \ref{am:cp},  $e_{u^*}(x)$ can be further identified as 
\begin{eqnarray}\nn
e_{u^*}(X) &=& f(M_{1} = m^*_1,M_{0} = m^*_0\mid X ) 
\\\nn
&=&c_\rho\big\{F(M_1 = m^*_1\mid X ), F(M_0 = m^*_0\mid X)\big\}\prod_{z\in\{0,1\}}f(M_z =m^*_z\mid X)
\\\nn
&=&c_\rho\big\{F(M_1 = m^*_1\mid X, Z = 1), F(M_0 = m^*_0\mid X,Z= 0)\big\}\prod_{z\in\{0,1\}}f(M_z =m^*_z\mid X,Z = z)
\\\nn
&&\quad\quad\quad\quad\quad\quad\quad\quad\quad\quad\quad\quad\quad\quad\quad\quad\quad\quad\quad\quad\quad\quad\quad\quad\quad\quad\quad\quad\,\,\,\,\,\text{(Assumption \ref{assump:TAignorability})}
\\\label{id:euxss}
& = & c_\rho\big\{F_{1m^*_1}(X), F_{0m^*_0}(X)\big\}\prod_{z\in\{0,1\}}f_{zm^*_z}(X).
\end{eqnarray}
\QEDB

\subsection{Proof of Theorem \ref{thm:idd}}
 By the law of the expectation (LOTE), we have
\begin{eqnarray}\nonumber
\tau_{u^*} & = & \E\Big[\big\{\mu_{1}(X,M_1) - \mu_{0}(X,M_0)\big\}\frac{k_{u^*}(U)}{\E \{k_{u^*}(U)\}}\Big]
\\\nonumber
& = & \frac{1}{\E \{k_{u^*}(U)\}}\cdot\E\Big[\E\Big[\big\{\mu_{1}(X,M_1) - \mu_{1}(X,M_0)\big\}k_{u^*}(U)\mid X\Big]\Big]
\\\label{id:res}
& = & \frac{1}{\E \{k_{u^*}(U)\}}\cdot\E\Bigg[\int_{\R^2}\big\{\mu_{1}( X,m_1) - \mu_{0}(X,m_0)\big\}k_{u^*}(u)e_{u}(X)du\Bigg],
\end{eqnarray}
We also have 
$\E\{k_{u^*}(U)\} = \E[\E\{k_{u^*}(U) \mid X\}]
=\E\{\int_{\R^2}k_{u^*}(u)e_{u}(X)du\}.
$
Combining above results and recalling the definition of $\lbt\,\cdot \,\rbt$ in \eqref{def:lu}, the identification formula follows, and $e_u(X)$ can be identified same as \eqref{id:euxss}.

\QEDB

\subsection{Proof of Theorem \ref{eif:tau}}
\subsubsection{Preliminary and auxiliary lemmas}
To obtain the EIFs, we consider some arbitrary one-dimensional parametric submodel of the whole data density, denoted by $\{f_\theta(V = v)\}_{\theta\in\mathbb{R}}$. Let $f_\theta(v)$ contain the truth $f(v)$ at $\theta = 0$, i.e., $f_0(v) = f(v)$. We use $s_\theta(\cdot)$ and $s(\cdot)$ to denote the corresponding score functions, e.g., $s_\theta(x) = \partial\log f_\theta(X = x)/\partial \theta$ and $s(x) = \partial\log f_\theta(X = x)/\partial \theta \mid _{\theta = 0}$. The submodel likelihood is factorized as
\beee\nn
&&f_{\theta}(V = v)
\\\nn
& =& f_{\theta}(X = x)f_{\theta}(Z = z\mid X = x)f_{\theta}(M = m\mid X = x,Z = z)f_{\theta}(Y = y\mid X = x,Z = z,M = m).\\
\label{np:decome}
\eee 
\par
Under the scenario of nonparametric estimation, the score function under submodel can be decomposed as
\beee\nn
s_\theta(v)& = &s_\theta(x) + s_\theta(z\mid x) + s_\theta(m\mid x,z) + s_\theta(y\mid x,z,m).
\eee
 Then the tangent space is 
\beee\label{tangent:1}
\Lambda&=&H_{1} \oplus H_{2} \oplus H_{3} \oplus H_{4}
\eee
where $\oplus$ is the set sum and,
\begin{equation}\nn
\begin{aligned}
H_{1} &=\{h(x)\mid  \mathbb{E}\{h(X)\}=0\}, \\
H_{2} &=\{h(x,z)\mid  \mathbb{E}\{h(X,Z) \mid X\}=0\}, \\
H_{3} &=\{h(x,z,m)\mid  \mathbb{E}\{h(X,Z,M) \mid X,Z\}=0\}, \\
H_{4} &=\{h(x,z,m,y)\mid  \mathbb{E}\{h(X,Z,M,Y) \mid X,Z,M\}=0\}.
\end{aligned}
\end{equation}
 Based on the semiparametric theory (see e.g. \citet{bickel1993efficient,jiang2020multiply}), we have that under the nonparametric scenario (Scenario (iii)) with a tangent space like \eqref{tangent:1}, EIF of target parameter $\tau_{u^*}$ is the unique function $\phi(\cdot)$ (up to the almost surely equivalence) satisfying,  
\begin{eqnarray}\label{EIF:def}
\frac{\partial}{\partial \theta}\tau_{u^*,\theta}\,\Big|_{\theta = 0}& =& \E \Big\{\phi(V)\cdot s(V)\Big\}. 
\end{eqnarray}
Under the scenario of semiparametric estimation with known or parametric $\mathcal{M}_{{\rm ps}}$ (Scenarios (i) and (ii)), due to the assumption that the distribution (or the parametric distribution model) of $M\mid X,T$ is already known, the corresponding tangent space is more restrictive in comparison with \eqref{tangent:1}. As a result, the $\phi(\cdot)$ satisfying \eqref{EIF:def} is no longer unique. Any $\phi(\cdot)$ satisfying \eqref{EIF:def} serves as an influence function (IF) of $\tau_{u^*}$, but is not necessary to be an EIF. To  show that  the $\phi(\cdot)$ we derive under  Scenarios (i) or (ii) are indeed EIFs, we further verify that the derived $\phi(\cdot)$ belong to the corresponding semiparametric tangent spaces; Technical details are deferred to Section \ref{semieif}. Different from the order of Theorem \ref{eif:tau}, during the proof, we first derive $\phi_{{\rm np}}(x,z,m,y)$ under Scenario (iii) in Section \ref{noneif}, based on which we further derive the  EIFs under Scenarios  (ii) and (i) in Sections \ref{noneif2} and \ref{semieif}, respectively. 
\par
We list  two technical lemmas that will be used in the main proof.
\par
\begin{lemma}\label{lm:MonXZ}
Suppose function $\xi=\xi(A_1,A_2)$, where $A_1,A_2$ are random variables belonging to $V$, and $A_1,A_2$ have no overlapping. We have
\beee\nn
\E\big\{\xi(A_1,A_2) \cdot s(A_1|A_2)\big\} &=& \E\big[\big[\xi(A_1,A_2) - \E\{\xi(A_1,A_2)\mid A_2\}\big] \cdot s(V)\big].
\eee
\end{lemma}
\textit{Proof of Lemma \ref{lm:MonXZ}.} By property of score function,  we have $s(A_1\mid A_2) = s(A_1,A_2) - s(A_2)$. Then
\begin{eqnarray}\label{lm:MonXZ:main}
\E\Big\{\xi(A_1,A_2) \cdot s(A_1\mid A_2)\Big\}  & = & \E\Big\{\xi(A_1,A_2) \cdot s(A_1,A_2) \Big\} - \E\Big\{\xi(A_1,A_2) \cdot s(A_2) \Big\}.
\end{eqnarray}
\par
If $(A_1,A_2)$ do not contain all the covariates in $V$, we let $A_3$ be the rest of the covariates in $V$ such that $A_3 = V\setminus(A_1,A_2)$. For the first term of \eqref{lm:MonXZ:main}, we note that 
\begin{eqnarray}\label{lms2:1}
\E\Big\{\xi(A_1,A_2) \cdot s(A_3\mid A_1,A_2) \Big\} &=& \E\Big[\E\big\{\xi(A_1,A_2) \cdot s(A_3\mid A_1,A_2) \mid A_1,A_2\big\}\Big]
\\\nonumber
& = &\E\Big[\xi(A_1,A_2) \cdot\E\big\{ s(A_3\mid A_1,A_2) \mid A_1,A_2\big\}\Big]
\\\nonumber
& = & \E\Big\{\xi(A_1,A_2)\cdot 0\Big\}
\\\nonumber
& = &0,
\end{eqnarray}
where the third equality is by the property of score function as $\E\{s(A_3\mid A_1,A_2)\mid A_1,A_2\} = 0$. Thus we have 
\begin{eqnarray}\nonumber
\E\Big\{\xi(A_1,A_2) \cdot s(A_1,A_2) \Big\} & = & \E\Big\{\xi(A_1,A_2) \cdot s(A_1,A_2) \Big\} + \E\Big\{\xi(A_1,A_2) \cdot s(A_3\mid A_1,A_2) \Big\} 
\\\nonumber
& = & \E\Big[\xi(A_1,A_2) \cdot\big\{ s(A_1,A_2) + s(A_3\mid A_1,A_2)\big\}\Big]
\\\label{lm:MonXZ:main:2}
& = & \E\Big\{\xi(A_1,A_2)\cdot s(V)\Big\},
\end{eqnarray}
where the first equality is by \eqref{lms2:1}. If $(A_1,A_2)$ contain all covariates in $V$, then $\E\big\{\xi(A_1,A_2) \cdot s(A_1,A_2) \big\} = \E\big\{\xi\cdot s(V)\big\}$ trivially.
\par
For the second term of \eqref{lm:MonXZ:main}, we have
\begin{eqnarray}\nonumber
\E\Big\{\xi(A_1,A_2) \cdot s(A_2) \Big\}
& = & \E\Big[s(A_2)\cdot\E\big\{\xi(A_1,A_2) \mid A_2\big\}\Big]\quad\text{(LOTE)}
\\\nonumber
& = & \E\Big[s(A_2)\cdot\E\big\{\xi(A_1,A_2) \mid A_2\big\}\Big] + \E\Big[s(A_1,A_3\mid A_2)\cdot\E\big\{\xi(A_1,A_2) \mid A_2 \big\}\Big]
\\\nonumber
& = &\E\Big[\big\{s(A_2) + s(A_1,A_3\mid A_2)\big\}\cdot \E\big(\xi \mid A_2 \big)\Big]
\\\label{lm:MonXZ:main:3}
& = & \E\Big\{\E\big(\xi \mid A_2\big)\cdot s(V)\Big\},
\end{eqnarray}
where the second equality holds because 
\begin{eqnarray*}
\E\Big[s(A_1,A_3\mid A_2)\cdot\E\big\{\xi(A_1,A_2) \mid A_2 \big\}\Big] & = &\E\Big[\E\Big[s(A_1,A_3\mid A_2)\cdot\E\big\{\xi(A_1,A_2) \mid A_2 \big\}\mid A_2\Big]\Big]
\\
& = & \E\Big[\E\big\{\xi(A_1,A_2) \mid A_2  \big\}\cdot\E\Big\{s(A_1,A_3\mid A_2)\mid A_2\Big\}\Big]
\\
& = & 0,
\end{eqnarray*}
by the property of score function as $\E\big\{s(A_1,A_3\mid A_2)\mid A_2\big\} = 0$. Note here if $(A_1,A_2)$ contain all the covariates in $V$, $A_3$ can be empty. Combing \eqref{lm:MonXZ:main}, \eqref{lm:MonXZ:main:2}, \eqref{lm:MonXZ:main:3}, the desired result follows immediately.
\par
\qed
\begin{lemma}\label{lm:dd}
Suppose $N_\theta$ and $D_{\theta}$ are differentiable with respect to  $\theta\in \R$. If $\partial N_\theta/\partial \theta\mid_{\theta = 0} = \E\{\phi_N(V)s(V)\}$ and $\partial D_\theta/\partial \theta\mid_{\theta = 0} = \E\{\phi_D(V)s(V)\}$, then
\bee
\frac{\partial}{\partial\theta}\frac{N_\theta}{D_\theta}\ \Big|_{\theta = 0} = \E\Bigg[\Big\{\frac{1}{D_0}\phi_N(V) - \frac{N_0}{D^2_0}\phi_D(V)\Big\}\cdot s(V)\Bigg].
\ee
\end{lemma}
\textit{Proof of Lemma \ref{lm:dd}.} 
See Lemma S2 in \citet{jiang2020multiply}.\qed

\subsubsection{Derivation of  $\phi_{{\rm np}}(x,z,m,y)$ under Scenario (iii)}\label{noneif}
For the nonparametric scenario, by verifying condition \eqref{EIF:def}, we  derive the EIFs of the denominator and numerator of $\tau_{u^*}$'s identification formula in \eqref{tau:id}, respectively. We then combine these EIFs together and derive the desired EIF of $\tau_{u^*}$ based on Lemma \ref{lm:dd}.
\par
\
\par
\noindent\textbf{Step 1 (EIF of $\E\{\int_{\R^2}k_{u^*}(u)e_{u}(X)du\}$):} Based on the submodel introduced before, we can write $\E\{\int_{\R^2}k_{u^*}(u)e_{u}(X)du\}$ as 
\begin{eqnarray}\label{denom:main}
\E\Big\{\int_{\R^2}k_{u^*}(u)e_{u}(X)du\Big\}&=& \E_{\theta}\Big\{\int_{\R^2}k_{u^*}(u)e^{(\theta)}_u(X)du\Big\}\ \Big|_{\theta = 0}.
\end{eqnarray}
 In \eqref{denom:main}, obtain the derivative with respect to $\theta$,
\begin{eqnarray*}
\frac{\partial}{\partial \theta}\E_{\theta}\Big\{\int_{\R^2}k_{u^*}(u)e^{(\theta)}_u(X)du\Big\} \Big|_{\theta = 0} &=& \E_{\theta}\Big\{\int_{\R^2}k_{u^*}(u)\frac{\partial}{\partial \theta}e^{(\theta)}_u(X)\ \Big|_{\theta = 0}du\Big\} 
\\\nn
&&+ \E\Big\{\int_{\R^2}k_{u^*}(u)e_{u}(X)du\cdot s(X)\Big\}.
\end{eqnarray*}
Now we take a careful look at $\partial e^{(\theta)}_u(x)/\partial \theta \mid_{\theta = 0}$. By definition \eqref{fm1m0x}, we have
\begin{eqnarray*}
\frac{\partial}{\partial \theta}e^{(\theta)}_u(x)\ \Big|_{\theta = 0} & = & \frac{\partial}{\partial \theta}\Big[c_\rho\Big\{F^{(\theta)}_{1m_1}(x), F^{(\theta)}_{0m_0}(x)\Big\}\prod_{z\in\{0,1\}}f_{zm_z}^{(\theta)}(x)\Big] \ \Big|_{\theta = 0}
\\
& = & \sum_{z' = 1,0} c_{\rho}^{(z')}\Big\{F_{1m_1}(x), F_{0m_0}(x)\Big\}\Big\{\prod_{z\in\{0,1\}}f_{zm_z}(x)\Big\}\cdot \frac{\partial}{\partial \theta}F^{(\theta)}_{z'm_{z'}}(x)  \ \Big|_{\theta = 0}
\\
&& +\sum_{z' = 1,0} c_{\rho}\{F_{1m_1}(x), F_{0m_0}(x)\}\cdot \frac{\partial}{\partial \theta}f_{z'm_{z'}}^{(\theta)}(x)\ \Big|_{\theta = 0}\cdot f_{1-z'm_{1 - z'}}(x).
\end{eqnarray*}
For $z' = 0,1$, since $F_{z'm_{z'}}^{(\theta)}(x) = \int_{-\infty}^{m_{z'}}f^{(\theta)}(M_{z'} = m \mid X = x) dm$, we have
\begin{eqnarray*}
&&\frac{\partial}{\partial \theta}F_{z'm_{z'}}^{(\theta)}(x) \ \Big|_{\theta = 0} 
\\
& =& \int_{\R} \mathbb{I}(m \leq m_{z'})\frac{\partial f^{(\theta)}(M_{z'} = m \mid X = x)}{\partial \theta}\ \Big|_{\theta = 0}dm
\\
&=& \int_{\R} \mathbb{I}(m \leq m_{z'})\frac{1}{f(M_{z'} = m \mid X = x)}\frac{\partial f^{(\theta)}(M_{z'} = m \mid X = x)}{\partial \theta}\ \Big|_{\theta = 0} f(M_{z'} = m \mid X = x)dm
\\
&=& \int_{\R} \mathbb{I}(m \leq m_{z'})\frac{1}{f(M = m \mid X, Z = z')}\frac{\partial f^{(\theta)}(M = m \mid X,Z= z')}{\partial \theta}\ \Big|_{\theta = 0} f(M = m \mid X = x, Z = z')dm
\\
& = &  \E\Big\{  \mathbb{I}(M \leq m_{z'})s(M\mid X, Z) \mid X, Z = z'\Big\},
\end{eqnarray*}
where the third equality is by $f(M_{z'} = m\mid X = x) = f(M = m\mid X = x,Z = z')$ under Assumption \ref{assump:TAignorability}. And similarly, 
\begin{eqnarray*}
&&\frac{\partial}{\partial \theta}f^{(\theta)}(M_{z'} = m_{z'}\mid X = x)\ \Big|_{\theta = 0} 
\\
& = &  \frac{1}{f(M = m_{z'}\mid X = x, Z = z')}\Bigg\{\frac{\partial}{\partial \theta}f^{(\theta)}(M = m_{z'}\mid X = x, Z = z')\ \Big|_{\theta = 0}\Bigg\}
 f(M = m_{z'}\mid X = x, Z = z')
\\
&=& s(M = m_{z'}\mid X = x,Z = z') f(M = m_{z'}\mid X = x, Z = z').
\end{eqnarray*}
\par
Combining the results above, we have
\begin{eqnarray}\nonumber
&&\frac{\partial}{\partial \theta}\E_{\theta}\Big\{\int_{\R^2}k_{u^*}(u)e^{(\theta)}_u(X)du\Big\} \Big|_{\theta = 0}
\\\nonumber
& = & \sum_{z' = 1,0} \E_{\theta}\Big[\int_{\R^2}k_{u^*}(u) c_{\rho}^{(z')}\big\{F_{1m_1}( X), F_{0m_0}( X)\big\}\Big\{\prod_{z\in\{0,1\}}f_{zm_z}( X)\Big\}\cdot \frac{\partial}{\partial \theta}F^{(\theta)}_{z'm_{z'}}(X) \ \Big|_{\theta = 0}du\Big] 
\\\nonumber
&& + \sum_{z' = 1,0}\E_{\theta}\Big[\int_{\R^2}k_{u^*}(u)c_{\rho}\{F_{1m_1}( X), F_{0m_0}( X)\}\cdot \frac{\partial}{\partial \theta}f_{z'm_{z'}}^{(\theta)}( X)\ \Big|_{\theta = 0}\cdot f_{1-z'm_{1 - z'}}( X)du\Big]
\\\nonumber
&&+ \E\Big\{\int_{\R^2}k_{u^*}(u)e_{u}(X)du\cdot s(X)\Big\}
\\\nonumber
&=& \sum_{z' = 1,0}\E\Big[ \int_{\R^2}k_{u^*}(u)c_{\rho}^{(z')}\big\{F_{1m_1}( X), F_{0m_0}(X)\big\}\big\{\prod_{z\in\{0,1\}}f_{zm_z}( X)\big\}
\\\nonumber
&&\cdot \E\big\{  \mathbb{I}(M \leq m_{z'})s(M\mid X, Z = z') \mid X, Z = z'\big\}du\Big]
\\\nonumber
&& + \sum_{z' = 1,0}\E\Big[\int_{\R^2}k_{u^*}(u)c_{\rho}\{F_{1m_1}( X), F_{0m_0}( X)\}\big\{\prod_{z\in\{0,1\}}f_{zm_z}(X)\big\} s(M = m_{z'}\mid X,Z = z')du\Big]
\\\label{main:thm1:low}
&&+\E\Big\{\int_{\R^2}k_{u^*}(u)e_{u}(X)du\cdot s(X)\Big\}.
\end{eqnarray}
By LOTE and Fubini's theorem, the first term on the right-hand side of \eqref{main:thm1:low} equals to
\beee\nn
&&\sum_{z' = 1,0}\E\Bigg[\E\Big[s(M\mid X, Z = z')\int_{\R^2}1(M \leq m_{z'})k_{u^*}(u)c_{\rho}^{(z')}\big\{F_{1m_1}(X), F_{0m_0}( X)\big\}
\\\nn
&&\cdot\prod_{z\in\{0,1\}}f_{zm_z}(X)  du\mid X, Z = z'\Big]\Bigg]
\\\nn
&=&\sum_{z' = 1,0}\E\Bigg[\E\Big[ \frac{1(Z = z')s(M\mid X, Z = z')}{\pi_{z'}(X)}
\\\nn
&&\cdot\int_{\R^2}1(M \leq m_{z'})k_{u^*}(u)c_{\rho}^{(z')}\big\{F_{1m_1}(X), F_{0m_0}( X)\big\}
\prod_{z\in\{0,1\}}f_{zm_z}(X)du\mid X\Big]\Bigg]
\\\nn
&=&\E\Bigg[\sum_{z' = 1,0}\frac{1(Z = z')s(M\mid X, Z = z')}{\pi_{z'}(X)}\M \int_{\R^2}1(M \leq m_{z'})k_{u^*}(u)c_{\rho}^{(z')}\big\{F_{1m_1}(X), F_{0m_0}( X)\big\}
\prod_{z\in\{0,1\}}f_{zm_z}(X)du\Bigg]
\\\nn
&=&\E\Big[s(M\mid X, Z)\cdot\frac{1}{\pi_{Z}(X)}\int_{\R^2}1(M \leq m_{Z})k_{u^*}(u)c_{\rho}^{(Z)}\big\{F_{1m_1}(X), F_{0m_0}( X)\big\}
\prod_{z\in\{0,1\}}f_{zm_z}(X)du\Big]
\\\label{eif:main:low:1}
&=&\E\Big\{\tilde{\xi}'_{2}(X,Z,M)\cdot s(V)\Big\} \quad (\text{By Lemma \ref{lm:MonXZ} with $A_1 = M,$ $A_2 = (X,Z)$}),
\eee
recalling 
\begin{eqnarray*}
\xi'_2(X,Z,M) & = & \frac{1}{\pi_{Z}(X)}\int_{\R^2}1(M \leq m_{Z})k_{u^*}(u)c_{\rho}^{(Z)}\big\{F_{1m_1}(X), F_{0m_0}( X)\big\}
\prod_{z\in\{0,1\}}f_{zm_z}(X)du,
\\
\tilde{\xi}'_2(X,Z,M) &= & \xi_2(X,Z,M) - \E\big\{\xi_2(X,Z,M) \mid X,Z\big\}.
\end{eqnarray*}
\par
Note here and in the following, for simplicity, we will omit the footnote $u^*$ in $\xi_{2,u^*}(x,z,m)$, $\xi_{2,u^*}'(x,z,m)$, $\tilde{\xi}_{2,u^*}'(z,x,m)$, etc. By definition, for both $z' = 0$ or $1$ we have
\beee\nn
{\gamma'_2}^{(z')}(M = m_{z'},X) &= &\int_{\R}k_{u^*}(u)c_{\rho}\{F_{1m_1}(X), F_{0m_0}( X)\}f_{1 - z'm_{1 - z'}}( X)dm_{1 - z'}.
\eee
The second term on the right-hand side of \eqref{main:thm1:low} equals to
\begin{eqnarray}\nonumber
&&\sum_{z' = 1,0}\E\Big[\int_{\R^2}k_{u^*}(u)c_{\rho}\{F_{1m_1}(X), F_{0m_0}( X)\}\prod_{z = 0,1}f_{zm_z}(X) \cdot s(M = m_{z'}\mid X,Z = z') du\Big]
\\\nonumber
&=&\sum_{z' = 1,0}\E\Big[\int_{\R}\int_{\R}k_{u^*}(u)c_{\rho}\{F_{1m_1}(X), F_{0m_0}( X)\}f_{1- z'm_{1 - z'}}(X)dm_{1 - z'}
\\\nonumber
&&\cdot s(M = m_{z'}\mid X,Z = z') f(M = m_{z'}\mid X, Z = z') dm_{z'}\Big]
\\\nonumber
&=&\sum_{z' = 1,0}\E\Big\{\int_{\R}{\gamma'_2}^{(z')}(X,M = m_{z'}) s(M = m_{z'}\mid X,Z = z') f(M = m_{z'}\mid X, Z = z') dm_{z'}\Big\}
\\\nonumber
&=&\sum_{z' = 1,0}\E\Big[\E\Big\{{\gamma'_2}^{(z')}(X,M)s(M\mid X,Z = z')\mid X,Z = z'\Big\}\Big]
\\\nonumber
&=&\sum_{z' = 1,0}\E\Big[\E\Big\{\frac{1(Z = z')}{\pi_{z'}(X)}{\gamma'_2}^{(z')}(X,M)s(M\mid X,Z )\mid X\Big\}\Big]
\\\nonumber
& = &\E\Big[\Big\{\sum_{z' = 1,0}\frac{1(Z = z')}{\pi_{z'}(X)}{\gamma'_2}^{(z')}(X,M)\Big\} s(M\mid X,Z )\Big]\quad\text{(LOTE)}
\\\nonumber
& = & \E\Big\{\frac{1}{\pi_{Z}(X)}{\gamma'_2}^{(Z)}(X,M) s(M\mid X,Z )\Big\}
\\\nonumber
& = & \E\{\xi'_3(X,Z,M)s(M\mid X,Z)\}
\\\label{eif:main:low:2}
& = & \E\{\tilde{\xi}'_3(X,Z,M)s(V)\}\quad (\text{Lemma \ref{lm:MonXZ} with $A_1 = M$, $A_2 = (X,Z)$}),
\end{eqnarray}
recalling $\xi'_3(X,Z,M) = \{\pi_{Z}(X)\}^{-1}{\gamma'_2}^{(Z)}(X,M)$ and $\tilde \xi'_3(X,Z,M) = \xi'_3(X,Z,M) - \E\big[\xi'_3(X,Z,M) \mid X,Z\big]$. \par For the third term on the right-hand side of \eqref{main:thm1:low}, since $s(X) + s(Z,M,Y\mid X) = s(V)$, we have
\begin{eqnarray}\nonumber
&&\E\Big\{\int_{\R^2}k_{u^*}(u)e_{u}(X)du\cdot s(X)\Big\}
\\\nn
& = & \E\Big[\int_{\R^2}k_{u^*}(u)e_{u}(X)du\cdot\{s(V) - s(Z,M,Y | X)\}\Big] 
\\\nonumber
& = & \E\Big\{\int_{\R^2}k_{u^*}(u)e_{u}(X)du\cdot s(V)\Big\} 
 -\E\Big[\int_{\R^2}k_{u^*}(u)e_{u}(X)du\cdot\underbrace{\E\Big\{s(Z,M,Y | X)\mid X\Big\}}_{ = 0} \Big]\quad(\text{LOTE})
\\\nonumber
& = &  \E\Big\{\int_{\R^2}k_{u^*}(u)e_{u}(X)du\cdot s(V)\Big\}
\\\nonumber
&=& \E\Big[\Big[\int_{\R^2}k_{u^*}(u)e_{u}(X)-\Big\{\E \int_{\R^2}k_{u^*}(u)e_{u}(X)du\Big\}\Big]\cdot s(V)\Big],
\\\label{eif:main:low:3}
\end{eqnarray}
where the last equality is by $\E\big[\E\{\int_{\R^2}k_{u^*}(u)e_{u}(X)du\}s(V)\big] = \E\{\int_{\R^2}k_{u^*}(u)e_{u}(X)du\}\E\{s(V)\} = 0$. Combining \eqref{main:thm1:low}, \eqref{eif:main:low:1}, \eqref{eif:main:low:2} and \eqref{eif:main:low:3}, we have 
\begin{eqnarray*}
&&\frac{\partial}{\partial \theta}\E_{\theta}\Big\{\int_{\R^2}k_{u^*}(u)e^{(\theta)}_u(X)du\Big\} \Big|_{\theta = 0} 
\\
& = & \E\Bigg[\Big[\tilde\xi'_2(X,Z,M) + \tilde\xi'_3(X,Z,M) + \int_{\R^2}k_{u^*}(u)e_{u}(X)du - \E \Big\{\int_{\R^2}k_{u^*}(u)e_{u}(X)du\Big\}\Big] s(V)\Bigg],
\end{eqnarray*}
and therefore 
\beee\label{eif:upper}
\tilde\xi'_2(x,z,m) + \tilde\xi'_3(x,z,m) + \int_{\R^2}k_{u^*}(u)e_{u}(x)du - \E \Bigg\{\int_{\R^2}k_{u^*}(u)e_{u}(X)du\Bigg\}
\eee is the EIF of $\E\{\int_{\R^2}k_{u^*}(u)e_{u}(X)du\}$.
\par
\
\par
\noindent\textbf{Step 2 (EIF of $\E[\int_{\R^2}\{\mu_{1}( X,m_1) - \mu_{0}(X,m_0)\}k_{u^*}(u)e_{u}(X)du]$):} We obtain the following derivates,
\begin{eqnarray}\nonumber
&&\frac{\partial}{\partial \theta}\E_{\theta}\Big[\int_{\R^2}\big\{\mu^{(\theta)}_{1}(X,m_1) - \mu^{(\theta)}_{0}(X,m_0)\big\}k_{u^*}(u)e^{(\theta)}_u(X)du\Big]\ \Big|_{\theta = 0}
\\\nonumber
& = &\E\Big[\int_{\R^2}\big\{\mu_{1}(X,m_1) - \mu_{0}(X,m_0)\big\}k_{u^*}(u)e_{u}(X)du\cdot s(X)\Big]
\\\nonumber
&& + \E\Big[\int_{\R^2}\Big\{\frac{\partial}{\partial \theta}\mu^{(\theta)}_{1}(X,m_1)\Big|_{\theta = 0} - \frac{\partial}{\partial \theta}\mu^{(\theta)}_{0}(X,m_0)\Big|_{\theta = 0}\Big\}k_{u^*}(u)e_{u}(X)du\Big]
\\\label{eif:up:main}
&& + \E\Big[\int_{\R^2}\big\{\mu_{1}(X,m_1) - \mu_{0}(X,m_0)\big\}k_{u^*}(u)\cdot\frac{\partial}{\partial \theta} e^{(\theta)}_u(X)\Big|_{\theta = 0}du\Big].
\end{eqnarray}
Similar to \eqref{eif:main:low:3}, the first term on the right-hand side of \eqref{eif:up:main} equals to 
\begin{eqnarray}\nn
&&\E\Big[\int_{\R^2}\big\{\mu_{1}(X,m_1) - \mu_{0}(X,m_0)\big\}k_{u^*}(u)e_{u}(X)du\cdot s(V)\Big]
\\\nonumber
& = & \E\Big[\Big\{\int_{\R^2}\big\{\mu_{1}(X,m_1) - \mu_{0}(X,m_0)\big\}k_{u^*}(u)e_{u}(X)du
\\\label{eif:up:1}
&&- \E\int_{\R^2}\big\{\mu_{1}(X,m_1) - \mu_{0}(X,m_0)\big\}k_{u^*}(u)e_{u}(X)du\Big\}\cdot s(V)\Big].
\end{eqnarray}
For the second term of \eqref{eif:up:main}, we first derive  
\begin{eqnarray*}
\frac{\partial}{\partial \theta}{\mu}^{(\theta)}_{z}(x,m_z)\Big|_{\theta = 0} & = & \frac{\partial}{\partial \theta}\E_\theta(Y\mid X = x, Z=z,M=m_z)\Big|_{\theta = 0} 
\\
& = &\E\big\{Ys(Y\mid X = x,Z=z,M = m_z)\mid X = x,Z=z, M=m_z\big\}
\\
& = &\E\big\{Ys(Y\mid X,Z,M)\mid X = x,Z=z, M=m_z\big\}.
\end{eqnarray*}
Then we have
\begin{eqnarray}\nonumber
&&\E\Big\{\int_{\R^2}\frac{\partial}{\partial \theta}\mu^{(\theta)}_{z}(X,m_z)\Big|_{\theta = 0}k_{u^*}(u)e_{u}(X)du\Big\}
\\\nonumber
& = & \E\Big[\int_{\R}\int_{\R}\E\big\{Ys(Y\mid X,Z,M)\mid X,Z=z, M=m_z\big\}\cdot k_{u^*}(u)e_{u}(X)dm_{1 - z}dm_z\Big]
\\\nonumber
& = & \E\Big[\int_{\R}\underbrace{\int_{\R} k_{u^*}(u)e_{(m_1,m_0)}(X)dm_{1 - z}}_{ = \gamma_1^{(z)}(X,M = m_z)}\cdot\E\big\{Ys(Y\mid X,Z,M)\mid X,Z=z, M=m_z\big\}dm_z\Big]
\\\nonumber
& = & \E\Big[\int_{\R}\frac{\gamma_1^{(z)}(X,m_z)}{f(M = m_z\mid X,Z = z)}\cdot\E\big\{Ys(Y\mid X,Z,M)\mid X,Z=z, M=m_z\big\}
 \cdot f(M = m_z\mid X,Z = z) dm_z\Big]
\\\nonumber
& = & \E\Big[\E\Big[\frac{\gamma_1^{(z)}(X,M)}{f(M \mid X,Z = z)}\cdot\E\big\{Ys(Y\mid X,Z=z,M )\mid X,Z=z, M\big\}\mid X, Z = z\Big]\Big]
\\\nonumber
& = &  \E\Big[\E\Big[\E\Big\{\frac{\gamma_1^{(z)}(X,M)}{f(M \mid X,Z = z)}\cdot Ys(Y\mid X,Z=z,M )\mid X, Z = z, M\Big\}\mid X,Z= z\Big]\Big]
\\\nonumber
& = &  \E\Big[\E\Big[\frac{\gamma_1^{(z)}(X,M)}{f(M \mid X,Z = z)}\cdot Ys(Y\mid X,Z=z,M )\mid X, Z = z\Big]\Big] \quad(\LOTE)
\\\nonumber
& = & \E\Big[\E\Big\{\frac{{1}(Z = z)}{\pi_z(X)}\frac{\gamma_1^{(Z)}(X,M)}{f(M \mid X,Z = z)}\cdot Ys(Y\mid X,Z=z,M )\mid X\Big\}\Big] \quad(\LOTE)
\\\nn
& = & \E\Big[\E\Big\{\frac{ \mathbb{I}(Z = z)\gamma_1^{(Z)}(X,M)}{\pi_Z(X)f(M \mid X,Z)}\cdot Ys(Y\mid X,Z,M )\mid X\Big\}\Big]
\\\label{id:res:3}
& = & \E\Big\{\frac{1(Z = z)\gamma_1^{(Z)}(X,M)}{\pi_Z(X)f_{ZM}(X)}\cdot Ys(Y\mid X,Z,M )\Big\}\quad\text{(LOTE)}.
\end{eqnarray}
By \eqref{id:res:3}, the second term on the right-hand side of \eqref{eif:up:main} is
\beee\nn
&&\E\Big[\Big\{\frac{1(Z = 1){\gamma_1}^{(Z)}(X,M)}{\pi_Z(X)f_{ZM}(X)}Y - \frac{1(Z = 0){\gamma_1}^{(Z)}(X,M)}{\pi_{Z}(X)f_{ZM}(X)}Y\Big\}s(Y\mid X,Z,M )\Big]
\\\nn
&=& \E\Big\{\frac{(-1)^{Z+1}{\gamma_1}^{(Z)}(X,M)}{\pi_{Z}(X)f_{ZM}(X)}Ys(Y\mid X,Z,M )\Big\},
\eee
recalling that
$
\xi_1(x,z,m)= \{\pi_{z}(x)f_{zm}(x)\}^{-1}{(-1)^{z+1}{\gamma_1}^{(z)}(x,m)}.
$
By Lemma \ref{lm:MonXZ} with $A_1 = Y$ and $A_2 = (X,Z,M)$, the second term on the right-hand side of \eqref{eif:up:main} then is, 
\begin{eqnarray}\label{eif:up:2}
\E\Big[\xi_1(X,Z,M)\{Y - \mu_{Z}(X,M)\}\cdot s(V)\Big].
\end{eqnarray}
where $\tilde{\xi}_1(X,Z,M,Y) = {\xi}_1(X,Z,M,Y) - \E\{{\xi}_1(X,Z,M,Y)\mid X,Z,M\}$.
\par
For the third term on the right-hand side of \eqref{eif:up:main}, similar arguments to \eqref{eif:main:low:1} and \eqref{eif:main:low:2} can directly show
\begin{eqnarray}\nonumber
&&\E\Big[\int_{\R^2}\big\{\mu_{1}( X,m_1) - \mu_{0}(X,m_0)\big\}k_{u^*}(u)\cdot\frac{\partial}{\partial \theta} e^{(\theta)}_u(X)\Big|_{\theta = 0}du\Big]
\\\nn
 & = & \E\Big[\big\{\tilde{\xi}_2(X,Z,M) + \tilde{\xi}_3(X,Z,M)\big\}\cdot s(V)\Big],
\end{eqnarray}
where recalling Section \ref{sec:fulleif}, we denote
\begin{itemize}
\item[(i)] $\tilde\xi_a (X,Z,M) = \xi_a(X,Z,M) - \E\{\xi_a(X,Z,M) \mid X,Z\}$ for $a = 2,3$;
\item[(ii)] $\xi_2(X,Z,M) = \frac{1}{\pi_{Z}(X)}\int_{\R^2} \mathbb{I}(M \leq m_{Z})\{\mu_{1}(X,m_1) - \mu_{0}(X,m_0)\} k_{u^*}(u)c_{\rho}^{(Z)}\big\{F_{1m_1}(X), F_{0m_0}(X)\big\}$ $\{\prod_{z\in\{0,1\}}f_{zm_z}(X)\}du$; 
\item[(iii)] $\xi_3(X,Z,M) = \frac{1}{\pi_{Z}(X)}{\gamma}_2^{(Z)}(X,M)$; 
\item[(iv)] $\gamma_2^{(1)}(X,M) = \int_{\R}\{\mu_{1}(X,M) - \mu_{0}(X,m_0)\}  k_{u^*}(M,m_0)c_{\rho}\{F_{1M}(X), F_{0m_0}(X)\}f_{0m_{0}}( X)dm_{0}$;
\item[(v)] $\gamma_2^{(0)}(X,M) = \int_{\R}\{\mu_{1}( X,m_1) - \mu_{0}(X,M)\} k_{u^*}(m_1,M)c_{\rho}\{F_{1m_1}(X), F_{0M}(X)\}f_{1m_{1}}( X)dm_{1}$.
\end{itemize}
Combining \eqref{eif:up:main}, \eqref{eif:up:1}, \eqref{eif:up:2} we have
\begin{eqnarray*}
&&\frac{\partial}{\partial \theta}\E_{\theta}\Big[\int_{\R^2}\big\{\mu^{(\theta)}_{1}(X,m_1) - \mu^{(\theta)}_{0}(X,m_0)\big\}k_{u^*}(u)e^{(\theta)}_u(X)du\Big]\ \Big|_{\theta = 0}
\\
&=&\E\Bigg[\Big[\tilde{\xi}_2(X,Z,M) + \tilde{\xi}_3(X,Z,M) + \xi_1(X,Z,M)\{Y - \mu_{Z}(X,M)\}
\\\nn
&&+ \int_{\R^2}\big\{\mu_{1}(X,m_1) - \mu_{0}(X,m_0)\big\}k_{u^*}(u)e_{u}(X)du
\\
&& - \E\int_{\R^2}\big\{\mu_{1}(X,m_1) - \mu_{0}(X,m_0)\big\}k_{u^*}(u)e_{u}(X)du\Big]\cdot s(V)\Bigg].
\end{eqnarray*}
Therefore the EIF of $\E\big[\int_{\R^2}\big\{\mu_{1}(X,m_1) - \mu_{0}(X,m_0)\big\}k_{u^*}(u)e_{u}(X)du\big]$ is 
\beee\nn
&&\tilde{\xi}_2(x,z,m) + \tilde{\xi}_3(x,z,m) + \xi_1(x,z,m)\{y - \mu_{z}(x,m)\} + \int_{\R^2}\big\{\mu_{1}(x,m_1) - \mu_{0}(x,m_0)\big\}k_{u^*}(u)e_{u}(x)du 
\\\label{eif:lower}
&&- \E\Bigg\{\int_{\R^2}\big\{\mu_{1}(X,m_1) - \mu_{0}(X,m_0)\big\}k_{u^*}(u)e_{u}(X)du\Bigg\}.
\eee
\par
\
\par
\noindent{\textbf{Step 3 (EIF of $\tau_{u^*}$):} Recalling identification formula \eqref{tau:id} and applying Lemma \ref{lm:dd} with derived EIFs in Step 1 and Step 2, we have
\begin{eqnarray}\nonumber
\frac{\partial }{\partial \theta}\tau_{u^*,\theta}&=&\scalemath{0.8}{\frac{1}{\E\big[\int_{\R^2}k_{u^*}(u)e_{u}(X)du\big]} \frac{\partial}{\partial \theta}\E_{\theta}\Big[\int_{\R^2}\big\{\mu^{(\theta)}_{1}( X,m_1) - \mu^{(\theta)}_{0}(X,m_0)\big\}k_{u^*}(u)e^{(\theta)}_u(X)du\Big]\ \Big|_{\theta = 0}}
\\\nonumber
&& \scalemath{0.8}{-  \frac{\E\big[\int_{\R^2}\big\{\mu_{1}( X,m_1) - \mu_{0}(X,m_0)\big\}k_{u^*}(u)e_{u}(X)du\big]}{\big\{\E\big[\int_{\R^2}k_{u^*}(u)e_{u}(X)du\big]\big\}^2}
\frac{\partial}{\partial \theta}\E_{\theta}\Big\{\int_{\R^2}k_{u^*}(u)e^{(\theta)}_u(X)du\Big\}\ \Big|_{\theta = 0}}
\\\nonumber
&=&\scalemath{0.8}{\frac{1}{\E\Big[\int_{\R^2}k_{u^*}(u)e_{u}(X)du\big]} \cdot \E\Bigg[\Big[\tilde{\xi}_2(X,Z,M) + \tilde{\xi}_3(X,Z,M) + \xi_1(X,Z,M)\{Y - \mu_{Z}(X,M)\}}
\\\nonumber
&& \scalemath{0.8}{+ \int_{\R^2}\big\{\mu_{1}( X,m_1) - \mu_{0}(X,m_0)\big\}k_{u^*}(u)e_{u}(X)du} -\scalemath{0.8}{ \E\int_{\R^2}\big\{\mu_{1}( X,m_1) - \mu_{0}(X,m_0)\big\}k_{u^*}(u)e_{u}(X)du\Big]\cdot s(V)\Bigg]}
\\\nonumber
&& \scalemath{0.8}{- \frac{\E\big[\int_{\R^2}\big\{\mu_{1}(X,m_1) - \mu_{0}(X,m_0)\big\}k_{u^*}(u)e_{u}(X)du\big]}{\big[\E\big\{\int_{\R^2}k_{u^*}(u)e_{u}(X)du\big\}\big]^2}\cdot \E\Bigg[\Big\{\tilde\xi'_2(X,Z,M) + \tilde\xi'_3(X,Z,M) }
\\\nonumber
&&\scalemath{0.8}{+ \int_{\R^2}k_{u^*}(u)e_{u}(X)du - \E \int_{\R^2}k_{u^*}(u)e_{u}(X)du\Big\}}\scalemath{0.8}{\cdot s(V)\Bigg]}
\\\nonumber
&=&\scalemath{0.8}{ \E\Bigg[\Big[\frac{1}{\E\big\{\int_{\R^2}k_{u^*}(u)e_{u}(X)du\big\}}   \cdot \Big[\tilde{\xi}_2(X,Z,M) + \tilde{\xi}_3(X,Z,M) + \xi_1(X,Z,M)\{Y - \mu_{Z}(X,M)\}}
\\\nonumber
&&\scalemath{0.8}{+ \int_{\R^2}\big\{\mu_{1}( X,m_1) - \mu_{0}(X,m_0)\big\}k_{u^*}(u)e_{u}(X)du
 - \E\int_{\R^2}\big\{\mu_{1}( X,m_1) - \mu_{0}(X,m_0)\big\}k_{u^*}(u)e_{u}(X)du\Big]}
\\\nonumber
&&\scalemath{0.8}{- \frac{\E\big[\int_{\R^2}\big\{\mu_{1}( X,m_1) - \mu_{0}(X,m_0)\big\}k_{u^*}(u)e_{u}(X)du\big]}{\big[\E\big\{\int_{\R^2}k_{u^*}(u)e_{u}(X)du\big\}\big]^2}\cdot \Big\{\tilde\xi'_1(X,Z,M) + \tilde\xi'_2(X,Z,M)}
\\\nonumber
&&\scalemath{0.8}{ + \int_{\R^2}k_{u^*}(u)e_{u}(X)du - \E \int_{\R^2}k_{u^*}(u)e_{u}(X)du\Big\}\Big]}\scalemath{0.8}{\cdot s(V)\Bigg],}
\\\nonumber
&=&\scalemath{0.8}{ \E\Bigg[\Bigg[\frac{1}{\E\big\{\int_{\R^2}k_{u^*}(u)e_{u}(X)du\big\}}   \cdot \Big[\tilde{\xi}_2(X,Z,M) + \tilde{\xi}_3(X,Z,M)  + \xi_1(X,Z,M)\{Y - \mu_{Z}(X,M)\} }
\\\label{eif:final}
& &
\scalemath{0.8}{+ \int_{\R^2}\big\{\mu_{1}( X,m_1) - \mu_{0}(X,m_0)\big\}k_{u^*}(u)e_{u}(X)du
\Big]}
\\\nn
&&\scalemath{0.8}{- \frac{\tau_{u^*}}{\big[\E\big\{\int_{\R^2}k_{u^*}(u)e_{u}(X)du\big\}\big]}\cdot \Big\{\tilde\xi'_2(X,Z,M) + \tilde\xi'_3(X,Z,M)}
\scalemath{0.8}{ + \int_{\R^2}k_{u^*}(u)e_{u}(X)du \Big\}\Bigg]}\scalemath{0.8}{\cdot s(V)\Bigg],}
\\\nn
&=&\scalemath{0.8}{\E\Bigg[\phi_{{\rm np}}(X,Z,M,Y)s(V)\Bigg],}
\end{eqnarray} 
after rearranging the terms and recalling the definitions in Section \ref{sec:fulleif} and \eqref{def:lu}. \eqref{eif:final} combining with \eqref{EIF:def} proves that $\phi_{{\rm np}}(x,z,m,y)
= \phi_{{\rm kn}}(x,z,m,y) + \tilde{\phi}_{\mathrm{np}}(x,z,m)$ is the full EIF of $\tau_{u^*}$. 

\subsubsection{Derivation of  $\phi_{{\rm p}}(x,z,m,y)$ under Scenario (ii)}\label{noneif2}
Under the model assumption that $\mathcal{M}_{{\rm ps}}$ is parametrically modeled and estimated, we first derive the corresponding tangent space $\Lambda$ and then project $\phi_{{\rm np}}(x,z,m,y)$ onto $\Lambda$ to obtain the new EIF $\phi_{{\rm p}}(x,z,m,y)$ under this specific scenario \citep[$\mathsection$ 25.3]{van2000asymptotic}.
\par Consider any regular  submodel with likelihood  $f_{\theta,\beta}(V = v)$, parametrized with some one-dimensional parameter $\theta\in \R$ and the principal score model parameter $\beta\in\R^{d_\beta}$; the true model takes place when  $f_{\theta = 0,\beta = \beta^*}(v) = f(v)$.  The likelihood is  decomposed as
\beee\nn
f_{\theta,\beta}(v) &= & f_{\theta}(X = x) f_{\theta}(Z = z\mid X = x)f_{}(M = m\mid X = x, Z = z,\beta)
\cdot f_{\theta}(Y = y\mid X = x,Z = z,M = m).
\eee
Thus the  score function of submodel is,
\beee\nn
s_{\beta^*}(v) = \begin{bmatrix}
s_{}(x) + s_{}(z\mid x) + s_{}(y\mid x,z,m)
\\\nn
s_{zm}(x\mid \beta^*)
\end{bmatrix} \in\R^{1 + d_\beta},
\eee
where we recall $s_{}(x) = \partial \log f_{\theta}(X= x)/\partial \theta \mid_{\theta = 0}$, $s_{zm}(x\mid \beta^*) = \partial \log f(M = m\mid X = x, Z = z,\beta)/\partial \beta\mid_{\beta = \beta^*}$, and similarly for others. The parametric submodel tangent space  can be represented as
\beee\nn
\Lambda_{\theta} = \underbrace{\Big\{\ell_1\{s_{}(x) + s_{}(z\mid x) + s_{}(y\mid x,z,m)\} \mid \ell_1 \in \R \Big\}}_{\Lambda_1}\oplus\underbrace{\Big\{ \ell_2^\T s_{zm}(x\mid \beta^*)\mid \ell_2 \in \R^{d_{\beta}}\Big\}}_{\Lambda_2}.
\eee
One has $\Lambda_1 \perp \Lambda_2$, as by LOTE,
\beee\nn
&&\E\Big[\ell_1\{s_{}(X) + s_{}(Z\mid X) + s_{}(Y\mid X,Z,M)\}\cdot \ell_2^\T s_{ZM}(X\mid \beta^*)\Big]
\\\nn
&=& \E\Big[\ell_1\{s_{}(X) + s_{}(Z\mid X)\}\cdot \E\big\{\ell_2^\T s_{ZM}(X\mid \beta^*) \mid X,Z\big\}\Big]
\\\nn
&& + \E\Big[\ell_2^\T s_{ZM}(X\mid \beta^*) \cdot \E\big\{s(Y\mid X,Z,M)\mid X,Z,M \big\}\Big]
\\\label{perpcheck}
& = &0,
\eee
for any $\ell_1\in\R$ and $\ell_2\in\R^{d_\beta}$. Then the semiparametric tangent space $\Lambda$ is the mean-square closure of $\Lambda_{\theta}$, over all possible regular submodels associated with parameter $\theta$. Note $\beta$ is associated with a known and fixed model, namely $f_{zm}(x\mid \beta)$. Thus we have $\Lambda =\bar{ \Lambda}_1 + \Lambda_2$, where $\bar{ \Lambda}_1$ is the mean-squre closure of ${\Lambda}_1$ over all regular parametric submodels associated with $\theta$. Recalling \eqref{svdecom}--\eqref{tangent}, we similarly have $\bar\Lambda_1 = H_1 \oplus H_2 \oplus H_4$ and thus
\beee\label{newlambda}
\Lambda &= & H_1 \oplus H_2 \oplus H_4\oplus \Lambda_2,
\eee
where 
\begin{equation}\nn
\begin{aligned}
H_{1} &=\{h(x)\mid  \mathbb{E}\{h(X)\}=0\}, \\
H_{2} &=\{h(x,z)\mid  \mathbb{E}\{h(X,Z) \mid X\}=0\},  \\
H_{4} &=\{h(x,z,m,y)\mid  \mathbb{E}\{h(X,Z,M,Y) \mid X,Z,M\}=0\},
\\
\Lambda_2 &=\Big\{ \ell_2^\T s_{zm}(x\mid \beta^*)\mid \ell_2 \in \R^{d_{\beta}}\Big\}.
\end{aligned}
\end{equation}
Similar to \eqref{perpcheck}, we can check that
\beee\label{perp:used}
H_1 \perp H_2\perp H_4 \perp \Lambda_2.
\eee
\par
To derive the semiparametric EIF, we  need to obtain the projection of $\phi_{{\rm np}}(x,z,m,y)$ onto the semiparametric tangent space $\Lambda$. We first decompose $\phi_{{\rm np}}(x,z,m,y)$ in \eqref{eif:final} as follows,
\beee\nn
\phi_{{\rm np}}(x,z,m,y) & = &\underbrace{\scalemath{1}{\frac{\tilde{\xi}_2(x,z,m) + \tilde{\xi}_3(x,z,m) - \tau_{u^*}\big\{\tilde\xi'_2(x,z,m) + \tilde\xi'_3(x,z,m)\big\} 
 }{\E\{\int_{\R^2}k_{u^*}(u)e_{u}(X)du\}}}}_{\phi_1(x,z,m)} + \underbrace{\frac{{\xi}_1(x,z,m)\{y - \mu_{z}(x,m)\}}{\E\{\int_{\R^2}k_{u^*}(u)e_{u}(X)du\}}}_{\phi_2(x,z,m,y)}
\\\nn
&& +\underbrace{\scalemath{1}{\frac{\int_{\R^2}\{\mu_{1}(x,m_1) - \mu_{0}(x,m_0)\}k_{u^*}(u)e_{u}(x)du -  
\tau_{u^*}\int_{\R^2}k_{u^*}(u)e_{u}(x)du}{\E\big\{\int_{\R^2}k_{u^*}(u)e_{u}(X)du\big\}}}}_{\phi_3(x)}.
\\
\label{phinpooo}
\eee Since the orthogonal projection is a linear operator, we only need to derive the projections of $\phi_1$--$\phi_3$ onto $\Lambda$, respectively, and sum them up to get the final projections. By definition, one has
\beee\nn
\E\{\phi_2(X,Z,M,Y)\mid X,Z,M\} 
&=& c_{k}^{-1}\Big[\xi_1(X,Z,M)\big\{\E(Y\mid X,Z,M) - \mu_{Z}(X,M)\big\}\Big]
\\\nn
& = & 0,
\eee
recalling $\xi_1$ is defined in Section \ref{sec:fulleif}; we denote $c_{k} = \E\{\int_{\R^2}k_{u^*}(u)e_{u}(X)du\}$ for simplicity. Therefore $\phi_2(x,z,m,y)\in H_4$, and by \eqref{perp:used}, we have the projection of $\phi_2(x,z,m,y)$ onto $\Lambda$ is itself. Similarly, we can check that $\E\{\phi_3(X)\} = \tau_{u^*} - \tau_{u^*} = 0$ by Theorem \ref{thm:idd}. Thus $\phi_3(x)\in H_1$. Similarly, the projection of $\phi_3(x)$ onto $\Lambda$ is also itself.
\par It is left to find the projection of $\phi_1(x,z,m)$. Recalling the definitions of $\tilde{\xi}_2,\tilde{\xi}_3,\tilde{\xi}_2',\tilde{\xi}_3'$ in Section \ref{sec:fulleif}, one can easily  show that
\beee\nn
\E\{\phi_1(X,Z,M)\mid X,Z\} =  0.
\eee
We claim that the projections of $\phi_1(X,Z,M)$ onto $H_1$, $H_2$, and $H_4$ are all zero, because $\phi_1(X,Z,M)\perp H_j$, $j = 1,2,4$. For example,  denote $h_1^{(1)}(x)\in H_1$ as the projection of $\phi_1(x,z,m)$ onto $\Lambda$. Then $h_1^{(1)}(x)$ minimizes the following function
\beee\nn
&&\E\big\{h_1^{(1)}(x) - \phi_1(X,Z,M)\big\}^2 
\\\nn
 & = &\E\{h_1^{(1)}(X)\}^2 + \E\{\phi^2_1(X,Z,M)\} - 2\E\big[h_1^{(1)}(X)\underbrace{\E\{\phi_1(X,Z,M)\mid X,Z\}}_{ = 0}\big]
\\\nn
& = & \E\{h_1^{(1)}(X)\}^2 + \E\{\phi^2_1(X,Z,M)\},
\eee
among all functions in $H_1$, which is clearly satisfied if and only if $h_1^{(1)}(x) = 0$, a.s. Thus $\phi_1(x,z,m)\perp H_1$. Observing that
\bee\nn
&\E\big\{h_2(X,Z)\phi_1(X,Z,M)\big\} = \E\big[h_2(X,Z)\E\{\phi_1(X,Z,M)\mid X,Z\}\big] = 0,
\\
&\E\big\{h_4(X,Z,M,Y)\phi_1(X,Z,M)\big\} = \E\big[\phi_1(X,Z,M)\E\{h_4(X,Z,M,Y)\mid X,Z,M\}\big] = 0,
\ee
for any $h_2(x,z)\in H_2$ and $h_4(x,z,m,y)\in H_4$, we thus similarly have $\phi_1(x,z,m)\perp H_2 \text{ and }\phi_1(x,z,m)\perp H_4$. So the projection of $\phi_1(x,z,m)$ onto $\Lambda$ is only the projection of $\phi_1(x,z,m)$ onto $\Lambda_2$. 
\par
Now we  obtain the projections of $\tilde{\xi}_2,\tilde{\xi}_2,\tilde{\xi}_2',\tilde{\xi}_3'$ onto $\Lambda_2$ one by one. For example, for $\tilde{\xi}_2$, this is equivalent to finding  $\tilde{\ell}_1^\T s_{zm}(x\mid\beta^*)\in \Lambda_2$ solving the following optimization problem,
\beee\nn
&&\arg\min_{\tilde{\ell}_1\in \R^{d_{\beta}}} \, \E\Big\{\tilde{\xi}_2(X,Z,M) - \tilde{\ell}_1^\T s_{ZM}(X\mid \beta^*)\Big\}^2
\\\nn
& = & \arg\min_{\tilde{\ell}_1\in \R^{d_\beta}} \,\tilde{\ell}_1^\T I_{\beta^*}\tilde{\ell}_1 - 2\tilde{\ell}_1^\T \E\big\{\tilde{\xi}_2(X,Z,M)s_{ZM}(X\mid \beta^*)\big\} + \E\{\tilde{\xi}_2^2(X,Z,M)\},
\eee
which can be solved by standard quadratic programming \citep{boyd2004convex}. In particular, the solution is
\beee\nn
\tilde{\ell}_{1,\diamond}  & = & I_{\beta^*}^{-1}\E\big\{\tilde{\xi}_2(X,Z,M)s_{ZM}(X\mid \beta^*)\big\}
\\\nn
& = & I_{\beta^*}^{-1}\E\big\{{\xi}_2(X,Z,M)s_{ZM}(X\mid \beta^*)\big\} - I_{\beta^*}^{-1}\E\big[\E({\xi}_2\mid X,Z)\underbrace{\E\{s_{ZM}(X\mid \beta^*)\mid X,Z\}}_{= 0}\big]
\\\nn
& = & I_{\beta^*}^{-1}\E\big\{{\xi}_2(X,Z,M)s_{ZM}(X\mid \beta^*)\big\}
\\\nn
& = & \scalemath{0.8}{I_{\beta^*}^{-1}\E\Bigg[\frac{1}{\pi_{Z}(X)}\int_{\R^2}1(M \leq m_{z})\{\mu_{1}(X,m_1) - \mu_{0}(X,m_0)\} k_{u^*}(u)}
\scalemath{0.8}{\cdot c_{\rho}^{(Z)}\big\{F_{1m_1}(X), F_{0m_0}(X)\big\}\Big\{\prod_{z\in\{0,1\}}f_{zm_z}(X)\Big\}du\cdot s_{ZM}(X\mid \beta^*)\Bigg]}
\\\nn
&= &\scalemath{0.8}{I_{\beta^*}^{-1}\sum_{z = 0,1}\E\Bigg[\E\Big[\int_{\R^2}1(M\leq m_z)\{\mu_{1}(X,m_1) - \mu_{0}(X,m_0)\} k_{u^*}(u) c_{\rho}^{(z)}\big\{F_{1m_1}(X), F_{0m_0}(X)\big\}}
\\\nn
&&\scalemath{0.8}{\cdot \Big\{\prod_{z\in\{0,1\}}f_{zm_z}(X)\Big\}du\cdot s_{ZM}(X\mid \beta^*)\mid X,Z = z\Big]\Bigg] \quad\text{(LOTE)}}
\\\nn
&= &\scalemath{0.8}{I_{\beta^*}^{-1}\sum_{z = 0,1}\E\Bigg[\int_{\R^2}\int_{-\infty}^{m_z}\{\mu_{1}(X,m_1) - \mu_{0}(X,m_0)\} k_{u^*}(u) c_{\rho}^{(z)}\big\{F_{1m_1}(X), F_{0m_0}(X)\big\}}
\\\nn
&&\scalemath{0.8}{\cdot\Big\{\prod_{z\in\{0,1\}}f_{zm_z}(X)\Big\}du\cdot f'_{zm}(X) \underbrace{f^{-1}_{zm}(X)f_{zm}(X)}_{=1}dmdu\Bigg]}
\\\nn
&= &\scalemath{0.8}{I_{\beta^*}^{-1}\sum_{z = 0,1}\mathbb{E}\Big[\int_{\R^2}k_{u^*}(u)\big\{\mu_{1}( X,m_1) - \mu_{0}(X,m_0)\big\} c_{\rho}^{(z)}\big\{{F}_{1m_1}( X),{F}_{0m_0}(X)\big\}{f}_{1m_1}( X){f}_{0m_0}( X)F'_{zm_z}(X)du\Big].}
\eee
Thus $\tilde{\ell}_{1,\diamond}^\T s_{zm}(x\mid\beta^*)$ is the   projection of $\tilde{\xi}_2$ onto $\Lambda_1$, which, combing with the previous discussions, is also the projection of $\tilde{\xi}_2$ onto the whole  $\Lambda$. With similar techniques, we have $\tilde{\ell}_{2,\diamond}^\T s_{zm}(x\mid\beta^*)$ is the projection of $\tilde{\xi}_3$ onto $\Lambda$, such that
\beee\nn
\tilde{\ell}_{2,\diamond}& = &I_{\beta^*}^{-1}\E\big\{{\xi}_3(X,Z,M)s_{ZM}(X\mid\beta^*)\big\}
\\\nn
& = & \sum_{z = 0,1}I_{\beta^*}^{-1}\E\Big[\E\big\{\gamma_{2}^{(z)}(X,M)s_{ZM}(X\mid\beta^*)
\mid X,Z = z\big\}\Big].
\eee
When $z = 1$, for the term on the right-hand side of above equation, we further have
\beee\nn
&&I_{\beta^*}^{-1}\E\Big[\E\big\{\gamma_{2}^{(1)}(X,M)s_{ZM}(X\mid\beta^*)
\mid X,Z = 1\big\}\Big]
\\\nn
& = &I_{\beta^*}^{-1}\E\Big[\E\Big\{\int_{\R}\{\mu_{1}(X,M) - \mu_{0}(X,m_0)\} k_{u^*}(M,m_0)c_{\rho}\{F_{1M}(X), F_{0m_0}(X)\}f_{0m_{0}}(X)dm_{0}
\\\nn
&&\cdot f^{-1}(M\mid X,Z = 1)f'_{1M}(X) \mid X,Z = 1\Big\}\Big]
\\\nn
& = &I_{\beta^*}^{-1}\E\Big[\int_{\R}\int_{\R}\{\mu_{1}(X,m) - \mu_{0}(X,m_0)\} k_{u^*}(m,m_0)c_{\rho}\{F_{1m}(X), F_{0m_0}(X)\}f_{0m_{0}}(X)f'_{1m}(X)dm_{0}
 dm\Big\}\Big]
\\\nn
& = & I_{\beta^*}^{-1}\mathbb{E}\Big[\int_{\R^2}k_{u^*}(u)\big\{\mu_{1}( X,m_1) - \mu_{0}(X,m_0)\big\}
c_{\rho}\big\{{F}_{1m_1}( X),{F}_{0m_0}(X)\big\} f_{0m_{0}}( X)f'_{1m_1}( X)du\Big]
\eee
We can similarly deal with the $z = 0$ case and deduce,
\beee\nn
\tilde{\ell}_{2,\diamond}& = &\scalemath{0.8}{I_{\beta^*}^{-1}\sum_{z = 0,1}\mathbb{E}\Big[\int_{\R^2}k_{u^*}(u)\big\{\mu_{1}( X,m_1) - \mu_{0}(X,m_0)\big\}
c_{\rho}\big\{{F}_{1m_1}( X),{F}_{0m_0}(X)\big\} f_{(1 - z)m_{1- z}}( X)f'_{zm_z}( X) du\Big].}
\eee Notice that $\tilde{\xi}_2',\tilde{\xi}_3'$ have very similar structures as $\tilde{\xi}_2,\tilde{\xi}_3$. Following the derivation strategy as above, we can show that  ${\tilde{\ell}_{1,\diamond}}'^\T s_{zm}(x\mid\beta^*),{\tilde{\ell}_{2,\diamond}}'^\T s_{zm}(x\mid\beta^*)$ are the projections of $\tilde{\xi}_2',\tilde{\xi}_3'$, respectively, such that
\beee\nn
{\tilde{\ell}_{1,\diamond}}' & = & \scalemath{0.8}{I_{\beta^*}^{-1}\sum_{z = 0,1}\mathbb{E}\Big[\int_{\R^2}k_{u^*}(u) c_{\rho}^{(z)}\big\{{F}_{1m_1}( X),{F}_{0m_0}(X)\big\}{f}_{1m_1}( X){f}_{0m_0}( X)F'_{zm_z}(X)du\Big],}
\\\nn
{\tilde{\ell}_{2,\diamond}}' & = & \scalemath{0.8}{I_{\beta^*}^{-1}\sum_{z = 0,1}\mathbb{E}\Big[\int_{\R^2}k_{u^*}(u)
c_{\rho}\big\{{F}_{1m_1}( X),{F}_{0m_0}(X)\big\} f_{(1 - z)m_{1- z}}( X)f'_{zm_z}( X) du\Big].}
\eee
Recalling that $\phi_1(x,z,m)$ is a linear combiniation of $\tilde{\xi}_2,\tilde{\xi}_3,\tilde{\xi}_2',\tilde{\xi}_3'$, we have the projection of $\phi_1(x,z,m)$ on $\Lambda$ is the corresponding linear combination of $\tilde{\xi}_2,\tilde{\xi}_3,\tilde{\xi}_2',\tilde{\xi}_3'$'s projections on $\Lambda$. In particular, the projection of $\phi_1(x,z,m)$ is 
\beee
\frac{\mathcal{A}_2^* - \tau_{u^*}\mathcal{A}_1^*}{\E\{\int_{\R^2}k_{u^*}(u)e_u(X)\}}I_{\beta^*}^{-1}s_{zm}(x\mid\beta^*),
\eee
where we recall that in Section \ref{sec:fulleif},
\beee\nonumber
{\mathcal{A}^*_1} & \scalemath{0.8}{=} & \scalemath{0.8}{ \sum_{z = 0,1}\mathbb{E}\Big[\int_{\R^2}k_{u^*}(u)c_{\rho}\big\{{F}_{1m_1}( X),{F}_{0m_0}( X)\big\} f_{(1 - z)m_{1- z}}( X)\{f'_{zm_z}( X)\}^\T du\Big]}
\\\nonumber
&&\scalemath{0.8}{+\sum_{z = 0,1}\mathbb{E}\Big[\int_{\R^2}k_{u^*}(u)\Big[c_{\rho}^{(z)}\big\{{F}_{1m_1}( X),{F}_{0m_0}(X)\big\}{f}_{1m_1}( X){f}_{0m_0}( X)\big\{F'_{zm_z}(X)\big\}^\T \Big]du\Big],}
\\\nonumber
{\mathcal{A}^*_2} & \scalemath{0.8}{=} &  \scalemath{0.8}{\sum_{z = 0,1}\mathbb{E}\Big[\int_{\R^2}k_{u^*}(u)\big\{\mu_{1}( X,m_1) - \mu_{0}(X,m_0)\big\}
c_{\rho}\big\{{F}_{1m_1}( X),{F}_{0m_0}(X)\big\} f_{(1 - z)m_{1- z}}( X)\{f'_{zm_z}( X)\}^\T du\Big]}
\\\nonumber
&&\scalemath{0.8}{+\sum_{z = 0,1}\mathbb{E}\Big[\int_{\R^2}k_{u^*}(u)\big\{\mu_{1}( X,m_1) - \mu_{0}(X,m_0)\big\} c_{\rho}^{(z)}\big\{{F}_{1m_1}( X),{F}_{0m_0}(X)\big\}{f}_{1m_1}( X){f}_{0m_0}( X)\big\{F'_{zm_z}(X)\big\}^\T du\Big].}
\eee
Summarizing the projections of $\phi_1(\cdot)$--$\phi_3(\cdot)$ on $\Lambda$ defined in \eqref{newlambda} as above, and recalling \eqref{phinpooo}, we conclude the projection of $\phi_{{\rm np}}(x,z,m,y)$ on $\Lambda$, i.e.,  the semiparametric EIF of $\tau_{u^*}$ when the parametric model of $\mathcal{M}_{{\rm ps}}$ is known, is
\beee\nn
\phi_{{\rm p}}(x,z,m,y) &=& \frac{\mathcal{A}^*_2 - \tau_{u^*}\mathcal{A}^*_1}{\mathbb{E}\big\{\int_{\R^2}k_{u^*}(u){e}_{u}(X)du\big\}} I_{\beta^*}^{-1}s_{zm}(x\mid\beta^*)  + \frac{{\xi}_1(x,z,m)\{y - \mu_{z}(x,m)\}}{\mathbb{E}\big\{\int_{\R^2}k_{u^*}(u){e}_{u}(X)du\big\}}
\\\nn
&&+ \frac{
\int_{\R^2}\big\{\mu_{1}(x,m_1) - \mu_{0}(x,m_0)\big\}k_{u^*}(u){e}_{u}(x)du - \tau_{u^*}\int_{\R^2}k_{u^*}(u)e_u(x)du}{\mathbb{E}\big\{\int_{\R^2}k_{u^*}(u){e}_{u}(X)du\big\}},
\eee
which shows $\phi_{{\rm p}}(x,z,m,y)
 =   \phi_{{\rm kn}}(x,z,m,y) + \tilde{\phi}_{\mathrm{p}}(x,z,m)$ by recalling \eqref{def:lu} and  \eqref{tphip}.

\subsubsection{Derivation of   $\phi_{{\rm kn}}(x,z,m,y)$ under Scenario (i)}\label{semieif}
 Now we assume the distribution of $M$ given $X,Z$ is  known and thus we have some model restriction. Since the model assumption becomes more restrictive, the $\phi(\cdot)$ satisfying \eqref{EIF:def} is not necessary to be an EIF but must be an influence function (IF) of $\tau_{u^*}$ and the IF can be  not unique \citep{bickel1993efficient}. Therefore, we can still derive an IF of $\tau_{u^*}$ in a similar yet simpler way as in Section \ref{noneif}.  In particular, we find that $\phi(\cdot) = \phi_{{\rm kn}}(\cdot)$ in \eqref{eifeuknown} satisfying  \eqref{EIF:def} and thus $\phi_{{\rm kn}}(\cdot)$ is an IF. Furthermore, based on the fundamental semiparametric statistics theory, we can verify that $\phi_{{\rm kn}}(\cdot)$ is actually the EIF by showing that it belongs to the semiparametric tangent space and thereby finish our proof; See  e.g. \citet[$\mathsection$ 25.3]{van2000asymptotic} for the relevant semiparametric statistics theory  used here.
\par
 We show that $\phi_{{\rm kn}}(\cdot)$ is an IF following the proof routine in Section \ref{noneif} with some simplification. In particular, since $f_{zm}(x)$ and $F_{zm}(x)$ are all known, we do not need to parametrize $e_u(x)$ as $e^{(\theta)}_u(x)$ and take it into account when deriving the EIFs of $$\E\Big[\int_{\R^2}\big\{\mu_{1}( X,m_1) - \mu_{0}(X,m_0)\big\}k_{u^*}(u)e_{u}(X)du\Big]$$ and $\E\big[\int_{\R^2}\big\{\mu_{1}( X,m_1) - \mu_{0}(X,m_0)\big\}k_{u^*}(u)e_{u}(X)du\big]$ in Section \ref{noneif}. Then we can show that the function 
$$
\int_{\R^2}k_{u^*}(u)e_{u}(x)du - \E \Bigg\{\int_{\R^2}k_{u^*}(u)e_{u}(X)du\Bigg\},
$$
which drops the terms $\tilde{\xi}'_2, \tilde{\xi}'_3$ in \eqref{eif:upper}, is the IF of 
$\E\big[\int_{\R^2}\big\{\mu_{1}( X,m_1) - \mu_{0}(X,m_0)\big\}k_{u^*}(u)e_{u}(X)du\big]
$ under Scenario (i). Similarly, the function
$$
 \tilde{\xi}_1(x,z,m,y) + \int_{\R^2}\big\{\mu_{1}(x,m_1) - \mu_{0}(x,m_0)\big\}k_{u^*}(u)e_{u}(x)du - \E\Bigg\{\int_{\R^2}\big\{\mu_{1}(X,m_1) - \mu_{0}(X,m_0)\big\}k_{u^*}(u)e_{u}(X)du\Bigg\},
$$
which drops the terms $\tilde{\xi}_2, \tilde{\xi}_3$ in \eqref{eif:lower}, is the IF of $$\E\Big[\int_{\R^2}\big\{\mu_{1}( X,m_1) - \mu_{0}(X,m_0)\big\}k_{u^*}(u)e_{u}(X)du\Big].$$ With simplified IFs of  $\E\big[\int_{\R^2}\big\{\mu_{1}( X,m_1) - \mu_{0}(X,m_0)\big\}k_{u^*}(u)e_{u}(X)du\big]$ and $\E\big[\int_{\R^2}\big\{\mu_{1}( X,m_1) - \mu_{0}(X,m_0)\big\}$ $k_{u^*}(u)e_{u}(X)du\big]$ as above and similar to \eqref{eif:final}, we can use Lemma \ref{lm:dd} to show that   $\phi(\cdot) = \phi_{{\rm kn}}(\cdot)$ in \eqref{eifeuknown} satisfying  \eqref{EIF:def} and thus $\phi_{{\rm kn}}(\cdot)$ is an IF of $\tau_{u^*}$.
\par
We now derive the semiparametric tangent space under Scenario (i). With known $\mathcal{M}_{{\rm ps}}$, the density function decomposition similar to \eqref{np:decome} is now 
\beee\nonumber
&&f_{\theta}(V = v) 
\\\nn
&=& f_{\theta}(X = x)f_{\theta}(Z = z\mid X = x)f(M= m\mid X = x,Z = z)f_{\theta}(Y = y\mid X = x,Z = z,M = m).
\eee Thus score function of some arbitrary parametric submodel is now decomposed as 
\beee\label{svdecom}
s_\theta(v)& = &s_\theta(x) + s_\theta(z\mid x) + s_\theta(y\mid x,z,m).
\eee
By considering the mean-square closure of all functions $s_\theta(v)$ over all possible submodels,  the tangent space is now degenerated to, 
\beee\label{tangent}
\Lambda&=&H_{1} \oplus H_{2}  \oplus H_{4},
\eee
where $\oplus$ is the set sum and,
\beee\nn
H_{1} &=&\{h(x)\mid  \mathbb{E}\{h(X)\}=0\}, \\\nn
H_{2} &=&\{h(x,z)\mid  \mathbb{E}\{h(X,Z) \mid X\}=0\}, \\\nn
H_{4} &=&\{h(x,z,m,y)\mid  \mathbb{E}\{h(X,Z,M,Y) \mid X,Z,M\}=0\}.
\eee
First, observing that $\E[\xi_1(X,Z,M)\{Y - \mu_{Z}(X,M)\}] = \E[\xi_1(X,Z,M)\E\{Y - \mu_{Z}(X,M)\mid X,Z,M\}] =  0$, we have 
\beee\nn
\frac{{\xi}_1(x,z,m)\{y  - \mu_{z}(x,m)\}
}{\E\big\{\int_{\R^2}k_{u^*}(u)e_{u}(X)du\big\}}&\in &H_4. 
\eee
Second,  observe that 
\beee\nn
&&\E \Bigg[\frac{ \int_{\R^2}\big\{\mu_{1}(X,m_1) - \mu_{0}(X,m_0)\big\}k_{u^*}(u)e_{u}(X)du
 - \tau_{u^*}\big\{\int_{\R^2}k_{u^*}(u)e_{u}(X)du\big\}}{\E\big\{\int_{\R^2}k_{u^*}(u)e_{u}(X)du\big\}}\Bigg]
\\\nn
& = & \frac{\E\big[\int_{\R^2}\big\{\mu_{1}( X,m_1) - \mu_{0}(X,m_0)\big\}k_{u^*}(u)e_{u}(X)du\big]}{\E\big\{\int_{\R^2}k_{u^*}(u)e_{u}(X)du\big\}}- \tau_{u^*}
\\
& = & 0,
\eee
by Theorem \ref{thm:idd}. Recalling the form of $\phi_{{\rm kn}}(\cdot)$ in 
\eqref{eifeuknown},  thus for $\psi_{{\rm kn}}(x)$ defined as follows, we have
\beee\nn
\psi_{{\rm kn}}(x) &= &\frac{ \int_{\R^2}\big\{\mu_{1}(x,m_1) - \mu_{0}(x,m_0)\big\}k_{u^*}(u)e_{u}(x)du
 - \tau_{u^*}\big\{\int_{\R^2}k_{u^*}(u)e_{u}(x)du\big\}}{\E\big\{\int_{\R^2}k_{u^*}(u)e_{u}(X)du\big\}}
\\\nn
&=&\phi_{{\rm kn}}(x,z,m,y) - \frac{{\xi}_1(x,z,m)\{y - \mu_{z }(x,m)\}
}{\E\big\{\int_{\R^2}k_{u^*}(u)e_{u}(X)du\big\}} \in H_1.
\eee
Summarizing above results, we conclude that IF $\phi_{{\rm kn}}(x,z,m,y)$ belongs to the semiparametric tangent space, as
\beee\nn
\phi_{{\rm kn}}(x,z,m,y) & =& \psi_{{\rm kn}}(x) + \frac{{\xi}_1(x,z,m)\{y - \mu_{z}(x,m)\}
}{\E\big\{\int_{\R^2}k_{u^*}(u)e_{u}(X)du\big\}} 
\\\nn
&\in& H_1  \oplus  H_4 \subseteq \Lambda.
\eee
This conclusion directly confirms that  $\phi_{{\rm kn}}(\cdot)$ is not only the IF, but also the EIF of $\tau_{u^*}$ under the semiparametric scenario with known $\mathcal{M}_{{\rm ps}}$ by the classic semiparametric statistics theory; See the discussions after Lemma 25.14 in \citet[$\mathsection$ 25.3]{van2000asymptotic}.
}

\subsection{Proof of Theorem \ref{thm:dbrb}}
With fixed $u^*$, we omit the footnote $u^*$ in $\bar{\xi}_{1,u^*}(\cdot)$ and directly write $\bar{\xi}_{1}(\cdot)$ for simplicity. By definition, we further write,
\beee\nonumber
&&\bar{{\xi}}_1(X,Z,M)\{Y - \bar{\mu}_{Z}(X,M)\} 
\\\nonumber
&=& \underbrace{\frac{\mathbb{I}(Z = 1){\gamma_1}^{(Z)}(X,M)}{\bar\pi_Z(X) f(M \mid X,Z )} Y}_{\bar{\tilde{\xi}}^{(1)}_1(X,Z,M,Y)} - \underbrace{\frac{\mathbb{I}(Z = 0){\gamma_1}^{(Z)}(X,M)}{\bar\pi_Z(X) f(M \mid X,Z )} Y}_{\bar{\tilde{\xi}}^{(0)}_1(X,Z,M,Y)}
\\\nonumber
&& -\underbrace{\frac{\mathbb{I}(Z = 1)\gamma_1^{(Z)}(X,M)}{\bar{\pi}_Z(X)f(M\mid X,Z)}\bar\mu_{Z}(X,M)}_{\bar{E}_{\xi_1}^{(1)}(X,Z,M)} + \underbrace{\frac{\mathbb{I}(Z = 0)\gamma_1^{(Z)}(X,M)}{\bar{\pi}_Z(X)f(M\mid X,Z)}\bar\mu_{Z}(X,M)}_{\bar{E}_{\xi_1}^{(0)}(X,Z,M)}
\eee We first consider the scenario when $\pi = \bar{\pi}$. Applying LOTE, one has
\beee\nn
\E\Big\{\bar{\tilde{\xi}}^{(1)}_1(X,Z,M,Y)\Big\} &=& \E\Bigg\{\frac{\mathbb{I}(Z = 1){\gamma_1}^{(Z)}(X,M)}{\pi_Z(X) f(M \mid X,Z )} Y\Bigg\}
\\\nn
&= &\E\Bigg[\E\Bigg\{\frac{\mathbb{I}(Z = 1){\gamma_1}^{(Z)}(X,M)}{\pi_Z(X) f(M \mid X,Z )} Y\mid X\Bigg\}\Bigg]
\\\nn
&=&  \E\Bigg[\pi_1(X)\E\Bigg\{\frac{{\gamma_1}^{(Z)}(X,M)}{\pi_1(X) f(M \mid X,Z )} Y\mid X, Z = 1\Bigg\} + \pi_0(X)\cdot0\Bigg]
\\\nn
& =&  \E\Bigg[\E\Bigg\{\frac{{\gamma_1}^{(1)}(X,M)}{ f(M \mid X,Z  = 1)} Y\mid X, Z = 1\Bigg\}\Bigg]
\\\nn
& =&  \E\Bigg[\E\Bigg\{\frac{{\gamma_1}^{(1)}(X,M)}{ f(M \mid X,Z  = 1)} \E\big(Y\mid X, Z = 1,M\big)\mid X,Z = 1\Bigg\}\Bigg]
\\\nn
& =&\E\Bigg\{\int_{\R}\frac{{\gamma_1}^{(1)}(M = m_1,X) \E\big(Y\mid X, Z = 1,M = m_1\big)}{ f(M = m_1 \mid X,Z  = 1)}f(M = m_1\mid X,Z= 1)dm_1\Bigg\}
\\\nn
& =&\E\Bigg\{\int_{\R}{{\gamma_1}^{(1)}(M = m_1,X) \mu_{1}(X,m_1)}dm_1\Bigg\}
\\\label{pf:dbr:main}
& = &\E\Bigg\{\int_{\R^2}{k_{u^*}(m_1,m_0)e_{(m_1,m_0)}(X)\mu_{1}(X,m_1)}du\Bigg\}.
\eee
Similarly, one has $\E\{\bar{\tilde{\xi}}^{(0)}_1(X,Z,M,Y)\} = \E\{\int_{\R^2}{k_{u^*}(m_1,m_0)e_{(m_1,m_0)}(X)\mu_{0}(X,m_0)}du\}$. With same techniques as above, we can show
\beee\nn
\E\Big\{\bar{E}_{\xi_1}^{(1)}(X,Z,M)\Big\} & =& \E\Bigg\{\frac{\mathbb{I}(Z = 1){\gamma_1}^{(Z)}(X,M)}{\pi_Z(X) f(M \mid X,Z )} \bar\mu_{Z}(X,M)\Bigg\}
\\\nn
& =& \E\Bigg[\E\Bigg\{\frac{{\gamma_1}^{(1)}(X,M)}{ f(M \mid X,Z  = 1)} \bar\mu_{1}(X,M)\mid X, Z = 1\Bigg\}\Bigg]
\\\nn
& =& \E\Bigg\{\int_{\R}{{\gamma_1}^{(1)}(M = m_1,X)}\bar\mu_{1}(X,m_1)dm_1\Bigg\}
\\\nn
& =& \E\Bigg\{\int_{\R^2}{k_{u^*}(u)e_{u}(X)\bar\mu_{1}(X,m_1)}du\Bigg\},
\eee
and, similarly $\E\big\{\bar{E}_{\xi_1}^{(0)}(X,Z,M)\big\} = \E\big\{\int_{\R^2}{k_{u^*}(u)e_{u}(X)\bar\mu_{0}(X,m_0)}du\big\}$. Summarizing all results above,
\beee\nn
\bar\tau_{u^*}& =& \frac{1}{\E\big\{ \int_{\R^2}k_{u^*}(u){e}_{u}(X)du\big\}}\Bigg[\E\Big[\int_{\R^2}\big\{\mu_{1}( X,m_1) - \mu_{0}(X,m_0)\big\}k_{u^*}(u){e}_{u}(X)du\Big]
\\\nn
&& - \E\Big[\int_{\R^2}\big\{\bar\mu_{1}( X,m_1) - \bar\mu_{0}(X,m_0)\big\}k_{u^*}(u){e}_{u}(X)du\Big] 
\\
&&+ \E\Big[\int_{\R^2}\big\{\bar\mu_{1}( X,m_1) - \bar\mu_{0}(X,m_0)\big\}k_{u^*}(u){e}_{u}(X)du\Big]\Bigg]
\\\nn
& = &\tau_{u^*},
\eee
where the last equality is from Theorem \ref{tau:id} and we  recall  \eqref{def:lu}.
\par
Next we consider the scenario when $\bar \mu = \mu$. With same technique as \eqref{pf:dbr:main}, we can show
\beee\nn
\E\Big\{\bar{\tilde{\xi}}^{(z)}_1(X,Z,M,Y)\Big\} &=& \E\Bigg\{\frac{\pi_z(X)}{\bar{\pi}_z(X)}\int_{\R^2}{k_{u^*}(u)e_{u}(X)\mu_{z}(X,m_z)}du\Bigg\}
\eee
and 
\beee\nn
\E\Big\{\bar{E}_{\xi_1}^{(z)}(X,Z,M)\Big\} &=& \E\Bigg\{\frac{\pi_z(X)}{\bar{\pi}_z(X)}\int_{\R^2}{k_{u^*}(u)e_{u}(X)\bar\mu_{z}(X,m_z)}du\Bigg\}
\\\nn
&=&\E\Bigg\{\frac{\pi_z(X)}{\bar{\pi}_z(X)}\int_{\R^2}{k_{u^*}(u)e_{u}(X)\mu_{z}(X,m_z)}du\Bigg\},
\eee
for both $z = 0,1$. This implies \beee\label{ebt3:bound}
\E\big\{\bar{\tilde{\xi}}_1(X,Z,M,Y)\big\}& =& 0
\eee Finally since $\bar{\mu} = \mu$, by Theorem \ref{thm:idd} we have 
\beee\nn
\bar{\tau}_{u^*} = \frac{ 0+\E\big[\int_{\R^2}\big\{\mu_{1}( X,m_1) - \mu_{0}(X,m_0)\big\}k_{u^*}(u){e}_{u}(X)du\big]}{\E\big\{\int_{\R^2}k_{u^*}(u){e}_{u}(X)du\big\}}
= {\tau}_{u^*}.
\eee
Combining with Proposition~\ref{po:approximate}, we further have $\lim_{h\rightarrow 0}\bar{\tau}_{u^*} = \lim_{h\rightarrow 0}{\tau}_{u^*} = \tau^*_{u^*} $\qed
\subsection{Regularity assumptions for Theorem \ref{thm:consist}}\label{sec:otherreg}
We first discuss Assumption \ref{am:emp} in the main paper. Assumption \ref{am:emp}  (i) allows either one of the models $\mathcal{M}_{{\rm tp}}$ or $\mathcal{M}_{{\rm om}}$, to be incorrectly estimated, while consistency and limiting distribution results in Theorem \ref{thm:consist} will still be valid. Assumption \ref{am:emp}  (ii)--(iii) are standard which assume that when the sample size grows, nuisance function estimators are consistent to their limits under  $\mathcal{L}_{\infty}$ norm and specify their $\mathcal{L}_2$ convergence rates to the truths. Assumption \ref{am:emp} (iv) restricts the functional space complexities of  estimators $\hat{\pi}$, $\hat{\mu}$ and their limits, which is a usual minimal condition for the use of empirical process techniques in theoretical analysis; See e.g., \citet{kennedy2016semiparametric,westling2020causal}.
\par
Now we  list other regularity conditions used for Theorem \ref{thm:consist} as follows.
\begin{assumption}\label{am:pi} There exists some $\epsilon > 0$ such that $\pi_z(X),\bar{\pi}_z(X) \in (\epsilon,1/\epsilon)$, uniformly for all $X\in\mathbb{R}$ and $z = 0,1$. 
\end{assumption}
\begin{assumption}\label{am:kernel}For some universal constant $C > 0$, we have $$\max\Bigg\{\sup_{h>0}\int_{\R^2}k_{u^*}(u)du,\,\sup_{z\in\{0,1\},h>0,m_z\in\R}h\int_{\R}k_{u^*}(u)dm_{1-z},\, \sup_{h>0}h^2\int_{\R^2}k^2_{u^*}(u)du\Bigg\} \leq C.$$
\end{assumption}
\begin{assumption}\label{am:eux}
(1) $\E\{e_{U}(X)\} < +\infty$; (2) $c(u,v)$ and all its first and second-order partial derivatives are bounded away from $+\infty$, uniformly over $(u,v)\in[-\infty,\infty]\times [-\infty,\infty]$; (3) $c^{(10)}(u,v) = c^{(01)}(u,v)$.
\end{assumption}
\begin{assumption}\label{am:fmm}
$\sup_{u\in\R^2}e_u = \sup_{(m_1,m_0)\in\R^2}f(M_1 = m_1,M_0 =m_0) < C_M$ for a universal constant $C_M > 0$.
\end{assumption}
\begin{assumption}\label{am:var:Y}
$\mathbb{Y}$ is compact and thus $\sup_{x,m\in\R\atop z\in\{0,1\}}\text{Var}(Y\mid X = x, Z = z, M = m) <+\infty$. 
\end{assumption}
\begin{assumption}\label{am:unibound}
For any $g_{zm_z}(\cdot\mid \beta)\in\big\{f_{zm_z}(\cdot\mid\beta),F_{zm_z}( \cdot\mid\beta)\big\}$ and any $\tilde{\beta}\in\R^{d_\beta}$, denote
$
g'_{zm_z}(\cdot\mid \tilde\beta) = \frac{\partial}{\partial \beta}g_{zm_z}( \cdot\mid\beta)\ \Big|_{\beta = \tilde \beta}$ and $g''_{zm_z}(\cdot\mid \tilde\beta) = \frac{\partial^2}{\partial \beta\partial \beta^\T}g_{zm_z}(\cdot\mid \beta)\ \Big|_{\beta = \tilde \beta}.$ There exist constants $c_0,C_0,\delta_0 > 0$ such that, $\inf_{m_z,x\in\R\atop z\in\{0,1\}} f_{zm_z}(x\mid \beta^*) \geq c_0$ and 
\bee\nonumber
\max&\Big\{\sup_{{m_z,x\in \R\atop z \in\{0,1\}}}\big| g_{zm_z}(x\mid\beta^*)\big|,\sup_{{m_z,x\in \R\atop z \in\{0,1\}}}\big\| g'_{zm_z}( x\mid\beta^*)\big\|,
\sup_{m_z,x\in \R\atop{z \in \{0,1\}\atop {\|\tilde \beta - \beta^*\|\leq \delta_0}}}\big\| g''_{zm_z}(x\mid \tilde{\beta})\big\|_2\Big\} \leq C_0.
\ee 
\end{assumption}

Assumption \ref{am:pi} is the standard overlap condition \citep{rosenbaum1983central}.  Assumptions \ref{am:kernel}--\ref{am:eux} can be satisfied by most of the commonly-used kernel functions and copula functions \citep{jaworski2010copula,fan1993local,wasserman2006all}. Assumptions \ref{am:fmm}--\ref{am:var:Y} are  standard density and moment conditions for the data distribution \citep{chen2015optimal,luedtke2016statistical,chu2022targeted}. The condition of a compact $\mathbb{Y}$ in Assumption \ref{am:var:Y} is only for controling the empirical process term (see Section \ref{pf:lm:emp}), and can be further relaxed if one exploits the standard sample splitting technique. Assumption \ref{am:unibound} is a standard regularity condition to bound the first and second-order derivatives of some  parametrized  likelihood \citep{van2000asymptotic,fan2004nonconcave}.
\subsection{Proof of Theorem \ref{thm:consist}}
Although $h$ is fixed for Loc.PCE, to further enable our nonparametric theoretical analysis of  Cont.PCE in Section \ref{hto0:p}, we consider a more general setting in the proof such that we allow $h\rightarrow 0$. Our results in Theorem~\ref{thm:consist} can be easily obtained by setting $h$ as fixed in our general results.   For the ease of exposition, in the following, we will omit the footnote $u^*$ in $\hat{\xi}_{1,u^*}(\cdot)$, $\bar{\xi}_{1,u^*}(\cdot)$, etc, and directly write $\hat{\xi}_{1}(\cdot)$, $\bar{\xi}_{1}(\cdot)$, etc. We will also occasionally omit random variables in $\hat\xi_1(X,Z,M)$ and directly write $\hat\xi_1$ for simplicity. On the other hand, for better illustration, we expand  the notation of $\lbt\cdot\rbt$ through its definition in \eqref{def:lu} during the proof, and only use $\lbt\cdot\rbt$ when wrapping up results.
\par
We first define 
\begin{eqnarray}\nonumber
\Delta_{1,n}&=& (\mathbb{P}_n - \mathbb{P})\Big\{\int_{\R^2}k_{u^*}(u){e}_{u}(X)du\Big\} 
+\mathbb{P}_n\Big\{\int_{\R^2}k_{u^*}(u)\hat{e}_{u}(X)du - \int_{\R^2}k_{u^*}(u){e}_{u}(X)du\Big\},
\\\nn
\Delta_{2,n}&=&\mathbb{P}_n\Big[\hat{\xi}_1(X,Z,M)\{Y - \hat{\mu}_{Z}(X,M)\}
+\int_{\R^2}\big\{\hat\mu_{1}(X,m_1) - \hat\mu_{0}(X,m_0)\big\}k_{u^*}(u)\hat{e}_{u}(X)du\Big]
\\\nn
&& -\mathbb{P}_n\Big[\bar{\xi}_1(X,Z,M)\{Y - \bar{\mu}_{Z}(X,M)\}
+\int_{\R^2}\big\{\bar\mu_{1}(X,m_1) - \bar\mu_{0}(X,m_0)\big\}k_{u^*}(u){e}_{u}(X)du\Big].
\end{eqnarray}
Then by Taylor expansion, we have for some $\delta_n\in[0,1]$,
\begin{eqnarray}\nonumber
\hat{\tau}_{u^*} - \tau_{u^*}& = & -\tau_{u^*}+  \frac{\mathbb{P}_n\big[\hat{{\xi}}_1\{Y - \hat{\mu}_{Z}(X,M)\}
+\int_{\R^2}\big\{\hat\mu_{1}( X,m_1) - \hat\mu_{0}(X,m_0)\big\}k_{u^*}(u)\hat{e}_{u}(X)du\big]}{\mathbb{P}_n\big\{\int_{\R^2}k_{u^*}(u)\hat{e}_{u}(X)du\big\}}
\\\nonumber
& = & - \tau_{u^*} +\frac{\mathbb{P}_n\big[\bar{{\xi}}_1\{Y - \bar{\mu}_{Z}(X,M)\}
+\int_{\R^2}\big\{\bar\mu_{1}( X,m_1) - \bar\mu_{0}(X,m_0)\big\}k_{u^*}(u){e}_{u}(X)du\big] + \Delta_{2,n}}{\mathbb{E}\big\{\int_{\R^2}k_{u^*}(u){e}_{u}(X)du\big\} + \Delta_{1,n}}
\\\nonumber
& = & - \tau_{u^*} +\frac{\mathbb{P}_n\big[\bar{{\xi}}_1\{Y - \bar{\mu}_{Z}(X,M)\}
+\int_{\R^2}\big\{\bar\mu_{1}( X,m_1) - \bar\mu_{0}(X,m_0)\big\}k_{u^*}(u){e}_{u}(X)du\big] + \Delta_{2,n}}{\mathbb{E}\big\{\int_{\R^2}k_{u^*}(u){e}_{u}(X)du\big\}} 
\\\nonumber
&& -  \frac{\mathbb{P}_n\big[\bar{{\xi}}_1\{Y - \bar{\mu}_{Z}(X,M)\}
+\int_{\R^2}\big\{\bar\mu_{1}( X,m_1) - \bar\mu_{0}(X,m_0)\big\}k_{u^*}(u){e}_{u}(X)du\big] + \Delta_{2,n}}{\big[\mathbb{E}\big\{\int_{\R^2}k_{u^*}(u){e}_{u}(X)du\big\}\big]^2} \Delta_{1,n}
\\\nonumber
&& +\frac{\mathbb{P}_n\big[\bar{{\xi}}_1\{Y - \bar{\mu}_{Z}(X,M)\}
+\int_{\R^2}\big\{\bar\mu_{1}( X,m_1) - \bar\mu_{0}(X,m_0)\big\}k_{u^*}(u){e}_{u}(X)du\big] + \Delta_{2,n}}{\big[\mathbb{E}\big\{\int_{\R^2}k_{u^*}(u){e}_{u}(X)du\big\}+ \delta_n\Delta_{1,n}\big]^3} \Delta^2_{1,n}
\\\nonumber
& =&\frac{(\mathbb{P}_n - \mathbb{P})\big[\bar{{\xi}}_1\{Y - \bar{\mu}_{Z}(X,M)\}
+\int_{\R^2}\big\{\bar\mu_{1}( X,m_1) - \bar\mu_{0}(X,m_0)\big\}k_{u^*}(u){e}_{u}(X)du\big] + \Delta_{2,n}}{\mathbb{E}\big\{\int_{\R^2}k_{u^*}(u){e}_{u}(X)du\big\}} 
\\\nonumber
&& - \frac{\mathbb{P}\big[\bar{{\xi}}_1\{Y - \bar{\mu}_{Z}(X,M)\}
+\int_{\R^2}\big\{\bar\mu_{1}( X,m_1) - \bar\mu_{0}(X,m_0)\big\}k_{u^*}(u){e}_{u}(X)du\big]}{\big[\mathbb{E}\big\{\int_{\R^2}k_{u^*}(u){e}_{u}(X)du\big\}\big]^2}\Delta_{1,n}
\\\nn
&& -  \frac{(\mathbb{P}_n - \mathbb{P})\big[\bar{{\xi}}_1\{Y - \bar{\mu}_{Z}(X,M)\}
+\int_{\R^2}\big\{\bar\mu_{1}( X,m_1) - \bar\mu_{0}(X,m_0)\big\}k_{u^*}(u){e}_{u}(X)du\big] + \Delta_{2,n}}{\big[\mathbb{E}\big\{\int_{\R^2}k_{u^*}(u){e}_{u}(X)du\big\}\big]^2} \Delta_{1,n}
\\\label{main:piece1}
&& +\frac{\mathbb{P}_n\big[\bar{{\xi}}_1\{Y - \bar{\mu}_{Z}(X,M)\}
+\int_{\R^2}\big\{\bar\mu_{1}( X,m_1) - \bar\mu_{0}(X,m_0)\big\}k_{u^*}(u){e}_{u}(X)du\big] + \Delta_{2,n}}{\big[\mathbb{P}\big\{\int_{\R^2}k_{u^*}(u){e}_{u}(X)du\big\}+ \delta_n\Delta_{1,n}\big]^3} \Delta^2_{1,n}
\end{eqnarray}
where the last equality is by the double robustness  introduced in Theorem \ref{thm:dbrb}. Next, in Sections \ref{sec:boundD1n} and \ref{sec:boundD2n}, we bound the relative error terms above, respectively.

\subsubsection{Bounding $\Delta_{1,n}$}\label{sec:boundD1n} For simplicity, we denote 
$$
\Delta_{1,n}^{(1)} = (\mathbb{P}_n - \mathbb{P})\Big\{\int_{\R^2}k_{u^*}(u){e}_{u}(X)du\Big\}\quad\text{and}\quad\Delta_{1,n}^{(2)} = \mathbb{P}_n\Big\{\int_{\R^2}k_{u^*}(u)\hat{e}_{u}(X)du - \int_{\R^2}k_{u^*}(u){e}_{u}(X)du\Big\}.
$$ 
Thus we have $\Delta_{1,n} = \Delta_{1,n}^{(1)} + \Delta_{1,n}^{(2)}$. 
\par 
Notice that, 
\begin{eqnarray}\nn
\mathbb{P}\Big\{\int_{\R^2}k_{u^*}(u){e}_{u}(X)du\Big\}^2 &\leq &\mathbb{P}\Big\{\int_{\R^2}k^2_{u^*}(u)du\cdot\int_{\R^2}{e}^2_{u}(X)du\Big\} \quad\text{(Cauchy--Schwartz inequality)}
\\\label{cherna}
&\precsim & h^{-2}\cdot\mathbb{P}\Big\{\int_{\R^2}{e}^2_{u}(X)du\Big\}\quad\text{(Assumption \ref{am:kernel})}
\\\nonumber
&=&h^{-2}\cdot\int_{\mathbb{R}^3}f^2(U = u\mid X = x)\cdot f(X = x)dxdu
\\\nonumber
&=&h^{-2}\cdot\int_{\mathbb{R}^3}f(U = u\mid X = x)\cdot f(X = x, U = u)dxdu
\\\nonumber
& = & h^{-2}\cdot\E\{e_{U}(X)\} \precsim h^{-2}. \quad (\text{Assumption \ref{am:eux}}).
\end{eqnarray} 
Then by a standard argument based on the Chebyshev's inequality, we  conclude 
\beee\nonumber
\mathrm{pr}\Bigg[\Big|\mathbb{P}_n\Big\{\int_{\R^2}k_{u^*}(u){e}_{u}(X)du\Big\} - \mathbb{P}\Big\{\int_{\R^2}k_{u^*}(u){e}_{u}(X)du\Big\}\Big| \geq kh^{-1}n ^{-1/2}\Bigg] &\precsim& k^{-2},
\eee
for any $k > 0$, which implies that,
\beee\nn
\Delta_{1,n}^{(1)} &= &(\mathbb{P}_n - \mathbb{P})\Big\{\int_{\R^2}k_{u^*}(u){e}_{u}(X)du\Big\} 
\\\label{d11n}
&= &\mathcal{O}_{\mathbb{P}}(h^{-1}n^{-1/2}).
\eee
\par
On the other hand, by Taylor expansion one has 
\begin{eqnarray}\label{delta11a}\Delta_{1,n}^{(2)}&=& \mathbb{P}_n\Bigg[\int_{\R^2}k_{u^*}(u)\Big[c_{\rho}\big\{{F}_{1m_1}( X\mid \hat{\beta}),{F}_{0m_0}(X\mid \hat{\beta})\big\}{f}_{1m_1}( X\mid\hat{\beta}){f}_{0m_0}(X\mid\hat{\beta}) - e_{u}(X)\Big]du\Bigg]
\\\label{ee:main}
&=&\scalemath{0.8}{\underbrace{\mathbb{P}_n\Big[\int_{\R^2}k_{u^*}(u)\Big[c_{\rho}\big\{{F}_{1m_1}( X\mid\beta^*),{F}_{0m_0}(X\mid \beta^*)\big\}{f}_{1m_1}( X\mid\hat{\beta}){f}_{0m_0}( X\mid\hat{\beta})- e_{u}(X)\Big]du\Big]}_{\Delta_{1,n}^{(2,1)}}}
\\\nonumber
& &\scalemath{0.8}{+ \underbrace{ \sum_{z = 0,1}\mathbb{P}_n\Big[\int_{\R^2}k_{u^*}(u)
\cdot \Big[c_{\rho}^{(z)}\big\{{F}_{1m_1}( X\mid \beta^*),{F}_{0m_0}( X\mid\beta^*)\big\}{f}_{1m_1}( X\mid\hat{\beta}){f}_{0m_0}( X\mid\hat{\beta})\big\{F_{zm_z}( X\mid\hat{\beta}) - F_{zm_z}( X\mid{\beta^*})\big\}\Big]du\Big]}_{\Delta_{1,n}^{(2,2)}}}
\\\nonumber
&&\scalemath{0.8}{ +\underbrace{\frac{1}{2}\sum_{z = 0,1}\mathbb{P}_n\Big[\int_{\R^2}k_{u^*}(u)
\Big[ c_{\rho}^{(zz)}\big\{\xi_{1,n},\xi_{0,n}\big\}{f}_{1m_1}( X\mid\hat{\beta}){f}_{0m_0}(X\mid \hat{\beta})\big\{F_{zm_z}(X\mid \hat{\beta}) - F_{zm_z}( X\mid{\beta^*})\big\}^2\Big]du\Big]}_{\Delta_{1,n}^{(2,3)}}}
\\\nonumber
&& \scalemath{0.8}{+ \underbrace{\mathbb{P}_n\Big[\int_{\R^2}k_{u^*}(u)
\Big[c_{\rho}^{(10)}\big\{\xi_{1,n},\xi_{0,n}\big\}{f}_{1m_1}( X\mid\hat{\beta}){f}_{0m_0}( X\mid\hat{\beta})\prod_{z = 0,1}\big\{F_{zm_z}(X\mid \hat{\beta}) - F_{zm_z}(X\mid {\beta^*})\big\}\Big]du\Big]}_{\Delta_{1,n}^{(2,4)}}},
\end{eqnarray}
where $\xi_{z,n} =\delta_{z,n} {F}_{zm_z}(X\mid\beta^*) + (1 - \delta_{z,n}){F}_{zm_z}(X\mid\hat{\beta})$, for some $\delta_{z,n}\in[0,1]$ and $z = 0,1$. We recall that $c^{(1)}(a,b) = \frac{\partial c(a,b)}{\partial a}$, $c^{(11)}(a,b) = \frac{\partial^2 c(a,b)}{\partial^2 a}$ and analogously for other notations, and in \eqref{ee:main} we use the symmetric property of the copula function under Assumption \ref{am:eux}. With straightforward algebra, we further have
\begin{eqnarray}\nonumber
&&{f}_{1m_1}( X\mid\hat{\beta}){f}_{0m_0}(X\mid \hat{\beta})- {f}_{1m_1}( X\mid{\beta^*}){f}_{0m_0}(X\mid {\beta^*})
\\\nn
&=&\big\{f_{1m_1}(X\mid \hat{\beta}) - {f}_{1m_1}(X\mid \beta^*)\big\}\big\{{f}_{0m_0}(X\mid \hat{\beta}) - {f}_{0m_0}(X\mid \beta^*)\big\}
\\\nn
&& + \big\{f_{1m_1}(X\mid \hat{\beta}) - {f}_{1m_1}(X\mid \beta^*)\big\}{f}_{0m_0}(X\mid \beta^*)
\\\label{ee:main:1}
&& +\big\{f_{0m_0}(X\mid \hat{\beta}) - {f}_{0m_0}( X\mid\beta^*)\big\}{f}_{1m_1}( X\mid\beta^*),
\end{eqnarray}
meanwhile by Taylor expansion we actually have for $z = 0,1$,
\bee\label{taylor:fa}
f_{zm_z}(X\mid \hat{\beta}) - f_{zm_z}(X\mid \beta^*)=\{f'_{zm_z}( X\mid\beta^*)\}^\T(\hat{\beta} - \beta^*)+ (\hat{\beta} - \beta^*)^\T f''_{zm_z}( X\mid\tilde{\beta}_{X,m_z})(\hat{\beta} - \beta^*),
\ee
where $\tilde{\beta}_{X,m_z} = \delta_{X,m_z}\hat{\beta} + (1 - \delta_{X,m_z})\beta^*$ for some $\delta_{X,m_z} \in [0,1]$. Combining \eqref{ee:main:1} and \eqref{taylor:fa}, the first term on the right-hand side of \eqref{ee:main} equals to 
\begin{eqnarray}\nonumber
\Delta_{1,n}^{(2,1)} &=&\mathbb{P}_n\Big[\int_{\R^2}k_{u^*}(u)c_{\rho}\big\{{F}_{1m_1}( X\mid\beta^*),{F}_{0m_0}( X\mid\beta^*)\big\}
\\\nn
&&\cdot\big\{{f}_{1m_1}( X\mid\hat{\beta}){f}_{0m_0}( X\mid\hat{\beta})
- {f}_{1m_1}(X\mid {\beta^*}){f}_{0m_0}( X\mid{\beta^*})\big\}du\Big]
\\\nonumber
&=& \mathbb{P}_n\Bigg[\int_{\R^2}k_{u^*}(u)c_{\rho}\big\{{F}_{1m_1}( X\mid\beta^*),{F}_{0m_0}( X\mid\beta^*)\big\}
\\\nonumber
&&\cdot\Big[(\hat{\beta} - \beta^*)^\T f'_{1m_1}(X\mid \beta^*)\{f'_{0m_0}( X\mid\beta^*)\}^\T(\hat{\beta} - \beta^*)
\\\nonumber
&& + (\hat{\beta} - \beta^*)^\T f'_{1m_1}(X\mid \beta^*)\cdot (\hat{\beta} - \beta^*)^\T f''_{0m_0}( X\mid\tilde{\beta}_{X,m_0})(\hat{\beta} - \beta^*)
\\\nonumber
&&+ (\hat{\beta} - \beta^*)^\T f'_{0m_0}( X\mid\beta^*)\cdot (\hat{\beta} - \beta^*)^\T f''_{1m_1}(X\mid \tilde{\beta}_{X,m_1})(\hat{\beta} - \beta^*)
\\\nonumber
&&+(\hat{\beta} - \beta^*)^\T f''_{0m_0}( X\mid\tilde{\beta}_{X,m_0})(\hat{\beta} - \beta^*)\cdot (\hat{\beta} - \beta^*)^\T f''_{1m_1}(X\mid \tilde{\beta}_{X,m_1})(\hat{\beta} - \beta^*)\Big]du\Bigg]
\\\nonumber
&&+\sum_{z = 0,1}\mathbb{P}_n\Bigg[\int_{\R^2}k_{u^*}(u)c_{\rho}\big\{{F}_{1m_1}( X\mid\beta^*),{F}_{0m_0}( X\mid\beta^*)\big\}
\\\nonumber
&&\cdot\Big[f_{(1 - z)m_{1- z}}( X\mid\beta^*)\{f'_{zm_z}(X\mid \beta^*)\}^\T(\hat{\beta} - \beta^*) 
\\\label{ee:main:2}
&&+ f_{(1 - z)m_{1- z}}( X\mid\beta^*) \cdot(\hat{\beta} - \beta^*)^\T f''_{zm_z}(X\mid \tilde{\beta}_{X,m_z})(\hat{\beta} - \beta^*) \Big]du\Bigg].
\end{eqnarray}
Since $\hat{\beta} - \beta^* = \mathcal{O}_\mathbb{P}(n^{-1/2})$ under a parametric model of $\mathcal{M}_{{\rm ps}}$, we have with probability approaching 1 (wpa1), $\|\tilde{\beta}_{x,m_z}- \beta^*\| \leq \delta_0$ uniformly for all $x,m_z\in \R$ when $n$ is sufficiently large. Denote $\|\cdot\|_2$  the matrix spectral norm. Then by Assumption \ref{am:unibound}, we have both $\sup_{m_z,x\in
\R}\|f''_{zm_z}( x\mid\tilde{\beta}_{x,m_z})\|_2$ and $\sup_{m_z,x\in
\R}\|f'_{zm_z}( x\mid\tilde{\beta}_{x,m_z})\|$ are $\mathcal{O}(1)$ for both $z = 0,1$. By the bounds of $k_{u^*}(\cdot,\cdot)$ and $c(\cdot,\cdot)$ given in Assumptions \ref{am:kernel} and \ref{am:eux}, we can bound the first and second terms in \eqref{ee:main:2} as following,
\begin{eqnarray}\nonumber
&&\mathbb{P}_n\Bigg[\int_{\R^2}k_{u^*}(u)c_{\rho}\big\{F_{1m_1}( X),F_{0m_0}( X)\big\}(\hat{\beta} - \beta^*)^\T f'_{1m_1}(X\mid \beta^*)\{f'_{0m_0}( X\mid\beta^*)\}^\T(\hat{\beta} - \beta^*)du\Bigg]
\\\nonumber
&\leq & \sup_{x\in\R}\Big|(\hat{\beta} - \beta^*)^\T\int_{\R^2}k_{u^*}(u)c_{\rho}\big\{{F}_{1m_1}( x),{F}_{0m_0}( x)\big\} f'_{1m_1}( x\mid\beta^*)\{f'_{0m_0}(x\mid \beta^*)\}^\T du\cdot (\hat{\beta} - \beta^*)\Big|
\\\nonumber
&\leq & \|\hat{\beta} - \beta^*\|^2\cdot \sup_{(u,v)\in \R^2}|c(u,v)| \cdot\prod_{z = 1,0} \sup_{m_z\in\mathbb{R},x\in\mathbb{R}^{d_X}}\|f_{zm_z}'(x\mid \beta^*)\|\underbrace{\sqrt{\int_{\R^2}k^2_{u^*}(u)du}}_{ \tiny{\mathcal{O}(h^{-1}) \text{ by Assumption~\ref{am:kernel}}}}
\\\nonumber
&= &\mathcal{O}_\mathbb{P}(h^{-1}n^{-1})
\end{eqnarray}
and similarly,
\begin{eqnarray}\nonumber
&&\mathbb{P}_n\Bigg[\int_{\R^2}k_{u^*}(u) c_{\rho}\big\{F_{1m_1}(X),F_{0m_0}( X)\big\}(\hat{\beta} - \beta^*)^\T f'_{1m_1}( X\mid\beta^*)
\cdot (\hat{\beta} - \beta^*)^\T f''_{0m_0}( X\mid\tilde{\beta}_{X,m_0})(\hat{\beta} - \beta^*)du\Bigg]
\\\nonumber
&\leq & \|\hat{\beta} - \beta^*\|^3\sup_{(u,v)\in \R^2}|c(u,v)| \sup_{m_1\in\R,x\in\mathbb{R}^{d_X}}\|f_{1m_1}'(x\mid \beta^*)\|
\sup_{m_0\in\R,x\in\R^{d_X}}\|f''_{0m_0}(X\mid \tilde{\beta}_{X,m_0})\|_2\cdot \sqrt{\int_{\R^2}k^2_{u^*}(u)du}
\\\nonumber
&=&\mathcal{O}_{\mathbb{P}}(h^{-1}n^{-3/2}).
\end{eqnarray}
With similar argument, we can finally show the first term in \eqref{ee:main} is
\begin{eqnarray}\nonumber
\Delta_{1,n}^{(2,1)} & = & \sum_{z = 0,1}\mathbb{P}_n\Bigg[\int_{\R^2}k_{u^*}(u)c_{\rho}\big\{{F}_{1m_1}( X),{F}_{0m_0}( X)\big\} f_{(1 - z)m_{1- z}}( X\mid\beta^*)\{f'_{zm_z}( X\mid \beta^*)\}^\T du\Bigg](\hat{\beta} - \beta^*)
\\\nonumber
&&+o_{\mathbb{P}}(h^{-1}n^{-1/2})
\\\nonumber
& = & \sum_{z = 0,1}\mathbb{P}\Bigg[\int_{\R^2}k_{u^*}(u)c_{\rho}\big\{{F}_{1m_1}( X),{F}_{0m_0}( X)\big\} f_{(1 - z)m_{1- z}}( X\mid\beta^*)\{f'_{zm_z}( X\mid\beta^*)\}^\T du\Bigg](\hat{\beta} - \beta^*)
\\\nonumber
&&+o_{\mathbb{P}}(h^{-1}  n^{-1/2});
\end{eqnarray}
Here in the last equality, by Chebyshev's inequality, we can easily show that,
\begin{eqnarray}\nonumber
&&(\mathbb{P}_n - \mathbb{P})\Bigg[\int_{\R^2}k_{u^*}(u)c_{\rho}\big\{{F}_{1m_1}( X),{F}_{0m_0}( X)\big\} f_{(1 - z)m_{1- z}}( X\mid\beta^*)\{f'_{zm_z}(X\mid \beta^*)\}^\T du\Bigg]\cdot(\hat{\beta} - \beta^*)
\\\nonumber
&=&\mathcal{O}_{\mathbb{P}}(h^{-1}n^{-1/2})\cdot\mathcal{O}_{\mathbb{P}}(n^{-1/2}) = \mathcal{O}_{\mathbb{P}}(h^{-1}n^{-1}).
\end{eqnarray}
 Note here $\int_{\R^2}k_{u^*}(u)c_{\rho}\big\{{F}_{1m_1}(X),{F}_{0m_0}(X)\big\} f_{(1 - z)m_{1- z}}( X\mid\beta^*)\{f'_{zm_z}( X\mid\beta^*)\}^\T du$ is a random variable with finite variance under Assumptions \ref{am:kernel}--\ref{am:eux}. With similar argument, we can finally show the second term on the right-hand side of \eqref{ee:main} is 
\begin{eqnarray}\nonumber
\Delta_{1,n}^{(2,2)}&=&\scalemath{0.8}{\sum_{z = 0,1}\mathbb{P}\Bigg[\int_{\R^2}k_{u^*}(u)\Big[c_{\rho}^{(z)}\big\{{F}_{1m_1}(X),{F}_{0m_0}(X)\big\}{f}_{1m_1}( X\mid{\beta}^*){f}_{0m_0}( X\mid{\beta}^*)\big\{F'_{zm_z}( X\mid{\beta}^*)\big\}^\T\Big]du\Bigg](\hat{\beta} - \beta^*)}
\\\label{delta11final}
&&+ o_{\mathbb{P}}(h^{-1}n^{-1/2}),
\end{eqnarray}
and $\Delta_{1,n}^{(2,3)}$--$\Delta_{1,n}^{(2,4)}$ on the right-hand side of \eqref{ee:main} are both $o_{\mathbb{P}}(h^{-1}n^{-1/2})$. Summarizing the results above, we finally conclude,
\begin{eqnarray}\nonumber
\Delta_{1,n}&=&\scalemath{0.8}{(\mathbb{P}_n - \mathbb{P})\Big\{\int_{\R^2}k_{u^*}(u){e}_{u}(X)du\Big\}}
\\\nonumber
&&\scalemath{0.8}{+ \sum_{z = 0,1}\mathbb{P}\Bigg[\int_{\R^2}k_{u^*}(u)c_{\rho}\big\{{F}_{1m_1}( X),{F}_{0m_0}( X)\big\} f_{(1 - z)m_{1- z}}( X\mid\beta^*)\{f'_{zm_z}( X\mid\beta^*)\}^\T du\Bigg](\hat{\beta} - \beta^*)}
\\\nonumber
&&\scalemath{0.8}{+\sum_{z = 0,1}\mathbb{P}\Bigg[\int_{\R^2}k_{u^*}(u)\Big[c_{\rho}^{(z)}\big\{{F}_{1m_1}( X),{F}_{0m_0}(X)\big\}{f}_{1m_1}( X\mid{\beta}^*){f}_{0m_0}( X\mid{\beta}^*)\big\{F'_{zm_z}(X\mid {\beta}^*)\big\}^\T\Big]du\Bigg]
(\hat{\beta} - \beta^*)}
\\\nonumber
&& + o_{\mathbb{P}}(h^{-1}n^{-1/2})
 \\\label{main:piece2}
 &=& \underbrace{{\Gamma}_{1,n}}_{\mathcal{O}_{\mathbb{P}}(h^{-1}n^{-1/2})} + o_{\mathbb{P}}(h^{-1}n^{-1/2})  = \mathcal{O}_{\mathbb{P}}(h^{-1}n^{-1/2}),
\end{eqnarray}
where the last equality follows from \eqref{d11n}, $\|\hat{\beta} - \beta^*\| = \mathcal{O}_{\mathbb{P}}(n^{-1/2})$, and the facts that for $z\in\{0,1\}$,
\begin{eqnarray}\nonumber
&&\Big\|\mathbb{P}\Big[\int_{\R^2}k_{u^*}(u)c_{\rho}\big\{{F}_{1m_1}( X),{F}_{0m_0}(X)\big\} f_{(1 - z)m_{1- z}}(X\mid \beta^*)\{f'_{zm_z}(X\mid \beta^*)\}^\T du\Big]\Big\|
\\\nonumber
&\leq& \sup_{(u,v)\in\R^2}c_{\rho}(u,v)\sup_{m_{1 - z}\in \R,\,x \in \R^{d_X}}\Big| f_{(1 - z)m_{1 - z}}( x\mid \beta^*) \Big|\sup_{m_z\in\R\, x\in \R^{d_X}}\big\|f'_{zm_z}(x\mid \beta^*)\big\|\cdot\sqrt{\int_{\R^2}k^2_{u^*}(u)du}
\\\label{unif:exp}
&\precsim& h^{-1}.
\end{eqnarray}
Under Assumptions \ref{am:eux} and \ref{am:unibound}, we similarly have,
$$
\Big\|\mathbb{P}\Big[\int_{\R^2}k_{u^*}(u)\big[c_{\rho}^{(z)}\big\{{F}_{1m_1}( X),{F}_{0m_0}( X)\big\}{f}_{1m_1}( X\mid{\beta}^*){f}_{0m_0}( X\mid{\beta}^*)\big\{F'_{zm_z}( X\mid{\beta}^*)\big\}^\T\big]du\Big]\Big\| \precsim h^{-1}.
$$

\par
\subsubsection{Bounding $\Delta_{2,n}$}\label{sec:boundD2n} We now bound $\Delta_{2,n}$. Decompose $\Delta_{2,n}$ as
\beee\nn
&&\Delta_{2,n} 
\\
&= &\underbrace{\scalemath{0.8}{\mathbb{P}_n\Bigg[
\int_{\R^2}k_{u^*}(u)\Big[\big\{\hat\mu_{1}( X,m_1) - \hat\mu_{0}(X,m_0)\big\}\big\{\hat{e}_{u}(X)-{e}_{u}(X)\big\}\Big]du\Bigg]}}_{ \Delta_{2,n}^{(1)}}
\\\nn
&&+\underbrace{\scalemath{0.8}{\mathbb{P}_n\Big[\hat{\xi}_1\{Y - \hat{\mu}_{Z}(X,M)\} - \bar{\xi}_1\{Y - \bar{\mu}_{Z}(X,M)\}\Big]+ \mathbb{P}_n\Bigg[
\int_{\R^2}k_{u^*}(u)\big[\{\hat\mu_{1}( X,m_1) - \hat\mu_{0}(X,m_0)\} - \{\bar\mu_{1}( X,m_1) - \bar\mu_{0m_0}(X)\}\big]{e}_{u}(X)du\Bigg]}}_{ \Delta_{2,n}^{(2)}}.
\eee
We bound the two terms above respectively. For $\Delta_{2,n}^{(1)}$, we further decompose it as
\begin{eqnarray}\nonumber
 \Delta_{2,n}^{(1)} &=&\underbrace{\mathbb{P}_n\Bigg[
 \int_{\R^2}k_{u^*}(u)\Big[\big\{\bar\mu_{1}( X,m_1) - \bar\mu_{0}(X,m_0)\big\}\{\hat{e}_{u}(X)-{e}_{u}(X)\}\Big]du\Bigg]}_{\Delta_{2,n}^{(1,1)}}
 \\\nonumber
& &+\underbrace{\mathbb{P}_n\Bigg[
\int_{\R^2}k_{u^*}(u)\big[\{\hat\mu_{1}( X,m_1) - \hat\mu_{0}(X,m_0)\}-\{\bar\mu_{1}( X,m_1) - \bar\mu_{0}(X,m_0)\}\big]\{\hat{e}_{u}(X)-{e}_{u}(X)\}du\Bigg]}_{\Delta_{2,n}^{(1,2)}}.
\end{eqnarray}
A similar argument to \eqref{delta11a}--\eqref{delta11final} that handles $\mathbb{P}_n\big[\int_{\R^2}k_{u^*}(u)\big\{\hat{e}_{u}(X)  - e_{u}(X)\big\}du\big]$ can similarly show that
\begin{eqnarray}\nonumber
&&\Delta_{2,n}^{(1,1)}
\\\nn
& = &\scalemath{0.75}{\sum_{z = 0,1}\mathbb{P}\Bigg[\int_{\R^2}k_{u^*}(u)\big\{\bar\mu_{1}( X,m_1) - \bar\mu_{0}(X,m_0)\big\}
c_{\rho}\big\{{F}_{1m_1}( X\mid\beta^*),{F}_{0m_0}(X\mid \beta^*)\big\}f_{(1 - z)m_{1- z}}( X\mid\beta^*)\{f'_{zm_z}( X\mid\beta^*)\}^\T du\Bigg]}\scalemath{0.75}{(\hat{\beta} - \beta^*)}
\\\nonumber
&&\scalemath{0.75}{+\sum_{z = 0,1}\mathbb{P}\Bigg[\int_{\R^2}k_{u^*}(u)\big\{\bar\mu_{1}( X,m_1) - \bar\mu_{0}(X,m_0)\big\} c_{\rho}^{(a)}\big\{{F}_{1m_1}( X\mid\beta^*),{F}_{0m_0}(X\mid \beta^*)\big\}}\scalemath{0.75}{{f}_{1m_1}( X\mid{\beta}^*){f}_{0m_0}( X\mid{\beta}^*)\big\{F'_{zm_z}(X\mid {\beta}^*)\big\}^\T du\Bigg]}\scalemath{0.75}{ (\hat{\beta} - \beta^*) }
\\\nonumber
&&\scalemath{0.75}{+ o_{\mathbb{P}}(h^{-1}n^{-1/2})}
\\\nonumber
&=& \scalemath{0.75}{\underbrace{{\Gamma}_{2,n}^{(1)}}_{\mathcal{O}_{\mathbb{P}}(h^{-1}n^{-1/2})} + o_{\mathbb{P}}(h^{-1}n^{-1/2}),}
\end{eqnarray}
under Assumptions \ref{am:emp} and \ref{am:kernel}. Note $\Delta_{2,n}^{(1,2)}$ is built by $\mathbb{P}_n\big[\int_{\R^2}k_{u^*}(u)\big\{\hat{e}_{u}(X)  - e_{u}(X)\big\}du\big] $ with an additional factor $\{\hat\mu_{1}( X,m_1) - \hat\mu_{0}(X,m_0)\}-\{\bar\mu_{1}( X,m_1) - \bar\mu_{0}(X,m_0)\}$ in the integrand. Thus we can similarly use  the Taylor expansion technique shown in \eqref{delta11a}--\eqref{delta11final} to deduce 
$$
\mathbb{P}_n\Bigg[\int_{\R^2}k_{u^*}(u)\big\{\hat{e}_{u}(X)  - e_{u}(X)\big\}du\Bigg] = \mathcal{O}_{\mathbb{P}}(h^{-1}n^{-1/2}),
$$ with a uniform bound such that 
\bee\nonumber
\sup_{x\in\R^{d_X}}\Big|\{\hat\mu_{1}(x,m_1) - \hat\mu_{0}(x,m_0)\}-\{\bar\mu_{1}(x,m_1) - \bar\mu_{0}(x,m_0)\}\Big| = o_{\mathbb{P}}(1),
\ee 
under Assumption \ref{am:emp}, thus we have $
\Delta_{2,n}^{(1,2)} = o_{\mathbb{P}}(h^{-1}n^{-1/2})$. In summary, one has
\beee\nn
\Delta_{2,n}^{(1)}& =& {\Gamma}_{2,n}^{(1)} + o_{\mathbb{P}}(h^{-1}n^{-1/2}).
\eee
For $\Delta_{2,n}^{(2)}$, we decompose
\begin{eqnarray}\nonumber
\Delta_{2,n}^{(2)} 
&=&\mathbb{P}_n\Bigg[\frac{(-1)^{Z+1}\hat{\gamma}_1^{(Z)}(X,M)}{\hat{\pi}_Z(X){f}_{ZM}( X\mid\hat{\beta})}\big\{Y - \hat{\mu}_{Z}(X,M)\big\}\Bigg] - \mathbb{P}_n\Bigg[\frac{(-1)^{Z+1}{\gamma}_1^{(Z)}(X,M)}{\hat{\pi}_Z(X){f}_{ZM}( X\mid{\beta}^*)}\big\{Y - \hat{\mu}_{Z}(X,M)\big\}\Bigg]
\\\nonumber
&&+\mathbb{P}_n\Bigg[\frac{(-1)^{Z+1}{\gamma}_1^{(Z)}(X,M)}{\hat{\pi}_Z(X){f}_{ZM}( X\mid{\beta}^*)}\big\{Y - \hat{\mu}_{Z}(X,M)\big\}\Bigg] - \mathbb{P}_n\Bigg[\frac{(-1)^{Z+1}{\gamma}_1^{(Z)}(X,M)}{\bar{\pi}_Z(X){f}_{ZM}( X\mid{\beta}^*)}\big\{Y - \bar{\mu}_{Z}(X,M)\big\}\Bigg]
\\\nonumber
&&+\mathbb{P}_n\Bigg[
\int_{\R^2}k_{u^*}(u)\big[\{\hat\mu_{1}( X,m_1) - \hat\mu_{0}(X,m_0)\} - \{\bar\mu_{1}( X,m_1) - \bar\mu_{0}(X,m_0)\}\big]{e}_{u}(X)du\Bigg]
\\\nonumber
&=&\mathbb{P}_n\Bigg[\frac{(-1)^{Z+1}}{{\pi}_Z(X)}\big\{Y - {\mu}_{Z}(X,M)\big\}\Big\{\frac{\hat{\gamma}_1^{(Z)}(X,M)}{{f}_{ZM}( X\mid\hat{\beta})} - \frac{{\gamma}_1^{(Z)}(X,M)}{{f}_{ZM}( X\mid\beta^*)}\Big\}\Bigg]
\\\nonumber
&&+\mathbb{P}_n\Bigg[\Big[\frac{(-1)^{Z+1}}{\hat{\pi}_Z(X)}\big\{Y - \hat{\mu}_{Z}(X,M)\big\}-\frac{(-1)^{Z+1}}{{\pi}_Z(X)}\big\{Y - {\mu}_{Z}(X,M)\big\}\Big]\Big\{\frac{\hat{\gamma}_1^{(Z)}(X,M)}{{f}_{ZM}( X\mid\hat{\beta})} - \frac{{\gamma}_1^{(Z)}(X,M)}{{f}_{ZM}( X\mid\beta^*)}\Big\}\Bigg] 
\\\nonumber
&&+\mathbb{P}_n\Bigg[\frac{(-1)^{Z+1}{\gamma}_1^{(Z)}(X,M)}{\hat{\pi}_Z(X){f}_{ZM}( X\mid{\beta}^*)}\big\{Y - \hat{\mu}_{Z}(X,M)\big\}+\int_{\R^2}k_{u^*}(u)\{\hat\mu_{1}( X,m_1) - \hat\mu_{0}(X,m_0)\} {e}_{u}(X)du\Bigg] 
\\\nonumber
&&- \mathbb{P}_n\Bigg[\frac{(-1)^{Z+1}{\gamma}_1^{(Z)}(X,M)}{\bar{\pi}_Z(X){f}_{ZM}( X\mid{\beta}^*)}\big\{Y - \bar{\mu}_{Z}(X,M)\big\}+\int_{\R^2}k_{u^*}(u)\{\bar\mu_{1}( X,m_1) - \bar\mu_{0}(X,m_0)\} {e}_{u}(X)du\Bigg]
\\
&=&\underbrace{\mathbb{P}_n\Bigg[\frac{(-1)^{Z+1}}{{\pi}_Z(X)}\big\{Y - {\mu}_{Z}(X,M)\big\}\Big\{\frac{\hat{\gamma}_1^{(Z)}(X,M)}{{f}_{ZM}( X\mid\hat{\beta})} - \frac{{\gamma}_1^{(Z)}(X,M)}{{f}_{ZM}( X\mid\beta^*)}\Big\}\Bigg]}_{\Delta_{2,n}^{(2,1)}}
\\\nonumber
&&+\underbrace{\mathbb{P}_n\Bigg[\Big[\frac{(-1)^{Z+1}}{\bar{\pi}_Z(X)}\big\{Y - \bar{\mu}_{Z}(X,M)\big\}-\frac{(-1)^{Z+1}}{{\pi}_Z(X)}\big\{Y - {\mu}_{Z}(X,M)\big\}\Big]\Big\{\frac{\hat{\gamma}_1^{(Z)}(X,M)}{{f}_{ZM}( X\mid\hat{\beta})} - \frac{{\gamma}_1^{(Z)}(X,M)}{{f}_{ZM}( X\mid\beta^*)}\Big\}\Bigg] }_{\Delta_{2,n}^{(2,2)}}
\\\nonumber
&&+\underbrace{\mathbb{P}_n\Bigg[\Big[\frac{(-1)^{Z+1}}{\hat{\pi}_Z(X)}\big\{Y - \hat{\mu}_{Z}(X,M)\big\}-\frac{(-1)^{Z+1}}{\bar{\pi}_Z(X)}\big\{Y - \bar{\mu}_{Z}(X,M)\big\}\Big]\Big\{\frac{\hat{\gamma}_1^{(Z)}(X,M)}{{f}_{ZM}( X\mid\hat{\beta})} - \frac{{\gamma}_1^{(Z)}(X,M)}{{f}_{ZM}( X\mid\beta^*)}\Big\}\Bigg] }_{\Delta_{2,n}^{(2,3)}}
\\\nonumber
&&+\mathbb{P}_n\Bigg[\frac{(-1)^{Z+1}{\gamma}_1^{(Z)}(X,M)}{\hat{\pi}_Z(X){f}_{ZM}( X\mid{\beta}^*)}\big\{Y - \hat{\mu}_{Z}(X,M)\big\}+\int_{\R^2}k_{u^*}(u)\big[\{\hat\mu_{1}( X,m_1) - \hat\mu_{0}(X,m_0)\} {e}_{u}(X)\big]du\Bigg] 
\\\nonumber
&&- \mathbb{P}_n\Bigg[\frac{(-1)^{Z+1}{\gamma}_1^{(Z)}(X,M)}{\bar{\pi}_Z(X){f}_{ZM}( X\mid{\beta}^*)}\big\{Y - \bar{\mu}_{Z}(X,M)\big\}+\int_{\R^2}k_{u^*}(u)\big[\{\bar\mu_{1}( X,m_1) - \bar\mu_{0}(X,m_0)\} {e}_{u}(X)\big]du\Bigg]
\\\nonumber
&=& \Delta_{2,n}^{(2,1)} + \Delta_{2,n}^{(2,2)} + \Delta_{2,n}^{(2,3)} + \mathbb{P}_n\{h(V\mid \hat{\pi},\hat{\mu})- h(V\mid \bar{\pi},\bar{\mu})\},
\end{eqnarray}
where we denote   
\beee\nn
h(V\mid {\pi},{\mu})&=& \frac{(-1)^{Z+1}\gamma_1^{(Z)}(X,M)}{\pi_Z(X)f_{ZM}( X)}\{Y - \mu_{Z}(X,M)\} + \int_{\R^2}k_{u^*}(u)\{\mu_{1}(X,m_1) - \mu_{0}(X,m_0)\}e_u(X)du.
\\
\label{def:h}
\eee
First, we study $\Delta_{2,n}^{(2,1)}$. By Taylor expansion, one has
\begin{eqnarray}\nonumber
&&\frac{\hat{\gamma}_1^{(1)}(X,M)}{{f}_{1M}(X\mid \hat{\beta})} - \frac{{\gamma}_1^{(1)}(X,M)}{{f}_{1M}(X\mid \beta^*)}
\\\nonumber
&=& \int_{\R}k_{u^*}(M,m_0)c_{\rho}\big\{F_{1M}( X\mid\hat{\beta}),F_{0m_0}( X\mid\hat{\beta})\big\}f_{0m_0}( X\mid\hat{\beta})dm_0
\\\nonumber
&&-\int_{\R}k_{u^*}(M,m_0)c_{\rho}\big\{F_{1M}( X\mid\beta^*),F_{0m_0}( X\mid\beta^*)\big\}f_{0m_0}( X\mid\beta^*)dm_0
\\\label{haha3}
&=&\Big[\underbrace{\{F'_{1M}( X\mid\beta^*)\}^\T \int_{\R}k_{u^*}(M,m_0)c_{\rho}^{(1)}\big\{F_{1M}(X),F_{0X}(m_0 )\big\}f_{0m_0}( X)dm_0 }_{\delta^{(1)}_{X,1,M}}
\\\nonumber
&&+\underbrace{\int_{\R}k_{u^*}(M,m_0)c_{\rho}^{(0)}\big\{F_{1M}( X),F_{0m_0}( X)\big\}\{F'_{0m_0}( X\mid\beta^*)\}^\T f_{0m_0}( X)dm_0}_{\delta^{(2)}_{X,1,M}}
\\\nonumber
&&+ \underbrace{\int_{\R}k_{u^*}(M,m_0)c_{\rho}\big\{F_{1M}(X),F_{0m_0}( X)\big\}\{f'_{0m_0}( X\mid \beta^*)\}^\T dm_0}_{\delta^{(3)}_{X,1,M}}\Big]\cdot (\hat{\beta} - \beta^*)
\\\nonumber
&&+(\tilde{\beta}_{X,1,M} - \beta^*)^\T I_{X,1,M}(\tilde{\beta}_{X,1,M} - \beta^*)
\\\nonumber
&=&\big(\delta^{(1)}_{X,1,M} + \delta^{(2)}_{X,1,M} + \delta^{(3)}_{X,1,M}\big)(\hat{\beta} - \beta^*) + (\tilde{\beta}_{X,1,M} - \beta^*)^\T I_{X,1,M}(\tilde{\beta}_{X,1,M} - \beta^*),
\end{eqnarray} 
 for some $\tilde{\beta}_{X,1,M} = \delta_{X,1,M}\hat{\beta} + (1 - \delta_{X,1,M})\beta^*$ for some $\delta_{X,1,M} \in (0,1)$. Here $\delta^{(1)}_{X,1,M}(\hat{\beta} - \beta^*),\delta^{(2)}_{X,1,M}(\hat{\beta } - \beta^*),\delta^{(3)}_{X,1,M}(\hat{\beta } - \beta^*)$  are the first three terms of \eqref{haha3}, and $I_{X,1,M}$ is the second order derivative matrix of $\gamma_1^{(1)}(X,M)/f_1(M\mid X,\beta)$ with respect to $\beta$, localized at $\beta = \tilde{\beta}_{X,1,M}$. In particular, one has
\beee\nn
I_{X,1,M} &=&F''_1(M\mid X,\tilde{\beta}_{X,1,M})
\\\nn
&&\cdot\int_{\R}k_{u^*}(M,m_0)c_{\rho}^{(1)}\big\{F_1(M\mid X,\tilde{\beta}_{X,1,M}),F_0(m_0\mid X,\tilde{\beta}_{X,1,M})\big\}f_0(m_0\mid X,\tilde{\beta}_{X,1,M})dm_0 
\\\nn
&&+F'_1(M\mid X,\tilde{\beta}_{X,1,M})\{F'_1(M\mid X,\tilde{\beta}_{X,1,M})\}^\T
\\\nn
&&\cdot\int_{\R}k_{u^*}(M,m_0)c_{\rho}^{(11)}\big\{F_1(M\mid X,\tilde{\beta}_{X,1,M}),F_0(m_0\mid X,\tilde{\beta}_{X,1,M})\big\} f_0(m_0\mid X,\tilde{\beta}_{X,1,M})dm_0 
\\\nn
&&+\dots
\\\nn
&&+\int_{\R}k_{u^*}(M,m_0)c_{\rho}\big\{F_1(M\mid X,\tilde{\beta}_{X,1,M}),F_0(m_0\mid X,\tilde{\beta}_{X,1,M})\big\}f''_0(m_0\mid X,\tilde{\beta}_{X,1,M})dm_0.
\eee
We can define $\delta^{(1)}_{X,0,M}(\hat{\beta} - \beta^*),\delta^{(2)}_{X,0,M}(\hat{\beta} - \beta^*),\delta^{(3)}_{X,0,M}(\hat{\beta } - \beta^*)$ and $I_{X,0,M}$ in a similar manner. We then write, 
\beee\label{d2ndecom}
\Delta_{2,n}^{(2,1)} & = &\sum_{j = 1,2,3}\mathbb{P}_n\Big[\frac{(-1)^{Z+1}}{{\pi}_Z(X)}\big\{Y - {\mu}_{Z}(X,M)\big\}\delta_{X,Z,M}^{(j)}\Big](\hat{\beta} - \beta^*)
\\\nonumber
&& + \mathbb{P}_n\Big[(\tilde{\beta}_{X,1,M} - \beta^*)^\T\frac{(-1)^{Z+1}}{{\pi}_Z(X)}\big\{Y - {\mu}_{Z}(X,M)\big\}I_{X,Z,M}(\tilde{\beta}_{X,1,M} - \beta^*)\Big].
\eee
For $j = 1,2,3$, similar to \eqref{unif:exp}, we can show that uniformly, 
$$
\sup_{x,z,m}\|\delta_{x,z,m}^{(j)}\| = \max\Big\{\sup_{x,m}\|\delta_{x,1,m}^{(j)}\|, \,\sup_{x,m}\|\delta_{x,0,m}^{(j)}\|\Big\} \precsim h^{-1}.
$$
To illustrate this, we use $\sup_{x,m}\|\delta_{x,1,m}^{(1)}\|$ as an example. In particular, we have
\beee\nonumber
&&\sup_{x,m}\|\delta_{x,1,m}^{(1)}\|
\\\nonumber
 & = & \sup_{x,m}\Big[\Big\|F'_{1m}( x\mid\beta^*)\}^\T \int_{\R}k_{u^*}(m,m_0)c_{\rho}^{(1)}\big\{F_{1m}(x),F_{0x}(m_0 )\big\}f_{0m_0}(x)dm_0\Big\|\Big]
\\\nonumber
&\leq & \sup_{x,m}\Big[\Big\|F'_{1m}( x\mid\beta^*)\Big\|\cdot\Big| \int_{\R}k_{u^*}(m,m_0)c_{\rho}^{(1)}\big\{F_{1m}(x),F_{0x}(m_0 )\big\}f_{0m_0}(x)dm_0\Big|\Big]
\\\nonumber
&\leq & \sup_{u,v}|c_{\rho}^{(1)}(u,v)|\cdot\sup_{x,m}|f_{0m}(x)|\cdot\sup_{x,m}\Big[\Big\|F'_{1m}( x\mid\beta^*)\Big\|\cdot\int_{\R}k_{u^*}(m,m_0)dm_0\Big]
\\\nonumber
& \precsim & h^{-1},
\eee
by Assumptions~\ref{am:kernel}-\ref{am:unibound}. Then we have
\beee\nn
&&\E\Big[\frac{(-1)^{Z+1}}{{\pi}_Z(X)}\big\{Y - {\mu}_{Z}(X,M)\big\}\delta_{X,Z,M,k}^{(j)}\Big]^2 \precsim h^{-1}\cdot\E\Big[\frac{(-1)^{Z+1}}{{\pi}_Z(X)}\big\{Y - {\mu}_{Z}(X,M)\big\}\Big]^2
\\\nn
&\precsim & \E\Big[{\big\{Y - {\mu}_{Z}(X,M)\big\}^2}\Big] \quad(\text{Assumption \ref{am:pi}})
\\\label{momentd111}
&\precsim & h^{-1}. \quad(\text{Assumption \ref{am:var:Y}}).
\eee
Here $\delta_{X,Z,M,k}^{(j)}$ represents the $k$th coordinate of $\delta_{X,Z,M}^{(j)}$. Finally by a similar argument as \eqref{cherna}--\eqref{d11n}, one can show the following by Chebyshev's inequality, 
\beee\nonumber
(\mathbb{P}_n - \mathbb{P})\Bigg[\Bigg|\frac{(-1)^{Z+1}}{{\pi}_Z(X)}\big\{Y - {\mu}_{Z}(X,M)\big\}\delta_{X,Z,M,k}^{(j)}\Bigg|\Bigg] &= &\mathcal{O}_{\mathbb{P}}(h^{-1}n^{-1/2}),
\eee
and by an entry-wise argument, we have,
\beee\label{S644}
\Bigg\|(\mathbb{P}_n - \mathbb{P})\Bigg[\frac{(-1)^{Z+1}}{{\pi}_Z(X)}\big\{Y - {\mu}_{Z}(X,M)\big\}\delta_{X,Z,M}^{(j)}\Bigg]\Bigg\|_2 &= &\mathcal{O}_{\mathbb{P}}(h^{-1}n^{-1/2}).
\eee In addition, with an entry-wise argument, we can also show that,
 \bee
 \mathbb{E} \Bigg[\Big\|\frac{(-1)^{Z+1}}{{\pi}_Z(X)}\big\{Y - {\mu}_{Z}(X,M)\big\}I_{X,Z,M}\Big]\Big\|_2\Bigg] \leq   \mathbb{E}\Bigg[ \sum_{1\leq l_1,l_2\leq d_\beta}\Bigg|\Bigg[\frac{(-1)^{Z+1}}{{\pi}_Z(X)}\big\{Y - {\mu}_{Z}(X,M)\big\}I_{X,Z,M}\Bigg]_{l_1l_2}\Bigg|\Bigg] \precsim h^{-1},
\label{ppdecom}
 \ee
and similarly as \eqref{momentd111}--\eqref{S644} we have
\beee\nonumber
(\mathbb{P}_n - \mathbb{P})\Bigg[\Bigg\|\frac{(-1)^{Z+1}}{{\pi}_Z(X)}\big\{Y - {\mu}_{Z}(X,M)\big\}I_{X,Z,M}\Big]\Bigg\|_2\Bigg] 
&=& \mathcal{O}_{\mathbb{P}}(h^{-1}n^{-1/2}).
 \eee
Summarizing the results above and by the fact that $\sup_{x,z,m} \|\tilde{\beta}_{x,z,m} - \beta^*\| =\sup_{x,z,m}\delta_{x,z,m}\|\hat{\beta} - \beta^*\| \leq \|\hat{\beta} - \beta^*\| = \mathcal{O}_{\mathbb{P}}(n^{-1/2})$, we  have
 \beee\nn
\Delta_{2,n}^{(2,1)} &=& \scalemath{0.8}{\sum_{j = 1,2,3}\mathbb{P}\Bigg[\frac{(-1)^{Z+1}}{{\pi}_Z(X)}\big\{Y - {\mu}_{Z}(X,M)\big\}\delta_{X,Z,M}^{(j)}\Bigg](\hat{\beta} - \beta^*)}
 \\\nn
 &&\scalemath{0.8}{+\sum_{j = 1,2,3}(\mathbb{P}_n - \mathbb{P})\Bigg[\frac{(-1)^{Z+1}}{{\pi}_Z(X)}\big\{Y - {\mu}_{Z}(X,M)\big\}\delta_{X,Z,M}^{(j)}\Bigg](\hat{\beta} - \beta^*)}
\\\nn
&& \scalemath{0.8}{+\Big\{(\mathbb{P}_n - \mathbb{P}) + \mathbb{P}\Big\}\Bigg[ \|\tilde{\beta}_{X,1,M} - \beta^*\|^2\Big(\frac{\tilde{\beta}_{X,1,M} - \beta^*}{\|\tilde{\beta}_{X,1,M} - \beta^*\|}\Big)^\T\frac{(-1)^{Z+1}}{{\pi}_Z(X)}\big\{Y - {\mu}_{Z}(X,M)\big\}
\cdot I_{X,Z,M}\Big(\frac{\tilde{\beta}_{X,1,M} - \beta^*}{\|\tilde{\beta}_{X,1,M} - \beta^*\|}\Big)\Bigg]}
\\\nn
&=&\scalemath{0.8}{\underbrace{\sum_{j = 1,2,3}\mathbb{P}\Bigg[\frac{(-1)^{Z+1}}{{\pi}_Z(X)}\big\{Y - {\mu}_{Z}(X,M)\big\}\delta_{X,Z,M}^{(j)}\Bigg](\hat{\beta} - \beta^*)}_{\Gamma^{(2,1)}_{2,n}} + o_{\mathbb{P}}(h^{-1}n^{-1/2})}
\eee
where in the last equality we use the  property of spectral norm such that by \eqref{ppdecom},
\beee\nn
&&\scalemath{0.8}{\Bigg|\Big\{(\mathbb{P}_n - \mathbb{P}) + \mathbb{P}\Big\}\Bigg[\|\tilde{\beta}_{X,1,M} - \beta^*\|^2\Big(\frac{\tilde{\beta}_{X,1,M} - \beta^*}{\|\tilde{\beta}_{X,1,M} - \beta^*\|}\Big)^\T\frac{(-1)^{Z+1}}{{\pi}_Z(X)}\big\{Y - {\mu}_{Z}(X,M)\big\}\cdot I_{X,Z,M}\Big(\frac{\tilde{\beta}_{X,1,M} - \beta^*}{\|\tilde{\beta}_{X,1,M} - \beta^*\|}\Big)\Bigg]\Bigg|}
\\\nn
&\leq&\scalemath{0.8}{\|\hat{\beta} - \beta^*\|^2\Big\{(\mathbb{P}_n - \mathbb{P}) + \mathbb{P}\Big\}\Bigg[ \sup_{\|v\| = 1}\Bigg| v^\T\frac{(-1)^{Z+1}}{{\pi}_Z(X)}\big\{Y - {\mu}_{Z}(X,M)\big\}I_{X,Z,M} \Bigg|\Bigg]}
\\\nn
&=&\scalemath{0.8}{\|\hat{\beta} - \beta^*\|^2    \Big\{(\mathbb{P}_n - \mathbb{P}) + \mathbb{P}\Big\}\Bigg[\Bigg\| \frac{(-1)^{Z+1}}{{\pi}_Z(X)}\big\{Y - {\mu}_{Z}(X,M)\big\}I_{X,Z,M} \Bigg\|_2\Bigg]}
\\\label{ppdecom22}
& =& \scalemath{0.8}{\mathcal{O}_{\mathbb{P}}(h^{-1}n^{-1}),}
\eee
and also for $j = 1,2,3$, by \eqref{S644},
\beee\nonumber
&&\Big\|(\mathbb{P}_n - \mathbb{P})\Big[\frac{(-1)^{Z+1}}{{\pi}_Z(X)}\big\{Y - {\mu}_{Z}(X,M)\big\}\delta_{X,Z,M}^{(j)}\Big](\hat{\beta} - \beta^*)\Big\| 
\\\nonumber
&\leq &\Big\|(\mathbb{P}_n - \mathbb{P})\Big[\frac{(-1)^{Z+1}}{{\pi}_Z(X)}\big\{Y - {\mu}_{Z}(X,M)\big\}\delta_{X,Z,M}^{(j)}\Big]\Big\|_2\Big\|\hat{\beta} - \beta^*\Big\|
\\\label{D2n121final}
& = &\mathcal{O}_{\mathbb{P}}(h^{-1}n^{-1}).
\eee
\par
{\color{black}Next we study $\Delta_{2,n}^{(2,2)}$.} When $\bar{\pi} = \pi$ and $\bar{\mu} = \mu$, it is easy to see $\Delta_{2,n}^{(2,2)} = 0$. When either $\bar{\pi} \neq \pi$ and $\bar{\mu} \neq \mu$, similar to \eqref{d2ndecom}, we have
\beee\nonumber
&&\Delta_{2,n}^{(2,2)}
\\\nonumber
 & = & \sum_{j = 1,2,3}\mathbb{P}_n\Bigg[\Big[\frac{(-1)^{Z+1}}{\bar{\pi}_Z(X)}\big\{Y - \bar{\mu}_{Z}(X,M)\big\}-\frac{(-1)^{Z+1}}{{\pi}_Z(X)}\big\{Y - {\mu}_{Z}(X,M)\big\}\Big]\delta_{X,Z,M}^{(j)}\Bigg](\hat{\beta} - \beta^*)
\\\nonumber
&& + \mathbb{P}_n\Bigg[(\tilde{\beta}_{X,1,M} - \beta^*)^\T\Big[\frac{(-1)^{Z+1}}{\bar{\pi}_Z(X)}\big\{Y - \bar{\mu}_{Z}(X,M)\big\}-\frac{(-1)^{Z+1}}{{\pi}_Z(X)}\big\{Y - {\mu}_{Z}(X,M)\big\}\Big]I_{X,Z,M}(\tilde{\beta}_{X,1,M} - \beta^*)\Bigg].
\eee
We then exploit the same technical argument as for $\Delta_{2,n}^{(2,1)}$ to bound $\Delta_{2,n}^{(2,2)}$. In particular, recalling $\|\delta^{(j)}_{X,Z,M}\|$ can be uniformly bounded by $\mathcal{O}(h^{-1})$, we have for any $j = 1,2,3$ and $k \in [d_\beta]$,
\beee\nonumber
&&\mathbb{E}\Bigg[\Big[\frac{(-1)^{Z+1}}{\bar{\pi}_Z(X)}\big\{Y - \bar{\mu}_{Z}(X,M)\big\}-\frac{(-1)^{Z+1}}{{\pi}_Z(X)}\big\{Y - {\mu}_{Z}(X,M)\big\}\Big]\delta_{X,Z,M}^{(j)}\Bigg]^2
\\\nonumber
&\precsim & h^{-2}\cdot\mathbb{E}\Bigg[\Big[\frac{(-1)^{Z+1}}{\bar{\pi}_Z(X)}\big\{Y - \bar{\mu}_{Z}(X,M)\big\}-\frac{(-1)^{Z+1}}{{\pi}_Z(X)}\big\{Y - {\mu}_{Z}(X,M)\big\}\Big]\Bigg]^2
\\\nonumber
&\leq & 2h^{-2}\mathbb{E}\Bigg[\Big[\frac{(-1)^{Z+1}}{\bar{\pi}_Z(X)}\big\{Y - \bar{\mu}_{Z}(X,M)\big\}\Big]\Bigg]^2 + 2h^{-2}\mathbb{E}\Bigg[\Big[\frac{(-1)^{Z+1}}{{\pi}_Z(X)}\big\{Y - {\mu}_{Z}(X,M)\big\}\Big]\Bigg]^2
\\\nonumber
&\leq & 4h^{-2}\mathbb{E}\Bigg[\Big[\frac{(-1)^{Z+1}}{\bar{\pi}_Z(X)}\big\{Y - {\mu}_{Z}(X,M)\big\}\Big]\Bigg]^2 + 4h^{-2}\mathbb{E}\Bigg[\Big[\frac{(-1)^{Z+1}}{\bar{\pi}_Z(X)}\big\{{\mu}_{Z}(X,M) - \bar{\mu}_{Z}(X,M)\big\}\Big]\Bigg]^2 
\\\label{momentd222}
&&+ 2h^{-2}\mathbb{E}\Bigg[\Big[\frac{(-1)^{Z+1}}{{\pi}_Z(X)}\big\{Y - {\mu}_{Z}(X,M)\big\}\Big]\Bigg]^2
\\\nn
&\precsim& h^{-2},
\eee
where the first and third term in \eqref{momentd222} can be bounded similar to \eqref{momentd111}, while the second term in  \eqref{momentd222} can be bounded by Assumptions \ref{am:pi} and \ref{am:emp},
\beee\nn
\mathbb{E}\Bigg[\Big[\frac{(-1)^{Z+1}}{\bar{\pi}_Z(X)}\big\{{\mu}_{Z}(X,M) - \bar{\mu}_{Z}(X,M)\big\}\Big]\Bigg]^2 & \leq & \epsilon^{-2} \cdot\big\{\E\big|\mu_{Z}(X,M)\big|^2 +\E\big| \bar{\mu}_{Z}(X,M)\big|^2 \big\}
\\\nn
& = &  \mathcal{O}(1).
\eee
Similar to \eqref{S644}, \eqref{momentd222} combining with Chebyshev's inequality and an entry-wise argument implies that
\beee\nn
\Bigg\|(\mathbb{P}_n - \mathbb{P})\Bigg[\Big[\frac{(-1)^{Z+1}}{\bar{\pi}_Z(X)}\big\{Y - \bar{\mu}_{Z}(X,M)\big\}-\frac{(-1)^{Z+1}}{{\pi}_Z(X)}\big\{Y - {\mu}_{Z}(X,M)\big\}\Big]\delta_{X,Z,M}^{(j)}\Bigg]\Bigg\|_2 &= &\mathcal{O}_{\mathbb{P}}(h^{-1}n^{-1/2}).
\eee
Then, a similar argument to \eqref{momentd111}--\eqref{D2n121final} can show that
 \beee\nn
 \Delta_{2,n}^{(2,2)} &=&\underbrace{\sum_{j = 1,2,3}\mathbb{P}\Bigg[\Big[\frac{(-1)^{Z+1}}{\bar{\pi}_Z(X)}\big\{Y - \bar{\mu}_{Z}(X,M)\big\}-\frac{(-1)^{Z+1}}{{\pi}_Z(X)}\big\{Y - {\mu}_{Z}(X,M)\big\}\Big]\delta_{X,Z,M}^{(j)}\Bigg](\hat{\beta} - \beta^*)}_{\Gamma^{(2,2)}_{2,n}} 
\\\nn
&&+ o_{\mathbb{P}}(h^{-1}n^{-1/2})
\eee
\par
Next we bound $\Delta_{2,n}^{(2,3)}$. Based on \eqref{haha}, we can show
\beee\nonumber
&&\Delta_{2,n}^{(2,3)}
\\\nonumber
& =& \mathbb{P}_n\Bigg[\Big[\frac{(-1)^{Z+1}}{\hat{\pi}_Z(X)}\big\{Y - \hat{\mu}_{Z}(X,M)\big\}-\frac{(-1)^{Z+1}}{\bar{\pi}_Z(X)}\big\{Y - \bar{\mu}_{Z}(X,M)\big\}\Big]\big(\delta^{(1)}_{X,1,M} + \delta^{(2)}_{X,1,M} + \delta^{(3)}_{X,1,M}\big)\Bigg](\hat{\beta} - \beta^*) 
\\\nn
&&+  \mathbb{P}_n\Bigg[\Big[\frac{(-1)^{Z+1}}{\hat{\pi}_Z(X)}\big\{Y - \hat{\mu}_{Z}(X,M)\big\}-\frac{(-1)^{Z+1}}{\bar{\pi}_Z(X)}\big\{Y - \bar{\mu}_{Z}(X,M)\big\}\Big](\tilde{\beta}_{X,1,M} - \beta^*)^\T I_{X,1,M}(\tilde{\beta}_{X,1,M} - \beta^*) \Bigg]
\\\label{d2n23decom}
& = & \mathbb{P}_n\Bigg\{\big(\kappa^{(1)}_{X,Z,Y} + \kappa^{(2)}_{X,Z,M}\big)\big(\delta^{(1)}_{X,1,M} + \delta^{(2)}_{X,1,M} + \delta^{(3)}_{X,1,M}\big)\Bigg\}(\hat{\beta} - \beta^*) 
\\\nn
&& +  \mathbb{P}_n\Bigg\{\big(\kappa^{(1)}_{X,Z,Y} + \kappa^{(2)}_{X,Z,M}\big)(\tilde{\beta}_{X,1,M} - \beta^*)^\T I_{X,1,M}(\tilde{\beta}_{X,1,M} - \beta^*) \Bigg\},
\eee
where we further define,
\beee\nn
\kappa_{X,Z,Y}^{(1)} &=&\frac{(-1)^{Z + 1}}{\hat{\pi}_Z(X)}Y - \frac{(-1)^{Z + 1}}{\bar{\pi}_Z(X)}Y,
\\\nn
 \kappa_{X,Z,M}^{(2)} &=& \frac{(-1)^{Z + 1}}{\bar{\pi}_Z(X)}\bar{\mu}_{Z}(X,M) -\frac{(-1)^{Z + 1}}{\hat{\pi}_Z(X)}\hat{\mu}_{Z}(X,M).
\eee
We argue that all terms in \eqref{d2n23decom} are $o_{\mathbb{P}}(h^{-1}n^{-1/2})$ and thus 
\beee\label{bd:D2n}
\Delta_{2,n}^{(2,3)} &=& o_{\mathbb{P}}(h^{-1}n^{-1/2}),
\eee is negligible. For example, we have
\beee\nn
\Big|\mathbb{P}_n\big(\kappa_{X,Z,Y}^{(1)}\delta_{X,1,M}^{(1)}\big)(\hat{\beta} - \beta^*)\Big| &\leq & \|\hat{\beta} - \beta^*\|\cdot \mathbb{P}_n\big(\big\|\kappa_{X,Z,Y}^{(1)}\delta_{X,1,M}^{(1)}\big\|\big)
\\\label{k1d1bound}
& = & o_{\mathbb{P}}(h^{-1}n^{-1/2}),
\eee
where $\|\hat{\beta} - \beta^*\| = \mathcal{O}_{\mathbb{P}}(n^{-1/2})$ and 
\beee\nn
&&\mathbb{P}_n\big(\big\|\kappa_{X,Z,Y}^{(1)}\delta_{X,1,M}^{(1)}\big\|\big)
\\\nn
& =&\mathbb{P}_n\Bigg[\Big\|\Big\{ \frac{\bar{\pi}_Z(X) -\hat{\pi}_Z(X) }{\hat{\pi}_Z(X)\bar{\pi}_Z(X)}\Big\}YF'_{1M}( X\mid\beta^*)\int_{\R}k_{u^*}(M,m_0)c_{\rho}^{(1)}\big\{F_{1M}(X),F_{0X}(m_0 )\big\}f_{0m_0}( X)dm_0\Big\|\Bigg]
\\\nn
&\leq & \|\bar{\pi}_Z(X) -\hat{\pi}_Z(X)\|_{\infty}\|\hat{\pi}^{-1}_Z(X)\|_{\infty}\|\bar{\pi}_Z^{-1}(X)\|_{\infty}\cdot \sup_{m_1,x}\|F'_{1m_1}(x\mid \beta^*)\|\cdot\sup_{u,v}|c_\rho^{(1)}(u,v)| \cdot\sup_{m_0,x}f_{0m_0}(x)
\\\label{bound1d23}
&&\cdot \Big\{\sup_{m_1}\int_{\R}k_{u^*}(u)dm_0\Big\}\mathbb{P}_n(|Y|)
\\\label{bound2d23}
& = & o_{\mathbb{P}}(1)\cdot \mathcal{O}_{\mathbb{P}}(1) \cdots \mathcal{O}(1)\cdot\mathcal{O}(h^{-1})\cdot \{(\mathbb{P}_n - \mathbb{E})(|Y|) + \mathbb{E}(|Y|)\}
\\\nn
& = & o_{\mathbb{P}}(h^{-1}),
\eee
where $\|\cdot\|_{\infty}$ denotes the sup-norm for function and random variable. For \eqref{bound1d23}, we have $\|\bar{\pi}_Z(X) -\hat{\pi}_Z(X)\|_{\infty} = o_{\mathbb{P}}(1)$ under Assumption \ref{am:emp}, $\|\bar{\pi}^{-1}_Z(X)\|_{\infty}\geq \epsilon^{-1}$ under Assumption \ref{am:pi} and thus $\|\hat{\pi}^{-1}_Z(X)\|_{\infty}$ is also bounded away from zero with probability approaching $1$, meanwhile all other terms are $\mathcal{O}_{\mathbb{P}}(1)$ under Assumptions \ref{am:kernel}--\ref{am:unibound}. For \eqref{bound2d23}, $\sup_{m_1}\int_{\R}k_{u^*}(u)dm_0 = \mathcal{O}(h^{-1})$ under Assumption~\ref{am:kernel}, $\mathbb{E}(|Y|) \leq \sqrt{\mathbb{E}(|Y|^2)} < +\infty$ under Assumption \ref{am:var:Y} and then $(\mathbb{P}_n - \mathbb{E})(|Y|) = \mathcal{O}_{\mathbb{P}}(n^{-1/2})$ follows by an application of  the Markov's inequality. Similarly, we can show 
$$
\big|\mathbb{P}_n\big(\kappa_{X,Z,Y}^{(1)}\delta_{X,1,M}^{(2)}\big)(\hat{\beta} - \beta^*)\big| = \mathcal{O}_{\mathbb{P}}(h^{-1}n^{-1/2})\quad \text{and} \quad \big|\mathbb{P}_n\big(\kappa_{X,Z,Y}^{(1)}\delta_{X,1,M}^{(3)}\big)(\hat{\beta} - \beta^*)\big| = \mathcal{O}_{\mathbb{P}}(h^{-1}n^{-1/2}).
$$ We can also show that
\beee\nn
\Big|\mathbb{P}_n\big(\kappa_{X,Z,M}^{(2)}\delta_{X,1,M}^{(1)}\big)(\hat{\beta} - \beta^*)\Big| &\leq & \|\hat{\beta} - \beta^*\| \mathbb{P}_n\big(\big\|\kappa_{X,Z,Y}^{(2)}\delta_{X,1,M}^{(1)}\big\|\big)
\\
&= & o_{\mathbb{P}}(h^{-1}n^{-1/2}),
\eee
where $\|\hat{\beta} - \beta^*\| = \mathcal{O}_{\mathbb{P}}(n^{-1/2})$, and similar to \eqref{k1d1bound},
\beee\nn
&&\mathbb{P}_n\big(\big\|\kappa_{X,Z,Y}^{(2)}\delta_{X,1,M}^{(1)}\big\|\big)
\\\nn
& =&\mathbb{P}_n\Bigg[\Bigg\|\Bigg[ \frac{(-1)^{Z + 1}}{\bar{\pi}_Z(X)}\bar{\mu}_{Z}(X,M) -\frac{(-1)^{Z + 1}}{\hat{\pi}_Z(X)}\hat{\mu}_{Z}(X,M)\Bigg]
 F'_{1M}( X\mid\beta^*)\int_{\R}k_{u^*}(M,m_0)
\\\nn
&&\cdot c_{\rho}^{(1)}\big\{F_{1M}(X),F_{0X}(m_0 )\big\}f_{0m_0}( X)dm_0\Bigg\|\Bigg]
\\\label{bound2d2333}
&\leq &\Bigg\| \frac{(-1)^{Z + 1}}{\bar{\pi}_Z(X)}\big\{\bar{\mu}_{Z}(X,M) -\hat{\mu}_{Z}(X,M)  \big\}+ \Bigg\{\frac{(-1)^{Z + 1}}{\bar{\pi}_Z(X)} -\frac{(-1)^{Z + 1}}{\hat{\pi}_Z(X)}\Bigg\}\hat{\mu}_{Z}(X,M)\Bigg\|_{\infty}
\\\nn
&&\cdot \sup_{m_1,x}\|F'_{1m_1}(x\mid \beta^*)\|\cdot\sup_{u,v}|c_\rho^{(1)}(u,v)| \cdot\sup_{m_0,x}f_{0m_i}(x)\cdot \sup_{m_1}\int_{\R}k_{u^*}(u)dm_0 \cdot\mathbb{P}_n(|Y|)
\\\nn
& = & o_{\mathbb{P}}(h^{-1}).
\eee
For \eqref{bound2d2333}, based on Assumptions \ref{am:pi} and \ref{am:emp}, we bound
\beee\nn
&&\Bigg\| \frac{(-1)^{Z + 1}}{\bar{\pi}_Z(X)}\big\{\bar{\mu}_{Z}(X,M) -\hat{\mu}_{Z}(X,M)  \big\}+ \Bigg\{\frac{(-1)^{Z + 1}}{\bar{\pi}_Z(X)} -\frac{(-1)^{Z + 1}}{\hat{\pi}_Z(X)}\Bigg\}\hat{\mu}_{Z}(X,M)\Bigg\|_{\infty}
\\\nn
&\leq & \Bigg\| \frac{(-1)^{Z + 1}}{\bar{\pi}_Z(X)}\big\{\bar{\mu}_{Z}(X,M) -\hat{\mu}_{Z}(X,M)  \big\}\Bigg\|_{\infty} + \Bigg\|\Bigg\{\frac{(-1)^{Z + 1}}{\bar{\pi}_Z(X)} -\frac{(-1)^{Z + 1}}{\hat{\pi}_Z(X)}\Bigg\}\hat{\mu}_{Z}(X,M)\Bigg\|_{\infty}
\\\nn
&\leq & \epsilon^{-1}\|\bar{\mu}_{Z}(X,M) -\hat{\mu}_{Z}(X,M)\|_{\infty} + \Big\|\frac{\hat{\pi}_Z(X) - \bar{\pi}_Z(X)}{\bar{\pi}_Z(X)\hat{\pi}_Z(X)}\Big\|_{\infty}\|\hat{\mu}_{Z}(X,M)\|_{\infty}
\\\nn
&\leq &  \epsilon^{-1}\|\bar{\mu}_{Z}(X,M) -\hat{\mu}_{Z}(X,M)\|_{\infty}  +  \|\bar{\pi}_Z(X) -\hat{\pi}_Z(X)\|_{\infty}\|\hat{\pi}^{-1}_Z(X)\|_{\infty}\|\bar{\pi}_Z^{-1}(X)\|_{\infty}
\\\nn
&&\cdot \Big\{\|\bar{\mu}_{Z}(X,M)\|_{\infty} + \|\hat{\mu}_{Z}(X,M) - \bar{\mu}_{Z}(X,M)\|_{\infty}\Big\}
\\\nn
& = & o_{\mathbb{P}}(1). 
\eee

By applying the similar technical arguments as above to each entry of $\mathbb{P}_n(\kappa^{(1)}_{X,Z,Y}I_{X,1,M})$ and $\mathbb{P}_n(\kappa^{(2)}_{X,Z,M}I_{X,1,M})$, we can show that $\mathbb{P}_n(\|\kappa^{(1)}_{X,Z,Y}I_{X,1,M}\|_2),\mathbb{P}_n(\|\kappa^{(2)}_{X,Z,M}I_{X,1,M}\|_2) = \mathcal{O}_{\mathbb{P}}(h^{-1})$ similar to \eqref{ppdecom}. Then by the basic property of spectral norm and similar to \eqref{ppdecom22}, one has 
\beee\nn
&&\Bigg|\mathbb{P}_n\Bigg\{\big(\kappa^{(1)}_{X,Z,Y} + \kappa^{(2)}_{X,Z,M}\big)(\tilde{\beta}_{X,1,M} - \beta^*)^\T I_{X,1,M}(\tilde{\beta}_{X,1,M} - \beta^*) \Bigg\}\Bigg|
\\\nn
& \leq & \|\hat{\beta} - \beta\|^2\Big\{\mathbb{P}_n\big(\|\kappa^{(1)}_{X,Z,Y}I_{X,1,M}\|_2\big) + \mathbb{P}_n\big(\|\kappa^{(2)}_{X,Z,M}I_{X,1,M}\|_2\big)\Big\}
\\\label{d2n23decom2}
& = & \mathcal{O}_{\mathbb{P}}(h^{-1}n^{-1}).
\eee
Combining \eqref{d2n23decom}--\eqref{d2n23decom2}, \eqref{bd:D2n} is thus shown.
\par
Summarizing the results above, we have
\bee\label{D2n:step}
&\Delta_{2,n} = \Gamma^{(1)}_{2,n} + \Gamma^{(2,1)}_{2,n} + \Gamma^{(2,2)}_{2,n} + o_{\mathbb{P}}(h^{-1}n^{-1/2}) + \mathbb{P}_n\{h(V\mid \hat{\pi},\hat{\mu}) - h(V\mid \bar{\pi},\bar{\mu})\}
\ee
We now study the last term on the right-hand side above. We decompose
\beee\nn
&&\mathbb{P}_n\{h(V\mid \hat{\pi},\hat{\mu}) - h(V\mid \bar{\pi},\bar{\mu})\} 
\\\nn
&=& (\mathbb{P}_n - \mathbb{P})\{h(V\mid \hat{\pi},\hat{\mu}) - h(V\mid \bar{\pi},\bar{\mu})\} + \mathbb{P}\{h(V\mid \hat{\pi},\hat{\mu}) - h(V\mid \bar{\pi},\bar{\mu})\}.
\eee
We first  bound $(\mathbb{P}_n - \mathbb{P})\{h(V\mid \hat{\pi},\hat{\mu}) - h(V\mid \bar{\pi},\bar{\mu})\}$ through Lemma \ref{lm:emp} using  empirical process theory. We defer a brief introduction of empirical process theory as well as the technical proof of Lemma \ref{lm:emp} to Section \ref{pf:lm:emp}.
\begin{lemma}\label{lm:emp}
Recall \eqref{def:h}. Under the conditions of Theorem \ref{thm:consist}, we have 
\beee\label{emp:con}
(\mathbb{P}_n - \mathbb{P})\{h(V\mid \hat{\pi},\hat{\mu}) - h(V\mid \bar{\pi},\bar{\mu})\}& =& o_{\mathbb{P}}(h^{-1}n^{-1/2}).
\eee
\end{lemma}
By Lemma \ref{lm:emp}, we have
\beee\nn
\mathbb{P}_n\big\{h(V\mid \hat{\pi},\hat{\mu}) - h(V\mid \bar{\pi},\bar{\mu})\big\} &= 
& o_{\mathbb{P}}(h^{-1}n^{-1/2}) + \mathbb{P}\big\{h(V\mid \hat{\pi},\hat{\mu}) - h(V\mid \bar{\pi},\bar{\mu})\big\}.
\eee
In addition, if either $\bar{\mu} = \mu$ or $\bar{\pi} = \pi$, one has
\beee\nn
\mathbb{P}\{h(V\mid \bar\pi,\bar\mu)\} &= &\mathbb{E}\Bigg[\bar{{\xi}}_1(X,Z,M)\{Y - \mu_{Z}(X,M)\}
+\int_{\R^2}\big\{\bar\mu_{1}( X,m_1) - \bar\mu_{0}(X,m_0)\big\}k_{u^*}(u){e}_{u}(X)du\Bigg]
\\\nn
& = & \bar{\tau}_{u^*} \cdot \E\Big\{\int_{\R^2}k_{u^*}(u)e_{u}(X)du\Big\}
\\\nn
& = & \tau_{u^*} \cdot \E\Big\{\int_{\R^2}k_{u^*}(u)e_{u}(X)du\Big\} \quad\text{(Theorem \ref{thm:dbrb})}
\\\nn
& = & \E\Big[\int_{\R^2}\big\{\mu_{1}( X,m_1) - \mu_{0}(X,m_0)\big\}k_{u^*}(u)e_{u}(X)du\Big]
\\\nn
& = & \E\Big[{{\xi}}_1(X,Z,M)\{Y - \mu_{Z}(X,M)\} + \int_{\R^2}\big\{\mu_{1}( X,m_1) - \mu_{0}(X,m_0)\big\}k_{u^*}(u)e_{u}(X)du\Big]
\\\nn
& = & \mathbb{P}\big\{h(V\mid \pi,\mu)\big\},
\eee
where the fourth equality is by Theorem \ref{thm:idd}, the fifth equality is by $\E\tilde{\xi}_1 = 0$ when $\tilde{\mu} = \mu$; See \eqref{ebt3:bound}. Therefore we have,
\beee\label{phh}
&&\mathbb{P}\big\{h(V\mid \hat{\pi},\hat{\mu}) - h(V\mid \bar{\pi},\bar{\mu})\big\} 
\\\nn
&=&\mathbb{P}\big\{h(V\mid \hat{\pi},\hat{\mu}) - h(V\mid {\pi},{\mu})\big\} 
 \\\nn
 &=& \mathbb{P}\Bigg[\frac{(-1)^{Z+1}{\gamma}_1^{(Z)}(X,M)}{\hat{\pi}_Z(X){f}_{Z}(M\mid X,{\beta}^*)}\big\{Y - \hat{\mu}_{Z}(X,M)\big\} + \int_{\R^2}k_{u^*}(u)\{\hat\mu_{1}(X,m_1) - \hat\mu_{0}(X,m_0)\}e_u(X)du
\\\nn
&& - \frac{(-1)^{Z+1}{\gamma}_1^{(Z)}(X,M)}{{\pi}_Z(X){f}_{Z}(M\mid X,{\beta}^*)}\big\{Y - {\mu}_{Z}(X,M)\big\} -\int_{\R^2}k_{u^*}(u)\{\mu_{1}(X,m_1) - \mu_{0}(X,m_0)\}e_u(X)du\Bigg] 
 \\\nn
 &= &\mathbb{P}\Bigg[\frac{(-1)^{Z+1}{\gamma}_1^{(Z)}(X,M)}{\hat{\pi}_Z(X){f}_{Z}(M\mid X,{\beta}^*)}\big\{Y - \hat{\mu}_{Z}(X,M)\big\} - \frac{(-1)^{Z+1}{\gamma}_1^{(Z)}(X,M)}{{\pi}_Z(X){f}_{Z}(M\mid X,{\beta}^*)}\big\{Y - {\mu}_{Z}(X,M)\big\}\Bigg]
\\\nn
& & + \mathbb{P}\Bigg[\int_{\R^2}k_{u^*}(u)\{\hat\mu_{1}(X,m_1) - \hat\mu_{0}(X,m_0)\}e_u(X)du-\int_{\R^2}k_{u^*}(u)\{{\mu}_{1}(X,m_1) - {\mu}_{0}(X,m_0)\}e_u(X)du\Bigg]. 
\\
\label{ep:11}
\eee
By LOTE, the first term in \eqref{ep:11} equals to,
\beee\nn
&&\sum_{z = 0,1}\mathbb{P}\Bigg[\mathbb{P}\Big[{\pi}_z(X)\frac{(-1)^{z+1}{\gamma}_1^{(z)}(X,M)}{\hat{\pi}_z(X){f}_{z}(M\mid X,{\beta}^*)}\big\{Y - \hat{\mu}_{z}(X,M)\big\} 
\\\nn
&&- \frac{(-1)^{z+1}{\gamma}_1^{(z)}(X,M)}{{f}_{z}(M\mid X,{\beta}^*)}\big\{Y - {\mu}_{z}(X,M)\big\} \mid X,Z = z\Big]\Bigg]
\\\nn
&=&\sum_{z = 0,1}\mathbb{P}\Bigg[\mathbb{P}\Bigg[\mathbb{P}\Big[{\pi}_z(X)\frac{(-1)^{z+1}{\gamma}_1^{(z)}(X,M)}{\hat{\pi}_z(X){f}_{z}(M\mid X,{\beta}^*)}\big\{Y - \hat{\mu}_{z}(X,M)\big\} 
\\\nn
&& - \frac{(-1)^{z+1}{\gamma}_1^{(z)}(X,M)}{{f}_{z}(M\mid X,{\beta}^*)}\big\{Y - {\mu}_{z}(X,M)\big\} \mid X,Z = z,M\Big]\mid X,Z = z\Bigg]\Bigg]
\\\nn
&=&\sum_{z = 0,1}\mathbb{P}\Bigg[\mathbb{P}\Bigg[{\pi}_z(X)\frac{(-1)^{z+1}{\gamma}_1^{(z)}(X,M)}{\hat{\pi}_z(X){f}_{z}(M\mid X,{\beta}^*)}\big\{\mu_{z}(X,M) - \hat{\mu}_{z}(X,M)\big\} 
\\\nn
&& - \frac{(-1)^{z+1}{\gamma}_1^{(z)}(X,M)}{{f}_{z}(M\mid X,{\beta}^*)}\big\{\underbrace{\mu_{z}(X,M) - {\mu}_{z}(X,M)}_{ = 0}\big\}  \mid X,Z = z\Bigg]\Bigg]
\\\nn
&=&\sum_{z = 0,1}\mathbb{P}\Bigg[\frac{{\pi}_z(X)(-1)^{z+1}}{\hat{\pi}_z(X)}\mathbb{P}\Big[\frac{{\gamma}_1^{(z)}(X,M)}{{f}_{z}(M\mid X,{\beta}^*)}\big\{\mu_{z}(X,M) - \hat{\mu}_{z}(X,M)\big\} \mid X,Z=z\Big]\Bigg]
\\\nn
&=&\sum_{z = 0,1}\mathbb{P}\Bigg[\frac{{\pi}_z(X)(-1)^{z+1}}{\hat{\pi}_z(X)}\int_{\R^2}k_{u^*}(u)\{\mu_{z}(X,m_z) - \hat{\mu}_{z}(X,m_z)\}e_u(X)du\Bigg],
\eee
which combining with \eqref{phh} yields 
\beee\nn
&&\mathbb{P}\{h(V\mid \hat{\pi},\hat{\mu}) - h(V\mid \bar{\pi},\bar{\mu})\}
\\\nn
&  = &\sum_{z = 0,1}(-1)^{z+1}\cdot\mathbb{P}\Bigg[\Big\{\frac{{\pi}_z(X)}{\hat{\pi}_z(X)} -1\Big\}\int_{\R^2}k_{u^*}(u)\{\mu_{z}(X,m_z) - \hat{\mu}_{z}(X,m_z)\}e_u(X)du\Bigg].
\eee
For either $z = 0\text{ or 1}$, by Cauchy--Schwartz inequality, we have
\beee\nn
&&\Bigg|\mathbb{P}\Bigg[\Big\{\frac{{\pi}_z(X)}{\hat{\pi}_z(X)} -1\Big\}\int_{\R^2}k_{u^*}(u)\{\mu_{z}(X,m_z) - \hat{\mu}_{z}(X,m_z)\}e_u(X)du\Bigg]\Bigg|
\\\nn
 &\leq & \Bigg|\mathbb{P}\Bigg\{\frac{{\pi}_z(X) - \hat{\pi}_z(X)}{\hat{\pi}_z(X)}\Bigg\}^2\Bigg|^{1/2}\cdot\sqrt{\int_{\R^2}k^2_{u^*}(u)du}\cdot \Bigg|\mathbb{P}\Bigg[\int_{\R^2}\{\mu_{z}(X,m_z) - \hat{\mu}_{z}(X,m_z)\}^2e^2_u(X)du\Bigg]\Bigg|^{1/2}
\\\label{Pdecome:bound1}
& = & \mathcal{O}_{\mathbb{P}}(h^{-1}r_\pi r_\mu).
\eee
To see \eqref{Pdecome:bound1},  firstly we note by Assumptions \ref{am:pi} and \ref{am:emp} and LOTE,
\beee\nn
\mathbb{P}\big[\{{\pi}_z(X) - \hat{\pi}_z(X)\}^2\big]&=&\mathbb{P}\Big[\mathbb{P}\big[\{{\pi}_Z(X) - \hat{\pi}_Z(X)\}^2\mid X, Z = z\big]\Big]
\\\nn
&\leq &\pi_z(X)^{-1}\cdot \mathbb{P}\Big[\pi_z(X)\mathbb{P}\big[\{{\pi}_Z(X) - \hat{\pi}_Z(X)\}^2\mid X, Z = z\big] 
\\\nn
&&
+\pi_{1-z}(X)\mathbb{P}\big[\{{\pi}_Z(X) - \hat{\pi}_Z(X)\}^2\mid X, Z =1- z\big]\Big]
\\\nn
&= &\pi_z(X)^{-1}\cdot\mathbb{P}\Big[\mathbb{P}\big[\{{\pi}_Z(X) - \hat{\pi}_Z(X)\}^2\mid X\big] 
\Big]
\\\nn
&= &\pi_z(X)^{-1}\cdot\mathbb{P}\big[\{{\pi}_Z(X) - \hat{\pi}_Z(X)\}^2\big] 
\\\nn
&=&\mathcal{O}_{\mathbb{P}}(r_\pi^2). 
\eee
and thus
\beee\nn
\Bigg|\mathbb{P}\Bigg\{\frac{{\pi}_z(X) - \hat{\pi}_z(X)}{\hat{\pi}_z(X)}\Bigg\}^2\Bigg| &\leq& \|\hat{\pi}^{-1}\|_{\infty}^2 \cdot \big|\mathbb{P}\{{\pi}_z(X) - \hat{\pi}_z(X)\}^2\big|= \mathcal{O}_{\mathbb{P}}(r_\pi^2),
\\\nn
\eee
because $\|\hat{\pi}^{-1}\|_{\infty}\precsim 1$ with probability approaching $1$ under Assumptions \ref{am:pi} and \ref{am:emp}, where $\bar{\pi}_z(X) \precsim 1$ and $\|\hat{\pi} - \bar{\pi}\|_{\infty} = o_{\mathbb{P}}(1)$. Secondly, we have 
\beee\nonumber
&&\mathbb{P}\Bigg[\int_{\R^2}k^2_{u^*}(u)du\int_{\R^2}\{\mu_{z}(X,m_z) - \hat{\mu}_{z}(X,m_z)\}^2e^2_u(X)du\Bigg]
\\\nn
& \precsim  & h^{-2}\cdot \mathbb{P}\Bigg[\int_{\R^2}\{\mu_{z}(X,m_z) - \hat{\mu}_{z}(X,m_z)\}^2e^2_u(X)du\Bigg] \quad\text{(Assumption \ref{am:kernel})}
\\\nn
&\precsim &h^{-2}\cdot\mathbb{P}\Bigg[\int_{\R^2}\{\mu_{z}(X,m_z) - \hat{\mu}_{z}(X,m_z)\}^2e_u(X)du\Bigg]
\\\nn
& = &h^{-2}\cdot \mathbb{P}\Bigg[\mathbb{P}\Big[\{\mu_{z}(X,M_z) - \hat{\mu}_{z}(X,M_z)\}^2\mid X\Big]\Bigg]
\\\nn
& = &h^{-2}\cdot \mathbb{P}\Bigg[\mathbb{P}\Big[\{\mu_{Z}(X,M) - \hat{\mu}_{Z}(X,M)\}^2\mid X, Z  = z\Big]\Bigg]\quad(\text{Assumption \ref{assump:TAignorability}})
\\\nn
& \leq  &h^{-2}\cdot \pi_{z}^{-1}(X)\cdot\mathbb{P}\Bigg[\pi_{z}(X)\mathbb{P}\Big[\{\mu_{Z}(X,M) - \hat{\mu}_{Z}(X,M)\}^2\mid X, Z = z\Big]
\\\nn
&& + \pi_{1-z}(X)\mathbb{P}\Big[\{\mu_{Z}(X,M) - \hat{\mu}_{Z}(X,M)\}^2\mid X, Z = 1-z\Big]\Bigg]
\\\nn
&=&h^{-2}\cdot\pi_{z}^{-1}(X)\cdot\mathbb{P}\Bigg[\mathbb{P}\Big[\{\mu_{Z}(X,M) - \hat{\mu}_{Z}(X,M)\}^2\mid X\Big]\Bigg]
\\\nn
&=&h^{-2}\cdot\pi_{z}^{-1}(X)\cdot\mathbb{P}\Big[\{\mu_{Z}(X,M) - \hat{\mu}_{Z}(X,M)\}^2\Big]
\\\nn
&= & \mathcal{O}(h^{-2}r_{\mu}^2)\quad \text{(Assumptions \ref{am:pi} and \ref{am:emp})},
\eee
where the second inequality is by $e_u(x)$ is uniformly bounded by some constants  for all $x \in \mathbb{X}$ and $u\in\R^2$, under Assumptions \ref{am:eux} and \ref{am:unibound}. Summarizing results above, \eqref{Pdecome:bound1} is thus shown.
We finally conclude, 
\bee\nn
&\mathbb{P}\{h(V\mid \hat{\pi},\hat{\mu}) - h(V\mid \bar{\pi},\bar{\mu})\} = \mathcal{O}_{\mathbb{P}}(h^{-1}r_\pi r_\mu)
\\
&\text{and}\quad\mathbb{P}_n\{h(V\mid \hat{\pi},\hat{\mu}) - h(V\mid \bar{\pi},\bar{\mu})\} =\mathcal{O}_{\mathbb{P}}(h^{-1}r_\pi r_\mu) + o_{\mathbb{P}}(h^{-1}n^{-1/2}),
\ee
which combining with \eqref{D2n:step} yields
\beee\nn
\Delta_{2,n} &=& \Gamma_{2,n}^{(1)} + \Gamma_{2,n}^{(2,1)}  + \Gamma_{2,n}^{(2,2)}+ \mathcal{O}_{\mathbb{P}}(h^{-1}r_\pi r_\mu) + o_{\mathbb{P}}(h^{-1}n^{-1/2})
\\\label{main:piece3}
& = & \underbrace{\Gamma_{2,n}}_{\mathcal{O}_{\mathbb{P}}(h^{-1}n^{-1/2})} + \mathcal{O}_{\mathbb{P}}(h^{-1}r_\pi r_\mu) + o_{\mathbb{P}}(h^{-1}n^{-1/2}),
\eee
where in summary, we define,
\beee\nn 
&&\Gamma_{2,n} 
\\\nn
& = &\scalemath{0.8}{\sum_{z = 0,1}\mathbb{P}\Bigg[\int_{\R^2}k_{u^*}(u)\big\{\bar\mu_{1}( X,m_1) - \bar\mu_{0}(X,m_0)\big\}
c_{\rho}\big\{{F}_{1m_1}( X\mid\beta^*),{F}_{0m_0}(X\mid \beta^*)\big\}}
 \scalemath{0.8}{f_{(1 - z)m_{1- z}}( X\mid\beta^*)\{f'_{zm_z}( X\mid\beta^*)\}^\T du\Bigg](\hat{\beta} - \beta^*)}
\\\nonumber
&&\scalemath{0.8}{+\sum_{z = 0,1}\mathbb{P}\Bigg[\int_{\R^2}k_{u^*}(u)\big\{\bar\mu_{1}( X,m_1) - \bar\mu_{0}(X,m_0)\big\} c_{\rho}^{(z)}\big\{{F}_{1m_1}( X\mid\beta^*),{F}_{0m_0}(X\mid \beta^*)\big\}}
\\\nn
&&\scalemath{0.8}{\quad\quad\quad\,\,\quad {f}_{1m_1}( X\mid{\beta}^*){f}_{0m_0}( X\mid{\beta}^*)\big\{F'_{zm_z}(X\mid {\beta}^*)\big\}^\T du\Bigg]}\scalemath{0.8}{\cdot (\hat{\beta} - \beta^*) }
\\\nn
&&\scalemath{0.8}{ + \sum_{j = 1,2,3}\mathbb{P}\Bigg[\Big[\frac{(-1)^{Z+1}}{\bar{\pi}_Z(X)}\big\{Y - \bar{\mu}_{Z}(X,M)\big\}\Big]\delta_{X,Z,M}^{(j)}\Bigg](\hat{\beta} - \beta^*)}. 
\eee

\subsubsection{Putting all  pieces together}
Combining \eqref{main:piece1}, \eqref{main:piece2} and \eqref{main:piece3}, we have that, in general,
\beee\nn
&&\hat{\tau}_{u^*} - \tau_{u^*} 
\\\nn
& = &\scalemath{0.8}{ (\mathbb{P}_n - \mathbb{P})\Bigg[\frac{\bar{{\xi}}_1\{Y - \mu_{Z}(X,M)\}
+\int_{\R^2}\big\{\bar\mu_{1}( X,m_1) - \bar\mu_{0}(X,m_0)\big\}k_{u^*}(u){e}_{u}(X)du }{\mathbb{E}\big\{\int_{\R^2}k_{u^*}(u){e}_{u}(X)du\big\}} \Bigg]}
\\\nn
&& \scalemath{0.8}{-\frac{\mathbb{E}\big[\bar{{\xi}}_1
\{Y - \mu_{Z}(X,M)\}+\int_{\R^2}\big\{\bar\mu_{1}( X,m_1) - \bar\mu_{0}(X,m_0)\big\}k_{u^*}(u){e}_{u}(X)du \big]}{\big[\mathbb{E}\big\{\int_{\R^2}k_{u^*}(u){e}_{u}(X)du\big\}\big]^2}
\Bigg[(\mathbb{P}_n - \mathbb{P})\Big\{\int_{\R^2}k_{u^*}(u){e}_{u}(X)du\Big\} + \mathcal{A}_1(\hat{\beta} - \beta^*)\Bigg]}
\\\nn
&& \scalemath{0.8}{+ \frac{\mathcal{A}_2}{\mathbb{E}\big\{\int_{\R^2}k_{u^*}(u)e_u(X)du\big\}}(\hat{\beta} - \beta^*)
+  \mathcal{O}_{\mathbb{P}}(h^{-1}r_\pi r_\mu) +o_{\mathbb{P}}(h^{-1}n^{-1/2})}
\\\label{rate:db2}
& = &\scalemath{0.8}{(\mathbb{P}_n - \mathbb{P})\Bigg[\frac{\bar{{\xi}}_1\{Y - \mu_{Z}(X,M)\}
+\int_{\R^2}\big\{\bar\mu_{1}( X,m_1) - \bar\mu_{0}(X,m_0)\big\}k_{u^*}(u){e}_{u}(X)du - \tau_{u^*}\int_{\R^2}k_{u^*}(u)e_u(X)du}{\mathbb{E}\big\{\int_{\R^2}k_{u^*}(u){e}_{u}(X)du\big\}} \Bigg]}
\\\nn
&& \scalemath{0.8}{+  \frac{\mathcal{A}_2 - \tau_{u^*}\mathcal{A}_1}{\mathbb{E}\big\{\int_{\R^2}k_{u^*}(u){e}_{u}(X)du\big\}}(\hat{\beta} - \beta^*)+  \mathcal{O}_{\mathbb{P}}(h^{-1}r_\pi r_\mu) +o_{\mathbb{P}}(h^{-1}n^{-1/2})}
\\\label{rate:db}
&= & \mathcal{O}_{\mathbb{P}}\big(h^{-1}n^{-1/2} + h^{-1}r_\pi r_\mu\big),
\eee
where the the second equality is by Theorem \ref{thm:dbrb} under Assumption \ref{am:emp}(1); The first two terms in \eqref{rate:db2} are $\mathcal{O}_{\mathbb{P}}(h^{-1}n^{-1/2})$ with a standard argument similar as before. We note that when $h$ is fixed, we have $\mathbb{E}\big\{\int_{\R^2}k_{u^*}(u){e}_{u}(X)du\big\} > 0$ is fixed, and if $h\rightarrow 0$,
\beee\nn
\mathbb{E}\Big\{\int_{\R^2}k_{u^*}(u){e}_{u}(X)du\Big\}& = &\mathbb{E}\Big[\Big\{k_{u^*}(U)\mid X\Big\}\Big]
\\\nn
& =& \mathbb{E}\big[k_{u^*}(U)\big]
\\\label{kubound} 
& = & e_{u^*} + \mathcal{O}(h^2) \succsim 1
\eee
by \eqref{euh}, which is also bounded away from $0$. We define $\mathcal{A}_1$ and $\mathcal{A}_2$ in \eqref{def:A1} and \eqref{def:A2}, $\delta_{X,1,M}^{(1)}$--$\delta_{X,1,M}^{(3)}$  in \eqref{haha}, and $\delta_{X,0,M}^{(1)}$--$\delta_{X,0,M}^{(3)}$ in \eqref{haha2}. 
\par
In the following, we focus on the fixed-$h$ setting for the asymptotic analysis of Loc.PCE. Then \eqref{rate:db} becomes
$$
\hat{\tau}_{u^*} - \tau_{u^*} = \mathcal{O}_{\mathbb{P}}\big(n^{-1/2} + r_\pi r_\mu\big),
$$
which implies \eqref{main:rate}. If moreover, $r_\pi r_\mu = o( n^{-1/2})$ and $\hat{\beta}$ is asymptotically linear, we further have,
{ \beee\nn
&&\sqrt{n}(\hat{\tau}_{u^*} - \tau_{u^*} )
\\\nn
& = &\scalemath{0.8}{ \sqrt{n}(\mathbb{P}_n - \mathbb{P})\Bigg[\frac{\bar{{\xi}}_1\{Y - \mu_{Z}(X,M)\}
+\int_{\R^2}\big\{\bar\mu_{1}( X,m_1) - \bar\mu_{0}(X,m_0)\big\}k_{u^*}(u){e}_{u}(X)du - \tau_{u^*}\int_{\R^2}k_{u^*}(u)e_u(X)du}{\mathbb{E}\big\{\int_{\R^2}k_{u^*}(u){e}_{u}(X)du\big\}} \Bigg]}
\\\nn
&& \scalemath{0.8}{+ \frac{\mathcal{A}_2 - \tau_{u^*}\mathcal{A}_1}{\mathbb{E}\big\{\int_{\R^2}k_{u^*}(u){e}_{u}(X)du\big\}}(\hat{\beta} - \beta^*) + o_{\mathbb{P}}(1)}
\\\nn
&=& \scalemath{0.8}{\sqrt{n}(\mathbb{P}_n - \mathbb{P})\Bigg[\frac{\bar{{\xi}}_1\{Y - \mu_{Z}(X,M)\}
+\int_{\R^2}\big\{\bar\mu_{1}( X,m_1) - \bar\mu_{0}(X,m_0)\big\}k_{u^*}(u){e}_{u}(X)du - \tau_{u^*}\int_{\R^2}k_{u^*}(u)e_u(X)du}{\mathbb{E}\big\{\int_{\R^2}k_{u^*}(u){e}_{u}(X)du\big\}}} 
\\\label{res:dist2}
&&\scalemath{0.8}{+ \frac{\mathcal{A}_2 - \tau_{u^*}\mathcal{A}_1}{\mathbb{E}\big\{\int_{\R^2}k_{u^*}(u){e}_{u}(X)du\big\}} \psi(X,Z,M)\Bigg] + o_{\mathbb{P}}(1)}
\\\label{res:dist}
&\leadsto & \mathcal{N}(0,\bar\sigma_{u^*}^2)
\eee}
where the second equality is by that $\psi(X,Z,M)$ is mean-zero and $ \|\{\mathcal{A}_2 - \tau_{u^*}\mathcal{A}_1\}[\mathbb{E}\{\int_{\R^2}k_{u^*}$ $(u){e}_{u}(X)du\}]^{-1}\|$ is bounded by some constant, which can be easily shown with a similar argument as the previous proof. Finally, \eqref{res:dist} follows from the central limit theorem (CLT) and Slutsky's theorem, as the random variable in $\sqrt{n}(\mathbb{P}_n - \mathbb{P})[\,\cdot\,]$ in \eqref{res:dist2} is mean-zero and has bounded variance under the conditions of Theorem \ref{thm:dbrb}. In particular, we have
\beee\nn
\bar\sigma_{u^*}^2 &=& \text{Var}\Bigg[\frac{\bar{{\xi}}_1\{Y - \mu_{Z}(X,M)\}
+\int_{\R^2}\big\{\bar\mu_{1}( X,m_1) - \bar\mu_{0}(X,m_0)\big\}k_{u^*}(u){e}_{u}(X)du - \tau_{u^*}\int_{\R^2}k_{u^*}(u)e_u(X)du}{\mathbb{E}\big\{\int_{\R^2}k_{u^*}(u){e}_{u}(X)du\big\}}
\\\nn
&&+ \frac{\mathcal{A}_2 - \tau_{u^*}\mathcal{A}_1}{\mathbb{E}\big\{\int_{\R^2}k_{u^*}(u){e}_{u}(X)du\big\}} \psi(X,Z,M)\Bigg]
\\\nn
&=& \E\Bigg[\frac{\mathcal{A}_2 - \tau_{u^*}\mathcal{A}_1}{\mathbb{E}\big\{\int_{\R^2}k_{u^*}(u){e}_{u}(X)du\big\}} \psi(X,Z,M) + \frac{\bar{{\xi}}_1\{Y - \mu_{Z}(X,M)\}
}{\mathbb{E}\big\{\int_{\R^2}k_{u^*}(u){e}_{u}(X)du\big\}}
\\\nn
&&+ \frac{\int_{\R^2}\big\{\bar\mu_{1}( X,m_1) - \bar\mu_{0}(X,m_0)\big\}k_{u^*}(u){e}_{u}(X)du - \tau_{u^*}\int_{\R^2}k_{u^*}(u)e_u(X)du}{\mathbb{E}\big\{\int_{\R^2}k_{u^*}(u){e}_{u}(X)du\big\}}\Bigg]^2.
\eee
\par
In Theorem \ref{eif:tau} (ii), we have derived the semiparametric EIF  under the setting that the parametric model $\mathcal{M}_{{\rm ps}}$ in \eqref{paraps} is specified and $\hat{\beta}$ is estimated. By the classic semiparametric theory (see e.g. \citet{bickel1993efficient}), the  corresponding semiparametric efficiency bound is
\beee\nn
\sigma_{u^*,\text{eff}}^2&=&\E\{\phi^2_{{\rm p}}(X,Z,M,Y)\} 
\\\nn
&=&
\E\Bigg[\frac{\bar{{\xi}}_1\{Y - \mu_{Z}(X,M)\}
+\int_{\R^2}\big\{\mu_{1}( X,m_1) - \mu_{0}(X,m_0)\big\}k_{u^*}(u){e}_{u}(X)du - \tau_{u^*}\int_{\R^2}k_{u^*}(u)e_u(X)du}{\mathbb{E}\big\{\int_{\R^2}k_{u^*}(u){e}_{u}(X)du\big\}}
\\\label{goal:rate}
&&+ \frac{\mathcal{A}^*_2 - \tau_{u^*}\mathcal{A}^*_1}{\mathbb{E}\big\{\int_{\R^2}k_{u^*}(u){e}_{u}(X)du\big\}} I_{\beta^*}^{-1}s_{ZM}(X\mid \beta^*)\Bigg]^2.
\eee
  We now show that, under the setting of Theorem \ref{thm:consist}, the asymptotic variance of our proposed estimator given in \eqref{sigmaform} achieves such bound and thus is semeparaetric efficient. Under the conditions of Theorem \ref{thm:consist}(III), we can check the following facts through definitions:
\begin{itemize}
\item[(i)] $\bar{\tilde{\xi}}_1= \tilde{\xi}_1$ because $\bar{\mu} = \mu$, $\bar{\pi}= \pi$; 
\item[(ii)] $\mathcal{A}_1 = \mathcal{A}_1^*$, $\mathcal{A}_2 = \mathcal{A}_2^*$, noting that the third term of $\mathcal{A}_2$ in \eqref{def:A2} under our scenario becomes,
\beee\nn
&& \sum_{j = 1,2,3}\mathbb{E}\Bigg[\Big[\frac{(-1)^{Z+1}}{{\pi}_Z(X)}\big\{Y - {\mu}_{Z}(X,M)\big\}\Big]\delta_{X,Z,M}^{(j)}\Bigg]
\\\nn
& = & \sum_{j = 1,2,3}\mathbb{E}\Bigg[\Big[\frac{(-1)^{Z+1}}{{\pi}_Z(X)}\underbrace{\E\big\{Y - {\mu}_{Z}(X,M)\mid X,Z,M\big\}}_{ = 0}\Big]\delta_{X,Z,M}^{(j)}\Bigg] = 0;
\eee
\item[(iii)] $\psi(X,Z,M) = I_{\beta^*}^{-1}s_{ZM}(X\mid \beta^*)$ by $\psi(x,z,m) = I_{\beta^*}^{-1}s_{zm}(x\mid \beta^*)$.\end{itemize}
 Based on above facts and comparing \eqref{sigmaform} with \eqref{goal:rate}, we can observe  that $\bar\sigma_{u^*}^2=\sigma_{u^*,\text{eff}}^2$, which implies that $\hat{\tau}_{u^*}$ is semiparametric efficient.

\qed
\subsection{Proof of Theorem \ref{hto0}}\label{hto0:p}
Combining \eqref{rate:db} and Proposition \ref{po:approximate}, we have,
\beee\nn
\hat{\tau}_{u^*} - \tau^*_{u^*} &=  &(\hat{\tau}_{u^*} - \tau_{u^*}) +(\tau_{u^*} - \tau^*_{u^*})
\\\nn
& = &\mathcal{O}_{\mathbb{P}}\big(h^{-1}n^{-1/2} + h^{-1}r_\pi r_\mu + h^2\big). 
\eee
If $r_\pi r_\mu=\mathcal{O}(n^{-1/2})$, one further has $\hat{\tau}_{u^*} - \tau^*_{u^*} = \mathcal{O}_{\mathbb{P}}(h^{-1}n^{-1/2}  + h^2)$. When choosing $h\asymp n^{-1/6}$, we then deduce the minimax optimal convergence rate $\hat{\tau}_{u^*} - \tau^*_{u^*} = \mathcal{O}_{\mathbb{P}}(n^{-1/3})$.
\par
By \eqref{rate:db2} and Proposition \ref{po:approximate}, when $h = o(n^{-1/6})$, $r_{\mu}r_\pi = o(n^{-1/2})$ and $\hat{\beta}$ is asymptotically linear, we further have,
\beee\nn
&&\sqrt{n}h(\hat{\tau}_{u^*} - \tau^*_{u^*})
\\\nn
&=&\sqrt{n}h(\hat{\tau}_{u^*} - \tau_{u^*}) + \sqrt{n}h({\tau}_{u^*} - \tau^*_{u^*})
\\\nn &=& \scalemath{0.9}{\sqrt{n}(\mathbb{P}_n - \mathbb{P})\Bigg[\frac{h\bar{{\xi}}_1\{Y - \mu_{Z}(X,M)\}
}{\mathbb{E}\big\{\int_{\R^2}k_{u^*}(u){e}_{u}(X)du\big\}} + \frac{
h\int_{\R^2}\big\{\bar\mu_{1}( X,m_1) - \bar\mu_{0}(X,m_0)\big\}k_{u^*}(u){e}_{u}(X)du}{\mathbb{E}\big\{\int_{\R^2}k_{u^*}(u){e}_{u}(X)du\big\}} - \frac{h\tau_{u^*}\int_{\R^2}k_{u^*}(u)e_u(X)du}{\mathbb{E}\big\{\int_{\R^2}k_{u^*}(u){e}_{u}(X)du\big\}} }
\\\label{nonp:conv2}
\\\label{nonp:conv}
&&\scalemath{0.9}{+\frac{h\mathcal{A}_2 - h\tau_{u^*}\mathcal{A}_1}{\mathbb{E}\big\{\int_{\R^2}k_{u^*}(u){e}_{u}(X)du\big\}}\psi(X,Z,M)\Bigg]+  \underbrace{\mathcal{O}_{\mathbb{P}}(\sqrt{n}r_\pi r_\mu) +o_{\mathbb{P}}(1) + \mathcal{O}(\sqrt{n}h^3)}_{o_{\mathbb{P}}(1)}.}
\eee
To apply central limit theorem, under regularity conditions, we check the $L_2$-norm bounds of all terms in \eqref{nonp:conv2} and \eqref{nonp:conv}. For example, for the first term in \eqref{nonp:conv2}, we have
\beee\nonumber
\E\Bigg[\frac{h\bar{{\xi}}_1\{Y - \mu_{Z}(X,M)\}
}{\mathbb{E}\big\{\int_{\R^2}k_{u^*}(u){e}_{u}(X)du\big\}}\Bigg]^2 & = &\E\Bigg[\frac{h^2\{\bar{\pi}_Z(X) f_{ZM}(X)\}^{-2}\big\{{{\gamma_1}^{(Z)}(Z,M)}\big\}^2\{Y - \mu_{Z}(X,M)\}^2
}{\big[\mathbb{E}\big\{\int_{\R^2}k_{u^*}(u){e}_{u}(X)du\big\}\big]^2}\Bigg]
\\\nn
& \precsim &\E\Big[{\big\{h{{\gamma_1}^{(Z)}(Z,M)}\big\}^2\{Y - \mu_{Z}(X,M)\}^2
}\Big]
\\\nn
& \precsim &\E\Big[{\{Y - \mu_{Z}(X,M)\}^2
}\Big] = \mathcal{O}(1),
\eee
where the first inequality is by \eqref{kubound}, Assumption \ref{am:pi} and Assumption \ref{am:unibound}, and the second inequality is by $\sup_{z,m}h{{\gamma_1}^{(z)}(z,m)} = \mathcal{O}(1)$ under Assumption \ref{am:kernel}. For the second term in \eqref{nonp:conv2}, we have
\beee\nn
\E\Bigg[\frac{h\int_{\R^2}\big\{\bar\mu_{1}( X,m_1) - \bar\mu_{0}(X,m_0)\big\}k_{u^*}(u){e}_{u}(X)du}{\mathbb{E}\big\{\int_{\R^2}k_{u^*}(u){e}_{u}(X)du\big\}}\Bigg]^2 & \precsim & \E\Bigg[{h\int_{\R^2}k_{u^*}(u)du}\Bigg]^2 
\\\nn
&= &h^2 \rightarrow 0.
\eee
Thus the second term in \eqref{nonp:conv2} is negligible, i.e., it does not contribute to the  asymptotic variance of $\sqrt{n}h(\hat{\tau}_{u^*} - \tau^*_{u^*})$. We can similarly analyze all terms in \eqref{nonp:conv2} and \eqref{nonp:conv} and show all terms have either bounded or vanishing $L_2$ norms; We omit details for simplicity. Finally by CLT, we can finally show that $$\sqrt{n}h(\hat{\tau}_{u^*} - \tau^*_{u^*})\leadsto \mathcal{N}(0,\tilde\sigma_{u^*}^2),$$ where
\beee\label{tildesigma}
\tilde\sigma_{u^*}^2 &=& \frac{\E\big[\mathcal{C} \psi(X,Z,M)  + h\bar{{\xi}}_1\{Y - \mu_{Z}(X,M)\}\big]^2}{\big[\mathbb{E}\big\{\int_{\R^2}k_{u^*}(u){e}_{u}(X)du\big\}\big]^2},
\\\nn
\mathcal{C} &=& \sum_{j = 1,2,3}\mathbb{E}\Bigg[\Big[\frac{(-1)^{Z+1}}{\bar{\pi}_Z(X)}\big\{Y - \bar{\mu}_{Z}(X,M)\big\}\Big]h\delta_{X,Z,M}^{(j)}\Bigg].
\eee
\qed
 \subsection{Proof of Theorem~\ref{thm:complex}.}\label{sec:pf:ni}
We focus on 
\bee\label{example:ni}
\lbt\hat{e}_{\dd\od}(X_i)\big\rbt_{u^*}  = \frac{1}{h^2}\int_{\R^2} \hat{e}_{u}(X_i)\mathcal{K}\left(\frac{m_1 - m_1^*}{h},\frac{m_0 - m_0^*}{h}\right)du
\ee as an example to investigate the computational complexity bound of calculating it. The argument is similar to derive the computational complexity in approximating  $\lbt\{\hat\mu_{1}( X_i,\dd) - \hat\mu_{0}(X_i,\od) \}\hat{e}_{\dd\od}(X_i)\rbt_{u^*}$.  We  first decompose  the domain of the integration in \eqref{example:ni}, to a 2D-square centered at $u^*$ with an $C_1h\sqrt{\log n}$ edge length for some $C_1 > 0$, and the rest of the area in $\R^2$. We denote the 2D domain by $\mathcal{S}_{u^*,C_1h\sqrt{\log n}}$. Then \eqref{example:ni} can be   decomposed as
\beee\label{example:ni:decom}
\lbt\hat{e}_{\dd\od}(X_i)\big\rbt_{u^*} & = &  \frac{1}{h^2}\int_{\mathcal{S}_{u^*,C_1h\sqrt{\log n}}} \hat{e}_{u}(X_i)\mathcal{K}\left(\frac{m_1 - m_1^*}{h},\frac{m_0 - m_0^*}{h}\right)du 
\\\nonumber
&& + \frac{1}{h^2}\int_{\R^2\setminus \mathcal{S}_{u^*,C_1h\sqrt{\log n}}} \hat{e}_{u}(X_i)\mathcal{K}\left(\frac{m_1 - m_1^*}{h},\frac{m_0 - m_0^*}{h}\right)du.
\eee
Recall that $\hat{e}_{u}(X_i)$ is uniformly bounded for all $u\in\R^2$. Let $\Psi(\cdot)$ be the Gaussian CDF. Then by the tail bound of $\Psi(\cdot)$  (see e.g., \citet{vershynin2018high}), we have for some sufficiently large and fixed $C_1 > 0$, $1 - \Psi(C_1\sqrt{\log n}/2) = \mathcal{O}(n^{-1})$, and thus
\beee\nonumber
&&\frac{1}{h^2}\int_{\R^2\setminus \mathcal{S}_{u^*,C_1h\sqrt{\log n}}} \hat{e}_{u}(X_i)\mathcal{K}\left(\frac{m_1 - m_1^*}{h},\frac{m_0 - m_0^*}{h}\right)du 
\\\nonumber
&\precsim& \frac{1}{h^2}\int_{\R^2\setminus \mathcal{S}_{u^*,C_1h\sqrt{\log n}}}  \mathcal{K}\left(\frac{m_1 - m_1^*}{h},\frac{m_0 - m_0^*}{h}\right)du
\\\nonumber
& = &1 - \frac{1}{h^2}\int_{  \mathcal{S}_{u^*,C_1h\sqrt{\log n}}}  \mathcal{K}\left(\frac{m_1 - m_1^*}{h},\frac{m_0 - m_0^*}{h}\right)du
\\\nonumber
& = &1 - \left\{2 \Psi\left(\frac{C_1\sqrt{\log n}}{2}\right) - 1\right\}^2
\\\nonumber
&\leq & 1 - (1 - n^{-1 })^2 = \mathcal{O}(n^{-1 }).
\eee
 Thus we have
  \bee\nonumber
\lbt\hat{e}_{\dd\od}(X_i)\big\rbt_{u^*} =  \frac{1}{h^2}\int_{\mathcal{S}_{u^*,C_1h\sqrt{\log n}}} \hat{e}_{u}(X_i)\mathcal{K}\left(\frac{m_1 - m_1^*}{h},\frac{m_0 - m_0^*}{h}\right)du + \mathcal{O}(n^{-1}),
\ee
and to approximate $\lbt\hat{e}_{\dd\od}(X_i)\big\rbt_{u^*}$ with an $o(n^{-1/2})$ accuracy, it is sufficient to numerically approximate the first term on the right-hand side of \eqref{example:ni:decom}. We can further write
\beee\nonumber
&&\frac{1}{h^2}\int_{\mathcal{S}_{u^*,C_1h\sqrt{\log n}}} \hat{e}_{u}(X_i)\mathcal{K}\left(\frac{m_1 - m_1^*}{h},\frac{m_0 - m_0^*}{h}\right)du
\\\nonumber
&=&\frac{1}{h^2}\int_{0}^{C_1h\sqrt{\log n}}\int_{0}^{C_1h\sqrt{\log n}} \hat{e}_{t_1 - C_1h\sqrt{\log n}/2 + m_1^*,t_0 - C_1h\sqrt{\log n}/2 +m_0^*}(X_i)
\\\nonumber
&&\quad \quad \quad \quad \quad \quad \quad \quad\quad \quad \quad \quad \cdot \mathcal{K}\left(\frac{t_1 - C_1h\sqrt{\log n}/2}{h},\frac{t_0 - C_1h\sqrt{\log n}/2}{h}\right)dt_1t_0
\\\nonumber
&=&C_1^2\log n\cdot \int_{0}^{1}\int_{0}^{1} \hat{e}_{C_1h\sqrt{\log n}t_1 - C_1h\sqrt{\log n}/2 + m_1^*,C_1h\sqrt{\log n}t_0 - C_1h\sqrt{\log n}/2 +m_0^*}(X_i)
\\\nonumber
&&\quad\quad\quad\quad\quad\quad\quad\quad\cdot\mathcal{K}\left({C_1\sqrt{\log n}t_1 - C_1\sqrt{\log n}/2} ,{C_1\sqrt{\log n}t_0 - C_1\sqrt{\log n}/2} \right)d {t_1} dt_0
\\\nonumber
&=:& C_1^2\log n\cdot \mathcal{I}_i.
\eee
 Denote the following $N\times N$ uniform grids on $[0,1]^2$: $\mathcal{H}_N = \{(h_1/N,h_0/N )\mid (h_1,h_0)\in[N]^2\}$. We will approximate $\mathcal{I}_i$  by the following Riemann sum,
\bee \nonumber
& \bar{\mathcal{I}}_i = \frac{1}{N^2}\sum_{(t_1,t_0)\in\mathcal{H}_N}\hat{e}_{C_1h\sqrt{\log n}t_1 - C_1h\sqrt{\log n}/2 + m_1^*,C_1h\sqrt{\log n}t_0 - C_1h\sqrt{\log n}/2 +m_0^*}(X_i)
\\\label{def:barIi}
&  \quad \quad \quad \quad \quad \quad \quad\quad\cdot\mathcal{K}\left({C_1\sqrt{\log n}t_1 - C_1\sqrt{\log n}/2} ,{C_1\sqrt{\log n}t_0 - C_1\sqrt{\log n}/2} \right),
\ee
which has a computational complexity of $\mathcal{O}(N^2r_n)$, recalling that calculating   $$\hat{e}_{C_1h\sqrt{\log n}t_1 - C_1h\sqrt{\log n}/2 + m_1^*,C_1h\sqrt{\log n}t_0 - C_1h\sqrt{\log n}/2 +m_0^*}(X_i)$$ is $\mathcal{O}(r_n)$, and calculating   $\mathcal{K}\left({C_1\sqrt{\log n}t_1 - C_1\sqrt{\log n}/2} ,{C_1\sqrt{\log n}t_0 - C_1\sqrt{\log n}/2} \right)$ is $\mathcal{O}(1)$ for the standard Gaussian kernel $\mathcal{K}$. We need the following direct deduction of the main theorem  in \citet[$\mathsection$5.5]{davis2007methods} to bound the approximation error.
\begin{lemma}[\citet{davis2007methods}]\label{lm:numerical}
Suppose $f(t_1,t_0)$ satisfies that
\bee\label{lm:condition:ni}
\int_{[0,1]^2}\left|\frac{\partial^2 f(m_1,m_0)}{\partial m_1\partial m_0}\right|dm_1 dm_0 + \int_{[0,1]}\left|\frac{\partial f(m_1,1)}{\partial m_1 }\right|dm_1   + \int_{[0,1]}\left|\frac{\partial f(1,m_0)}{ \partial m_0}\right| dm_0 \leq V,
\ee
for some $V > 0$. Then we have
\bee\nonumber
\left|\frac{1}{N^2}\sum_{(t'_1,t'_0)\in\mathcal{H}_N}f(t'_1,t'_0) - \int_{[0,1]^2}f(t_1,t_0)dt_1dt_0\right| \leq \frac{2V}{{N}}.
\ee
\end{lemma}
Now let 
\bee\nonumber
 f(t_1,t_0) 
&= \hat{e}_{C_1h\sqrt{\log n}t_1 - C_1h\sqrt{\log n}/2 + m_1^*,C_1h\sqrt{\log n}t_0 - C_1h\sqrt{\log n}/2 +m_0^*}(X_i)\\
&\quad\quad \cdot\mathcal{K}\left({C_1\sqrt{\log n}t_1 - C_1\sqrt{\log n}/2} ,{C_1\sqrt{\log n}t_0 - C_1\sqrt{\log n}/2} \right).
\ee
Under \eqref{regular:int:condition} and with a standard Gaussian kernel $\mathcal{K}$, it can be shown by some direct calculations that, for any    $C_1$,  $h$ and $n$, condition~\eqref{lm:condition:ni} is satisfied with $V = C_V$, for some constant $C_V > 0$ only depending on $C$ in \eqref{regular:int:condition}. Then by Lemma~\ref{lm:numerical}  with $N = (\log n)^{1 +\epsilon/2}\sqrt{n}$, for any fixed $\epsilon > 0$,
\bee\nonumber
|C_1^2\log n\cdot \mathcal{I}_i - C_1^2\log n\cdot \bar{\mathcal{I}}_i| = \mathcal{O}\left\{(\log n)^{-\epsilon/2}n^{-1/2}\right\} = o(n^{-1/2}).
\ee
In summary, $C_1^2\log n\cdot \bar{\mathcal{I}}_i$ is a numerical approximation of $\lbt\hat{e}_{\dd\od}(X_i)\big\rbt_{u^*}$ with an approximation error bounded by $o(n^{-1/2})$, and it has a computational complexity of $\mathcal{O}(N^2r_n) = \mathcal{O}\{(\log n)^{2 + \epsilon}nr_n\}$. Similar result also holds for $\lbt\{\hat\mu_{1}( X_i,\dd) - \hat\mu_{0}(X_i,\od) \}\hat{e}_{\dd\od}(X_i)\rbt_{u^*}$. Thus, to numerically approximate   all  $\{\lbt\hat{e}_{\dd\od}(X_i)\big\rbt_{u^*}\}_{i = 1}^n$ and  $\{\lbt\{\hat\mu_{1}( X_i,\dd) - \hat\mu_{0}(X_i,\od) \}\hat{e}_{\dd\od}(X_i)\rbt_{u^*}\}_{i = 1}^n$, with approximation errors  bounded by $o(n^{-1/2})$ uniformly for all the corresponding numerical integrations, the overall computational complexity can be bounded by $\mathcal{O}\{(\log n)^{2 + \epsilon}n^2r_n\}$.
\par
Next, we study the numerical integration for $\hat{\gamma}_{1,u^*}^{(Z_i)}(M_i,X_i)$. Without loss of generality, assume $Z_i = 0$, and then
$$
\hat{\gamma}_{1,u^*}^{(Z_i)}(M_i,X_i)=\int_{\R} k_{u^*}(m_1,M_i)\hat{e}_{(m_1,M_i
)}(X_i)dm_{1} = \frac{1}{h^2}\int_\R\mathcal{K}\left(\frac{m_1 - m_1^*}{h},\frac{M_i - m_0^*}{h}\right)\hat{e}_{(m_1,M_i
)}(X_i)dm_{1}.
$$
Following a similar argument as before, we only  need to focus on the numerical approximation for 
\bee\nonumber
\frac{1}{h^2}\int_{m_1^*-C_2 h\sqrt{\log (n/h)}}^{m_1^*+C_2 h\sqrt{\log (n/h)}}\mathcal{K}\left(\frac{m_1 - m_1^*}{h},\frac{M_i - m_0^*}{h}\right)\hat{e}_{(m_1,M_i
)}(X_i)dm_{1},
\ee
for some sufficiently large but fixed $C_2 > 0$. Since $h\succsim n^{-1/2}$, we further have that we only need to  focus on the numerical approximation for
\beee\nonumber
&&\frac{1}{h^2}\int_{m_1^*-C_3 h\sqrt{\log n}}^{m_1^*+C_3 h\sqrt{\log n}}\mathcal{K}\left(\frac{m_1 - m_1^*}{h},\frac{M_i - m_0^*}{h}\right)\hat{e}_{(m_1,M_i
)}(X_i)dm_{1}
\\
&=&\frac{C_3\sqrt{\log n}}{h}\int_{ -1 }^{ 1 }\mathcal{K}\left(C_3\sqrt{\log n}t_1,\frac{M_i - m_0^*}{h}\right)\hat{e}_{(C_3h\sqrt{\log n}t_1 + m_1^*,M_i
)}(X_i)dt_1
\\
&=:& \frac{C_3\sqrt{\log n}}{h}{\mathcal{I}_i'},
\eee
for some sufficiently large $C_3 > 0$. Let 
$$
f(t_1) = \mathcal{K}\left(C_3\sqrt{\log n}t_1,\frac{M_i - m_0^*}{h}\right)\hat{e}_{(C_3h\sqrt{\log n}t_1 + m_1^*,M_i
)}(X_i),
$$
be the integrand of $\mathcal{I}_i'$. It is easy to verify that under the regularity conditions, we have
\bee\nonumber
\left|\frac{d}{dt_1}f(t_1) \right| = \mathcal{O}\ (\sqrt{\log n}),
\ee
uniformly for all $t_1\in[-1,1]$. To approximate $\mathcal{I}_i' = \int_{-1}^1 f(t_1)dt_1$, we use the following Riemann sum:
\bee\nonumber
\bar{\mathcal{I}_i'} = \frac{1}{N}\sum_{t_1 = -1 + \frac{2h}{N},\, h
\in[N]}f(t_1),
\ee
which has a computational complexity of $\mathcal{O}(Nr_n)$, where similarly as \eqref{def:barIi}, the computational complexity to calculate each $f(t_1)$ is $\mathcal{O}(r_n)$. By the result from  \citet[$\mathsection$2.1]{davis2007methods} which provides the error bound for the 1D numerical integration, we have
\bee\nonumber
\left|\frac{C_3\sqrt{\log n}}{h}{\mathcal{I}_i'} - \frac{C_3\sqrt{\log n}}{h}\bar{\mathcal{I}_i'}\right| = \mathcal{O}\left( \frac{{\log n}}{hN}\right).
\ee
By taking $N = (\log n)^{1 +\epsilon}{n}$, we have
\bee\nonumber
\left|\frac{C_3\sqrt{\log n}}{h}{\mathcal{I}_i'} - \frac{C_3\sqrt{\log n}}{h}\bar{\mathcal{I}_i'}\right| = o(n^{-1/2}),
\ee
recalling that $h \succsim n^{-1/2}$. Thus we can use ${C_3\sqrt{\log n}}\bar{\mathcal{I}_i'}/{h}$ to approximate $\hat{\gamma}_{1,u^*}^{(Z_i)}(M_i,X_i)$ with computational complexity bounded by $\mathcal{O}\{(\log n)^{1 +\epsilon}{n}r_n\}$. To numerically  approximate all $\{\hat{\gamma}_{1,u^*}^{(Z_i)}(M_i,X_i)\}_{i = 1}^n$, with approximation errors  bounded by $o(n^{-1/2})$ uniformly for all corresponding numerical integrations, the overall computational complexity can be bounded by $\mathcal{O}\{(\log n)^{1 + \epsilon}n^2r_n\}$.
\qed 
\subsection{Proof of Theorem~\ref{prop:linear:triple}.}\label{sec:pf:linear} When $(\bar{\pi},\bar{e}) = (\pi,e)$, we have $\check{\tau}_{u^*} = \bar\tau_{u^*}$, and by Theorem~\ref{thm:dbrb}, we have $\lim_{h\rightarrow 0}\check{\tau}_{u^*} = \lim_{h\rightarrow 0}\bar{\tau}_{u^*} = \tau^*_{u^*}$.  When $\bar{\mu} = \mu$, we can show that
\bee\nonumber
\E\Big[\check\xi_{1,u^*}(X,Z,M)\{Y - \bar{\mu}_{Z}(X,M)\} 
 \Big] = \E\Big[\check\xi_{1,u^*}(X,Z,M)\E\{Y - {\mu}_{Z}(X,M)\mid X,Z,M\} 
 \Big]  = 0,
\ee
which implies that
\bee\nonumber
\check\tau_{u^*} = \frac{\E \big\lbt\big\{\mu_{1}(X,\dd) - \mu_{0}(X,\od)\big\}\bar{e}_{\dd\od}(X)\big\rbt_{u^*} }{\E\big\lbt{\bar e}_{\dd\od}(X)\big\rbt_{u^*}}.
\ee
Since $\bar{e}_u(x)$ satisfies the same regularity condition as ${e}_u(x)$, similar to the proof of Proposition~\ref{po:approximate}, we have
\bee\nonumber
\lim_{h\rightarrow 0}\check\tau_{u^*}  &= \lim_{h\rightarrow 0}\frac{\E \big\lbt\big\{\mu_{1}(X,\dd) - \mu_{0}(X,\od)\big\}\bar{e}_{\dd\od}(X)\big\rbt_{u^*} }{\E\big\lbt{\bar e}_{\dd\od}(X)\big\rbt_{u^*}} 
\\
&= \frac{\E\big[\big\{\mu_{1}(X,m^*_1) - \mu_{0}(X,m^*_0)\big\}\bar e_{u^*}(X)\big]}{\bar e_{u^*}},
\ee
where we define $\bar{e}_{u^*} = \E \{\bar{e}_{u^*}(X)\}$. We have $\bar{e}_{u^*} = \E \{\bar{e}_{u^*}(X)\} > 0$ by the regularity conditions in Proposition~\ref{po:approximate} which now apply for $\bar{e}_u(x)$. Since $\mu_z(x,m)$ follows the linear model, we further have
\bee\nonumber
\lim_{h\rightarrow 0}\check\tau_{u^*}  = \frac{\E\left[\{\lambda_{d_X + 2} + \lambda_{d_X + 3}(m_1^* - m_0^*)\}\bar{e}_{u^*}(X) \right]}{\bar{e}_{u^*}}
 = \lambda_{d_X + 2} + \lambda_{d_X + 3}(m_1^* - m_0^*).
\ee 
By Theorem~\ref{thm:iddd}, we can also  show 
$$
\tau_{u^*}^* =\frac{\E\big[\big\{\lambda_{d_X + 2} + \lambda_{d_X + 3}(m_1^* - m_0^*)\big\}e_{u^*}(X)\big]}{e_{u^*}} = \lambda_{d_X + 2} + \lambda_{d_X + 3}(m_1^* - m_0^*), $$ and thus $\lim_{h\rightarrow 0}\check\tau_{u^*} = \tau_{u^*}^*$.

\qed
 
\section{Proof of Lemma \ref{lm:emp}}\label{pf:lm:emp}
\subsection{Preliminary}\label{emp:pre}
We first introduce some basic concepts and results in the empirical process theory, e.g., the stochastic equicontinuity,  covering number, entropy integral; See  \citet{vaart1997weak} for a more detailed introduction. Based on the application of  empirical process theory, we   prove Lemma \ref{lm:emp}. Denote $\mathbb{G}_n = \sqrt{n}(\mathbb{P}_n - \mathbb{P})$. Let $\mathcal{E}(g)$ be an operator of function $g$ where $g$ is in some metric space $\mathcal{G}$ equipped with semimetric $d(\cdot,\cdot)$. In this paper, we typically consider the uniform difference metric, i.e., for any $g_1,g_2\in\mathcal{G}$,
\beee\nonumber
d(g_1,g_2) &=& \sup_{v\in\mathbb{V}}|g_1(v) - g_2(v)|.
\eee
 We now introduce some notations to measure the complexity of the functional space $\mathcal{G}$. Let $G$ be an envelope function for $\mathcal{G}$ such that $G(v) \geq |g(v)|$ for any $g \in \mathcal{G}$ and $v\in\mathbb{V}$. Let $N(\epsilon,\mathcal{G},\|\cdot\|)$ denote the covering number of $\mathcal{G}$, which is the minimal number of $\epsilon$-balls needed to cover $\mathcal{G}$; the $\epsilon$-ball is under the distance induced by norm $\|\cdot\|$. We then define the uniform entropy integral,
\beee\nonumber
J(\delta,\mathcal{G},\mathcal{L}_2) & =& \sup_{Q}\int_{0}^{\delta}\sqrt{\log N\{\epsilon\|G\|_{Q,2},\mathcal{G},\|\cdot\|_{Q,2}\}}d\epsilon,
\eee
where $Q$ can be any random measure, $\|\cdot\|_{Q,2}$ is the $\mathcal{L}_2$-norm under $Q$, such that $\|g\|_{Q,2} = \sqrt{\int_{\mathbb{V}} g^2 dQ}$.
 A series of empirical process $\{\mathbb{G}_n\mathcal{E}(g)\mid n \geq 1,\, g\in\mathcal{G}\}$ is \textit{stochastically equicontinuous} if for every $\epsilon,\zeta > 0$, there exists a $\delta > 0$ such that
\beee\nonumber
{\lim\sup}_{n\rightarrow +\infty}\mathbb{P}\Bigg\{\sup_{d(g_1,g_2)<\delta\atop g_1,g_2\in\mathcal{G}}\Big|\mathbb{G}_n\mathcal{E}(g_1) - \mathbb{G}_n\mathcal{E}(g_2)\Big|>\epsilon\Bigg\} & <& \zeta.
\eee
Above property implies that for any stochastically equicontinuous $\{\mathbb{G}_n \mathcal{E}(g)\mid n\geq 1,\, g \in\mathcal{G}\}$,  
\beee\label{sqbound}
\mathbb{G}_n\{\mathcal{E}(g_1) - \mathcal{E}(g_2)\} & = & o_{\mathbb{P}}(1)
\eee
whenever $d(g_1,g_2) = o_{\mathbb{P}}(1)$ and $g_1,g_2\in\mathcal{G}$ \citep{pollard2012convergence}. 
\par
We also denote 
$
\|\mathcal{E}  \|_{\mathcal{G}} = \sup_{g\in \mathcal{G}}|\mathcal{E}\{g(V)\}|.
$
It can be seen that $\|\mathcal{E}  \|_{\mathcal{G}}$ is generally a random variable as it depends on  $V$. To show a series of empirical process $\{\mathbb{G}_n\mathcal{E}(g)\mid n\ge 1, g\in\mathcal{G}\}$ is stochastically equicontinuous, one can  use Theorem 2.11.1 in \cite{vaart1997weak}, which boils down the verification of stochastic equicontinuity, to checking the following two Lindeberg conditions and one condition of  $\mathcal{G}$'s complexity,
\begin{enumerate}
\item[(1).] $\E\big\{\|\mathcal{E}\|_{\mathcal{G}}^2\cdot 1\big(\|\mathcal{E}\|_{\mathcal{G}} > \epsilon \sqrt{n}\big)\big\}\rightarrow 0$ for every $\epsilon > 0$,
\item[(2).] $\sup_{d(g_1,g_2 ) < \delta_n}\E\big[\{\mathcal{E}(g_1) - \mathcal{E}(g_2)\}^2\big] \rightarrow 0$ for every sequence $\delta_n \rightarrow 0$,
\item[(3).] $\int_{0}^{\delta_n}\sqrt{\log N(\epsilon,\mathcal{G},\|\cdot\|_{\mathbb{P}_n,2})}d
\epsilon \rightarrow 0$ in probability, for every sequence $\delta_n \rightarrow 0$.
\end{enumerate}
Now we prove Lemma \ref{lm:emp}. Using the new notation, we decompose our target term to bound, namely $(\mathbb{P}_n - \mathbb{P})\{h(V\mid \hat{\pi},\hat{\mu}) - h(V\mid \bar{\pi},\bar{\mu})\}$, as 
\beee\nn
(\mathbb{P}_n - \mathbb{P})\{h(V\mid \hat{\pi},\hat{\mu}) - h(V\mid \bar{\pi},\bar{\mu})\}&=& h^{-1}n^{-1/2}\mathbb{G}_n\{h_1(V\mid \hat{\pi},\hat{\mu}) - h_1(V\mid \bar{\pi},\bar{\mu})\}
\\\nn
&&  + n^{-1/2}\mathbb{G}_n\{h_2(V\mid \hat{\mu}) - h_2(V\mid \bar{\mu})\}
\\\label{decom:Gnh}
&& + n^{-1/2}\mathbb{G}_n\{h_3(V\mid \hat{\mu}) - h_3(V\mid \bar{\mu})\},
\eee
where we define 
\beee\nn
h_1(V\mid \pi,\mu)&= &\frac{(-1)^{Z + 1}h\gamma_1^{(Z)}(X,M)}{\pi_Z(X)f_{ZM}(X)}\big\{Y - \mu_{Z}(X,M)\big\},
\\\nn
h_2(V\mid \mu) &= & \int_{\R^2}k_{u^*}(u)\mu_{1}(X,m_1)e_u(X)du,
\\\nn
h_3(V\mid \mu) &=& -\int_{\R^2}k_{u^*}(u)\mu_{0}(X,m_0)e_u(X)du.
\eee
In the following sections, we show the root-$n$-rates of all three terms on the right-hand side of \eqref{decom:Gnh}, and thereby show the desired result \eqref{emp:con}.
\subsection{Bounding $\mathbb{G}_n\{h_1(V\mid \hat{\pi},\hat{\mu}) - h_1(V\mid \bar{\pi},\bar{\mu})\}$}\label{Sec5.2}
We use a similar proof strategy as \citet[Appendix 5]{kennedy2017non} to bound the first term of \eqref{decom:Gnh}. Denote
\beee\nn
\mathcal{E}_1(g) &=& \frac{(-1)^{Z + 1}{h}\gamma_1^{(Z)}(X,M)}{f_{ZM}(X)} g(V).
\eee
The following lemma gives a sufficient condition, under which $\{\mathbb{G}_n \mathcal{E}_1(g) \mid n\geq 1\}$ is stochastically equicontinuous.
\begin{lemma}\label{lm:emp:1} Suppose Assumptions \ref{assump:TAignorability}--\ref{am:emp} and other regularity conditions in Section \ref{sec:otherreg} hold. Suppose $g$ belongs to $\mathcal{G}$, such that $\mathcal{G}$ has:
\begin{itemize}
\item[(i).] A finite uniform entropy integral, i.e., $J(\delta,\mathcal{G},\mathcal{L}_2) < +\infty$ for every $\delta > 0$;
\item[(ii).] A uniformly bounded envelope function $G(v)$ such that $\sup_{v\in\mathbb{V}}G(v)< C_G$ with some universal constant $C_G > 0$, and $G(v) > |g(v)|$ for any $g \in\mathcal{G}$ and any $v\in\mathbb{V}$.
\end{itemize}  
Then  $\{\mathbb{G}_n \mathcal{E}_1(g) \mid n\geq 1\}$ is stochastically equicontinuous. 
\end{lemma}
\noindent\textit{Proof of Lemma \ref{lm:emp:1}.} To show $\{\mathbb{G}_n \mathcal{E}_1(g) \mid n\geq 1\}$ is stochastically equicontinuous, we only need to verify aforementioned Conditions (1)--(3) in Section \ref{emp:pre}. We first verify Lindeberg Condition (1). We have for any $c > 0$,
\beee\nn
\ \lim_{n\rightarrow +\infty}\text{pr}\Big\{\|\mathcal{E}_1\|^2_{\mathcal{G}}\cdot 1\big(\|\mathcal{E}_1\|_{\mathcal{G}} > \epsilon\sqrt{n}\big) \geq c\Big\} &\leq & \lim_{n\rightarrow +\infty}\text{pr}\Big(\|\mathcal{E}_1\|_{\mathcal{G}} > c\sqrt{n}\Big)
\\\nn
&\leq&\lim_{n\rightarrow +\infty}\text{pr}\Bigg\{\frac{{h}\gamma_1^{(Z)}(X,M)}{f_{ZM}(X)}G(V) > c\sqrt{n}\Bigg\}
\\\label{lincon11}
& = & 0,
\eee
where the last equality is for some universal constant $ C_{\mathcal{E}_1} > 0$, we have a.s.,
\beee\nn
\Bigg| \frac{{h}\gamma_1^{(Z)}(X,M)}{f_{ZM}(X)} G(V) \Bigg| < C_{\mathcal{E}_1},
\eee
where $G(V)$ is bounded by $C_G$, $f_{ZM}(X)$ is bounded away from $0$ under Assumption \ref{am:unibound}, and when $Z =  0$,
\beee\nn
{h}\gamma_1^{(Z)}(X,M) & = & {h}\int_{\R}k_{u^*}(m_1,M)e_{(m_1,M)}(X)dm_1
\\\nn
&\precsim  &{h}\int_{\R}k_{u^*}(m_1,M)dm_1 \quad\text{(Assumptions \ref{am:eux} and \ref{am:unibound})}
\\\nn
&\leq & \sup_{m_0\in\R, h > 0} h\int_{\R}k_{u^*}(m_1,m_0)dm_1
\\\nn
&\precsim  & 1 \quad(\text{Assumption \ref{am:kernel}}),
\eee
similar for $Z = 1$ and thus $h\gamma_1^{(Z)}(X,M)$ is uniformly bounded by some constant. By \eqref{lincon11},
\beee\label{lincon12}
\|\mathcal{E}_1\|^2_{\mathcal{G}}\cdot 1\big(\|\mathcal{E}_1\|_{\mathcal{G}} > \epsilon\sqrt{n}\big) = o_{\mathbb{P}}(1).
\eee
On the other hand,  under the regularity conditions,
\beee\nn
\E \|\mathcal{E}_1\|^2_{\mathcal{G}} &\leq& \E\Bigg\{\frac{h\gamma_1^{(Z)}(X,M)}{f_{ZM}(X)} G(V)\Bigg\}^2 
\\\nn
&\leq & C_G^2 c_0^{-2}h^2 \E\Big\{\gamma_1^{(Z)}(X,M)\Big\}^2
\\\nn
&= & \sum_{z = 0,1}C_G^2 c_0^{-2}h^2\E\Bigg[ \pi_{z}(X)\cdot\E\Big[\Big\{\gamma_1^{(z)}(X,M)\Big\}^2\mid X,Z = z\Big]\Bigg]
\\\label{bound:Engamma}
& \precsim & 1,
\eee
where the last inequality follows because we have when $z = 0$,
\beee\nn
\E\Bigg[ \pi_{z}(X)\cdot\E\Big[\Big\{\gamma_1^{(z)}(X,M)\Big\}^2\mid X,Z = z\Big]\Bigg]  &\leq&  \epsilon \cdot \E\Bigg[ \E\Big[\Big\{\gamma_1^{(0)}(X,M)\Big\}^2\mid X,Z = 0\Big]\Bigg]
\\\nn
&\leq & \epsilon \cdot \E\Bigg[ \E\Big\{ \int_{\R} k^2_{u^*}(m_1,M)e_{(m_1,M)}^2(X)dm_{1 } \mid X,Z = 0\Big\}\Bigg]
\\\nn
&\precsim  & \E\Bigg[ \E\Big\{ \int_{\R} k^2_{u^*}(m_1,M)dm_{1 } \mid X,Z = 0\Big\}\Bigg]
\\\nn
& = & \E\Bigg[\int_{\R^2} k^2_{u^*}(u) f_{0m_0}(X)du\Bigg]
\\\label{lincon13}
&\precsim & h^{-2},
\eee
where the third inequality is because that $e_u(x)$ is uniformly bounded under regularity conditions and the last inequality is by $h^2\int_{\R^2}k_{u^*}^2(u)du\precsim 1$ and $f_{0m_0}(x)$ is uniformly upper bounded; Similar result also holds for $ z = 1$. Combining \eqref{lincon12} and \eqref{lincon13}, and noting $\|\mathcal{E}_1\|^2_{\mathcal{G}}\cdot 1\big(\|\mathcal{E}_1\|_{\mathcal{G}} > \epsilon\sqrt{n}\big) \leq \|\mathcal{E}_1\|^2_{\mathcal{G}}$, we can invoke dominated convergence theorem and show Lindeberg Condition (1) holds for $\mathcal{E}_1(g)$.
\par
We next verify Lindeberg Condition (2) when $d(g_1,g_2) = \|g_1 - g_2\|_{\infty}$, where $\|\cdot\|_{\infty}$ is the sup-norm. We have,
\beee\nn
\sup_{d(g_1,g_2)<\delta_n}\E\Big[\{\mathcal{E}_1(g_1) - \mathcal{E}_1(g_2)\}^2\Big]
& = & \E\Bigg[\frac{\{h\gamma_1^{(Z)}(X,M)\}^2}{f^2_{ZM}(X)} \Big\{g_1(V) - g_2(V)\Big\}^2\Bigg]
\\\nn
& \leq  & c_0^{-2}h^2\|g_1 - g_2\|_{\infty}^2\E\Big[\{\gamma_1^{(Z)}(X,M)\}^2\Big] 
\\\nn
&\precsim & \delta_n \rightarrow 0,
\eee
where the last inequality follows from $\|g_1 - g_2\|_{\infty} \leq \delta_n$ and $\E\big[\{\gamma_1^{(Z)}(X,M)\}^2\big]  \precsim h^{-2}$ as shown in \eqref{bound:Engamma}.
\par
We finally verify Condition (3). By  \citet[$\mathsection$ 2.11.1.1 \& Lemma 2.11.6]{vaart1997weak},  Condition (3) is satisfied if $\mathcal{G}$ has a finite uniform entropy integral and for some random measure $\nu_n$,
\beee\label{entropycon1}
\frac{1}{n}\Big\{\mathcal{E}_1(g_1) - \mathcal{E}_1(g_2)\Big\}^2 &\leq &\int_{\mathbb{V}}\big\{g_1(v) - g_2(v)\big\}^2d \nu_n;
\eee
see also the proof of Lemma 1 in \citet{kennedy2017non}. Noticing that
\beee\nn
\frac{1}{n}\Big\{\mathcal{E}_1(g_1) - \mathcal{E}_1(g_2)\Big\}^2 = \frac{\{h\gamma_1^{(Z)}(X,M)\}^2}{nf^2_{ZM}(X)} \Big\{g_1(V) - g_2(V)\Big\}^2,
\eee
\eqref{entropycon1} is satisfied by the following random measure 
$
\nu_n= \frac{\{h\gamma_1^{(Z)}(X,M)\}^2}{nf^2_{ZM}(X)} \delta_V,
$
where $\delta_V$ is a (random) Dirac measure centered at $V$. We thus have verified Condition (3). With Conditions (1)--(3) being satisfied, we have  $\{\mathbb{G}_n \mathcal{E}_1(g) \mid n\geq 1\}$ is stochastically equicontinuous.

\qed

Define
$$
g^{\natural}(V\mid \pi,\mu)=\frac{Y - \mu_{Z}(X,M)}{\pi_Z(X)}.
$$ We then have
\beee\label{G1decom}
\mathbb{G}_n\{h_1(V\mid \hat{\pi},\hat{\mu}) - h_1(V\mid \bar{\pi},\bar{\mu})\}& = & \mathbb{G}_n\Big[\mathcal{E}_1\big\{g^\natural(V\mid \hat{\pi},\hat{\mu})\big\} - \mathcal{E}_1\big\{g^\natural(V\mid \bar{\pi},\bar{\mu}\big\}\Big].            
\eee 
For the series of empirical process $\{\mathbb{G}_n \mathcal{E}_1(g^{\natural})\mid n\geq 1,\, g^\natural \in\mathcal{G}^\natural\}$ where $g^{\natural}$ can be all possible  $g^{\natural}(\cdot\mid \hat{\pi},\hat{\mu})$ and $g^{\natural}(\cdot\mid \bar{\pi},\bar{\mu})$ under Assumption \ref{am:emp}. We first prove that there exists some metric space $\mathcal{G}^{\natural}$ satisfying the condition in Lemma \ref{lm:emp:1}, and thus $\{\mathbb{G}_n \mathcal{E}_1(g^{\natural})\mid n\geq 1,\, g^\natural \in\mathcal{G}^\natural\}$ is stochastically equicontinuous. Under Assumption \ref{am:unibound}, we have functional classes $\mathcal{G}_{\mu}$ and $\mathcal{G}_{\pi}$ containing $\bar\mu,\hat\mu$ and $\bar\pi,\bar\pi$, respectively, and both $\mathcal{G}_{\mu}$ and $\mathcal{G}_{\pi}$ have uniformly bounded envelop functions and finite uniform entropy integrals. We then construct 
\beee\nonumber
\mathcal{G}^\natural &=& (\mathcal{Y} \ominus \mathcal{G}_{\mu}) \circ\mathcal{G}_{\pi}^{-1},
\eee
where we denote $\mathcal{Y}$ is the set of a single function outputting $Y$ from $Z$, and for two functional sets, e.g., $\mathcal{G}$ and $\mathcal{G}'$, we define the operations,  $\mathcal{G} \ominus \mathcal{G}' = \{g - g'\mid g\in\mathcal{G}, g'\in\mathcal{G}'\}$,  $\mathcal{G}^{-1} = \{g^{-1}\mid g\in\mathcal{G}\}$, and $\mathcal{G}\circ\mathcal{G}' = \{gg'\mid g\in\mathcal{G}, g'\in\mathcal{G}\}$. Under Assumptions \ref{am:var:Y} and \ref{am:unibound} such that  $\mu_{z}(x,m)$ and $\pi_z(x)^{-1}$ are uniformly bounded by some constant and $\mathbb{Y}$ is compact, it can be seen that $\mathcal{G}^\natural$ has an envelope function with a uniform constant upper bound; thus Condition (ii) in Lemma \ref{lm:emp:1} is verified. The Condition (i) in Lemma \ref{lm:emp:1} follows from the basic property of uniform entropy integral. Under Assumption \ref{am:unibound}, both $\mathcal{G}_{\mu}$ and $\mathcal{G}_{\pi}^{-1}$ have bounded entropy integrals and bounded envelop functions and so does the single-function  space $\mathcal{Y}$. Therefore, by  \citet[Theorem 3]{andrews1994empirical}, one has the uniform entropy integral of $\mathcal{G}^\natural$ is bounded. Thus Lemma \ref{lm:emp:1} verifies $\{\mathbb{G}_n \mathcal{E}_1(g^{\natural})\mid n\geq 1,\, g^\natural \in\mathcal{G}^\natural\}$ is stochastically equicontinuous.
\par
On the other hand, recalling we define $d(\cdot,\cdot)$ as the uniform difference metric, we have
\beee\nn
d\Big\{g^\natural(\cdot\mid \hat{\pi},\hat{\mu}), g^\natural(\cdot\mid \bar{\pi},\bar{\mu})\Big\} & = & \sup_{v\in\mathbb{V}}\Big|\frac{y - \hat{\mu}_{z}(x,m)}{\hat{\pi}_z(x)} - \frac{y - \bar{\mu}_{z}(x,m)}{\bar{\pi}_z(x)}\Big|
\\\nn
&\leq & \sup_{v\in\mathbb{V}}\Big|\frac{y }{\hat{\pi}_z(x)\bar{\pi}_z(x)}\{\bar{\pi}_z(x) - \hat{\pi}_z(x)\}\Big| + \sup_{v\in\mathbb{V}}\Big|\frac{1}{\hat{\pi}_z(x)}\{\hat{\mu}_{z}(x,m) - \bar{\mu}_{z}(x,m)\}\Big| 
\\\nn
&&+ \sup_{v\in\mathbb{V}}\Big|\frac{\bar{\mu}_{z}(x,m)}{\hat{\pi}_z(x)\bar{\pi}_z(x)}\{\bar{\pi}_z(x) - \hat{\pi}_z(x)\}\Big|
\\\nn
& = & o_{\mathbb{P}}(\| \hat{\pi} - \bar{\pi}\|_{\infty} + \|\hat{\mu} - \bar{\mu}\|_{\infty}) 
\\\label{g1bound1}
& = & o_{\mathbb{P}}(1),
\eee
under Assumptions \ref{am:var:Y} and \ref{am:unibound}. Finally, by \eqref{sqbound}, \eqref{G1decom} and \eqref{g1bound1}, we conclude $h^{-1}n^{-1/2}\mathbb{G}_n\{h_1(V\mid \hat{\pi},\hat{\mu}) - h_1(V\mid \bar{\pi},\bar{\mu})\} = o_{\mathbb{P}}(h^{-1}n^{-1/2})$. 
\subsection{Bounding $\mathbb{G}_n\{h_j(V\mid \hat{\mu}) - h_j(V\mid \bar{\mu})\}$ for $j = 2,3$}\label{Sec5.3}
When bounding  $\mathbb{G}_n\{h_2(V\mid \hat{\mu}) - h_2(V\mid \bar{\mu})\}$ and $\mathbb{G}_n\{h_3(V\mid \hat{\mu}) - h_3(V\mid \bar{\mu})\}$, the arguments will be similar. Therefore, we  focus on bounding the former for brevity. Define,
\beee\nn
\mathcal{E}_2(\mu) & = & \int_{\R^2}k_{u^*}(u)\mu_{1}(X,m_1)e_u(X)du.
\eee
We then write 
\beee\label{targetemp2}
n^{-1/2}\mathbb{G}_n\{h_2(V\mid \hat{\mu}) - h_2(V\mid \bar{\mu})\}
&=&n^{-1/2}\mathbb{G}_n\{\mathcal{E}_2(\hat\mu) - \mathcal{E}_2(\bar\mu)\}.
\eee
By \citet[Theorem 19.14]{van2000asymptotic} and Assumption \ref{am:emp}, we have the functional space $\mathcal{G}_{\mu_1}$ which contains all possible functions $\hat{\mu}_{1}(x,m_1),\bar{\mu}_{1}(x,m_1)$, is $P$-Donsker, because $\mathcal{G}_{\mu_1}$ has a finite uniform entropy integral when $\delta = 1$ and a uniformly bounded (and thus also mean-squared-average bounded) envelop function; See the definition of $P$-Donsker in \citet[$\mathsection$ 19.2]{van2000asymptotic}. Next, we consider the functional space,
\beee\nn
\mathcal{E}_2(\mathcal{G}_{\mu_1}) &=& \Big\{\mathcal{E}_2(\mu)(x) = \int_{\R^2}k_{u^*}(u)\mu_{1}(x,m_1)e_u(x)du\mid \mu_{1}(x,m_1)\in\mathcal{G}_{\mu_1}\Big\},
\eee
which contains functions for all possible $\mathcal{E}_2(\hat{\mu})$ and $\mathcal{E}_2(\bar{\mu})$. 
\par
Now we show that $\mathcal{E}_2(\mathcal{G}_{\mu_1}) $ is  $P$-Donsker. Since $e_u(x)$ is a fixed function and uniformly bounded (Assumptions \ref{am:eux} and \ref{am:unibound}), we have $\mathcal{G}_{\mu_1 e} = \{\mu_{1}(x,m_1)e_u(x)\mid \mu_{1}(x,m_1)\in\mathcal{G}_{\mu_1}\}$ is also $P$-Donsker by \citet[Theorem 19.14]{van2000asymptotic}. Define $R(u)$ as a fixed probability measure over $u$ with probability density function $k_{u^*}(u)/\int_{\R^2}k_{u^*}(u)du$ and clearly $C_{\mathcal{E}}= \int_{\R^2}k_{u^*}(u)du < +\infty$ is a fixed constant under our regularity conditions. Then noticing that
\beee\nn
\int_{\R^2}k_{u^*}(u)\mu_{1}(x,m_1)e_u(x)du &= & C_{\mathcal{E}}\int_{\R^2}\mu_{1}(x,m_1)e_u(x)R(du),
\eee
we can apply  \citet[Lemma 5.2]{van2006estimating} with $r = s =2$ and deduce,
\beee\label{Nbound1}
\sup_{Q}N\{\epsilon \|G_{\mathcal{E}}\|_{Q,2},\mathcal{E}_2(\mathcal{G}_{\mu_1}),\|\cdot\|_{Q,2}\} &\leq& \sup_{Q}N\{\epsilon \|G_{\mu_1e}\|_{Q,2}/2,\mathcal{G}_{\mu_1 e},\|\cdot\|_{Q,2}\},
\eee
where $G_{\mu_1e}(x,u)$ can be any envelope function of $\mathcal{G}_{\mu_1 e}$ and, correspondingly, define
\beee
G_{\mathcal{E}}(x) &=& C_{\mathcal{E}}\Big\{\int_{\R^2}G^2_{\mu_1 e}(x,u)dR(du)\Big\}^{1/2},
\eee
which serves as an envelope function of $\mathcal{E}_2(\mathcal{G}_{\mu_1})$ by H\"older's inequality.
\par
On the other hand, noticing that for any $\nu(u,x) = \mu_{1}(x,m_1)e_u(x)$, $\nu'(v) = \mu_{1}'(x,m_1)e_u(x)$ for any $\mu,\mu'\in \mathcal{G}_{\mu_1}$, we have
\beee\nn
\big\|\nu  - \nu'\big\|_{Q,2} &\leq& \Big\{\sup_{x,u}e_u(x)\Big\}\cdot \big\|\mu_{1 }(\cdot)  - \mu'_{1 }(\cdot)\big\|_{Q,2}
\\\nn
&\leq& C_{e}\big\|\mu_{1 }(\cdot)  - \mu'_{1 }(\cdot)\big\|_{Q,2}
\eee
for some fixed constant $C_{e} > 0$ such that
\beee\label{boundeux}
\sup_{x,u}e_u(x) \leq C_e,
\eee as $e_u(x)$ is uniformly bounded. Thus for any $(C_e^{-1}c)$-balls covering $\mathcal{G}_{\mu_1}$ with functional centers, $\mu^{(1)}_{1}(x,m_1),\mu^{(2)}_{1}(x,m_1)$, etc, the corresponding $c$-balls with functional centers, $e_u(x)\mu^{(1)}_{1}(x,m_1),$ $e_u(x)\mu^{(2)}_{1}(x,m_1)$, etc, cover $\mathcal{G}_{\mu_1e}$ for any $c > 0$. Thus we have
\beee\label{basiccovering}
N\{c,\mathcal{G}_{\mu_1 e},\|\cdot\|_{Q,2}\} & \leq & N\{C_e^{-1}c,\mathcal{G}_{\mu_1},\|\cdot\|_{Q,2}\},
\eee
for any $c > 0$ and any random measure $Q$. Let $G_{\mu_1}(x,m_1)$ be an envelope function of $\mathcal{G}_{\mu_1}$ satisfying Assumption \ref{am:emp}. By \eqref{boundeux}, we can  take $G_{\mu_1 e}(x,u)= C_e G_{\mu_1}(x,m_1)$ as  the envelop function of $\mathcal{G}_{\mu_1 e}$. Taking $c = \epsilon\|G_{\mu_1 e}\|_{Q,2}/2 = \epsilon C_e\|G_{\mu_1}\|_{Q,2}/2$ in \eqref{basiccovering} and considering the supremum over all possible $Q$, we have
\beee\label{Nbound2}
\sup_{Q}N\{\epsilon\|G_{\mu_1 e}\|_{Q,2}/2,\mathcal{G}_{\mu_1 e},\|\cdot\|_{Q,2}\} & \leq & \sup_{Q}N\{\epsilon\|G_{\mu_1}\|_{Q,2}/2,\mathcal{G}_{\mu_1},\|\cdot\|_{Q,2}\}.
\eee
Combing \eqref{Nbound1} and \eqref{Nbound2}, we have
\beee\nn
\sup_{Q}N\{\epsilon \|G_{\mathcal{E}}\|_{Q,2},\mathcal{E}_2(\mathcal{G}_{\mu_1}),\|\cdot\|_{Q,2}\} &\leq &\sup_{Q}N\{\epsilon\|G_{\mu_1}\|_{Q,2}/2,\mathcal{G}_{\mu_1},\|\cdot\|_{Q,2}\}
\eee
and therefore,
\beee\nn
J\big\{1,\mathcal{E}_2(\mathcal{G}_{\mu_1}),\mathcal{L}_2\big\} & =& \sup_{Q}\int_{0}^{1}\sqrt{\log N\{\epsilon \|G_{\mathcal{E}}\|_{Q,2},\mathcal{E}_2(\mathcal{G}_{\mu_1}),\|\cdot\|_{Q,2}\}}d\epsilon
\\\nn
&\leq &\sup_{Q}\int_{0}^{1}\sqrt{\log N\{\epsilon\|G_{\mu_1}\|_{Q,2}/2,\mathcal{G}_{\mu_1},\|\cdot\|_{Q,2}\}}d\epsilon
\\\label{finalep1}
&\leq &2 \sup_{Q}\int_{0}^{1/2}\sqrt{\log N\{\epsilon\|G_{\mu_1}\|_{Q,2},\mathcal{G}_{\mu_1},\|\cdot\|_{Q,2}\}}d\epsilon
\\\nn
& < & +\infty,
\eee
under Assumption \ref{am:emp}. In addition,  we have
\beee\nn
\E\Big\{ G^2_{\mathcal{E}}(X) \Big\}&=& \E\Big\{ C^2_{\mathcal{E}}C^2_e\int_{\R^2} G^2_{\mu_1}(m_1,X)dR(du)\Big\}
\\\nn
& \leq & C^2_{\mathcal{E}}C^2_e \Big\{\sup_{m_1,x} G_{\mu_1}(m_1,x)\Big\}^2
\\\label{finalep111}
& < & +\infty,
\eee
as $G_{\mu_1}(m_1,x)$ is uniformly bounded. Summarizing \eqref{finalep1}--\eqref{finalep111} and by  \citet[Theorem 19.14]{van2000asymptotic}, we conclude that $\mathcal{E}_2(\mathcal{G}_{\mu_1})$ is $P$-Donsker.
\par
With the Donsker property, a standard application of the continuous mapping theorem \citep[Theorem 19.24]{van2000asymptotic} implies that
\beee\label{targetemp}
\mathbb{G}_n\{\mathcal{E}_2(\hat\mu) - \mathcal{E}_2(\bar\mu)\} = o_{\mathbb{P}}(1)
\eee
for any $\mathcal{E}_2(\hat\mu)$ and $\mathcal{E}_2(\bar\mu)$ in $\mathcal{E}_2(\mathcal{G}_{\mu_1})$,  as long as $\|\mathcal{E}_2(\hat\mu) - \mathcal{E}_2(\bar\mu)\|_{\infty} = o_{\mathbb{P}}(1)$; we omit this standard argument for simplicity and refer the interested reader to \citet[$\mathsection$ 4.2]{kennedy2016semiparametric} for details. Thus \eqref{targetemp} gives the desired bound for \eqref{targetemp2}. Therefore, to finish the proof, it is only left to show $\|\mathcal{E}_2(\hat\mu) - \mathcal{E}_2(\bar\mu)\|_{\infty}= o_{\mathbb{P}}(1)$ which can be bounded as follows under Assumption \ref{am:emp},
\beee\nn
\|\mathcal{E}_2(\hat\mu) - \mathcal{E}_2(\bar\mu)\|_{\infty}&=& \sup_{x\in\mathbb{X}}\Bigg|\int_{\R^2}k_{u^*}(u)\big\{\hat\mu_{1}(x,m_1) - \bar\mu_{1}(x,m_1)\big\}e_u(x)du \Bigg|
\\\nn
&\leq & \|\hat\mu_{1}(\cdot,m_1) - \bar\mu_{1}(\cdot,m_1)\|_{\infty}C_eC_{\mathcal{E}}
\\\nn
& = & o_{\mathbb{P}}(1).
\eee 
 Summarizing above results, we have $n^{-1/2}\mathbb{G}_n\{h_2(V\mid \hat{\mu}) - h_2(V\mid \bar{\mu})\} = o_{\mathbb{P}}(n^{-1/2})$, and symmetrically we can show $n^{-1/2}\mathbb{G}_n\{h_3(V\mid \hat{\mu}) - h_3(V\mid \bar{\mu})\} = o_{\mathbb{P}}(n^{-1/2})$. Combining the bounds in Sections \ref{Sec5.2} and \ref{Sec5.3} along with \eqref{decom:Gnh}, Lemma \ref{lm:emp} follows. 
\qed
\section{Additional Figures for simulation in Section~\ref{sec:cd}}\label{sec:add:simu}
\begin{figure}
\centering
\begin{subfigure}[b]{1\textwidth}
\centering
\includegraphics[width=1\textwidth]{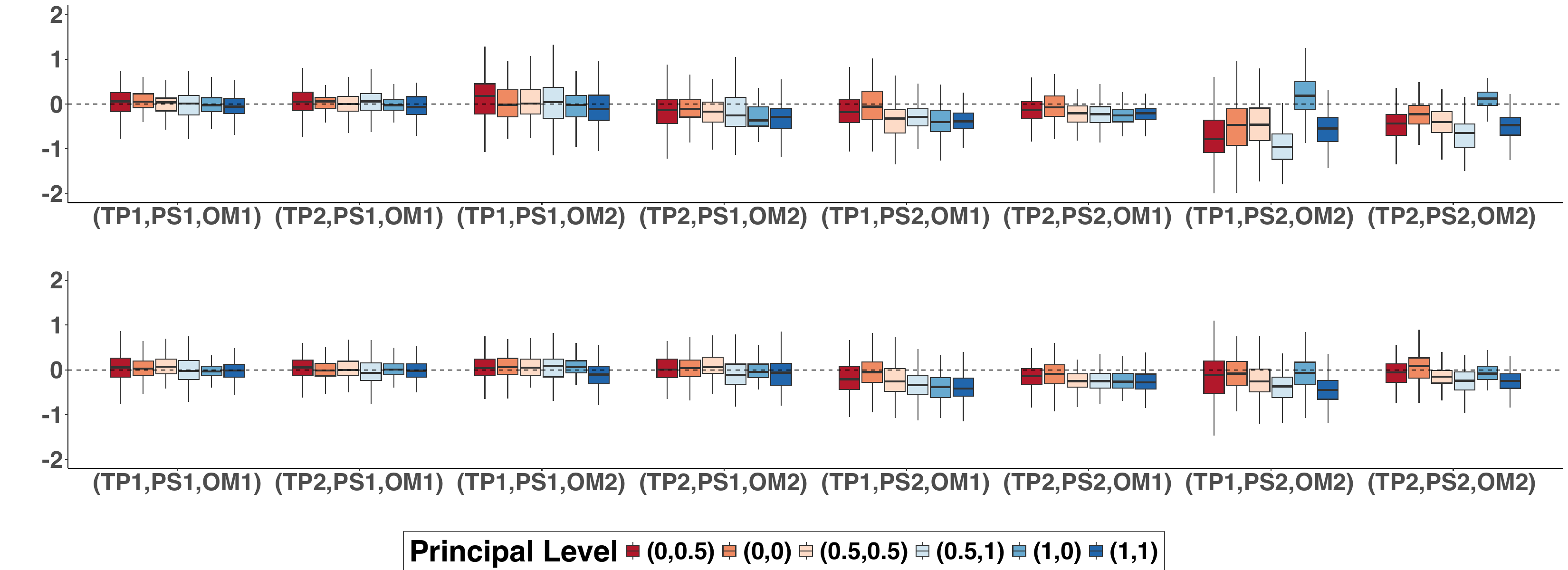}
\caption{$n = 500$}
\end{subfigure}
\begin{subfigure}[b]{1\textwidth}
\centering
\includegraphics[width=1\textwidth]{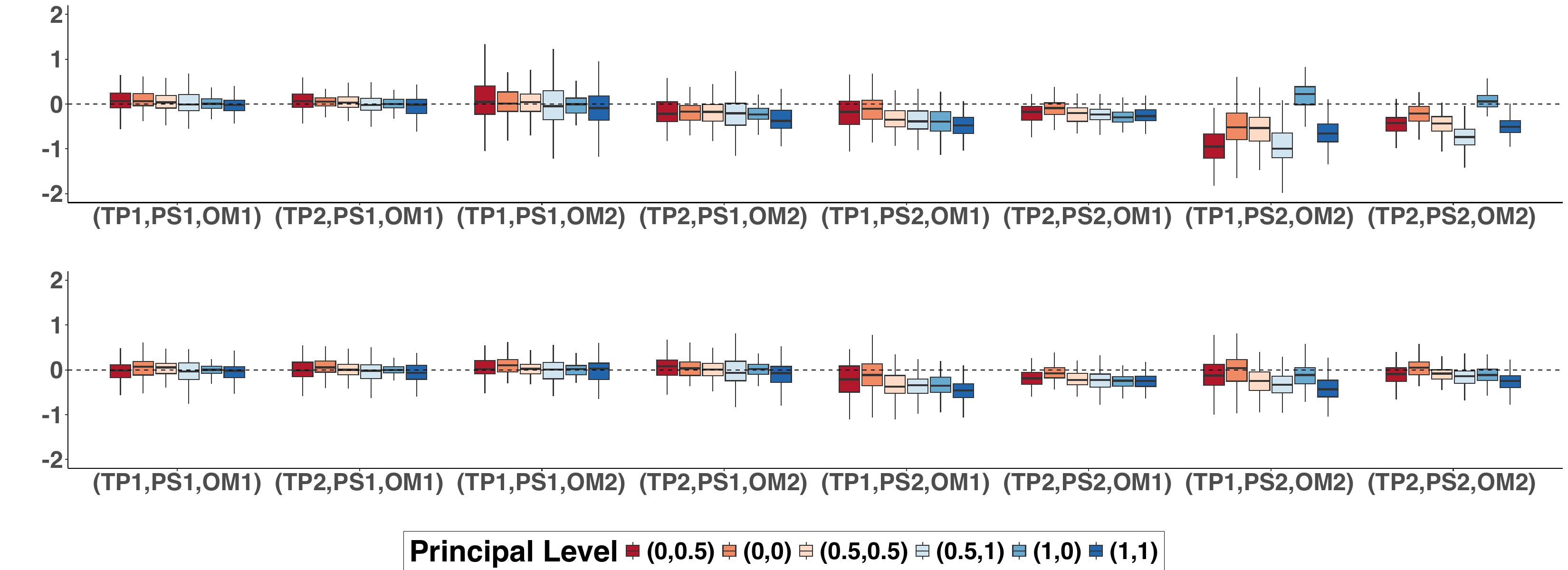}
\caption{$n = 1000$}
\end{subfigure}
\begin{subfigure}[b]{1\textwidth}
\centering
\includegraphics[width=1\textwidth]{2000.pdf}
\caption{$n = 2000$}
\end{subfigure}

\caption{\small The box-plots of $\hat{\tau}_{u}-\tau^*_{u}$ over $100$ MC rounds when $n = 500,1000$ and $2000$ for the simulation in Section~\ref{sec:cd}.  X-axis represents different data generation mechanisms. The top and bottom panels in each subfigure present the results when nuisance functions are estimated through  parametric and semiparametric strategies, respectively.}
\label{fig:completesimu}
\end{figure}
{\color{black}We report the complete simulation results for our numerical experiments in Section~\ref{sec:cd}. In Figures~\ref{fig:completesimu}(a)--(c), we present the histograms of $\hat{\tau}_u-\tau_u^*$ over 100 Monte Carlo iterations under different data generation settings, with $n = 500,1000$, and $2000$, respectively. In each subfigure, the top and bottom panels present the results when nuisance functions are estimated through  parametric and semiparametric strategies, respectively. X-axis represents different data generation mechanisms. For example, (TP2,PS1,OM1) means  data are generated from $(\mathcal{F}_{{\rm tp}},\mathcal{F}_{{\rm ps}},\mathcal{F}_{{\rm ps}}) = (\mathcal{F}_{{\rm tp}}^{(2)},\mathcal{F}_{{\rm ps}}^{(1)},\mathcal{F}_{{\rm ps}}^{(1)})$. Different colors represent different principal levels $u$.
\par
Under the data generation settings when $\hat{\tau}_u$ can consistently estimate $\tau_u^*$ as indicated by Theorem~\ref{hto0}, i.e., the data generation settings are (TP1,PS1,OM1),  (TP2,PS1,OM1), (TP1,PS1,  OM2) when nuisance functions are trained by the parametric strategy, and the data generation settings are (TP1,PS1,OM1),  (TP2,PS1,OM1), (TP1,PS1,OM2), (TP2,PS1,OM2) when nuisance functions are trained by the semiparametric strategy, we can observe the corresponding estimation biases of $\hat{\tau}_u - \tau_u^*$ are close to zero for all $n$, and in general,  the estimation variances and biases become smaller as $n$ increases, which verifies the nonparametric consistency of $\hat{\tau}_u$ (cf. Theorem~\ref{hto0}). In addition, when all nuisance functions can be  consistently esitmated, i.e., the data generation setting is (TP1,PS1,OM1) with parametric nuisance training and the data generation settings are  (TP1,PS1,OM1),  (TP2,PS1,OM1), (TP1,PS1,OM2), (TP2,PS1,OM2) with semiparametric nuisance training, the estimation errors of $\hat{\tau}_u$ become particularly small due to the rate-double robustness (cf. Theorem~\ref{hto0}).}

\end{document}